\documentclass[12pt]{article}
\usepackage[T1]{fontenc}
\usepackage[utf8]{inputenc}
\usepackage[letterpaper]{geometry}
\usepackage{lmodern}
\usepackage{jheppub}

\usepackage{parskip}
\usepackage{physics}
\usepackage{amssymb}
\usepackage{xcolor}
\usepackage{mathtools}
\usepackage{hhline}
\usepackage{array}
\usepackage{arydshln}
\usepackage{float}
\usepackage{subcaption}
\usepackage{etoc}

\usepackage{cleveref}
\crefformat{section}{#2§#1#3}
\crefformat{subsection}{#2§#1#3}
\crefformat{appendix}{#2App.~#1#3}
\crefformat{figure}{#2Fig.~#1#3}

\usepackage{ytableau}
\ytableausetup{boxsize=0.4em,centertableaux}

\usepackage{tikz}
\usetikzlibrary{arrows,decorations.markings,cd,3d,decorations.pathmorphing}

\makeatletter
\g@addto@macro\bfseries{\boldmath}
\makeatother

\newcommand{\Z}{\mathbb{Z}}

\newcommand{\bZ}{\mathbb{Z}}
\newcommand{\bR}{\mathbb{R}}
\newcommand{\bC}{\mathbb{C}}
\newcommand{\bQ}{\mathbb{Q}}

\newcommand{\bS}{\mathbb{S}}
\newcommand{\cT}{\mathcal{T}}

\newcommand{\su}{\mathfrak{su}}

\newcommand{\cC}{\mathcal{C}}
\newcommand{\cD}{\mathcal{D}}

\newcommand{\cN}{\mathcal{N}}

\newcommand{\cS}{\mathcal{S}}
\newcommand{\cM}{\mathcal{M}}
\newcommand{\cA}{\mathcal{A}}

\newcommand{\cO}{\mathcal{O}}
\newcommand{\cZ}{\mathcal{Z}}
\newcommand{\cR}{\mathcal{R}}
\newcommand{\cK}{\mathcal{K}}
\newcommand{\cG}{\mathcal{G}}
\newcommand{\nn}{\nonumber}
\renewcommand{\bar}{\overline}
\renewcommand{\tilde}{\widetilde}
\renewcommand{\mod}{\text{ mod }}
\newcommand{\id}{\text{id}}
\newcommand{\Sym}{\text{Sym}}
\newcommand{\Rep}{\text{Rep}}
\newcommand{\Spin}{\text{Spin}}
\newcommand{\SO}{SO}
\newcommand{\SU}{SU}
\newcommand{\pt}{\text{pt}}
\newcommand{\Hom}{\text{Hom}}
\newcommand{\End}{\text{End}}
\newcommand{\Arf}{\text{Arf}}
\newcommand{\Diag}{\text{Diag}}
\newcommand{\NS}{\mathrm{NS}}
\newcommand{\R}{\mathrm{R}}
\newcommand{\1}{\mathbf{1}}
\newcommand{\bimod}[3]{{\vphantom{#2}}_{#1} {#2}_{#3}}

\colorlet{good}{green!60!black}
\colorlet{bad}{red!70!black}

\title{Twisted scattering from Chern--Simons theory}

\preprint{%
  \begin{tabular}[t]{@{}r@{}}
    \texttt{IFT-UAM/CSIC-26-155} \\
    \texttt{Imperial-TP-2026-RM-1}
  \end{tabular}%
}

\author{Guillermo Arias-Tamargo$^{a,b}$,}
\author{Philip Boyle Smith$^{c,d}$,}
\author{Rishi Mouland$^a$,}
\author{and Max Velásquez Cotini Hutt$^{a,e}$}

\affiliation[a]{Abdus Salam Centre for Theoretical Physics, \\ Imperial College London, London SW7 2AZ, UK}
\affiliation[b]{Instituto de Física Teórica IFT-UAM/CSIC,\\
C/ Nicolás Cabrera 13-15, Campus de Cantoblanco, 28049 Madrid, Spain}
\affiliation[c]{SISSA, via Bonomea 265, 34136 Trieste, Italy}
\affiliation[d]{INFN, Sezione di Trieste, via Valerio 2, 34127 Trieste, Italy}
\affiliation[e]{Department of Applied Mathematics and Theoretical Physics, \\ University of Cambridge, CB3 0WA, UK\\}

\emailAdd{guillermo.arias@ift.csic.es}
\emailAdd{philip.boyle.smith@sissa.it}
\emailAdd{r.mouland@imperial.ac.uk}
\emailAdd{mv623@cam.ac.uk}

\abstract{We consider boundary conditions of two-dimensional Dirac fermions which preserve a non-anomalous continuous symmetry which acts differently on the left- and right-movers. Such boundaries are equivalent to topological defects of Weyl fermions preserving the symmetry. We give a general construction of such defects for conformally-embedded symmetries using Chern--Simons theory. They always exist and can be realised as half-space gauging defects for some finite categorical symmetry. Using this technology, we compute the twist lines to which fermions become attached upon scattering through the defect, the boundary central charge of the corresponding folded boundary state, and its mod-2 valued SPT class. As applications, we first consider abelian symmetries and find agreement with known results in the literature. We then study a novel example where an $SU(2) \times SU(4) \times U(1)^2$ symmetry is preserved. In contrast to the abelian case, the twist lines are intrinsically non-invertible and belong to either a $\text{TY}(\bZ_3 \times \bZ_3)$ or $\text{Rep}(\bS_3) \times \bZ_3$ symmetry depending on whether scattering is to the left or the right.}

\begin{document}

\maketitle

\newpage

\section{Introduction}\label{sec:int}

A long-standing problem in Quantum Field Theory (QFT) is to determine when a given QFT can be put on a manifold with boundary while preserving certain symmetries.

There are many facets to this question: one can consider QFTs in various dimensions, one can make statements about all QFTs or only concrete examples, one can pose the question for continuous or discrete symmetries, invertible or non-invertible symmetries, and one may or may not demand that the boundary is simple (has a unique topological point operator).

The discrete case is largely solved. If a QFT has a perturbative gravitational anomaly, then it has no boundaries at all \cite{hellerman}. Otherwise, it is a commonly assumed folklore theorem that it has at least one simple boundary. One can then render this boundary symmetric (but non-simple) by summing over a symmetry orbit of boundaries. So the relevant question is whether a boundary exists that is both simple \emph{and} symmetric. If the symmetry has an anomaly, then the answer is no \cite{Thorngren:2020yht}, while if the symmetry is anomaly-free, the answer may be yes or no depending on the dynamical details of the theory and has no clear connection to anomalies \cite{Wei:2025zyd}. This story has an extension to non-invertible discrete symmetries in two dimensions \cite{Choi_2023}, however we will mostly be interested in invertible symmetries in this work.

The continuous case is less solved. The key departure from the discrete case comes when we try to sum over the symmetry orbit of boundaries. Now instead of a discrete sum over $g \in G$, we would have to perform a path integral $\int [Dg]$ over a $G$-valued symmetry parameter over the whole boundary in order to guarantee local current conservation, and it is far from obvious how to regulate this path integral or whether it is even defined. So it is not clear that symmetric boundaries exist, even non-simple ones. In fact, it is known that if the symmetry has a perturbative anomaly, then symmetric boundaries do not exist \cite{jensen}. So the question now is: do symmetric boundaries exist when there is no perturbative anomaly?

In this paper, we will be interested in giving a positive answer to this question. Specifically, we will consider $N$ free Dirac fermions in $d=1+1$ dimensions, and pick a connected Lie group symmetry $G$ with no perturbative anomaly. The challenge is then to construct a symmetric boundary preserving $G$.

Our reason for focusing on free fermions is twofold. First, this theory is a rich source of interesting symmetries $G$, by virtue of the fact one can pick any two $N$-dimensional representations $L, R$ of $G$ with matching anomalies $k(L) = k(R)$ to get such a symmetry. Second and more importantly, symmetric boundaries of free fermions describe the effective physics of four-dimensional massless fermions interacting with a magnetic monopole \cite{vanBeest:2023dbu,vanBeest:2023mbs}. Thus our ability to construct symmetric boundaries is directly related to our ability to define 't Hooft lines in chiral gauge theories---an important probe into their dynamics. Indeed, we will spend much of our effort in \cref{sec:non-abelian} analysing a single such chiral gauge theory where the relevant symmetric boundary was not previously known \cite{tongscgp}. However most of our techniques generalise with little modification to arbitrary (rational) CFTs.

\begin{figure}
    \centering
    \begin{tikzpicture}
        \filldraw[color=black, fill=blue!15,use as bounding box] (0,-1) ellipse (3.5 and 2.8);
        \filldraw[color=black, fill=blue!10] (0,-0.5) ellipse (2.5 and 1.8);
        \filldraw[color=black, fill=blue!5] (0,0) ellipse (1.5 and 0.8);
        \node[] at (0,0) {\begin{tabular}{c} equal rank \end{tabular}};
        \node[] at (0,-1.55) {\begin{tabular}{c} conformally \\ embedded \end{tabular}};
        \node[] at (0,-3.05) {\begin{tabular}{c} non-conformally \\ embedded \end{tabular}};
    \end{tikzpicture}
    \caption{The hierarchy of difficulties of symmetries $G$.}
    \label{fig:venndiagram}
\end{figure}
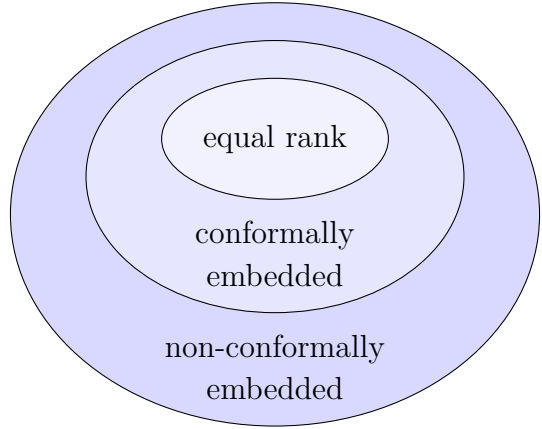

The possible symmetries $G$ we might want to preserve form a hierarchy of difficulties shown in \cref{fig:venndiagram}. The easiest is the equal-rank case, when $\text{rank}(G) = N$, which has already been solved in the literature. Here $G$ contains a $U(1)^N$ subgroup, and one can explicitly construct all $U(1)^N$-preserving boundary states \cite{Han_2017,BoyleSmith:2019jnh}, a subset of which will also preserve $G$ \cite{BoyleSmith:2020nuf}. These boundaries have also been constructed from non-invertible duality defects \cite{Arias-Tamargo:2026urw, Wei:2026fsn} and quantum mechanical rotor systems coupled to the fermions, which flow to the relevant boundary at low energies \cite{Polchinski:1984uw, Loladze:2024ayk, Loladze:2025jsq}. More generally, if $G$ is a subgroup of an existing anomaly-free rank-$N$ symmetry, then it can be extended to the equal-rank case and solved this way too. A larger but harder class of examples are symmetries with respect to which the theory is rational (or in other words, $G$ embeds conformally into the full chiral algebra of the theory), which is when $c(G) = N$ with $c(G)$ the Sugawara central charge of the symmetry. The goal of this paper is to solve the problem in this situation. This leaves the generic case $c(G) < N$ as an open problem; we will comment more on it in \cref{sec:outlook}.

\subsection*{An example: the 3450 model}

To give a flavour of why constructing symmetric boundaries is non-trivial, even in the equal-rank case, we need look no further than the 3450 model (also known as Pythagoras's Theory), which has $N=2$ Dirac fermions and symmetry $G=U(1)$. We give the left-moving and right-moving fermions charges
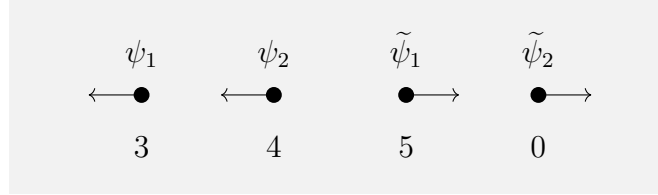
\[
    \begin{tikzpicture}[scale=0.7]
        \newcommand{\x}{2.5}
        \fill[black!5] (-\x,\x*0.75) rectangle (4*\x,-\x*0.75);
        \draw[->] (0   ,0) node[shape=circle,fill=black,inner sep=1pt] {} node[above, yshift=0.5em] {$\psi_1$}         node[below,yshift=-1em] {3} -- ++(-1,0);
        \draw[->] (  \x,0) node[shape=circle,fill=black,inner sep=1pt] {} node[above, yshift=0.5em] {$\psi_2$}         node[below,yshift=-1em] {4} -- ++(-1,0);
        \draw[->] (2*\x,0) node[shape=circle,fill=black,inner sep=1pt] {} node[above, yshift=0.5em] {$\tilde{\psi}_1$} node[below,yshift=-1em] {5} -- ++( 1,0);
        \draw[->] (3*\x,0) node[shape=circle,fill=black,inner sep=1pt] {} node[above, yshift=0.5em] {$\tilde{\psi}_2$} node[below,yshift=-1em] {0} -- ++( 1,0);
        \fill (0,0) circle (0.15);
        \fill (\x,0) circle (0.15);
        \fill (2*\x,0) circle (0.15);
        \fill (3*\x,0) circle (0.15);
    \end{tikzpicture}
\]
All perturbative anomalies vanish thanks to $3^2 + 4^2 = 5^2 + 0^2$. Yet, a left-moving particle of charge $3$ incident on a boundary at $x=0$ seemingly has no right-moving counterpart to scatter into. Nonetheless, a boundary can preserve this $U(1)$, by scattering this incoming particle into an outgoing excitation created by a \textit{twist operator}, that lives at the end of a topological line attaching it to the boundary, as shown in the left-hand panel below.
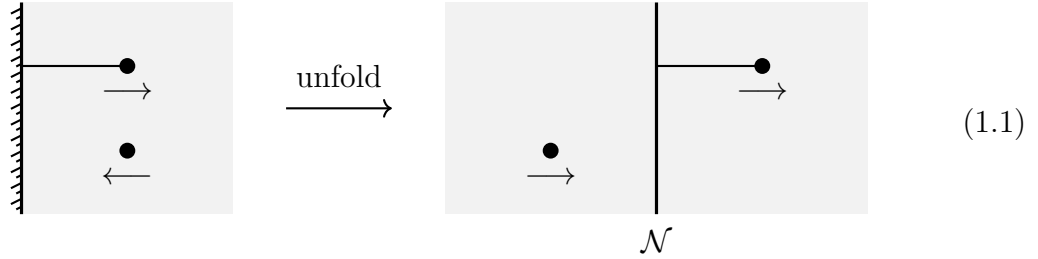
\begin{equation}\label{eq:2d scattering intro}
    \begin{tikzpicture}[scale=0.7,baseline=(current bounding box.center)]
        \fill[black!5] (6,0) rectangle (10,-4);
        \draw[very thick] (6,0) -- (6,-4);
        \fill (8,-1.2) circle (0.15);
        \fill (8,-2.8) circle (0.15);
        \draw[thick] (6,-1.2) -- (8,-1.2);
        \node[below] at (8,-1.4) {$\longrightarrow$};
        \node[below] at (8,-3) {$\longleftarrow$};
        \foreach \x in {0,...,12} \draw[thick] (6-.2,-4+.3*\x+.04+.15) -- (.2+6-.2,-4+.3*\x+.1+.04+.15);
        \foreach \x in {-1,...,11} \draw[thick] (6-.1,-4+.3*\x+.09+.15+.15) -- (.2+6-.2,-4+.3*\x+.1+.04+.15+.15);
        \draw[thick,->] (11,-2) -- (13,-2);
        \node[above] at (12,-1.8) {unfold};
        \begin{scope}[shift={(12,0)}]
            \fill[black!5] (2,0) rectangle (6,-4);
            \fill[black!5] (6,0) rectangle (10,-4);
            \draw[very thick] (6,0) -- (6,-4);
             \fill (8,-1.2) circle (0.15);
            \fill (4,-2.8) circle (0.15);
            \draw[thick] (6,-1.2) -- (8,-1.2);
            \node[below] at (8,-1.4) {$\longrightarrow$};
            \node[below] at (4,-3) {$\longrightarrow$};
            \node[below] at (6,-4.1) {$\cN$};
        \end{scope}	
    \end{tikzpicture}
\end{equation}
A useful equivalent perspective is found by `unfolding' this picture, shown in the right-hand panel. The boundary becomes a topological line $\cN$---now in the theory of $N$ right-moving complex Weyl fermions---that must preserve a symmetry $G$ which acts differently on the fermions on either side of it, but crucially with the same $G$ anomaly on both sides. From this point on, we will mostly work in the unfolded picture.

The reason this example was solvable was because the symmetry was extendable to an equal-rank symmetry. To do this, we could pick a second $U(1)$ with charges $-4, 3, 0, 5$, and the combined $U(1)^2$ would still have no perturbative anomaly.

With the above choice of $U(1)^2$ extension, the twist lines that appear in the scattering form an invertible $\bZ_5$ symmetry, generated by the basic twist line
\begin{equation}\label{eq:basictwist}
    (\tilde{\psi}_1, \tilde{\psi}_2) \to (e^{2 \pi i / 5} \tilde{\psi}_1, e^{2 \pi i \cdot 2 / 5} \tilde{\psi}_2)
\end{equation}
This reflects the general fact that in the equal-rank case, the twist lines always form an invertible finite abelian group \cite{Wei:2026fsn,Arias-Tamargo:2026urw}. We reproduce this result in \cref{sec:abelian}. This twisted scattering phenomenon has also been argued to be a consequence of an anomaly localised to the defect \cite{Antinucci:2026uuh} and the physical properties of such a state have been elaborated in \cite{Tachikawa:2026cxd}.

Although the 3450 model was somewhat contrived to illustrate the above issues, the exact same physics can be seen in the effective theory of the Callan--Rubakov effect, which has $N$ Dirac fermions, equal-rank symmetry $G = SU(N) \times U(1)$, and the fermions transform as $\square_{+1}$ and $\square_{-1}$ \cite{Callan:1982ac,Callan:1982au,Callan:1982ah,Rubakov:1982fp}. Here the twist lines form the group $\bZ_{N/\gcd(N,2)}$.

\subsection*{A harder example}

To construct examples that cannot be solved using known techniques, we must go beyond the equal-rank case. The simplest example that we know of appeared in \cite{Tong:2021phe} and has $N=9$ Weyl fermions of the same chirality with symmetry
\begin{equation}
    G = SU(2) \times SU(4) \times U(1) \times U(1)
\end{equation}
The fermions to the left $\psi_{1,2}$ and to the right $\tilde{\psi}_{1,2}$ of the defect $\cN$ transform as
\begin{align}
\begin{array}{c|cccc}
    &\, \psi_{1}\, &\, \psi_{2}\,  &\, \tilde\psi_1 \, & \tilde\psi_2 \\ \hline
    SU(2) 	& \ydiagram{2} 		& 1  	& \ydiagram{1}   & 1  \\
    SU(4)		& 1	 & \ydiagram{1,1}	& \ydiagram{1} & 1 \\
    U(1) & 2 & 0 & 1 & -2 \\
    U(1)				& 0  	&  2		& 1  & 4
\end{array}
\end{align}
Overall, both the fermions to the left and to the right transform in 9-dimensional representations of $G$, and it can be checked that they have matching perturbative anomalies. The Sugawara central charge is $c(G) = 9$, which places us in the conformally-embedded case. However $\text{rank}(G) = 6 < 9$, and $G$ is moreover not extendable to a symmetry of rank 9, so we are outside the equal-rank case.

This example is especially interesting because it exhibits an \emph{$N$-ality problem}. The fermions on the right carry excitations charged under the centre of $SU(2)$, while the fermions on the left do not. Thus if we send in a $\tilde{\psi}_1$ particle going left, there is seemingly nothing that can come out, just like in the 3450 model. Unlike the 3450 model, however, it has been argued \cite[§5.3]{vanBeest:2023mbs} that an $N$-ality problem under a non-abelian group is more severe than a $U(1)$ charge conservation problem, and requires (among other loopholes) non-invertible twist lines. We will solve this example in \cref{sec:non-abelian}, and see this expectation borne out: the twist lines for scattering to the left belong to a Tambara--Yamagami category $\text{TY}(\bZ_3 \times \bZ_3, \begin{psmallmatrix} 1 & 0 \\ 0 & -1 \end{psmallmatrix}, +1)$, while the twist lines for scattering to the right belong to $\text{Rep}(\bS_3) \times \bZ_3$.

Note that this example is not purely academic: we will see in \cref{sec:mpole} that it sits in an infinite sequence of examples with $G = SU(N) \times SU(N+2) \times U(1)^2$ for which the corresponding $G$-symmetric boundary conditions precisely model the low energy scattering of massless fermions off 't Hooft lines in a corresponding family of four-dimensional abelian gauge theories.

\subsection{Summary of results}

Our goal is to construct topological defects $\cN$ of $N$ right-moving Weyl fermions preserving a connected Lie group symmetry $G$ that acts differently to the left and the right of the defect, with matching perturbative anomaly $k$ on both sides. As discussed above, we specialise to the case that $G_k$ is conformally embedded in the full current algebra $SO(2N)_1$. The desired outcome is shown in \cref{fig:intro_slab_2d} below, with the two realisations $U_g^{L,R}$, $g\in G$ of the symmetry forming a topological junction with $\cN$.

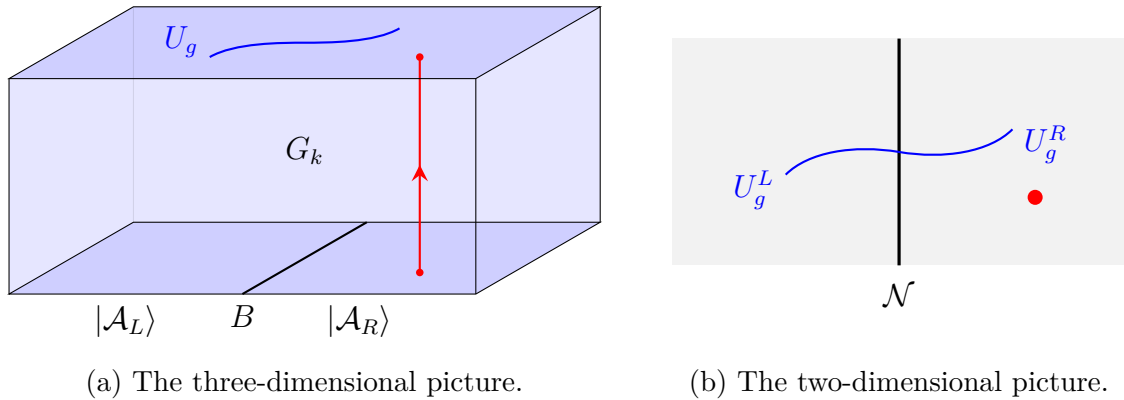
\begin{figure}[H]
    \begin{subfigure}[t]{0.55\textwidth}
        \centering
        \begin{tikzpicture}[z={(30:2cm)},scale=0.95]
            \newcommand{\x}{6.5}
            \newcommand{\y}{3}

            \fill[blue,opacity=0.1] (0,0,0) -- ++(0,-\y,0) -- ++(\x,0,0) -- ++(0,0,1) -- ++(0,\y,0) -- ++(-\x,0,0) -- cycle;
            \fill[fill=blue,opacity=0.1] (0,0,0) -- ++(\x,0,0) -- ++(0,0,1) -- ++(-\x,0,0) -- cycle;
            \filldraw[black,fill=blue,fill opacity=0.1] (0,-\y,0) -- ++(\x,0,0) -- ++(0,0,1) -- ++(-\x,0,0) -- cycle;
            \draw[black] (0,0,0) -- ++(0,-\y,0);
            \draw[black] (\x,0,0) -- ++(0,-\y,0);
            \draw[black] (\x,0,1) -- ++(0,-\y,0);
            \draw[black,opacity=0.1] (0,0,1) -- ++(0,-\y,0);
            \draw[thick,black] (\x/2,-\y,0) -- ++(0,0,1);
            \draw[thick,blue] (\x*0.35,0,0.3) node[left,yshift=0.2cm] {$U_g$} .. controls +(0,0,0.2) and +(\x*-0.05,0,0) .. (\x*0.5,0,0.5) .. controls +(\x*0.05,0,0) and +(0,0,-0.2) .. (\x*0.65,0,0.7);

            \node at (\x/2,-\y/2,0.5) {$G_k$};
            \node[below] at (\x*0.25,-\y,0) {$\ket{\cA_L}$};
            \node[below] at (\x*0.75,-\y,0) {$\ket{\cA_R}$};
            \node[below] at (\x*0.5,-\y,0) {$B$};

            \begin{scope}[every node/.style={shape=circle,inner sep=1pt}]
                \draw[thick,red,decoration = {markings, mark=at position .5 with {\arrowreversed[scale=1.5,rotate=0]{stealth}}}, postaction=decorate] (\x*0.8,0,0.3) node[fill=red] {} -- ++(0,-\y,0) node[fill=red] {};
            \end{scope}

            \draw[black] (0,0,0) -- ++(\x,0,0) -- ++(0,0,1) -- ++(-\x,0,0) -- cycle;
        \end{tikzpicture}
        \caption{The three-dimensional picture.}
        \label{fig:intro_slab_3d}
    \end{subfigure}
    \begin{subfigure}[t]{0.45\textwidth}
        \centering
        \raisebox{1em}{
        \begin{tikzpicture}[z={(30:2cm)}]
            \newcommand{\x}{6}
            \newcommand{\y}{3}
            \fill[black!5] (0,0) rectangle (\x,-\y);
            \draw[very thick] (\x/2,0) -- ++(0,-\y);
            \draw[thick,blue] (\x*0.25,-\y*0.6) node[left,yshift=-0.2cm] {$U_g^L$} .. controls +(\x*0.05,\y*0.1) and +(\x*-0.1,0.1) .. (\x*0.5,-\y*0.5) .. controls +(\x*0.1,-0.1) and +(-\x*0.05,\y*-0.1) .. (\x*0.75,-\y*0.4) node[right,yshift=-0.2cm] {$U_g^R$};
            \node[fill=red,shape=circle,inner sep=2pt] at (\x*0.8,-\y*0.7) {};
            \node[below] at (\x/2,-\y-0.1) {$\cN$};
        \end{tikzpicture}
        }
        \caption{The two-dimensional picture.}
        \label{fig:intro_slab_2d}
    \end{subfigure}
    \caption{TQFT construction of the $G$-preserving defect $\cN$.}
    \label{fig:intro_slab}
\end{figure}

To do this, we use a three-dimensional formulation of the theory. Let us begin with just one realisation of the symmetry $G$, with associated current algebra $G_k$. Any purely right-moving CFT with this current algebra admits a realisation involving the $G_k$ Chern--Simons theory \cite{Witten:1988hf,Elitzur:1989nr}. We take a slab of this Chern--Simons theory, and on the top we place the standard gapless boundary condition of $G_k$ hosting right-moving modes of central charge $c(G_k)$. This boundary condition can be thought of as the right-moving half of the corresponding $G_k$ WZW model.\footnote{Unlike the full WZW model, this theory is typically not a well-defined CFT on its own, unless $G_k = (E_8)_1$ or $\SO(n)_1$.} The gapless excitations here are our fermions, as is familiar from non-abelian bosonisation. This construction is similar to the SymTFT \cite{Freed:2022qnc,Apruzzi:2021nmk}, and we will elaborate more on this later.

The bottom boundary generically harbours gapless modes with central charge $c(\cT)-c(G_k)$, but specialising to $G_k$ conformally embedded ensures that this vanishes, and so this boundary is topological.\footnote{In this simplified introduction we have suppressed a gravitational SPT of $((E_8)_1)^n$ if $\cT$ is bosonic with $c(\cT) = 8n$, and $\SO(n)_1$ if $\cT$ is fermionic with $c(\cT) = n/2$, since it plays no role here. We will restore these SPTs from \cref{sec:theory} onwards.}  The realisation of $G$ carried by our $N$ Weyl fermions---equivalently, the conformal embedding of $G_k$ inside the full current algebra $SO(2N)_1$---is encoded in the boundary condition at this topological boundary.

This can be understood in terms of the anyons (i.e.\ Wilson lines) of the bulk Chern--Simons theory. All anyons are able to end on the top boundary, and each such endpoint corresponds to a point operator in the two-dimensional theory. In contrast, only a subset $\cA$ of the anyons are allowed to end on the bottom boundary, and as such, we denote the bottom boundary by $\ket{\cA}$.

\emph{Local} operators in two dimensions correspond to anyons belonging to $\cA$, which begin on one boundary and end on the other, as shown in red in \cref{fig:intro_slab}. Charge under $G$ is carried by the top endpoint, and is measured by topological lines $U_g$ that are supported in the top boundary, and which generically cannot venture into the bulk. We can also consider an anyon that lies outside $\cA$. While such an anyon can end on the upper boundary, it cannot end on the lower one, and instead must trail off to infinity. The corresponding two-dimensional picture is then an operator attached to a topological line, which thus defines a twist operator.

The scenario of interest is when our theory of $N$ Weyl fermions admits two $G$ symmetries $G^{L,R}$ with the same anomaly; that is, two inequivalent conformal embeddings of $G_k$ inside $SO(2N)_1$. Each comes with an associated topological boundary condition $\ket{\cA_{L,R}}$. To construct $\cN$ in three dimensions, we take one boundary condition $\ket{\cA_L}$ in one region of the bottom boundary, and $\ket{\cA_R}$ in the other, separated by a topological interface $B$. By construction, $G$ acts differently on the local fermions of the two-dimensional theory on each side of $B$. Upon collapsing the slab, we find the desired defect $\cN$ shown in \cref{fig:intro_slab_2d}.

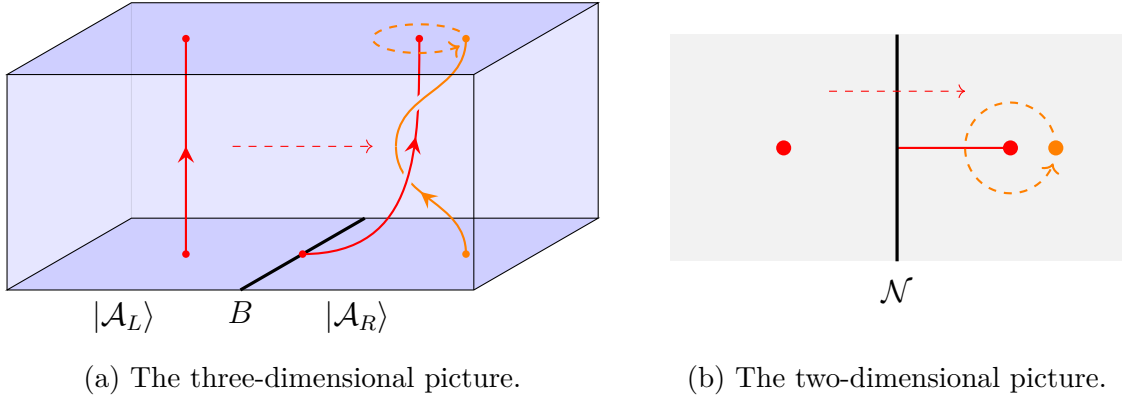
\begin{figure}[t]
    \begin{subfigure}[t]{0.55\textwidth}
        \centering
        \begin{tikzpicture}[z={(30:2cm)},scale=0.95]
            \newcommand{\x}{6.5}
            \newcommand{\y}{3}

            \fill[blue,opacity=0.1] (0,0,0) -- ++(0,-\y,0) -- ++(\x,0,0) -- ++(0,0,1) -- ++(0,\y,0) -- ++(-\x,0,0) -- cycle;
            \fill[fill=blue,opacity=0.1] (0,0,0) -- ++(\x,0,0) -- ++(0,0,1) -- ++(-\x,0,0) -- cycle;
            \filldraw[black,fill=blue,fill opacity=0.1] (0,-\y,0) -- ++(\x,0,0) -- ++(0,0,1) -- ++(-\x,0,0) -- cycle;
            \draw[black] (0,0,0) -- ++(0,-\y,0);
            \draw[black] (\x,0,0) -- ++(0,-\y,0);
            \draw[black] (\x,0,1) -- ++(0,-\y,0);
            \draw[black,opacity=0.1] (0,0,1) -- ++(0,-\y,0);
            \draw[very thick,black] (\x/2,-\y,0) -- ++(0,0,1);

            \node[below] at (\x*0.25,-\y,0) {$\ket{\cA_L}$};
            \node[below] at (\x*0.75,-\y,0) {$\ket{\cA_R}$};
            \node[below] at (\x*0.5,-\y,0) {$B$};

            \begin{scope}[every node/.style={shape=circle,inner sep=1pt}]
                \draw[thick,orange,decoration = {markings, mark=at position .4 with {\arrowreversed[scale=1.5,rotate=0]{stealth}}}, postaction=decorate] (\x*0.7,-\y/2,0.5) .. controls +(0,-\y*0.25) and +(0,\y*0.25) .. (\x*0.85,-\y,0.5) node[fill=orange] {};

                \begin{scope}
                    \clip (\x/2,-\y+0.1,1) rectangle (\x-0.1,-0.1,0);
                    \draw[blue!10,line width=4pt] (\x*0.75,0,0.5) .. controls+(0,-\y/2) and +(\x*0.25,0) .. (\x*0.5,-\y,0.5);
                \end{scope}

                \draw[thick,red,decoration = {markings, mark=at position .5 with {\arrowreversed[scale=1.5,rotate=0]{stealth}}}, postaction=decorate] (\x*0.25,0,0.5) node[fill=red] {} -- ++(0,-\y,0) node[fill=red] {};
                \draw[thick,red,decoration = {markings, mark=at position .35 with {\arrowreversed[scale=1.5,rotate=-4]{stealth}}}, postaction=decorate] (\x*0.75,0,0.5) node[fill=red] {} .. controls+(0,-\y/2) and +(\x*0.25,0) .. (\x*0.5,-\y,0.5) node[fill=red] {};
                \draw[->,dashed,red] (\x*0.35,-\y/2,0.5) -- ++(\x*0.3,0,0);

                \begin{scope}
                    \clip (\x/2,-\y+0.1,1) rectangle (\x-0.1,-0.1,0);
                    \draw[blue!10,line width=4pt] (\x*0.85,0,0.5) .. controls +(0,-\y*0.25) and +(0,\y*0.25) .. (\x*0.7,-\y/2,0.5);
                \end{scope}

                \draw[thick,orange] (\x*0.85,0,0.5) node[fill=orange] {} .. controls +(0,-\y*0.25) and +(0,\y*0.25) .. (\x*0.7,-\y/2,0.5);
            \end{scope}

            \draw[thick,dashed,->,orange] ({\x*0.75+\x*0.1*cos(30)},{0.2*sin(30)},0.5) arc[start angle=30,end angle=330,x radius=\x*0.1,y radius=0.2];

            \draw[black] (0,0,0) -- ++(\x,0,0) -- ++(0,0,1) -- ++(-\x,0,0) -- cycle;
        \end{tikzpicture}
        \caption{The three-dimensional picture.}
        \label{fig:intro_braiding_3d}
    \end{subfigure}
    \begin{subfigure}[t]{0.45\textwidth}
        \centering
        \raisebox{1em}{
        \begin{tikzpicture}[z={(30:2cm)}]
            \newcommand{\x}{6}
            \newcommand{\y}{3}
            \fill[black!5] (0,0) rectangle (\x,-\y);
            \begin{scope}[every node/.style={fill=red,shape=circle,inner sep=2pt}]
                \draw[thick,red] (\x/2,-\y/2) -- (\x*0.75,-\y/2) node{};
                \node at (\x*0.25,-\y/2) {};
                \draw[very thick] (\x/2,0) -- ++(0,-\y);
                \draw[->,dashed,red] (\x*0.35,-\y*0.25) -- ++(\x*0.3,0);
            \end{scope}
            \draw[thick,dashed,->,orange] ([shift=(20:\x*0.1)]\x*0.75,-\y/2) arc[start angle=20,end angle=340,radius=\x*0.1];
            \fill[orange] (\x*0.85,-\y/2) circle (0.1);
            \node[below] at (\x/2,-\y-0.1) {$\cN$};
        \end{tikzpicture}
        }
        \caption{The two-dimensional picture.}
        \label{fig:intro_braiding_2d}
    \end{subfigure}
    \caption{Twisted scattering occurs when an anyon ending on $\ket{\cA_L}$ can no longer end on $\ket{\cA_R}$. The twist line monodromy can be probed by braiding anyons.}
    \label{fig:intro_braiding}
\end{figure}

We can then consider scattering through $\cN$ from the three-dimensional perspective, as shown in \cref{fig:intro_braiding_3d}. An incoming particle is described by a vertical anyon that ends on $\ket{\cA_L}$, and travels right. There is no interface on the upper boundary: the right-moving gapless excitation described by the top endpoint of the anyon sails straight through. Since charge under $G$ is measured on this gapless boundary, the symmetry $G$ is manifestly preserved. All non-triviality comes from what's happening downstairs. It may be that this anyon can also end on $\ket{\cA_R}$, in which case a local operator can emerge on the other side, and the scattering is boring. But if the anyon \emph{cannot} end on $\ket{\cA_R}$, the lower endpoint will get snagged on $B$ and left behind, dragging the anyon out into a twist line. Collapsing the slab, we find the classic picture: the outgoing excitation is created by a twist operator living at the end of a topological line, the other end of which is anchored on $\cN$.

The properties of this twist operator are encoded in two dimensions by the monodromies of local operators as they encircle it, as shown in \cref{fig:intro_braiding_2d}, which are in turn captured in three dimensions by the braiding of anyons, shown in \cref{fig:intro_braiding_3d}.

The question then becomes how to construct $B$. The assumption that the bottom boundaries are topological is key, as it brings this problem into the realm of generalised symmetries where it can be solved. The topological boundaries $\ket{\cA_{L,R}}$ are characterised by the sets of bulk anyons $\cA_{L,R}$ which can end on them. An interface $B$ is then a line on which both $\cA_L$ and $\cA_R$ know how to end in a mutually consistent way. Mathematicians have studied such objects \cite{ostrik2001}, and refer to them as \emph{$(\cA_L, \cA_R)$-bimodules}.\footnote{We thank Francesco Benini and Antonio Santaniello for helping us out with this point.} In particular, they always exist, and there exists a finite algorithm to construct them all, which yields a classification of all symmetric defects $\cN$. Moreover, as a byproduct of this technology, we will be able to explicitly compute the (generically non-invertible) twist lines generated by scattering from the data of the bimodule $B$.

We will also give a purely two-dimensional construction of $\cN$ by identifying a particular symmetry such that when gauged in a region of spacetime, the interface between the original and gauged theory is precisely $\cN$. That is, one side of the scattering setup \eqref{eq:2d scattering intro} can be viewed as a gauged version of the other. Indeed, in the equal-rank case this was achieved in \cite{Arias-Tamargo:2026urw, Wei:2026fsn} where the symmetry to be gauged is a finite abelian group. In the conformally embedded case, generically we must gauge a finite categorical symmetry, rather than a conventional group symmetry.

Boundary conditions of two-dimensional CFTs also admit a number of other important properties such as their Affleck--Ludwig central charge $g$ \cite{affleckludwig} and, for fermionic CFTs, an SPT class valued in $\mho^3_\Spin(\pt) \cong \bZ_2$ relating to the Arf topological phase \cite{BoyleSmith:2019jnh,Arias-Tamargo:2026urw}. We will also show how this information can be recovered from the three-dimensional picture in \cref{sec:SPTs_general}.

\subsection{The plan of the paper}

In \cref{sec:theory} we put some mathematical meat on the bones of this construction. Everything we do applies to a generic pair of conformal embeddings $G_k^{L,R} \subseteq SO(2N)_1$, for any $N$. The methods generalise immediately to generic RCFTs. The essential tools will be those of a \emph{super-modular tensor category}, which is the appropriate language to describe the fusion and braiding of anyons in Chern--Simons theory, as well as the construction of the topological boundary conditions $\ket{\cA_{L,R}}$. We demonstrate how to identify which anyons can end on the lower boundary specified by each conformal embedding. We find that an interface $B$ always exists, and show how to classify possible distinct choices thereof, and compute their quantum dimension, corresponding to the boundary central charge. Moreover, we show how the braiding of anyons determines the way that twist operators generated by scattering behave with respect to regular local operators. We also uncover the purely two-dimensional realisation of $\cN$ as a generalised half-space gauging defect. Finally, we address a number of subtleties stemming from the fact we are working with fermions, and thus need to keep track of what changes when we consider different spin structures.

We then apply this construction to a number of examples. In \cref{sec:abelian} we reproduce the known $G$-symmetric defects in the case $G=U(1)^N$, which serves as a useful proving ground for our new perspective. In \cref{sec:non-abelian} we apply our method to a family of examples with $G$ non-abelian and with $\rank(G)<N$, but for which the embedding $G_k\subseteq SO(2N)_1$ is conformal. Finally, we return to the generic non-conformally embedded case in \cref{sec:outlook} and offer outlook on the interpretation of our results in the context of scattering chiral fermions off 't Hooft lines in four-dimensional gauge theories.

\paragraph{Note added.} As this work neared completion, we learnt about the related work \cite{yunqinetal}. We thank the authors for agreeing to coordinate submission.

\section{Three-dimensional description of scattering}\label{sec:theory}

In this section, we develop the general framework introduced above to construct symmetric defects by using the auxiliary three-dimensional picture. In \cref{sec:CS_basics} we recapitulate the basic facts of the correspondence between two-dimensional rational CFTs (RCFTs) and three-dimensional Chern--Simons theory, which will be central to our construction; as well as the algorithm to construct all possible interfaces $B$ from the bulk data. In \cref{sec:scattering_from_bulk} we specialise this general technology to our case of interest, resulting in the construction of the symmetric defects $\cN$ for the theory of Weyl fermions, and discuss scattering through it. In \cref{sec:2d_descrip} we explain how to determine the symmetry that needs to be gauged in half-space (à la \cite{Arias-Tamargo:2026urw, Wei:2026fsn}) to arrive at a purely two-dimensional description of $\cN$. In \cref{sec:monodromies} we elaborate on the consistency of these two pictures by seeing how the monodromy due to the twist line in two dimensions arises from the braiding of anyons in three dimensions. Finally in \cref{sec:SPTs_general} we discuss a collection of fermionic subtleties that arise.

\subsection{RCFTs and Chern--Simons theories}\label{sec:CS_basics}

An RCFT comes with an associated chiral algebra, $G_k$, and the Hilbert space of the theory decomposes into representations of this algebra. More specifically, only the integrable representations of $G_k$ are used, of which there are finitely many. Primary states are therefore labelled by representations $a$, and descendants are found by acting with the current modes generating the algebra. A full RCFT consists of a holomorphic and anti-holomorphic sector, paired together in a consistent fashion. One ingredient of this pairing is a matrix $Z_{ab}\in\bZ_{\geq0}$ which is invariant under modular transformations. This specifies a modular invariant partition function. Denoting by $\chi_a(\tau)$ the character associated with the representation $a$, the partition function then takes the form
\begin{equation}\label{eq:PF}
    \cZ(\tau) = \sum_{a,b} Z_{ab} \chi_a(\tau) \overline{\chi}_b(\overline{\tau})
\end{equation}
Crucially, in an RCFT the sum over $a,b$ is finite. One choice of modular invariant for any $G_k$ is the diagonal theory $Z_{ab}=\delta_{ab}$ (the WZW model) where each representation appears once, paired with its conjugate. Specifying the full RCFT requires more data than $Z_{ab}$ alone \cite{Moore:1988qv, Fuchs:2002cm}.

Through the state-operator correspondence, primary operators $\cO_a$ of the RCFT are also labelled by integrable representations. The Sugawara construction allows us to determine the conformal dimensions $h_a$ of all the primaries, as well as the central charge $c$ of the theory, from purely algebraic data of $G_k$. Since there are only finitely-many primaries, the OPE is constrained to the form
\begin{equation}\label{eq:OPE}
    \cO_a (z) \cO_b (0) \sim \sum_{c \in a \otimes b} C_{ab}^c z^{h_c - h_a - h_b} \cO_c (0) + \text{descendants}
\end{equation}
where the $C_{ab}^c$ are constants. The representations $c$ occurring in the OPE are determined by tensor product of the representations $a$ and $b$, also known as the fusion rules of the algebra.

Let us now turn to the Chern--Simons description of the theory \cite{Witten:1988hf, Elitzur:1989nr}. We consider a three-dimensional manifold $\cM = \Sigma \times [0,1]$, where $\Sigma$ is the two-dimensional manifold on which the RCFT is defined. A Chern--Simons theory is also defined by a chiral algebra $G_k$, such that there is a simple line called an anyon or Wilson line associated with each representation $a$. The theory is topological and the anyons can be moved around freely in $\cM$. The anyon corresponding to the trivial representation is often denoted $\1$.

The relation to two-dimensional RCFT comes since the Chern--Simons theory admits boundary conditions which realise the gapless holomorphic $G_k$ chiral algebra on one boundary and the anti-holomorphic $G_k$ chiral algebra on the other. That is, an anyon $a$ extended between the two boundaries gives rise to a local excitation in the $a$ representation of the holomorphic and anti-holomorphic copies of $G_k$. Since this can be performed for all anyons $a$, the boundary degrees of freedom are precisely those of the diagonal modular invariant RCFT described above:
\begin{equation}\label{eq:diagonalWZW}
\begin{tikzpicture}[z={(30:2cm)},baseline=(current bounding box.center)]
    \newcommand{\x}{4}
    \newcommand{\y}{3}

    \fill[blue,opacity=0.1] (0,0,0) -- ++(0,-\y,0) -- ++(\x,0,0) -- ++(0,0,1) -- ++(0,\y,0) -- ++(-\x,0,0) -- cycle;
    \filldraw[black,fill=blue,fill opacity=0.1] (0,-\y,0) -- ++(\x,0,0) -- ++(0,0,1) -- ++(-\x,0,0) -- cycle;

    \draw[black] (0,0,0) -- ++(0,-\y,0);
    \draw[black] (\x,0,0) -- ++(0,-\y,0);
    \draw[black] (\x,0,1) -- ++(0,-\y,0);
    \draw[black,opacity=0.1] (0,0,1) -- ++(0,-\y,0);

    \draw[thick,red,decoration = {markings, mark=at position .5 with {\arrowreversed[scale=1.5]{stealth}}}, postaction=decorate] (\x*0.25,0,0.5) node[style={fill=red,shape=circle,inner sep=1pt}]{} -- ++(0,-\y,0) node[style={fill=red,shape=circle,inner sep=1pt}]{};
    \node[left,red] at (\x*0.25,-\y/2,0.5) {$a$};

    \filldraw[black,fill=blue,fill opacity=0.1] (0,0,0) -- ++(\x,0,0) -- ++(0,0,1) -- ++(-\x,0,0) -- cycle;

    \node at (\x/2,-\y*0.5,0.5) {$G_k$};
    \node[left] at (0,0,0) {$\ket{G_k}$};
    \node[left] at (0,-\y,0) {$\ket{\overline{G}_k}$};

    \draw[->,thick] (\x*0.4,0,0.5) -- (\x*0.6,0,0.5);
    \draw[->,thick] (\x*0.6,-\y,0.5) -- (\x*0.4,-\y,0.5);
\end{tikzpicture}
\end{equation}
The black arrows indicate that the upper boundary corresponds to the holomorphic sector and the lower boundary to the anti-holomorphic sector. Other modular invariants can be constructed by inserting a surface defect $S$ which lies horizontally across the centre of the bulk which transforms an anyon $a$ into the anyon $b$ with which it is paired in the partition function \eqref{eq:PF} \cite{Kapustin:2010if} (see also \cite{Komargodski:2020mxz} for a recent discussion):
\begin{equation}
\begin{tikzpicture}[z={(30:2cm)},baseline=(current bounding box.center)]
    \newcommand{\x}{4}
    \newcommand{\y}{4}

    \fill[blue,opacity=0.1] (0,0,0) -- ++(0,-\y,0) -- ++(\x,0,0) -- ++(0,0,1) -- ++(0,\y,0) -- ++(-\x,0,0) -- cycle;
    \filldraw[black,fill=blue,fill opacity=0.1] (0,-\y,0) -- ++(\x,0,0) -- ++(0,0,1) -- ++(-\x,0,0) -- cycle;
    \filldraw[black,fill=orange,fill opacity=0.1] (0,-\y/2,0) -- ++(\x,0,0) -- ++(0,0,1) -- ++(-\x,0,0) -- cycle;

    \draw[black] (0,0,0) -- ++(0,-\y,0);
    \draw[black] (\x,0,0) -- ++(0,-\y,0);
    \draw[black] (\x,0,1) -- ++(0,-\y,0);
    \draw[black,opacity=0.1] (0,0,1) -- ++(0,-\y,0);

    \draw[thick,red,decoration = {markings, mark=at position .5 with {\arrowreversed[scale=1.5]{stealth}}}, postaction=decorate] (\x*0.25,0,0.5) node[style={fill=red,shape=circle,inner sep=1pt}]{} -- ++(0,-\y/2,0) node[style={fill=red,shape=circle,inner sep=1pt}]{};
    \draw[thick,red,decoration = {markings, mark=at position .5 with {\arrowreversed[scale=1.5]{stealth}}}, postaction=decorate] (\x*0.25,-\y/2,0.5) node[style={fill=red,shape=circle,inner sep=1pt}]{} -- ++(0,-\y/2,0) node[style={fill=red,shape=circle,inner sep=1pt}]{};
    \node[left,red] at (\x*0.25,-\y/4,0.5) {$a$};
    \node[left,red] at (\x*0.25,-\y*0.75,0.5) {$b$};

    \filldraw[black,fill=blue,fill opacity=0.1] (0,0,0) -- ++(\x,0,0) -- ++(0,0,1) -- ++(-\x,0,0) -- cycle;

    \node at (\x/2,-\y*0.25,0.5) {$G_k$};
    \node[left] at (0,0,0) {$\ket{G_k}$};
    \node[left] at (0,-\y,0) {$\ket{\overline{G}_k}$};
    \node[orange] at (\x*0.6,-\y/2,0.5) {$S$};

    \draw[->,thick] (\x*0.4,0,0.5) -- (\x*0.6,0,0.5);
    \draw[->,thick] (\x*0.6,-\y,0.5) -- (\x*0.4,-\y,0.5);
\end{tikzpicture}
\end{equation}
Folding the bulk along this surface gives a product theory $G_k \boxtimes \overline{G_k} \cong G_k \boxtimes G_{-k}$. Here, and throughout, we use a bar to denote orientation reversal. We caution that $G_{-k}$ does not refer to the category of representations of the chiral algebra $G_{-k}$ with negative level. Instead it refers to the Chern--Simons theory with negative level. We will discuss orientation-reversal at the level of the anyon data further below. The surface defect $S$ becomes a topological boundary of the folded theory while the upper boundary then carries both the holomorphic and anti-holomorphic modes. This reflects the correspondence between this setup and the SymTFT construction \cite{Freed:2022qnc,Apruzzi:2021nmk}.\footnote{In particular, for a two-dimensional theory with symmetry $\cC$, the SymTFT construction asks us to construct a three-dimensional TQFT described by the Drinfel'd centre $\cZ(\cC)$. This TQFT admits a `physical' boundary where the dynamical modes live and a `symmetry' boundary which is topological. From this perspective, the reason for which $G_k \boxtimes G_{-k}$ appears in the construction above is because $\cZ(\cC) \cong \cC \boxtimes \overline{\cC}$ when $\cC$ is modular \cite{mueger2001}, which is the case when $\cC = \Rep(G_k)$ is the category of representations of a chiral algebra.}

Let us elaborate briefly on the structures exhibited by the Chern--Simons anyons. The set of anyons is the set of (integrable) representations of $G_k$, which we denote $\cC=\Rep(G_k)$. Bringing two anyons $a$ and $b$ together, they can be understood as a composite object which is expressible as a sum of other anyons. This \emph{fusion} is written
\begin{equation}\label{eq:fusion}
    a \otimes b = \bigoplus_c N_{ab}^c c
\end{equation}
where $N_{ab}^c \in \bZ_{\geq0}$. The coefficient $N_{ab}^c$ describes the number of distinct junctions between incoming lines $a$ and $b$ and an outgoing line $c$. These junctions then form an $N_{ab}^c$-dimensional vector space denoted $\Hom(a\otimes b,c)$. We will denote junctions by Greek letters $\mu$, $\nu$, etc.

The fusion of the bulk anyons is the same as that of the boundary primary operators discussed above, encoding the structure of the OPE. The associativity of the OPE is encoded in a map (morphism) $\alpha_{a,b,c} : (a \otimes b) \otimes c \to a \otimes (b \otimes c)$ whose components in a given choice of basis of the junction vector spaces are known as \emph{$F$-symbols}. Diagrammatically,
\begin{equation}\label{eq:F_def}
    \begin{tikzpicture}[baseline=-1em]
    	\draw [thick, decoration = {markings, mark=at position .7 with {\arrow[scale=1.5]{stealth}}}, postaction=decorate] (-1, -1) node[below] {\small $a$} to (0,0) node[right] {\footnotesize $\nu$};
    	\draw [thick, decoration = {markings, mark=at position .7 with {\arrow[scale=1.5]{stealth}}}, postaction=decorate] (-0, -1) node[below] {\small $b$} to (.5,-.5) node[right] {\footnotesize $\mu$};
    	\draw [thick, decoration = {markings, mark=at position .7 with {\arrow[scale=1.5]{stealth}}}, postaction=decorate] (1,-1) node[below] {\small $c$} to (.5,-.5);
    	\draw [thick, decoration = {markings, mark=at position .7 with {\arrow[scale=1.5]{stealth}}}, postaction=decorate] (.5, -.5) node[above] {\small $e$} to (0,0);
    	\draw [thick, decoration = {markings, mark=at position .7 with {\arrow[scale=1.5]{stealth}}}, postaction=decorate] (0,0) to (-.5,.5) node[above] {\small $d$};
    \end{tikzpicture}
    = \sum_{f,\rho,\sigma} (F^{abc}_{def})^{\mu\nu}_{\rho\sigma}
    \,
    \begin{tikzpicture}[baseline=-1em]
    	\draw [thick, decoration = {markings, mark=at position .7 with {\arrow[scale=1.5]{stealth}}}, postaction=decorate] (-1, -1) node[below] {\small $a$} to (-.5,-.5) node[left] {\footnotesize $\rho$};
    	\draw [thick, decoration = {markings, mark=at position .7 with {\arrow[scale=1.5]{stealth}}}, postaction=decorate] (-0, -1) node[below] {\small $b$} to (-.5,-.5);
    	\draw [thick, decoration = {markings, mark=at position .7 with {\arrow[scale=1.5]{stealth}}}, postaction=decorate] (1,-1) node[below] {\small $c$} to (0,0) node[right] {\footnotesize $\sigma$};
    	\draw [thick, decoration = {markings, mark=at position .7 with {\arrow[scale=1.5]{stealth}}}, postaction=decorate] (-.5, -.5) node[above] {\small $f$} to (0,0);
    	\draw [thick, decoration = {markings, mark=at position .7 with {\arrow[scale=1.5]{stealth}}}, postaction=decorate] (0,0) to (-.5,.5) node[above] {\small $d$};
    \end{tikzpicture}
\end{equation}
These must satisfy a consistency condition known as the pentagon identity \cite[§2.1]{etingof}.

Two important quantities associated to an anyon $a$ are its quantum dimension $d_a$ and topological spin $\theta_a$, defined by the expectation value of the following configurations of anyons:
\begin{equation}
    d_a =
    \begin{tikzpicture}[baseline=-0.5ex]
        \draw [thick, decoration = {markings, mark=at position .47 with {\arrowreversed[scale=1.5,rotate=10]{stealth}}}, postaction=decorate] (0,0) circle [radius=.5];
        \node at (-.8,0) {\small $a$};
    \end{tikzpicture}
    \qquad \qquad \theta_a = \frac{1}{d_a} \,
    \begin{tikzpicture}[baseline=-0.5ex]
        \draw [thick] (0,0) ++(45:.5) arc (45:315:.5);
        \draw [thick] (0,0) ++(-45:.5) to +(45:1);
        \draw [thick, decoration = {markings, mark=at position .46 with {\arrowreversed[scale=1.5,rotate=10]{stealth}}}, postaction=decorate] (0,0) ++(-45:1) ++(45:.5) arc (-135:135:.5);
        \node at (2.20,0) {\small $a$};
        \draw [thick] (0,0) ++(45:.5) to +(-45:.3); \draw [thick] (0,0) ++(45:.5) ++ (-45:.7) to +(-45:.3);
    \end{tikzpicture}
\end{equation}
The latter is related to the conformal dimension of the local operator $\cO_a$ of the RCFT by
\begin{equation}\label{eq:theta=exp(h)}
    \theta_a = \exp(2\pi i h_a)
\end{equation}
These diagrams should be interpreted within the so-called \emph{string calculus}, which uses pictures to represent morphisms between objects. This roughly corresponds to the evaluation of defect networks in the physical theory, up to some details relating to the Frobenius--Schur indicator \cite{fs}.

Finally, the Chern--Simons theory admits a \emph{braiding}, which is described by a (non-degenerate) map $c_{a,b} : a \otimes b \to b \otimes a$. Consistency with the associativity structure imposes conditions known as the hexagon identities on $c_{a,b}$ \cite[§8.1]{etingof}. Moreover, consistency between the braiding and the topological spins imposes the balancing equation
\begin{equation}\label{eq:balance}
    c_{b,a} \circ c_{a,b} \eval_c = \frac{\theta_c}{\theta_a \theta_b}
\end{equation}
where $c \in a \otimes b$. Here, $c_{b,a} \circ c_{a,b}$ is an endomorphism of $a \otimes b$ and $\eval_c$ indicates a restriction to the fusion channel $c$. In a chosen basis for the junction vector spaces, the braiding can be represented by an \emph{$R$-matrix} as
\begin{equation}\label{eq:R_def}
    \begin{tikzpicture}[baseline=-0.1cm]
        \draw [thick, decoration = {markings, mark=at position .7 with {\arrow[scale=1.5]{stealth}}}, postaction=decorate] (0,-.5) node[below] {\small $c$} to (0,0) node[right] {\footnotesize $\mu$};
        \draw [thick] (0,0) arc [radius=.3, start angle=240, end angle=120];
        \draw [thick, decoration = {markings, mark=at position .75 with {\arrow[scale=1.5]{stealth}}}, postaction=decorate] (0,0.52) to +(30:.5) node[right] {\small $b$};
        \draw [thick] (0,0) arc [radius=.3, start angle=-60, end angle=42];
        \draw [thick, decoration = {markings, mark=at position .8 with {\arrow[scale=1.5]{stealth}}}, postaction=decorate] (0,0.52) ++(150:.1) to +(150:.4) node[left] {\small $a$};
    \end{tikzpicture}
    =
    \sum_{\nu} (R^{ab}_c)_{\mu\nu}
    \begin{tikzpicture}[baseline=-0.4cm]
        \draw [thick, decoration = {markings, mark=at position .5 with {\arrow[scale=1.5]{stealth}}}, postaction=decorate] (0,-.8) node[below] {\small $c$} to (0,0);
        \draw [thick, decoration = {markings, mark=at position .7 with {\arrow[scale=1.5]{stealth}}}, postaction=decorate] (0,0) node[below right] {\footnotesize $\nu$} to (-.6,.5) node[left] {\small $a$};
        \draw [thick, decoration = {markings, mark=at position .7 with {\arrow[scale=1.5]{stealth}}}, postaction=decorate] (0,0)--(.6,.5) node[right] {\small $b$};
    \end{tikzpicture}
\end{equation}
whence \eqref{eq:balance} becomes
\begin{equation}\label{eq:ribbon}
    \sum_\lambda (R^{ab}_c)_{\mu\lambda} (R^{ba}_c)_{\lambda\nu} = \frac{\theta_c}{\theta_a\theta_b} \delta_{\mu\nu}
\end{equation}

We will often be interested in theories where there is never more than one type of junction between any three lines; that is, $N_{ab}^c \in \{0,1\}$ in \eqref{eq:fusion}. In this case, we will drop the junction labels $\mu$, $\nu$, etc.\ from the $F$-symbols and $R$-matrices.

As discussed previously, we will often make use of a \emph{folding} operation by which we take a Chern--Simons theory, pick a surface in it, take the theory on one side of the surface and fold it over to the other side. The result is a product (or `stacked') theory denoted $\cC \boxtimes \overline{\cC}$. In terms of the defining data introduced above, the orientation-reversed theory $\overline{\cC}$ is the same as $\cC$ but with its braiding reversed: $\overline{c}_{X,Y}=c_{Y,X}^{-1}$. In the mathematics literature this is often written $\overline{\cC}=\cC^{\text{rev}}$.

Overall, the Chern--Simons anyons have the structure of a \emph{modular tensor category}. Given that the two-dimensional theory we seek to describe is fermionic, we will be forced to keep track of fermionic subtleties in the corresponding Chern--Simons theory. This will require us to invoke the technology of fermionic Chern--Simons theories (such as $U(1)_k$ for odd $k$) which are described by super-modular tensor categories. We will discuss this in more detail in \cref{sec:SPTs_general}.

\subsubsection*{Boundaries of Chern--Simons theories}

Boundaries between different Chern--Simons theories can be inferred from embeddings of chiral algebras \cite{Moore:1989yh}. In particular, suppose that $G_k \to H_{k'}$ is an embedding. The anyons of the $H_{k'}$ theory are labelled by representations of $H_{k'}$, and can be branched to representations of $G_k$. In terms of Chern--Simons theories this means there is an interface between the $H_{k'}$ and $G_k$ theories, which is equivalent to a boundary of the folded theory $G_k \boxtimes H_{-k'}$:
\begin{equation}
    \begin{tikzpicture}[z={(30:2cm)}]
        \newcommand{\x}{4}
        \newcommand{\y}{4}

        \shade[top color=green,bottom color=white,middle color=green,opacity=0.1] (0,-\y/2,0) -- ++(0,-\y/2,0) -- ++(\x,0,0) -- ++(0,0,1) -- ++(0,\y/2,0) -- ++(-\x,0,0) -- cycle;
        \shade[top color=white,bottom color=blue,middle color=blue,opacity=0.1] (0,0,0) -- ++(0,-\y/2,0) -- ++(\x,0,0) -- ++(0,0,1) -- ++(0,\y/2,0) -- ++(-\x,0,0) -- cycle;
        \draw[black] (0,-\y/2,0) -- ++(\x,0,0) -- ++(0,0,1) -- ++(-\x,0,0) -- cycle;

        \draw[black] (0,0,0) -- ++(0,-\y,0);
        \draw[black] (\x,0,0) -- ++(0,-\y,0);
        \draw[black] (\x,0,1) -- ++(0,-\y,0);
        \draw[black,opacity=0.1] (0,0,1) -- ++(0,-\y,0);

        \node at (\x/2,-\y*0.85,0.5) {$H_{k'}$};
        \node at (\x/2,-\y*0.25,0.5) {$G_k$};

        \node[left] at (0,-\y/2,0) {$\ket{H_{k'}/G_k}$};
    \end{tikzpicture}
    \hspace{0.4cm}
    \begin{tikzpicture}[z={(30:2cm)}]
        \newcommand{\x}{4}
        \newcommand{\y}{4}

        \node at (-\x*0.2,-\y*0.3,0) {$=$};

        \shade[top color=white,bottom color=teal,middle color=teal,opacity=0.1] (0,0,0) -- ++(0,-\y/2,0) -- ++(\x,0,0) -- ++(0,0,1) -- ++(0,\y/2,0) -- ++(-\x,0,0) -- cycle;
        \draw[black] (0,-\y/2,0) -- ++(\x,0,0) -- ++(0,0,1) -- ++(-\x,0,0) -- cycle;
        \filldraw[black,fill=teal,fill opacity=0.1] (0,-\y/2,0) -- ++(\x,0,0) -- ++(0,0,1) -- ++(-\x,0,0) -- cycle;

        \draw[black] (0,0,0) -- ++(0,-\y/2,0);
        \draw[black] (\x,0,0) -- ++(0,-\y/2,0);
        \draw[black] (\x,0,1) -- ++(0,-\y/2,0);
        \draw[black,opacity=0.1] (0,0,1) -- ++(0,-\y/2,0);

        \node at (\x/3,-\y*0.25,0.5) {$G_k \boxtimes H_{-k'}$};

        \node[right] at (\x*1.05,-\y/2,0) {$\ket{H_{k'}/G_k}$};
        \node at (0,-\y*0.85,0) {};
    \end{tikzpicture}
\end{equation}
The boundary degrees of freedom are described by the coset $H_{k'}/G_k$ \cite{DiFrancesco:1997nk}. Of particular importance for us is the case where the embedding is conformal, meaning that the central charges of $G_k$ and $H_{k'}$ are equal. In this case the coset theory is gapped ($c=0$) and the corresponding boundary becomes topological, so can be freely deformed.

There is a simple description of topological boundaries in terms of the bulk anyons: the boundary is defined by specifying which lines can end on it. The set of lines which can end on the boundary, which is fixed by the embedding $G_k\to H_{k'}$, form an algebra $\cA$ in $G_k \boxtimes H_{-k'}$. An algebra is a collection of anyons,
\begin{equation}
    \cA = \bigoplus_a n_a a \qc n_a \in \bZ_{\geq0}
\end{equation}
with an associative multiplication $m: \cA \otimes \cA \to \cA$ and a unit map $\eta: \1 \to \cA$ satisfying a compatibility condition \cite{etingof}. Additional constraints must be placed on $\cA$ to ensure that these lines can end consistently. Of particular importance is the constraint that $\cA$ is braided commutative, meaning that $m\circ c_{\cA,\cA} = m$. Additionally we must impose that $\cA$ is maximal, in the sense that all anyons not in $\cA$ braid non-trivially with at least one line in $\cA$. This constraint is equivalent to equating the dimension of the algebra, $\dim(\cA) = \sum_a n_a d_a$, to the total quantum dimension of $\cC$, $\cD(\cC) = \sqrt{\sum_{a\in\cC} d_a^2}$:
\begin{equation}\label{eq:bosonic_lagr}
    \dim(\cA) = \cD(\cC)
\end{equation}
Such an algebra is called a \emph{Lagrangian algebra}. The maximality constraint is what tells us that $\cA$ describes a boundary of the theory, rather than an interface to another smaller theory containing those anyons which braid trivially with $\cA$. We denote this boundary by $\ket{\cA}$.

One way to understand the boundary $\ket{\cA}$ is via \emph{anyon condensation}. We take the theory $G_k \boxtimes H_{-k'}$ and divide it into two halves by choosing a surface and then gauge (i.e.\ condense) the algebra $\cA$ in the lower half.\footnote{The fact that the algebra $\cA$ can be consistently gauged is part of the conditions imposed above. Namely, a Lagrangian algebra is automatically a symmetric special Frobenius algebra.} The gauged algebra $\cA$ is often referred to as the condensate. Gauging $\cA$ corresponds to inserting a (sufficiently dense) mesh of the lines contained in it, giving
\begin{equation}
    \begin{tikzpicture}[z={(30:2cm)},baseline=(current bounding box.center)]
        \newcommand{\x}{4}
        \newcommand{\y}{6}

        \shade[top color=teal,bottom color=white,middle color=teal,opacity=0.1] (0,-\y/2,0) -- ++(0,-\y/2,0) -- ++(\x,0,0) -- ++(0,0,1) -- ++(0,\y/2,0) -- ++(-\x,0,0) -- cycle;
        \shade[top color=white,bottom color=teal,middle color=teal,opacity=0.1] (0,0,0) -- ++(0,-\y/2,0) -- ++(\x,0,0) -- ++(0,0,1) -- ++(0,\y/2,0) -- ++(-\x,0,0) -- cycle;
        \draw[black] (0,-\y/2,0) -- ++(\x,0,0) -- ++(0,0,1) -- ++(-\x,0,0) -- cycle;

        \draw[black] (0,0,0) -- ++(0,-\y,0);
        \draw[black] (\x,0,0) -- ++(0,-\y,0);
        \draw[black] (\x,0,1) -- ++(0,-\y,0);
        \draw[black,opacity=0.1] (0,0,1) -- ++(0,-\y,0);

        \node at (\x/3,-\y*0.25,0.5) {$G_k \boxtimes H_{-k'}$};

        \node[left] at (0,-\y/2,0) {$\ket{\cA}$};


        \draw[red,thick] (3,-3,0.3)--(2,-5,0.5);
        \draw[red,thick] (3,-5,1)--(3,-4,-1);
        \draw[red,thick] (3,-5,1)--(1,-3,0.8);
        \draw[red,thick] (1,-3,0.8)--(3,-3,0.5);
        \draw[red,thick] (1,-3,0.8)--(0.4,-3,0.3);
        \draw[red,thick] (3,-3,0.5)--(3.8,-3,0.8);
        \draw[red,thick] (3,-3,0.5)--(3,-5,1);
        \draw[red,thick] (2,-5,0.5)--(0.4,-3,0.3);
        \draw[red,thick] (3,-3,0.3)--(3,-4,-1);
        \draw[red,thick] (3,-4,-1)--(0.3,-4,0);
        \draw[red,thick] (2,-5,0.5)--(3.5,-5,0.2);
        \draw[red,thick] (3,-3,0.3)--(0.5,-3.5,0.6);
        \draw[red,thick] (3,-5,1)--(3.9,-4.5,0.8);

    \end{tikzpicture}
\end{equation}
The  theory in the lower half is guaranteed to be completely trivial. Indeed, when performing the gauging we must project out any lines which do not braid trivially with $\cA$. This is because in the gauged theory $\cA$ should be transparent to all surviving lines; but $\cA$ is Lagrangian, and hence maximal, so every line not in $\cA$ must be projected out.

Those lines which are not in $\cA$ cannot end on $\ket{\cA}$, so instead become non-trivial topological lines on the boundary. On the other hand, since $\cA$ is condensed on $\ket{\cA}$, a line $a\in\cA$ approaching $\ket{\cA}$ from the upper half can be absorbed into the mesh of $\cA$ lines on the boundary, and end on it:
\begin{equation}\label{eq:lines_end}
\begin{tikzpicture}[z={(30:2cm)},baseline=(current bounding box.center)]
    \newcommand{\x}{4}
    \newcommand{\y}{5}

    \shade[top color=white,bottom color=teal,middle color=teal,opacity=0.1] (0,0,0) -- ++(0,-\y/2,0) -- ++(\x,0,0) -- ++(0,0,1) -- ++(0,\y/2,0) -- ++(-\x,0,0) -- cycle;
    \draw[black] (0,-\y/2,0) -- ++(\x,0,0) -- ++(0,0,1) -- ++(-\x,0,0) -- cycle;
    \filldraw[black,fill=teal,fill opacity=0.1] (0,-\y/2,0) -- ++(\x,0,0) -- ++(0,0,1) -- ++(-\x,0,0) -- cycle;

    \draw[black] (0,0,0) -- ++(0,-\y/2,0);
    \draw[black] (\x,0,0) -- ++(0,-\y/2,0);
    \draw[black] (\x,0,1) -- ++(0,-\y/2,0);
    \draw[black,opacity=0.1] (0,0,1) -- ++(0,-\y/2,0);

    \node at (\x/2,0,0.5) {$G_k \boxtimes H_{-k'}$};

    \node[left] at (0,-\y/2,0) {$\ket{\cA}$};

    \draw[thick,red,decoration = {markings, mark=at position .5 with {\arrowreversed[scale=1.5]{stealth}}}, postaction=decorate] (\x/4,0,0.2) -- (\x/4,-\y/2,0.2) node[style={fill=red,shape=circle,inner sep=1pt}]{};
    \node[left,red] at (\x/4,-\y/4,0.2) {$a \in \cA$};

    \draw[thick,red,decoration = {markings, mark=at position .5 with {\arrowreversed[scale=1.5]{stealth}}}, postaction=decorate] (\x*0.75,0,0.2) -- (\x*0.75,-\y/2,0.2);
    \draw[thick,red] (\x*0.75,-\y/2,0.2) -- (\x,-\y/2,0.2);
    \node[left,red] at (\x*0.75,-\y/4,0.2) {$b \notin \cA$};
\end{tikzpicture}
\end{equation}

\subsubsection*{Boundary lines \& boundary interfaces}

We now expand on what becomes of the other lines $a\notin\cA$ when they are pushed to the boundary $\ket{\cA}$. On the boundary, $\cA$ is condensed and so acts trivially. As such, any two bulk anyons $a$ and $a'$ which differ by fusion with an anyon in $\cA$ should be identified. Indeed, this is part of the procedure of gauging the algebra $\cA$ discussed above. Therefore, the bulk anyons are grouped into multiplets on the boundary $\ket{\cA}$ which are closed under fusion with $\cA$. That is, on the boundary, each distinct topological line is in fact a collection of bulk lines. These multiplets are called $\cA$-modules over $G_k \boxtimes H_{-k'}$.

The topological lines on the boundary can themselves be fused together, giving the set of boundary lines the structure of a fusion category. The category of $\cA$ modules over $G_k \boxtimes H_{-k'}$ is denoted $\cS \coloneqq (G_k \boxtimes H_{-k'})_\cA$. This is the symmetry of the boundary itself.

We note that while each boundary line is closed under fusion with $\cA$, it may be decomposable in the sense that it can be written as a direct sum of modules containing fewer lines which are themselves closed under fusion with $\cA$. That is, there may be a subset of the bulk anyons in the module which are themselves closed under fusion with $\cA$. In order to describe the symmetry category $\cS$ of the boundary, we must identify the indecomposable modules, which are then the simple topological lines on the $\ket{\cA}$ boundary.

Thankfully, the simple boundary lines can be found systematically by \emph{induction}. That is, given a bulk anyon $a\in G_k\boxtimes H_{-k'}$, we consider the set of lines
\begin{equation}\label{eq:induced_module}
    \text{Ind}_{\cA}(a) \coloneqq a \otimes \cA
\end{equation}
This can be equipped with a right $\cA$ action simply by using the algebra multiplication $m:\cA \otimes \cA \to \cA$, giving it the structure of an $\cA$-module. The induced modules are generically decomposable in the sense described above. It is a fact that all simple modules can be found by decomposing induced modules from bulk anyons $a$ \cite{Fuchs:2002cm}.

Now suppose that we have a pair of Lagrangian algebras, $\cA_L$ and $\cA_R$. Each describes a topological boundary of the Chern--Simons theory, so we can consider a setup of the form
\begin{equation}
\begin{tikzpicture}[z={(30:2cm)},baseline=(current bounding box.center)]
    \newcommand{\x}{8}
    \newcommand{\y}{5}

    \shade[top color=white,bottom color=teal,middle color=teal,opacity=0.1] (0,0,0) -- ++(0,-\y/2,0) -- ++(\x,0,0) -- ++(0,0,1) -- ++(0,\y/2,0) -- ++(-\x,0,0) -- cycle;
    \draw[black] (0,-\y/2,0) -- ++(\x,0,0) -- ++(0,0,1) -- ++(-\x,0,0) -- cycle;
    \filldraw[black,fill=teal,fill opacity=0.1] (0,-\y/2,0) -- ++(\x,0,0) -- ++(0,0,1) -- ++(-\x,0,0) -- cycle;

    \draw[black] (0,0,0) -- ++(0,-\y/2,0);
    \draw[black] (\x,0,0) -- ++(0,-\y/2,0);
    \draw[black] (\x,0,1) -- ++(0,-\y/2,0);
    \draw[black,opacity=0.1] (0,0,1) -- ++(0,-\y/2,0);

    \draw[very thick,black] (\x/2,-\y/2,0) -- ++(0,0,1);

    \node at (\x/2,-\y*0.25,0.5) {$G_k \boxtimes H_{-k'}$};

    \node[below] at (\x*0.25,-\y/2,0) {$\ket{\cA_L}$};
    \node[below] at (\x*0.75,-\y/2,0) {$\ket{\cA_R}$};
    \node[below] at (\x*0.5,-\y/2,0) {$B$};
\end{tikzpicture}
\end{equation}
On the $\ket{\cA_L}$ boundary, $\cA_L$ is condensed and the boundary lines are given by $\cA_L$-modules, so there is a symmetry $\cS_L = (G_k\boxtimes H_{-k'})_{\cA_L}$. Similarly, on the $\ket{\cA_R}$ boundary, $\cA_R$ is condensed and the boundary lines are given by $\cA_R$-modules, giving a symmetry $\cS_R = (G_k\boxtimes H_{-k'})_{\cA_R}$. We now ask what structure describes a topological interface $B$ between these two topological boundaries.

On the left of $B$, $\cA_L$ is condensed so the lines in $\cA_L$ act trivially. Therefore, $B$ must be able to absorb lines in $\cA_L$ from the left. The interface $B$ should therefore be comprised of a set of bulk anyons which is closed under fusion with $\cA_L$ from the left. Similarly, it must be closed under fusion with $\cA_R$ from the right. That is, $B$ must admit a left action of $\cA_L$ and a right action of $\cA_R$, and these two actions should commute. Mathematically, this means that $B$ is simultaneously a (left) $\cA_L$-module and a (right) $\cA_R$ module or, equivalently, $B$ is a $(\cA_L,\cA_R)$-bimodule. The set of all possible bimodules is denoted $\bimod{\cA_L}{(G_k\boxtimes H_{-k'})}{\cA_R}$.

While a given interface $B$ is closed under the actions of $\cA_L$ and $\cA_R$, it may be decomposable; that is, a subset of the lines contained in $B$ may themselves be closed under the left action of $\cA_L$ and right action of $\cA_R$. We will be primarily interested in the simple interfaces, which are indecomposable. Then any other interface can be written as a sum of simple ones.

Much like modules, all simple bimodules can be found by decomposing induced bimodules \cite{Fuchs:2002cm}. An induced bimodule is found by taking a bulk anyon $a \in G_k \boxtimes H_{-k'}$ and fusing with both condensates:
\begin{equation}
    \text{Ind}_{\cA_L,\cA_R}(a) \coloneqq \cA_L \otimes a \otimes \cA_R
\end{equation}
This has a natural left action of $\cA_L$ given simply by the multiplication of the algebra $m_L$, and similarly a natural right action of $\cA_R$ from its multiplication $m_R$. These actions commute, giving the structure of a bimodule. Generically an induced bimodule is decomposable, and the full set of simple interfaces $B$ can be found systematically by decomposing them. Note that this process is constructive, and bulk anyons always exist, so at least one choice of $B$ is guaranteed to exist.

\subsection{Twisted scattering from the bulk}\label{sec:scattering_from_bulk}

We now move on to our main aim of constructing symmetric boundary conditions of fermionic CFTs. Our starting point is $N$ left-moving and $N$ right-moving Weyl fermions, charged under a chiral symmetry $G$. Our aim is to construct a boundary condition preserving $G$. We begin by unfolding the two-dimensional theory along its boundary, giving a chiral theory of $N$ Weyl fermions propagating on a space without boundary, but with a defect separating the left and right halves. Since the boundary was symmetric, the resulting defect is also $G$-symmetric. Constructing a symmetric boundary condition is equivalent to constructing a symmetric defect.

The action of $G$ on the left- and right-movers individually is anomalous, but the vanishing of the anomaly of the combined $G$ symmetry acting on both chiralities implies that the anomalies of the left and right $G$ symmetries are equal. This means that the levels of the chiral algebras generated by the $G$ currents on the left and right are equal. We denote this level by $k$.

To begin with, let us ignore the defect and consider a single realisation of the $G$ symmetry on the fermions. Since the theory we seek to describe is chiral (i.e.\ purely holomorphic) the methods of the previous section require slight variation. A chiral CFT is generically not modular invariant, and can only be realised as a relative field theory. That is, it naturally lives on the boundary of a three-dimensional TQFT which matches its gravitational anomaly via inflow (see, e.g.\ \cite{Freed:2012bs, Gannon:2026ttf}). In particular, the free fermion theory is rational with respect to the chiral algebra $\SO(2N)_1$ and can be realised as a gapless boundary of $\SO(2N)_1$ Chern--Simons theory:
\begin{equation}
    \begin{tikzpicture}[z={(30:2cm)},baseline=(current bounding box.center)]
    \newcommand{\x}{4}
    \newcommand{\y}{4}

    \shade[top color=green,bottom color=white,middle color=green,opacity=0.1] (0,0,0) -- ++(0,-\y/2,0) -- ++(\x,0,0) -- ++(0,0,1) -- ++(0,\y/2,0) -- ++(-\x,0,0) -- cycle;

    \filldraw[black,fill=green,fill opacity=0.1] (0,0,0) -- ++(\x,0,0) -- ++(0,0,1) -- ++(-\x,0,0) -- cycle;

    \draw[black] (0,0,0) -- ++(0,-\y/2,0);
    \draw[black] (\x,0,0) -- ++(0,-\y/2,0);
    \draw[black] (\x,0,1) -- ++(0,-\y/2,0);
    \draw[black,opacity=0.1] (0,0,1) -- ++(0,-\y/2,0);

    \node at (\x/3,-\y*0.35,0.5) {$SO(2N)_1$};

    \draw[->,thick] (\x*0.4,0,0.5) -- (\x*0.6,0,0.5);
\end{tikzpicture}
\end{equation}
The $\SO(2N)_1$ Chern--Simons theory has two simple lines denoted $\1$ and $V$. We note that the affine algebra $\widehat{\mathfrak{so}}(2N)_1$ has four integrable representations. Our convention is that $\SO(2N)_1$ contains only the lines $\1, V$ corresponding to the trivial and vector representations, whereas $\Spin(2N)_1$ additionally hosts the spinor and co-spinor lines $S$ and $C$. See \cref{app:MTCs} for more details. In the NS-NS sector, for example, the partition function of the two-dimensional free fermion theory can be written simply as $\chi_\1 + \chi_V$.\footnote{Neveu--Schwarz (NS) and Ramond (R) boundary conditions refer to anti-periodic and periodic boundary conditions for fermions, respectively.}

The symmetry $G_k$ we are interested in is a subalgebra of the chiral algebra $\SO(2N)_1$ so, from the discussion of \cref{sec:CS_basics}, there is a boundary between the two Chern--Simons theories:
\begin{equation}\label{eq:coset_CFT}
    \begin{tikzpicture}[z={(30:2cm)},baseline=(current bounding box.center)]
    \newcommand{\x}{4}
    \newcommand{\y}{4}

    \shade[top color=green,bottom color=white,middle color=green,opacity=0.1] (0,-\y/2,0) -- ++(0,-\y/2,0) -- ++(\x,0,0) -- ++(0,0,1) -- ++(0,\y/2,0) -- ++(-\x,0,0) -- cycle;
    \fill[blue,opacity=0.1] (0,0,0) -- ++(0,-\y/2,0) -- ++(\x,0,0) -- ++(0,0,1) -- ++(0,\y/2,0) -- ++(-\x,0,0) -- cycle;
    \draw[black] (0,-\y/2,0) -- ++(\x,0,0) -- ++(0,0,1) -- ++(-\x,0,0) -- cycle;
    \filldraw[black,fill=blue,fill opacity=0.1] (0,0,0) -- ++(\x,0,0) -- ++(0,0,1) -- ++(-\x,0,0) -- cycle;
    \node[left] at (0,0,0) {$\ket{G_k}$};
    \draw[black] (0,0,0) -- ++(0,-\y,0);
    \draw[black] (\x,0,0) -- ++(0,-\y,0);
    \draw[black] (\x,0,1) -- ++(0,-\y,0);
    \draw[black,opacity=0.1] (0,0,1) -- ++(0,-\y,0);

    \node at (\x/3,-\y*0.85,0.5) {$\SO(2N)_1$};
    \node at (\x/3,-\y*0.25,0.5) {$G_k$};

    \node[left] at (0,-\y/2,0) {$\ket{\SO(2N)_1/G_k}$};

    \draw[->,thick] (\x*0.4,0,0.5) -- (\x*0.6,0,0.5);
\end{tikzpicture}
\end{equation}
Here, the upper boundary of the $G_k$ Chern--Simons theory is the standard gapless boundary condition which realises the right-moving $G_k$ chiral algebra and the boundary between the two Chern--Simons theories is described by the coset degrees of freedom $\SO(2N)_1/G_k$.

This can be thought of as the right-moving half of the diagonal $G_k$ WZW model \eqref{eq:diagonalWZW}, with the necessary introduction of the fermionic SPT phase $\SO(2N)_1$ to cancel the gravitational anomaly.

The interface between the two Chern--Simons theories is equivalent to a boundary of $G_k \boxtimes \SO(2N)_{-1}$. Here we place the (generically gapless) boundary condition that realises the right-moving chiral algebra one obtains from doing the coset construction $\SO(2N)_1 / G_k$. However if $G_k$ has central charge $N$ then it is embedded conformally in $\SO(2N)_1$ and the coset is gapped ($c=0$). That is, the coset describes a TQFT and the boundary described by the coset becomes topological. The two-dimensional CFT resulting from interval compactification of the $G_k$ portion of the picture is then also rational with respect to the $G_k$ chiral algebra.  In this work, we restrict to this conformally embedded case; we will make some comments on the extension to the generic case in the conclusion.

As described in \cref{sec:CS_basics}, topological boundaries of Chern--Simons theory are described by a set of anyons $\cA$ that can end on the boundary. Equivalently, $\cA$ is condensed on the boundary. The algebra $\cA$ which we must condense can be derived from the branching rules associated with the conformal embedding $G_k \to \SO(2N)_1$. This is because each anyon in $\SO(2N)_1$ corresponds to an (integrable) representation of $\SO(2N)_1$. Taking this line through the topological boundary corresponds simply to viewing that anyon as a representation of $G_k$, which is given by the branching rules. Suppose an anyon $L$ in $\SO(2N)_1$ branches to an anyon $L'$ in $G_k$ (which may be a sum of simple lines), then the line $(L',\overline{L})$ in $G_k \boxtimes \SO(2N)_{-1}$ can end on the boundary of the folded theory and is in the condensate $\cA$. Since $\SO(2N)_1$ contains only two lines, $\1$ and $V$, branching the $\1 \oplus V$ line gives rise to the full set of $G_k$ anyons making up $\cA$. We note in passing that, as the theory resulting from our three-dimensional setup is fermionic, the algebra will also be fermionic (i.e.\ will contain lines with $\theta_a = -1$). The consequences of this are discussed in full in \cref{sec:SPTs_general}, but for the time being we will neglect these fermionic subtleties.

In order to describe chiral symmetries, we require two conformal embeddings of $G_k$ in $\SO(2N)_1$,
\begin{equation}\label{eq:two_embeddings}
    L, R : G_k \to SO(2N)_1
\end{equation}
The two embeddings allow for the left- and right-moving fermions to see different realisations $G^L$ and $G^R$ of the same symmetry group $G$. To each conformal embedding is associated a topological boundary described by condensation of a Lagrangian algebra $\cA_{L,R}$ in $G_k\boxtimes \SO(2N)_{-1}$. We will denote these boundaries $\ket{\cA_{L,R}}$. From the discussion in \cref{sec:CS_basics}, an interface between these topological boundaries is described by an $(\cA_L,\cA_R)$-bimodule $B$, and such an interface always exists. Choosing a particular interface, we assemble the setup:
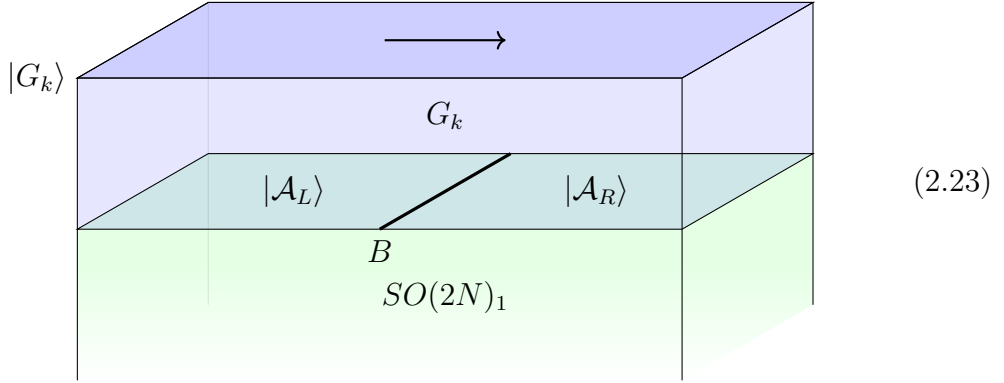
\begin{equation}\label{eq:3d_setup}
\begin{tikzpicture}[z={(30:2cm)},baseline=(current bounding box.center)]
    \newcommand{\x}{8}
    \newcommand{\y}{4}

    \shade[top color=green,bottom color=white,middle color=green,opacity=0.1] (0,-\y/2,0) -- ++(0,-\y/2,0) -- ++(\x,0,0) -- ++(0,0,1) -- ++(0,\y/2,0) -- ++(-\x,0,0) -- cycle;
    \fill[blue,opacity=0.1] (0,0,0) -- ++(0,-\y/2,0) -- ++(\x,0,0) -- ++(0,0,1) -- ++(0,\y/2,0) -- ++(-\x,0,0) -- cycle;
    \draw[black] (0,-\y/2,0) -- ++(\x,0,0) -- ++(0,0,1) -- ++(-\x,0,0) -- cycle;
    \draw[black] (0,0,0) -- ++(\x,0,0) -- ++(0,0,1) -- ++(-\x,0,0) -- cycle;
    \filldraw[black,fill=blue,fill opacity=0.1] (0,0,0) -- ++(\x,0,0) -- ++(0,0,1) -- ++(-\x,0,0) -- cycle;
    \node[left] at (0,0,0) {$\ket{G_k}$};
    \draw[black] (0,0,0) -- ++(0,-\y,0);
    \draw[black] (\x,0,0) -- ++(0,-\y,0);
    \draw[black] (\x,0,1) -- ++(0,-\y,0);
    \draw[black,opacity=0.1] (0,0,1) -- ++(0,-\y,0);
    \draw[very thick,black] (\x/2,-\y/2,0) -- ++(0,0,1);

    \node at (\x/2,-\y*0.85,0.5) {$\SO(2N)_1$};
    \node at (\x/2,-\y*0.25,0.5) {$G_k$};

    \node at (\x*0.25,-\y/2,0.5) {$\ket{\cA_L}$};
    \node at (\x*0.75,-\y/2,0.5) {$\ket{\cA_R}$};
    \node[below] at (\x*0.5,-\y/2,0) {$B$};

    \draw[->,thick] (\x*0.4,0,0.5) -- (\x*0.6,0,0.5);
\end{tikzpicture}
\end{equation}
Ultimately, the topological interface $B$ in this setup is the key object. Upon compactifying the $G_k$ Chern--Simons theory, $B$ describes a topological defect of $N$ Weyl fermions preserving the chiral symmetry acting as $G^L$ to the left of the defect and $G^R$ to the right of the defect. The topological defects in the various regions of this picture will be the central players in all that follows.

We remark that the Chern--Simons theories themselves host only finitely-many topological line operators. They do \emph{not}, therefore, describe the continuous $G$ (or $\SO(2N)$) global symmetry of the two-dimensional CFT which results from performing interval compactification on the $G_k$ Chern--Simons theory. Rather, the $G$ symmetry lines $U_g$ ($g\in G$) are supported on the upper boundary of the picture. There, they implement an anomalous $G$ symmetry with level $k$. Upon closing the sandwich, $U_g$ acts as $G^L$ if placed to the left of $B$ and as $G^R$ if placed to the right of $B$. We can also place $U_g$ transversely to $B$, in which case we get topological junctions that change $G^L$ defects into $G^R$ defects as they cross $B$. The lines in the $G_k$ Chern--Simons bulk describe only a finite $\Rep(G_k)$ symmetry, so the continuous $U_g$ lines \emph{cannot} be pulled off of the two-dimensional boundary into the three-dimensional bulk.

The solution of the scattering problem then follows from basic properties discussed in \cref{sec:CS_basics}. We begin with a local operator on the left of the interface. In the Chern--Simons realisation, this corresponds to the endpoint of a bulk anyon $a \in \cA_L$. The scattering proceeds by shifting this anyon to the right. Upon encountering the interface $B$, the lower endpoint of the anyon will get snagged if $a \notin \cA_R$ since it then cannot end on the $\ket{\cA_R}$ boundary, leaving the following simplified picture of the scattering:
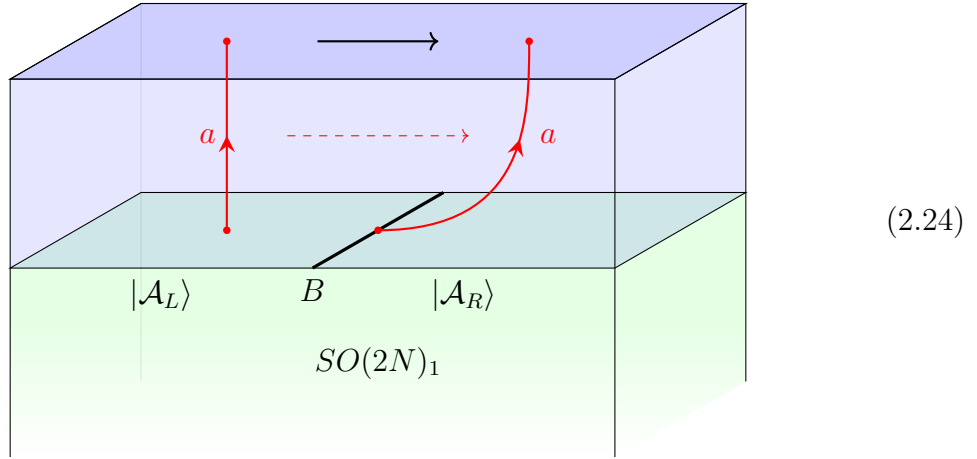
\begin{equation}\label{eq:3d_scattering}
\begin{tikzpicture}[z={(30:2cm)},baseline=(current bounding box.center)]
    \newcommand{\x}{8}
    \newcommand{\y}{5}

    \shade[top color=green,bottom color=white,middle color=green,opacity=0.1] (0,-\y/2,0) -- ++(0,-\y/2,0) -- ++(\x,0,0) -- ++(0,0,1) -- ++(0,\y/2,0) -- ++(-\x,0,0) -- cycle;
    \fill[blue,opacity=0.1] (0,0,0) -- ++(0,-\y/2,0) -- ++(\x,0,0) -- ++(0,0,1) -- ++(0,\y/2,0) -- ++(-\x,0,0) -- cycle;
    \draw[black] (0,-\y/2,0) -- ++(\x,0,0) -- ++(0,0,1) -- ++(-\x,0,0) -- cycle;
    \draw[black] (0,0,0) -- ++(\x,0,0) -- ++(0,0,1) -- ++(-\x,0,0) -- cycle;
    \filldraw[black,fill=blue,fill opacity=0.1] (0,0,0) -- ++(\x,0,0) -- ++(0,0,1) -- ++(-\x,0,0) -- cycle;
    \draw[black] (0,0,0) -- ++(0,-\y,0);
    \draw[black] (\x,0,0) -- ++(0,-\y,0);
    \draw[black] (\x,0,1) -- ++(0,-\y,0);
    \draw[black,opacity=0.1] (0,0,1) -- ++(0,-\y,0);
    \draw[very thick,black] (\x/2,-\y/2,0) -- ++(0,0,1);

    \node at (\x/2,-\y*0.85,0.5) {$\SO(2N)_1$};
    \node[below] at (\x*0.25,-\y/2,0) {$\ket{\cA_L}$};
    \node[below] at (\x*0.75,-\y/2,0) {$\ket{\cA_R}$};
    \node[below] at (\x*0.5,-\y/2,0) {$B$};

    \draw[->,thick] (\x*0.4,0,0.5) -- (\x*0.6,0,0.5);

    \begin{scope}[every node/.style={fill=red,shape=circle,inner sep=1pt}]
        \draw[thick,red,decoration = {markings, mark=at position .5 with {\arrowreversed[scale=1.5]{stealth}}}, postaction=decorate] (\x*0.25,0,0.5) node{} -- ++(0,-\y/2,0) node{};
        \draw[thick,red,decoration = {markings, mark=at position .35 with {\arrowreversed[scale=1.5]{stealth}}}, postaction=decorate] (\x*0.75,0,0.5) node{} .. controls+(0,-1) and +(2,0) .. (\x*0.5,-\y/2,0.5) node{};
        \draw[->,dashed,red] (\x*0.35,-\y*0.25,0.5) -- ++(\x*0.3,0,0);
    \end{scope}
    \node[left,red] at (\x*0.25,-\y/4,0.5) {$a$};
    \node[right,red] at (\x*0.75,-\y/4,0.5) {$a$};
\end{tikzpicture}
\end{equation}
To reveal the two-dimensional avatar of this process we need only collapse the $G_k$ Chern--Simons theory, giving
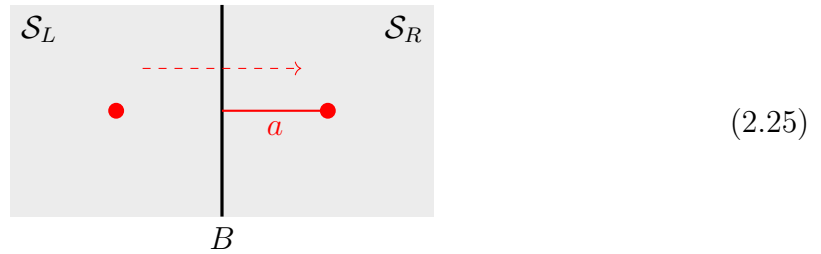
\begin{equation}\label{eq:2d_projec}
    \begin{tikzpicture}[scale=0.7,baseline=(current bounding box.center)]
    \fill[gray!15] (-8,-2) rectangle (0,2);

    \draw[very thick] (-4,-2) -- (-4,2);
    \node[below] at (-4,-2) {$B$};

    \fill[red] (-6,0) circle (0.15);

    \draw[thick,red] (-4,0) -- (-2,0);
    \fill[red] (-2,0) circle (0.15);
    \node[below,red] at (-3,0) {$a$};
    \draw[->,dashed,red] (-6+0.5,0.8) -- (-2-0.5,0.8);

    \node[below right] at (-8,2) {$\cS_L$};
    \node[below left] at (0,2) {$\cS_R$};
    \end{tikzpicture}
\end{equation}
This picture gives a simple explanation of the scattering in the conformally embedded case: the twist line attached to the outgoing excitation is simply the anyon $a\in\cA_L$, corresponding to the incoming state, pushed against the boundary $\ket{\cA_R}$.

In the case that the bulk anyon $a \in \cA_L \cap \cA_R$ then it can end on both boundaries. Hence, when scattered it may continue through $B$ unharmed. In this case there is no twist line attached to the outgoing state of the scattering. In fact, if the anyons have non-invertible fusion, this is not guaranteed and it is possible that the local operator living at the end of a bulk anyon $a \in \cA_L \cap \cA_R$ may nonetheless get attached to a twist line when scattered through a symmetry-preserving defect. The precise details of the scattering, including whether a twist line is attached, is fixed by the choice of bimodule $B$. A simple example is the Kramers--Wannier defect $\cN$ of the Ising CFT, which acts non-invertibly on the spin operator $\sigma$ despite the fact that a local operator ($\sigma$ itself) exists on the other side of the defect with the correct charges to preserve the $\bZ_2$ symmetry.

The figures above contain all the necessary ingredients to solve the scattering problem in full for any conformally embedded symmetry $G$, and we will see them at play in concrete examples later. Let us recapitulate. We have two topological boundaries $\ket{\cA_{L,R}}$ of $G_k \boxtimes SO(2N)_{-1}$ described by the condensation of the anyons in the algebras $\cA_{L,R}$. Let's consider $\ket{\cA_L}$ first. Since the algebra is condensed, any bulk anyons $a$ and $a'$ which differ by fusion with $\cA_L$ are identified on the boundary so, as described in \cref{sec:CS_basics}, the distinct lines on the boundary form $\cA_L$-modules. These boundary lines can be fused together, giving a fusion category on the $\ket{\cA_L}$ boundary which we denote $\cS_L \coloneqq (G_k \boxtimes SO(2N)_{-1})_{\cA_L}$. This category of boundary lines can be found by decomposing the induced modules of the bulk anyons. Similarly, the boundary lines on $\ket{\cA_R}$ are $\cA_R$-modules and form a fusion category $\cS_R \coloneqq (G_k \boxtimes SO(2N)_{-1})_{\cA_R}$.

Between the topological boundaries $\ket{\cA_L}$ and $\ket{\cA_R}$ sits the topological interface $B$ which, following \cref{sec:CS_basics}, is an $(\cA_L,\cA_R)$-bimodule. Such an interface $B$ is a collection of lines in $G_k\boxtimes \SO(2N)_{-1}$ which is closed under the left action of $\cA_L$ and the right action of $\cA_R$. The simple interfaces can also be found by decomposing all the induced bimodules.

\subsection{Symmetric boundaries from gauging}\label{sec:2d_descrip}

The construction above gives an economical description of the scattering in three dimensions. We now seek to realise the topological interface $B$, which describes the symmetric boundary state of the CFT, in purely two-dimensional terms. In particular, we will find that $B$ can be described simply by gauging a particular symmetry of the theory.

More concretely, we seek to construct the interface by beginning with the two-dimensional theory which comes from compactifying the $G_k$ Chern--Simons theory with the $\ket{\cA_L}$ boundary condition. This theory has symmetry $\cS_L$. We then isolate a particular finite symmetry $a_L$ in $\cS_L$ and, finally, we split the spacetime in two halves and gauge the symmetry to the right of the dividing line:
\begin{equation}\label{eq:halfspace}
    \begin{tikzpicture}[scale=0.8,baseline=(current bounding box.center)]
    \fill[gray!15] (-8,-2) rectangle (0,2);

    \draw[very thick] (-4,-2) -- (-4,2);
    \node[below] at (-4,-2) {$B$};

    \node[above] at (-6,2) {$\cS_L$};
    \node[above] at (-2,2) {$\bimod{a_L}{(\cS_L)}{a_L}=
    \cS_R$};

    \draw[thick,red] (-4,1) -- (-2,2);
    \draw[thick,red] (-4,0.2) -- (-1.4,0.5);
    \draw[thick,red] (-4,-0.5) -- (-1.4,0.5);
    \draw[thick,red] (-4,-1.3) -- (-2.5,-2);

    \draw[thick,red] (-3,1.5) -- (-2.7,0.35);
    \draw[thick,red] (-1.4,0.5) -- (0,0);
    \draw[thick,red] (-0.35,0.125) -- (-1.3,2);
    \draw[thick,red] (-2.5,1.75) -- (-1.65/2,2.125/2);
    \draw[thick,red] (-2.7,0) -- (-1.8,-2);
    \draw[thick,red] (-4.5/2,-1) -- (0,-0.4);
    \draw[thick,red] (-4.5/8,-1.1/2) -- (0,-1.6);
    \draw[thick,red] (-22.5/16,-0.8) -- (-1.5,-2);

    \node[right,red] at (-1.3,0) {$a_L$};

    \end{tikzpicture}
\end{equation}
The symmetry $a_L$ is chosen such that the gauged theory on the right is precisely the one found above by collapsing the $G_k$ Chern--Simons theory with boundary condition $\ket{\cA_R}$. Hence, this half-space gauging constructs an interface between the original theory $\cS_L$ and the gauged theory $\cS_R$. When performing the gauging, we must choose boundary conditions describing how the $a_L$ lines end on the interface, and different choices yield different defects. As we will shortly show, it is always possible to choose boundary conditions such that the interface is precisely the $B$ which we started with from our three-dimensional construction. That is, the symmetry-preserving boundary of the two-dimensional CFT can always be described as a gauging interface.

The same arguments also hold in the opposite direction. Namely, there is a symmetry $a_R$ in $\cS_R$ with the property that gauging $a_R$ in half of space gives rise to $\cS_L$, separated from the theory on the right by the interface $B$.

The first step is to identify the all-important symmetries $a_L$ and $a_R$. To describe them in detail, we must first recall some facts about gauging categorical symmetries in two dimensions \cite{Bhardwaj:2017xup}. Given a theory with a global symmetry $\cS_L$, we can gauge the symmetry generated by a set of lines $a_L$ in $\cS_L$ by inserting a (sufficiently dense) mesh of the lines in $a_L$. The trivalent vertices of the mesh are defined by a multiplication map $\mu_L:a_L\otimes a_L\to a_L$, describing how two lines in $a_L$ meet to form a third. We demand that the result of the gauging does not depend on the precise mesh chosen, which constrains the choices of $a_L$ which can be gauged. We will call such a set of lines $a_L$ a \emph{gaugeable algebra}.\footnote{Specifically, $a_L$ must be a symmetric special Frobenius algebra in $\cS_L$ \cite{Fuchs:2002cm,Bhardwaj:2017xup}.}

Given such an algebra $a_L$, we can gauge it in half of space as described above in \eqref{eq:halfspace}. Since $a_L$ is condensed on the right, if we bring a line in $a_L$ to the interface from the left it will be absorbed into the mesh. The way in which the line is absorbed is given by a boundary condition for the $a_L$ lines. Similarly to the discussion of \cref{sec:CS_basics}, this boundary condition is described by an $a_L$-module over $\cS_L$. That is, the interface is made up of a set of lines in $\cS_L$ which is closed under fusion with $a_L$. Let us denote a particular interface by $m$. To uniquely specify it, we must specify not only the algebra $a_L$, but also an $a_L$-module. Hence, the full set of interfaces which can lie between the ungauged and gauged theories make up the category $(\cS_L)_{a_L}$ of $a_L$-modules over $\cS_L$.

While lines in $a_L$ are absorbed into the interface $m$, if we instead fuse a line $M^L \in \cS_L$ which is \emph{not} in $a_L$ into the interface it will generically change it from $m$ to another interface $m'$. That is, the set of interfaces $(\cS_L)_{a_L}$ is acted on from the left by $\cS_L$. Mathematically, this endows $(\cS_L)_{a_L}$ with the structure of a module category over $\cS_L$.
\begin{equation}
    \begin{tikzpicture}[scale=0.7,baseline=(current bounding box.center)]
    \fill[gray!15] (-8,-2) rectangle (0,2);

    \draw[very thick] (-4,-2) -- (-4,2);
    \node[below] at (-4,-2) {$m$};
    \node[above] at (-4,2) {$m'$};

    \draw[thick, red] (-6,-2) -- (-4,0);
    \node[above left, red] at (-5,-1) {$M^L$};

    \node[below right] at (-8,2) {$\cS_L$};
    \node[below left] at (0,2) {$\cS_R$};
    \end{tikzpicture}
\end{equation}

More generally, given any topological interface of the theory, we must be able to fuse lines in $\cS_L$ into it, and this process may change this interface into another. The orbit of interfaces reached by fusion with $\cS_L$ make up a module category, which we will generically denote $\cM$. The discussion above shows that $(\cS_L)_{a_L}$ is an example of such a module category.

We have seen that given a gaugeable algebra $a_L$ we can find a topological boundary $m$ of $\cS_L$ by gauging $a_L$ in half space and choosing an $a_L$-module to describe the interface. Importantly, this process can be reversed. Given an object $n$ in an arbitrary $\cS_L$-module category $\cM$, we can always find a gaugeable algebra $a_L$ in $\cS_L$ such that $n$ is found by gauging $a_L$ in half of space with a particular choice of boundary condition \cite{Bhardwaj:2017xup}.

The construction proceeds as follows. From a particular boundary condition $n\in\cM$, fusion with a generic line in $\cS_L$ will transform $n$ to some other boundary $n'$. We ask for the set of lines in $\cS_L$ which leave $n$ invariant. That is, we find lines $M^L \in \cS_L$ such that
\begin{equation}
    \begin{tikzpicture}[scale=0.7,baseline=(current bounding box.center)]
    \fill[gray!15] (-8,-2) rectangle (0,2);

    \draw[very thick] (-4,-2) -- (-4,2);
    \node[below] at (-4,-2) {$n$};
    \node[above] at (-4,2) {$n$};

    \draw[thick, red] (-6,-2) -- (-4,0);
    \node[above left, red] at (-5,-1) {$M^L$};

    \node[below right] at (-8,2) {$\cS_L$};
    \node[below left] at (0,2) {$\cS_R$};
    \end{tikzpicture}
\end{equation}
Fusion of any two such lines gives another line which also leaves $n$ invariant, and the set of lines with this property can be given the structure of a gaugeable algebra. The resulting algebra is denoted by
\begin{equation}
    a_L = \underline{\End}_{\cS_L}(n)
\end{equation}
and is known as the \emph{internal end algebra} of $n$. We then perform the steps above and gauge this algebra in half of space. The interface between the ungauged and gauged theory is described by an $a_L$-module, and the set of such $a_L$-modules forms the module category $(\cS_L)_{a_L}$. The key result of \cite[§7.10]{etingof} is that $(\cS_L)_{a_L}$ is equivalent to $\cM$. Therefore, there exists a choice of boundary conditions (i.e.\ an $a_L$-module) for which the half-space gauging defect is precisely the module $n$ we began with. Furthermore, the relevant choice of $a_L$-module is the \emph{regular} $a_L$-module. As an object in $\cS_L$, the regular module is given by $a_L$ itself and its right-action is simply given by the $a_L$ algebra multiplication. Choosing a different $a_L$-module corresponds to a different object $n'$ in the module category $\cM$ we began with.

When any finite symmetry is gauged, the resulting theory has a dual `quantum' symmetry. For a finite abelian group $G$, the quantum symmetry is the Pontryagin dual $\hat{G} = \Hom(G,U(1))$. In the case of a categorical symmetry, gauging an algebra $a_L$ in $\cS_L$ gives a quantum symmetry category $\bimod{a_L}{(\cS_L)}{a_L}$ of $(a_L,a_L)$-bimodules in $\cS_L$ \cite{Bhardwaj:2017xup}. As indicated in \eqref{eq:halfspace}, the internal end construction ensures that $\bimod{a_L}{(\cS_L)}{a_L} = \cS_R$.

This construction of the defect via gauging $a_L$ is also useful to find the Affleck--Ludwig central charge of the boundary $n$. In terms of the categorical data, it is simply given by the quantum dimension, $g = d_{n}$, which can be established from the fusion rules. For the construction above, since $n$ corresponds to the regular $a_L$-module, it satisfies $n \otimes_{a_L} \bar{n} = a_L$. Here $\otimes_{a_L}$ denotes the relevant tensor product for $a_L$-modules: the relative tensor product over $a_L$ \cite{Bhardwaj:2017xup}. Intuitively, bringing parallel $n$ and $\bar{n}$ lines close to each other gives a slab of the gauged theory between the two lines. The gauged theory has a mesh of $a_L$ lines, so fusing $n$ and $\bar{n}$ squashes the mesh to a single vertical $a_L$ line, which is the result of the fusion.\footnote{Mathematically, the fusion rule can be seen by noting that, as an object in $\cS_L$, the regular module is $n=a_L$. We then use the property $a_L \otimes_{a_L} a_L = a_L$.} Therefore, the Affleck--Ludwig central charge of the symmetry-preserving boundary is 
\begin{equation}
    g=\sqrt{\dim{a_L}}
\end{equation}

All the same operations can be performed instead starting from the theory on the right with symmetry $\cS_R$. Namely, interfaces $n$ can be fused with lines in $\cS_R$, generically transforming them into other interfaces $n'$, and the set of interfaces specifies a topological boundary condition for the theory on the right (a module category over $\cS_R$). Choosing a particular interface $n$, we identify an algebra $a_R$ in $\cS_R$ by finding all those lines in $\cS_R$ which leave $n$ invariant,
\begin{equation}
    a_R = \underline{\End}_{\cS_R}(n)
\end{equation}
Gauging $a_R$ in half space with a boundary condition described by the regular $a_R$-module then similarly gives rise to the interface described by $n$.

One important remark is that this construction began with a particular choice of $n$. Choosing a different interface $n'$ gives different algebras $a_L'$ and $a_R'$ but the module category which results from them is equivalent, $(\cS_L)_{a_L} \cong (\cS_L)_{a_L'}$. Different algebras with this property are said to be Morita equivalent. This is a manifestation of the fact that, for categorical symmetries, multiple different gaugeable algebras can correspond to the same gauging operation. That is, gauging $a_L$ and $a_L'$ gives rise to the same resulting theory and the same set of interfaces.

This technology can be readily applied to our situation once we establish how to probe these structures from our three-dimensional setup. In that context, we have an interface $B$ which is an $(\cA_L,\cA_R)$-bimodule over $G_k \boxtimes \SO(2N)_{-1}$. In order to apply the internal end construction, we must understand how this object $B$ can be understood as the object we denoted $n$ above, which sits in an $\cS_L$-module category $\cM$. To do so, we must determine how a line in $\cS_L$ can act on the interface $B$. Given this we can simply apply the internal end construction to find an algebra $a_L$ in $\cS_L$ which realises $B$ as a gauging interface. Here we will be somewhat brief, leaving technical details to the later sections.

Recall that $B$ can absorb bulk lines in the algebra $\cA_L$ from the left and $\cA_R$ from the right as it is an $(\cA_L,\cA_R)$-bimodule. The set of such interfaces is therefore $\bimod{\cA_L}{(G_k \boxtimes \SO(2N)_{-1})}{\cA_R}$. We seek to define a left action of the boundary lines $\cS_L$ on the set of interfaces, making it into a $\cS_L$-module category which we denoted $\cM$ above. Recall that the boundary lines $M^L\in\cS_L$ are $\cA_L$-modules, so are equipped with a right action of $\cA_L$. Using the braiding of the Chern--Simons theory, we can use this right action to define a left action of $\cA_L$ as well. One simply braids the lines in $\cA_L$ past $M^L$ and then uses the right action \cite[§8.8]{etingof} (see also \cite{Fuchs:2002cm}). Thus $M^L$ can be promoted to an $(\cA_L,\cA_L)$-bimodule. There is a natural way to fuse an $(\cA_L,\cA_L)$-bimodule with the $(\cA_L,\cA_R)$-bimodule $B$, giving another $(\cA_L,\cA_R)$-bimodule $B'$, known as the relative tensor product over $\cA_L$. This defines the left action of $\cS_L$ on the set of interfaces $\bimod{\cA_L}{(G_k \boxtimes \SO(2N)_{-1})}{\cA_R}$, so this set forms a module category over $\cS_L$. This module category was denoted $\cM$ above. This gives the requisite structure to perform the internal end construction and compute the algebras $a_L$ and $a_R$ associated with $B$.

\subsection{Monodromies \& braids}\label{sec:monodromies}

We now describe a simple observable property of the twisted sector operators which emerge from the scattering as in \eqref{eq:2d_projec}. One simple way to probe the twist line is to insert a local operator, $\cO_b$, on the right of the defect, and move it around the twisted sector operator which emerged by scattering a local operator on the left through the defect:
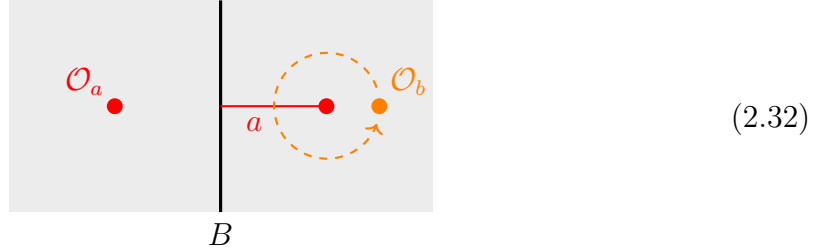
\begin{equation}\label{eq:loopdeloop}
    \begin{tikzpicture}[scale=0.7,baseline=(current bounding box.center)]
    \fill[gray!15] (-8,-2) rectangle (0,2);

    \draw[very thick] (-4,-2) -- (-4,2);
    \node[below] at (-4,-2) {$B$};

    \fill[red] (-6,0) circle (0.15);
    \node[red,above left] at (-6,0) {$\cO_a$};

    \draw[thick,red] (-4,0) -- (-2,0);
    \node[red, below left] at (-3,0) {$a$};

    \fill[red] (-2,0) circle (0.15);

    \draw[thick,dashed,->,orange] (-1.05,0.35) arc[start angle=20,end angle=340,radius=1];

    \fill[orange] (-1,0) circle (0.15);
    \node[orange,above right] at (-1,0) {$\cO_b$};

    \end{tikzpicture}
\end{equation}
The result of this manipulation is computable in two-dimensions, without appealing to the Chern--Simons description. The OPE of the two $G_k$ primary operators $\cO_a$ and $\cO_b$ is completely determined and takes the form \eqref{eq:OPE}. Circling the local operator on the right around the twisted sector operator coming out of the scattering corresponds to taking $z\to e^{2\pi i} z$, giving rise to monodromy phases in the OPE. From \eqref{eq:OPE}, we see that the monodromy operation yields
\begin{equation}\label{eq:monodromies}
    \cO_a (e^{2\pi i}z) \cO_b (0) \sim \sum_{c \in a \otimes b} C_{ab}^c e^{2\pi i (h_c-h_a-h_b)} z^{h_c - h_a - h_b} \cO_c (0) + \text{descendants}
\end{equation}
The conformal dimensions of any descendant differs from its corresponding primary by an integer, so does not change the phase produced in each summand.

Let us try to interpret \eqref{eq:monodromies}. Consider first the simplest case where $G$ is abelian, so the fusion of any two anyons results in a single anyon. In this case, the OPE has only a single term and \eqref{eq:OPE} and \eqref{eq:monodromies} imply
\begin{equation}\label{eq:abelian_overall_phase}
    \cO_a(e^{2\pi i}z) \cO_b(0) = C^{a\otimes b}_{ab}e^{2\pi i (h_{a\otimes b} - h_a - h_b)} z^{h_{a\otimes b} - h_a - h_b} \cO_{a\otimes b}(0) + \text{descendants}
\end{equation}
We see that the looping operation in \eqref{eq:loopdeloop}, which drags $\cO_b$ across the twist line, simply has the effect of producing a phase $\exp(2\pi i (h_{a\otimes b} - h_a - h_b))$ in the correlation function. In a rational CFT, all conformal dimensions are rational, so this phase is a root of unity. This is consistent with the twist line acting on $\cO_b$ as the generator of a finite invertible global symmetry. Indeed, we will confirm in \cref{sec:abelian} that for abelian $G$ the twist lines form a finite group which can be derived simply by studying the phases arising in the OPEs of local operators on the left and right.

The generic case where $G$ is non-abelian is harder to interpret. There are generically many fusion channels in the OPE \eqref{eq:OPE} and \eqref{eq:monodromies} shows that in each fusion channel we see a generically different phase. That is, moving $\cO_b$ past the twist line does not produce an overall phase in the correlation function as in \eqref{eq:abelian_overall_phase}. This is inconsistent with the twist line being the generator of a finite invertible symmetry. This simple observation demands explanation. From the two-dimensional perspective alone, the determination of the twist line seems out of reach. For example, from the form of the OPE monodromies \eqref{eq:monodromies} it is not clear whether the twist line could be a sum of invertible lines, or a simple non-invertible line.

The three-dimensional picture described previously captures all these features, demystifying the two-dimensional observations above. Consider the Chern--Simons setup corresponding to \eqref{eq:loopdeloop}. Since the initial operator $\cO_a$ scattered against the interface was local on the left, we must have $a\in\cA_L$. Similarly, since the probe operator $\cO_b$ is local on the right, $b\in\cA_R$. Dragging $\cO_b$ in a loop around $\cO_a$ corresponds to the corresponding anyons becoming \textit{braided},
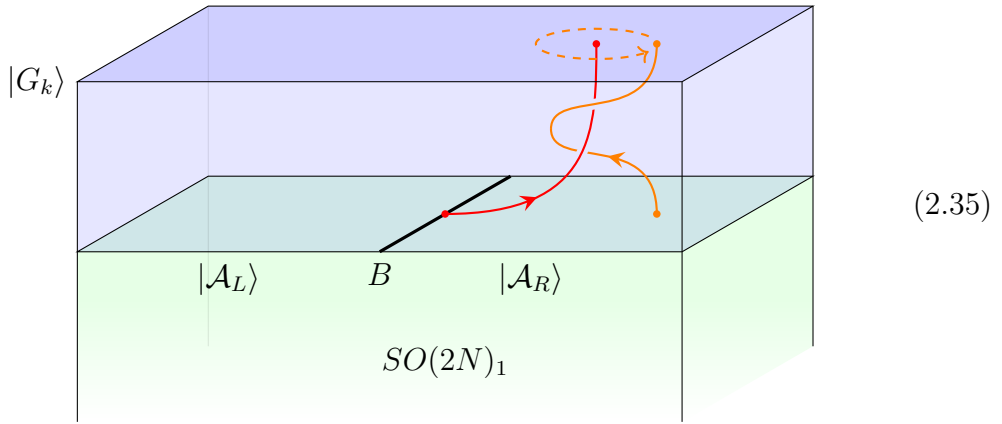
\begin{equation}\label{eq:3d_loopdeloop}
\begin{tikzpicture}[z={(30:2cm)},baseline=(current bounding box.center)]
    \newcommand{\x}{8}
    \newcommand{\y}{2.25}

    \shade[top color=green!10,bottom color=white,middle color=green!10] (0,-\y,0) -- ++(0,-\y,0) -- ++(\x,0,0) -- ++(0,0,1) -- ++(0,\y,0) -- ++(-\x,0,0) -- cycle;
    \fill[blue,opacity=0.1] (0,0,0) -- ++(0,-\y,0) -- ++(\x,0,0) -- ++(0,0,1) -- ++(0,\y,0) -- ++(-\x,0,0) -- cycle;
    \fill[fill=blue,opacity=0.1] (0,0,0) -- ++(\x,0,0) -- ++(0,0,1) -- ++(-\x,0,0) -- cycle;
    \draw[black] (0,-\y,0) -- ++(\x,0,0) -- ++(0,0,1) -- ++(-\x,0,0) -- cycle;
    \draw[black] (0,0,0) -- ++(0,-2*\y,0);
    \draw[black] (\x,0,0) -- ++(0,-2*\y,0);
    \draw[black] (\x,0,1) -- ++(0,-2*\y,0);
    \draw[black,opacity=0.1] (0,0,1) -- ++(0,-2*\y,0);
    \draw[very thick,black] (\x/2,-\y,0) -- ++(0,0,1);

    \node[left] at (0,0,0) {$\ket{G_k}$};
    \node at (\x/2,-\y*1.65-0.5,0.5) {$\SO(2N)_1$};
    \node[below] at (\x*0.25,-\y,0) {$\ket{\cA_L}$};
    \node[below] at (\x*0.75,-\y,0) {$\ket{\cA_R}$};
    \node[below] at (\x*0.5,-\y,0) {$B$};

    \begin{scope}[every node/.style={shape=circle,inner sep=1pt}]
        \draw[thick,orange,decoration = {markings, mark=at position .45 with {\arrowreversed[scale=1.5,rotate=0]{stealth}}}, postaction=decorate] (\x*0.675,-\y/2,0.5) .. controls +(0,-\y*0.25) and +(0,\y*0.5) .. (\x*0.85,-\y,0.5) node[fill=orange] {};

        \begin{scope}
            \clip (\x/2,-\y+0.1,1) rectangle (\x-0.1,-0.1,0);
            \draw[blue!10,line width=4pt] (\x*0.75,0,0.5) .. controls+(0,-\y/2) and +(\x*0.25,0) .. (\x*0.5,-\y,0.5);
        \end{scope}

        \draw[thick,red,decoration = {markings, mark=at position .65 with {\arrowreversed[scale=1.5,rotate=-4]{stealth}}}, postaction=decorate] (\x*0.75,0,0.5) node[fill=red] {} .. controls+(0,-\y/2) and +(\x*0.25,0) .. (\x*0.5,-\y,0.5) node[fill=red] {};

        \begin{scope}
            \clip (\x/2,-\y+0.1,1) rectangle (\x-0.1,-0.1,0);
            \draw[blue!10,line width=4pt] (\x*0.85,0,0.5) .. controls +(0,-\y*0.5) and +(0,\y*0.25) .. (\x*0.675,-\y/2,0.5);
        \end{scope}

        \draw[thick,orange] (\x*0.85,0,0.5) node[fill=orange] {} .. controls +(0,-\y*0.5) and +(0,\y*0.25) .. (\x*0.675,-\y/2,0.5);
    \end{scope}

    \draw[thick,dashed,->,orange] ({\x*0.75+\x*0.1*cos(30)},{0.2*sin(30)},0.5) arc[start angle=30,end angle=330,x radius=\x*0.1,y radius=0.2];

    \draw[black] (0,0,0) -- ++(\x,0,0) -- ++(0,0,1) -- ++(-\x,0,0) -- cycle;
\end{tikzpicture}
\end{equation}

Our aim is to demonstrate that the properties of the Chern--Simons anyons also produce a phase $\exp(2\pi i(h_c-h_a-h_b))$ in each fusion channel, giving an interpretation of the various phases appearing in \eqref{eq:monodromies}. We begin by considering the insertion of the $b$ anyon before looping it around $a$,
\begin{equation}\label{eq:parallel}
\begin{tikzpicture}[z={(30:2cm)},baseline=(current bounding box.center)]
    \newcommand{\x}{8}
    \newcommand{\yt}{2.25} 
    \newcommand{\yb}{2.25} 

    \shade[top color=green!10,bottom color=white,middle color=green!10] (0,-\yt,0) -- ++(0,-\yb,0) -- ++(\x,0,0) -- ++(0,0,1) -- ++(0,\yb,0) -- ++(-\x,0,0) -- cycle;
    \fill[blue,opacity=0.1] (0,0,0) -- ++(0,-\yt,0) -- ++(\x,0,0) -- ++(0,0,1) -- ++(0,\yt,0) -- ++(-\x,0,0) -- cycle;
    \draw[black] (0,-\yt,0) -- ++(\x,0,0) -- ++(0,0,1) -- ++(-\x,0,0) -- cycle;
    \draw[black] (0,0,0) -- ++(\x,0,0) -- ++(0,0,1) -- ++(-\x,0,0) -- cycle;
    \filldraw[black,fill=blue,fill opacity=0.1] (0,0,0) -- ++(\x,0,0) -- ++(0,0,1) -- ++(-\x,0,0) -- cycle;
    \node[left] at (0,0,0) {$\ket{G_k}$};
    \draw[black] (0,0,0) -- ++(0,-\yt-\yb,0);
    \draw[black] (\x,0,0) -- ++(0,-\yt-\yb,0);
    \draw[black] (\x,0,1) -- ++(0,-\yt-\yb,0);
    \draw[black,opacity=0.1] (0,0,1) -- ++(0,-\yt-\yb,0);
    \draw[very thick,black] (\x/2,-\yt,0) -- ++(0,0,1);

    \node at (\x/2,-\yt-\yb*0.65-0.5,0.5) {$\SO(2N)_1$};
    \node[below] at (\x*0.25,-\yt,0) {$\ket{\cA_L}$};
    \node[below] at (\x*0.75,-\yt,0) {$\ket{\cA_R}$};
    \node[below] at (\x*0.5,-\yt,0) {$B$};

    \newcommand{\redpath}{(\x*0.75,0,0.5) .. controls+(0,-1) and +(2,0) .. (\x*0.5,-\yt,0.5)}

    \draw[thick,red,decoration = {markings, mark=at position .35 with {\arrowreversed[scale=1.5]{stealth}}}, postaction=decorate] \redpath;

    \draw[thick,orange,decoration = {markings, mark=at position .5 with {\arrowreversed[scale=1.5]{stealth}}}, postaction=decorate] (\x*0.84,0,0.5) -- (\x*0.84,-\yt,0.5);

    \node[left,red] at (\x*0.73,-\yt/2,0.5) {$a$};
    \node[right,orange] at (\x*0.84,-\yt/2,0.5) {$b$};

    \begin{scope}[every node/.style={shape=circle,inner sep=1pt}]
        \node[fill=red] at (\x*0.75,0,0.5) {};
        \node[fill=red] at (\x*0.5,-\yt,0.5) {};
        \node[fill=orange] at (\x*0.84,0,0.5) {};
        \node[fill=orange] at (\x*0.84,-\yt,0.5) {};
    \end{scope}
\end{tikzpicture}
\end{equation}
The anyons satisfy a completeness relation\footnote{We follow conventions of \cite[Appendix E]{Kitaev:2005hzj}.}
\begin{equation}\label{eq:complete}
    \begin{tikzpicture}[baseline=0.35cm-0.5ex]
        \draw [thick, decoration = {markings, mark=at position .56 with {\arrow[scale=1.5]{stealth}}}, postaction=decorate] (0,-.5) to (0,1.2) node[above] {\small $a$};
        \draw [thick, decoration = {markings, mark=at position .56 with {\arrow[scale=1.5]{stealth}}}, postaction=decorate] (.8,-.5) to (.8,1.2) node[above] {\small $b$};
    \end{tikzpicture}
    = \sum_{c,\mu} \sqrt{\frac{d_c}{d_ad_b}}
    \quad
    \begin{tikzpicture}[baseline=0.35cm-1ex]
        \draw [thick, decoration = {markings, mark=at position .7 with {\arrow[scale=1.5]{stealth}}}, postaction=decorate] (-.5, -.5) node[below] {\small $a$} to (0,0) node[right] {\footnotesize $\mu$};
        \draw [thick, decoration = {markings, mark=at position .7 with {\arrow[scale=1.5]{stealth}}}, postaction=decorate] (.5, -.5) node[below] {\small $b$} to (0,0);
        \draw [thick, decoration = {markings, mark=at position .7 with {\arrow[scale=1.5]{stealth}}}, postaction=decorate] (0,0) to node[midway, left] {\small $c$} (0,.7) node[right] {\footnotesize $\mu$};
        \draw [thick, decoration = {markings, mark=at position .7 with {\arrow[scale=1.5]{stealth}}}, postaction=decorate] (0, .7) to (-.5,1.2) node[above] {\small $a$};
        \draw [thick, decoration = {markings, mark=at position .7 with {\arrow[scale=1.5]{stealth}}}, postaction=decorate] (0, .7) to (.5,1.2) node[above] {\small $b$};
    \end{tikzpicture}
\end{equation}
so the two anyons in \eqref{eq:parallel} can be replaced with a sum over intermediate fusion channels $c \subset a \otimes b$.

We now turn to the setup \eqref{eq:3d_loopdeloop}, after dragging $\cO_b$ around $\cO_a$. We can manipulate the anyons here using the $R$-matrix. In particular, it is useful to note the relation
\begin{equation}\label{eq:swap}
    \begin{tikzpicture}[baseline=-0.35cm+1ex]
        \draw [thick, decoration = {markings, mark=at position .9 with {\arrow[scale=1.5]{stealth}}}, postaction=decorate] (-.6,-.6) node[left] {\small $b$} to (.6,.6) node[right] {\small $b$};
        \draw [thick] (.6,-.6) node[right] {\small $a$} to (.1,-.1);
        \draw [thick, decoration = {markings, mark=at position .75 with {\arrow[scale=1.5]{stealth}}}, postaction=decorate] (-.1,.1) to (-.6,.6) node[left] {\small $a$};
    \end{tikzpicture}
    = \sum_{c,\mu,\nu} \sqrt{\frac{d_c}{d_a d_b}} (R^{ab}_c)_{\mu\nu}
    \begin{tikzpicture}[baseline=0.35cm-1ex]
        \draw [thick, decoration = {markings, mark=at position .7 with {\arrow[scale=1.5]{stealth}}}, postaction=decorate] (-.5, -.5) node[left] {\small $b$} to (0,0) node[right] {\footnotesize $\mu$};
        \draw [thick, decoration = {markings, mark=at position .7 with {\arrow[scale=1.5]{stealth}}}, postaction=decorate] (.5, -.5) node[right] {\small $a$} to (0,0);
        \draw [thick, decoration = {markings, mark=at position .7 with {\arrow[scale=1.5]{stealth}}}, postaction=decorate] (0,0) to node[midway, left] {\small $c$} (0,.7) node[right] {\footnotesize $\nu$};
        \draw [thick, decoration = {markings, mark=at position .7 with {\arrow[scale=1.5]{stealth}}}, postaction=decorate] (0, .7) to (-.5,1.2) node[left] {\small $a$};
        \draw [thick, decoration = {markings, mark=at position .7 with {\arrow[scale=1.5]{stealth}}}, postaction=decorate] (0, .7) to (.5,1.2) node[right] {\small $b$};
    \end{tikzpicture}
\end{equation}
which follows from the definition \eqref{eq:R_def} using the completeness relation \eqref{eq:complete} and the orthogonality relation for the anyons,
\begin{equation}\label{eq:orthog}
    \begin{tikzpicture}[baseline=0.5cm-0.5ex]
        \draw [thick, decoration = {markings, mark=at position .8 with {\arrow[scale=1.5]{stealth}}}, postaction=decorate] (0,-.5) node[left] {\small $d$} to (0,0) node[right] {\footnotesize $\nu$};
        \draw [thick, decoration = {markings, mark=at position .56 with {\arrow[scale=1.5, rotate=10]{stealth}}}, postaction=decorate] (0,0) arc (250:110:.5) node[midway, left] {\small $a$};
        \draw [thick, decoration = {markings, mark=at position .56 with {\arrow[scale=1.5, rotate=-10]{stealth}}}, postaction=decorate] (0,0) arc (-70:70:.5) node[midway, right] {\small $b$};
        \draw [thick, decoration = {markings, mark=at position .8 with {\arrow[scale=1.5]{stealth}}}, postaction=decorate] (0,.93) node[right] {\footnotesize $\mu$} to +(0,.5) node[left] {\small $c$};
    \end{tikzpicture}
    = \delta_{cd} \delta_{\mu\nu} \sqrt{\frac{d_a d_b}{d_c}}
    \quad
    \begin{tikzpicture}[baseline=0.47cm-0.5ex]
        \draw [thick, decoration = {markings, mark=at position .56 with {\arrow[scale=1.5]{stealth}}}, postaction=decorate] (0,-.5) to node[midway, right] {\small $c$} (0,1.43);
    \end{tikzpicture}
\end{equation}
We can then manipulate the braided anyons in \eqref{eq:3d_loopdeloop} as follows. First we use \eqref{eq:swap} on the upper braid:
\begin{equation}
    \raisebox{-3em}{\begin{tikzpicture}

    \draw [thick, decoration = {markings, mark=at position .9 with {\arrow[scale=1.5]{stealth}}}, postaction=decorate] (.5,-.5) node[right] {\small $a$} .. controls +(0,0.3) and +(0,-0.3) .. (-.5,.5) node[left] {\small $a$};

    \draw [thick, decoration = {markings, mark=at position .9 with {\arrow[scale=1.5]{stealth}}}, postaction=decorate,preaction={draw=white, line width=4pt}] (-.5,-.5) node[left] {\small $b$} .. controls +(0,0.3) and +(0,-0.3) .. (.5,.5) node[right] {\small $b$};

    \draw [thick, decoration = {markings, mark=at position .9 with {\arrow[scale=1.5]{stealth}}}, postaction=decorate] (.5,-1.5) node[right] {\small $b$} .. controls +(0,0.3) and +(0,-0.3) .. (-.5,-.5);

    \draw [thick, decoration = {markings, mark=at position .9 with {\arrow[scale=1.5]{stealth}}}, postaction=decorate,preaction={draw=white, line width=4pt}] (-.5,-1.5) node[left] {\small $a$} .. controls +(0,0.3) and +(0,-0.3) .. (.5,-.5);

    \end{tikzpicture}}
    =
    \sum_{c,\mu,\nu} \sqrt{\frac{d_c}{d_a d_b}} (R^{ab}_c)_{\mu\nu}
    \raisebox{-3.5em}{\begin{tikzpicture}

    \draw [thick] (-.5, -.5) node[left] {\small $b$} .. controls +(0,.15) .. (0,0) node[right] {\footnotesize $\mu$};

    \draw [thick] (.5, -.5) node[right] {\small $a$} .. controls +(0,0.15) .. (0,0);

    \draw [thick, decoration = {markings, mark=at position .7 with {\arrow[scale=1.5]{stealth}}}, postaction=decorate] (0,0) to node[midway, left] {\small $c$} (0,.7) node[right] {\footnotesize $\nu$};

    \draw [thick, decoration = {markings, mark=at position .7 with {\arrow[scale=1.5]{stealth}}}, postaction=decorate] (0, .7) to (-.5,1.2) node[left] {\small $a$};

    \draw [thick, decoration = {markings, mark=at position .7 with {\arrow[scale=1.5]{stealth}}}, postaction=decorate] (0, .7) to (.5,1.2) node[right] {\small $b$};

    \draw [thick, decoration = {markings, mark=at position .9 with {\arrow[scale=1.5]{stealth}}}, postaction=decorate] (.5,-1.5) node[right] {\small $b$} .. controls +(0,0.3) and +(0,-0.3) .. (-.5,-.5);

    \draw [thick, decoration = {markings, mark=at position .9 with {\arrow[scale=1.5]{stealth}}}, postaction=decorate,preaction={draw=white, line width=4pt}] (-.5,-1.5) node[left] {\small $a$} .. controls +(0,0.3) and +(0,-0.3) .. (.5,-.5);
    
    \end{tikzpicture}}
\end{equation}
Using \eqref{eq:swap} again on the lower braid, we have
\begin{equation}
    \sum_{c,\mu,\nu} \sqrt{\frac{d_c}{d_a d_b}} (R^{ab}_c)_{\mu\nu} \raisebox{-3.5em}{\begin{tikzpicture}

    \draw [thick] (-.5, -.5) node[left] {\small $b$} .. controls +(0,.15) .. (0,0) node[right] {\footnotesize $\mu$};

    \draw [thick] (.5, -.5) node[right] {\small $a$} .. controls +(0,0.15) .. (0,0);

    \draw [thick, decoration = {markings, mark=at position .7 with {\arrow[scale=1.5]{stealth}}}, postaction=decorate] (0,0) to node[midway, left] {\small $c$} (0,.7) node[right] {\footnotesize $\nu$};

    \draw [thick, decoration = {markings, mark=at position .7 with {\arrow[scale=1.5]{stealth}}}, postaction=decorate] (0, .7) to (-.5,1.2) node[left] {\small $a$};

    \draw [thick, decoration = {markings, mark=at position .7 with {\arrow[scale=1.5]{stealth}}}, postaction=decorate] (0, .7) to (.5,1.2) node[right] {\small $b$};

    \draw [thick, decoration = {markings, mark=at position .9 with {\arrow[scale=1.5]{stealth}}}, postaction=decorate] (.5,-1.5) node[right] {\small $b$} .. controls +(0,0.3) and +(0,-0.3) .. (-.5,-.5);

    \draw [thick, decoration = {markings, mark=at position .9 with {\arrow[scale=1.5]{stealth}}}, postaction=decorate,preaction={draw=white, line width=4pt}] (-.5,-1.5) node[left] {\small $a$} .. controls +(0,0.3) and +(0,-0.3) .. (.5,-.5);
    \end{tikzpicture}}
    =
    \sum_{\substack{c,\mu,\nu\\d,\rho,\sigma}} \sqrt{\frac{d_c}{d_a d_b}} \sqrt{\frac{d_d}{d_ad_b}} (R^{ab}_c)_{\mu\nu} (R^{ba}_d)_{\rho\sigma}
    \raisebox{-4em}{\begin{tikzpicture}

    \draw [thick, decoration = {markings, mark=at position .6 with {\arrow[scale=1.5]{stealth}}}, postaction=decorate] (-.5, -.5) node[left] {\small $b$} .. controls +(0,.15) .. (0,0) node[right] {\footnotesize $\mu$};

    \draw [thick, decoration = {markings, mark=at position .6 with {\arrow[scale=1.5]{stealth}}}, postaction=decorate] (.5, -.5) node[right] {\small $a$} .. controls +(0,0.15) .. (0,0);

    \draw [thick, decoration = {markings, mark=at position .7 with {\arrow[scale=1.5]{stealth}}}, postaction=decorate] (0,0) to node[midway, left] {\small $c$} (0,.7) node[right] {\footnotesize $\nu$};

    \draw [thick, decoration = {markings, mark=at position .7 with {\arrow[scale=1.5]{stealth}}}, postaction=decorate] (0, .7) to (-.5,1.2) node[left] {\small $a$};

    \draw [thick, decoration = {markings, mark=at position .7 with {\arrow[scale=1.5]{stealth}}}, postaction=decorate] (0, .7) to (.5,1.2) node[right] {\small $b$};

    \draw [thick] (-.5, -.5) node[left] {\small $b$} .. controls +(0,-.15) .. (0,-1) node[right] {\footnotesize $\sigma$};

    \draw [thick] (.5, -.5) node[right] {\small $a$} .. controls +(0,-0.15) .. (0,-1);

    \draw [thick, decoration = {markings, mark=at position .7 with {\arrow[scale=1.5]{stealth}}}, postaction=decorate] (0,-1.7) to node[midway, left] {\small $d$} (0,-1);

    \draw [thick, decoration = {markings, mark=at position .7 with {\arrow[scale=1.5]{stealth}}}, postaction=decorate] (-.5, -2.2) node[left] {\small $a$} to (0,-1.7) node[right] {\footnotesize $\rho$};

    \draw [thick, decoration = {markings, mark=at position .7 with {\arrow[scale=1.5]{stealth}}}, postaction=decorate] (.5, -2.2) node[right] {\small $b$} to (0,-1.7);

    \end{tikzpicture}}
\end{equation}
Then using \eqref{eq:orthog} and the balance relation \eqref{eq:ribbon}, we find
\begin{equation}\label{eq:MTC_monodromy phase}
    \raisebox{-3em}{\begin{tikzpicture}

    \draw [thick, decoration = {markings, mark=at position .9 with {\arrow[scale=1.5]{stealth}}}, postaction=decorate] (.5,-.5) node[right] {\small $a$} .. controls +(0,0.3) and +(0,-0.3) .. (-.5,.5) node[left] {\small $a$};

    \draw [thick, decoration = {markings, mark=at position .9 with {\arrow[scale=1.5]{stealth}}}, postaction=decorate,preaction={draw=white, line width=4pt}] (-.5,-.5) node[left] {\small $b$} .. controls +(0,0.3) and +(0,-0.3) .. (.5,.5) node[right] {\small $b$};

    \draw [thick, decoration = {markings, mark=at position .9 with {\arrow[scale=1.5]{stealth}}}, postaction=decorate] (.5,-1.5) node[right] {\small $b$} .. controls +(0,0.3) and +(0,-0.3) .. (-.5,-.5);

    \draw [thick, decoration = {markings, mark=at position .9 with {\arrow[scale=1.5]{stealth}}}, postaction=decorate,preaction={draw=white, line width=4pt}] (-.5,-1.5) node[left] {\small $a$} .. controls +(0,0.3) and +(0,-0.3) .. (.5,-.5);

    \end{tikzpicture}} = \sum_{c,\mu} \sqrt{\frac{d_c}{d_ad_b}} \frac{\theta_c}{\theta_a\theta_b} \,\raisebox{-3.2em}{\begin{tikzpicture}
    \draw [thick, decoration = {markings, mark=at position .7 with {\arrow[scale=1.5]{stealth}}}, postaction=decorate] (-.5, -.5) node[below] {\small $a$} to (0,0) node[right] {\footnotesize $\mu$};
    \draw [thick, decoration = {markings, mark=at position .7 with {\arrow[scale=1.5]{stealth}}}, postaction=decorate] (.5, -.5) node[below] {\small $b$} to (0,0);
    \draw [thick, decoration = {markings, mark=at position .7 with {\arrow[scale=1.5]{stealth}}}, postaction=decorate] (0,0) to node[midway, left] {\small $c$} (0,.7) node[right] {\footnotesize $\mu$};
    \draw [thick, decoration = {markings, mark=at position .7 with {\arrow[scale=1.5]{stealth}}}, postaction=decorate] (0, .7) to (-.5,1.2) node[above] {\small $a$};
    \draw [thick, decoration = {markings, mark=at position .7 with {\arrow[scale=1.5]{stealth}}}, postaction=decorate] (0, .7) to (.5,1.2) node[above] {\small $b$};
    \end{tikzpicture}}
    =\sum_{c,\mu} e^{2\pi i (h_c - h_a - h_b)} \sqrt{\frac{d_c}{d_ad_b}}  \,\raisebox{-3.2em}{\begin{tikzpicture}
    \draw [thick, decoration = {markings, mark=at position .7 with {\arrow[scale=1.5]{stealth}}}, postaction=decorate] (-.5, -.5) node[below] {\small $a$} to (0,0) node[right] {\footnotesize $\mu$};
    \draw [thick, decoration = {markings, mark=at position .7 with {\arrow[scale=1.5]{stealth}}}, postaction=decorate] (.5, -.5) node[below] {\small $b$} to (0,0);
    \draw [thick, decoration = {markings, mark=at position .7 with {\arrow[scale=1.5]{stealth}}}, postaction=decorate] (0,0) to node[midway, left] {\small $c$} (0,.7) node[right] {\footnotesize $\mu$};
    \draw [thick, decoration = {markings, mark=at position .7 with {\arrow[scale=1.5]{stealth}}}, postaction=decorate] (0, .7) to (-.5,1.2) node[above] {\small $a$};
    \draw [thick, decoration = {markings, mark=at position .7 with {\arrow[scale=1.5]{stealth}}}, postaction=decorate] (0, .7) to (.5,1.2) node[above] {\small $b$};
    \end{tikzpicture}}
\end{equation}
where in the final step we have used \eqref{eq:theta=exp(h)}. Comparing with \eqref{eq:complete} before the monodromy operation was performed, indeed we see that in each fusion channel $c\subset a\otimes b$ we acquire the same phase $e^{2\pi i (h_c - h_a - h_b)}$ as computed by the two-dimensional OPE. Our three-dimensional construction therefore gives a clear resolution of how the various monodromy phases arise in the OPE: they are the direct consequence of the braiding of the corresponding anyons.

\subsection{Fermionic symmetries \& SPT classes}\label{sec:SPTs_general}

Crucially, the two-dimensional theory we are studying is fermionic and, as always when fermions are involved, there are some additional subtleties which need to be taken account of. Fermionic theories, by definition, enjoy a privileged $\bZ_2$ symmetry describing fermion parity, generated by a topological line $(-1)^F$, and the theory depends on a choice of spin structure. We must find a way to keep track of these features in the three-dimensional construction described previously. Before doing so, we take a slight digression and give some general remarks about symmetries of fermionic field theories in two dimensions.

\subsubsection*{Fermionic symmetry categories}

Much as the symmetries of bosonic QFTs in two dimensions are described by fusion categories, those of fermionic QFTs are described by fusion \emph{super}categories. We point the reader to \cite{Bhardwaj:2024ydc} and references therein for more details; here we review only the essential ingredients.

The salient difference which the prefix `super' introduces is a $\bZ_2$ grading of the junctions between topological lines. Given two lines $a$, $b$, the junctions between them form a vector space $\Hom(a,b)$. The vector space $\Hom(a,a)$ therefore encodes the set of topological local operators living on a line $a$. A simple (or indecomposable) line in an ordinary fusion category has $\Hom(a,a) \cong \bC$, and any line with $\Hom(a,a)\cong\bC^n$ with $n>1$ can be decomposed into a sum of simple lines. In a superfusion category, on the other hand, this vector space is replaced by a super vector space, and a simple line then has either
\begin{equation}
    \Hom(a,a) \cong \bC^{1|0} \quad \text{($m$-type)} \qq{or} \Hom(a,a) \cong \bC^{1|1} \quad \text{($q$-type)}
\end{equation}
Simple lines with $\Hom(a,a) \cong \bC^{1|0}$ are called $m$-type, and they do not host any fermionic topological local operators. Those with $\Hom(a,a)\cong\bC^{1|1}$, known as $q$-type lines, do admit a local fermionic operator. The trivial line, for example, is $m$-type: $\Hom(\1,\1) \cong \bC^{1|0}$. We stress that whether a line is $m$-type or $q$-type is not fixed by the species of line itself, but by the junctions it admits.

The standard features of a fusion category are then carried over to this fermionic setting. Namely, lines can be moved together and fused, satisfying fusion rules of the form \eqref{eq:fusion}. The associativity of the fusion is again described by a set of $F$-symbols as in \eqref{eq:F_def}. The consistency condition on the $F$-symbols is modified slightly by the fact that the junctions may be bosonic or fermionic. This has the effect of introducing some additional signs into the pentagon identity \cite{Aasen:2017ubm, Gu:2010na}.

Not all fusion supercategories have the necessary structures to describe fermionic QFTs. We now discuss the particular subset of them which do. Firstly, in order to describe the symmetries of a fermionic QFT, the category must include a privileged line which we denote $\pi$. The $\pi$ line's purpose in life is to encode the spin structure. The $\pi$ line must generate a $\bZ_2$ symmetry: $\pi\otimes\pi=\1$. Furthermore, there must be a junction between $\pi$ and the trivial line $\1$, so $\Hom(\1,\pi)$ must be non-trivial. In other words, the $\pi$ line must have a topological endpoint. It is then straightforward to show that a simple line $a$ is $q$-type if and only if there is a bosonic junction between $a$ and $a\otimes \pi$, and that the quantum dimension of such a line is $d_a \geq \sqrt{2}$ \cite{Bhardwaj:2024ydc}. For an $m$-type line on the other hand, only a fermionic junction exists between $a$ and $a\otimes\pi$, and $d_a\geq 1$.

The way in which the $\pi$ line keeps track of the spin structure is through its relation with fermion parity. Fermion parity is a (non-anomalous) $\bZ_2$ symmetry generated by a line $(-1)^F$. The $\pi$ line has the property that it measures charge $-1$ on states in the $(-1)^F$-twisted sector:
\[
    \raisebox{-1.6em}{\begin{tikzpicture}[scale=0.7]
    \draw[thick,black, decoration = {markings, mark=at position .5 with {\arrow[scale=1.5]{stealth}}}, postaction=decorate] (-3,0) -- (-5,0);
    \draw[thick,black, decoration = {markings, mark=at position .7 with {\arrow[scale=1.5]{stealth}}}, postaction=decorate] (-2,0) -- (-3,0);

    \fill (-2,0) circle (0.15);

    \draw[thick,purple, decoration = {markings, mark=at position .9 with {\arrowreversed[scale=1.5]{stealth}}}, postaction=decorate] (-2,0) circle[radius=1];

    \node[above right] at (-5,0) {$(-1)^F$};
    \node[above right,purple] at (-2,1) {$\pi$};

    \end{tikzpicture}} \quad = \quad (-1) \quad \raisebox{0em}{\begin{tikzpicture}[scale=0.7]
    \draw[thick,black, decoration = {markings, mark=at position .5 with {\arrow[scale=1.5]{stealth}}}, postaction=decorate] (-2,0) -- (-5,0);

    \fill (-2,0) circle (0.15);

    \node[above right] at (-5,0) {$(-1)^F$};

    \end{tikzpicture}}
\]
The $(-1)^F$-twisted sector is equivalent to the Ramond (R) sector, while the $(-1)^F$-untwisted sector corresponds to the Neveu--Schwarz (NS) sector. Therefore, the $\pi$ line measures charge $-1$ on the R sector and $+1$ on the NS sector.

Crucially, however, there is not a unique possible $(-1)^F$ line. Specifically, there is always a two-fold choice of the generator of fermion parity. Given such a choice, the line $\pi \otimes (-1)^F$ is equally valid. We will return in a moment to the relation between the two choices. The necessary structure to describe the symmetries of a fermionic theory is thus a superfusion category with a $\pi$ line and a choice of $(-1)^F$; we will refer to such objects as \emph{fermionic symmetry categories}, following \cite{Bhardwaj:2024ydc}.

By way of example, let us introduce a particularly simple, yet particularly important, fermionic symmetry category known as $\Arf$. This theory hosts two simple lines, $\1$ and $\pi$, with $\pi \otimes \pi = \1$. The choice of fermion parity line is $(-1)^F = \pi$. (The same supercategory with $(-1)^F$ chosen to be $\1$ is simply the symmetry of a trivial fermionic theory). The Arf theory is a fermionic SPT phase and, while it contains only the minimal structure, it plays a crucial role, as we will soon see.

Suppose now that we have a pair of fermionic theories, $A$ and $B$, each with an associated fermionic symmetry category, $\cS_A$ and $\cS_B$. If we stack these theories on top of one another---that is, consider the two decoupled theories as a single larger theory---then what is the fermionic symmetry category describing the full theory? One obvious guess is that the full symmetry should be described by the (Deligne) product $\cS_A \boxtimes \cS_B$ of the two original symmetry categories. If the theories were bosonic, this would indeed be correct. However, as discussed above, a fermionic theory must have a \emph{single} $\pi$ line and a choice of $(-1)^F$. The product, however, contains two $\pi$ lines, $(\pi_A,\1_B)$ and $(\1_A,\pi_B)$, and we have not specified a choice of $(-1)^F$. The resolution is, of course, to associate the two $\pi$ lines: $(\pi_A,\1_B) \sim (\1_A,\pi_B)$. When taking products of fermionic theories, we will assume throughout that this quotient is taken and will not write it explicitly.\footnote{In the mathematics literature this product of fermionic theories is often denoted $\boxtimes_{\text{sVec}}$ and called the relative Deligne product over $\text{sVec}$. While taking the quotient by $(\pi_A,\pi_B)$ is straightforward at the level of objects, finding consistent $F$-symbols for the quotient category is non-trivial and requires use of the braiding. This can be done systematically using results from \cite{Eliens:2013epa}.} We also assign the generator of fermion parity to simply be the product $(-1)^F = (-1)^F_A (-1)^F_B$.

Let us now return to the observation above that there are always two choices of $(-1)^F$ since, given a particular choice, we could equally use $(-1)^F \otimes \pi$. The remarks of the previous paragraph tell us that these two fermionic symmetry categories are related by stacking with the Arf SPT. That is, Arf does nothing more than keep track of our allocation of $(-1)^F$.

In down-to-earth terms, let's see what this means on a space $T^2$. A choice of spin structure on $T^2$ involves choosing periodic (R) or anti-periodic (NS) boundary conditions for the fermions around each $S^1$ factor. There are, therefore, four spin structures: NS-NS, NS-R, R-NS, and R-R. In any given theory, we must choose the action of $(-1)^F$. The choice of $(-1)^F$ is unambiguous in the NS-NS sector as this follows from the action of fermion parity on the local fields of the theory. Modular transformations then relate the NS-NS sector to the NS-R and R-NS sectors, allowing us to fix the action of $(-1)^F$ there also. The R-R sector, on the other hand, is unfixed. That is, the choice of $(-1)^F$ boils down to introducing an extra sign in the R-R sector.

The Arf SPT can equivalently be described as a topological theory which measures the number of zero modes of the Dirac operator modulo 2, in the sense that its partition function can be written $(-1)^{\Arf(\rho)}$ where $\Arf(\rho)$ is the Arf invariant evaluated on the spin structure $\rho$ \cite{Karch:2019lnn}. On $T^2$, only the R-R spin structure has a zero mode. Therefore, the partition function of a theory with fermionic symmetry $\cS$ and the same theory stacked with Arf, $\cS \boxtimes \Arf$, is precisely an extra sign in the R-R sector.

\subsubsection*{Fermionic TQFTs}

We now discuss how the fermionic features discussed above arise from the three-dimensional perspective. In order to account for these features, we will have recourse to analyse fermionic (spin) TQFTs in our construction. Fermionic Chern--Simons theories (such as $U(1)_k$ for odd $k$) depend on a choice of spin structure to be unambiguously defined, and are not quite described by the modular tensor categories introduced above. This is for the simple reason that a modular tensor category is required to have a \emph{non-degenerate} braiding. The non-degeneracy refers to the fact that the only anyon which braids trivially with all others is the trivial line $\1$. A line with this property is said to be transparent. The novel feature of fermionic Chern--Simons theories is that they contain a \emph{transparent fermion}. That is, they contain an anyon $a$ with spin $\theta_a = -1$ (so it cannot be the identity line) but nevertheless braids trivially with all other lines. As such, the braiding is degenerate.

All is not lost, however, as the degeneracy is very mild. For brevity, we will refer to TQFTs with non-degenerate braiding as \emph{bosonic}, which are described by modular tensor categories, and those which have a transparent fermion as \emph{fermionic}, which are described by super-modular tensor categories.\footnote{In the mathematics literature such categories are sometimes called `slightly degenerate' \cite{davydov2011, Johnson2021}.}  A particularly simple example of a fermionic TQFT is $U(1)_3$ Chern--Simons theory. Denoting the lines by their charge $q$, the fusion is simply $q \otimes q' = (q+q')$. We may expect to have only three independent anyons with charges $0,1,2$. However, the spin is given by $\theta_q = \exp(2\pi i q^2/6)$, so an anyon with charge 3 has $\theta_3 = -1$, and thus cannot be associated with the trivial line $q=0$. The braiding of anyons $q$ and $q'$ is a phase $\exp(2\pi iqq'/3)$, which gives $1$ for all $q'$ for $q=3$. That is, $q=3$ is a transparent fermion. There are, therefore, six independent lines, $q=0,\dots,5$ and the braiding is degenerate. See \cref{app:MTCs} for more details of the $U(1)_k$ Chern--Simons theories.

Given a fermionic Chern--Simons theory, we can enhance it to a larger theory which is bosonic by summing over spin structures. This is nothing more than bosonisation in the language of three-dimensional TQFTs \cite{lorentzfrac}. For example, $U(1)_1$ is fermionic and contains two lines $q=0,1$, with $q=1$ being the transparent fermion. Its bosonised version is $U(1)_4$ which has twice as many lines, $q'=0,1,2,3$, and only the $q'=0$ line is transparent. We can return to the fermionic $U(1)_1$ theory by condensing an algebra $\cA=0\oplus 2$ in $U(1)_4$. This requires some further discussion, however, since the charge 2 line is a fermion ($\theta_2=-1$). In order for a gauging to be consistent, as discussed in \cref{sec:CS_basics}, the algebra $\cA$ to be gauged must satisfy certain constraints. Amongst them is the requirement that all lines in $\cA$ have topological spin $\theta_a=1$. Since $\cA$ is trivial in the gauged theory, it should be invariant under a $2\pi$ rotation. Thus the algebra $\cA$ cannot be gaugeable in the ordinary sense.

Instead, we do the following. We can consider the larger theory $U(1)_4 \boxtimes \SO(0)_{-1}$. The $\SO(0)_{-1}$ theory has two lines, $\1$ and $V$, and $V$ is a transparent fermion, so the product theory is fermionic. In this fermionic theory, we consider an algebra $\tilde{\cA} = (0,\1) \oplus (2,V)$. That is, to the boson $q'=0$ in the algebra $\cA$ that we want to gauge we attach the trivial line and to the fermion $q'=2$ we attach the transparent fermion $V$. As a result, all the lines in $\tilde{\cA}$ are bosons, and the algebra is gaugeable. The catch, however, is that in the bigger theory $U(1)_4 \boxtimes \SO(0)_{-1}$ the algebra $\tilde{\cA}$ is not maximal. That is, there are lines which braid trivially with the whole of $\tilde{\cA}$. Indeed, the line $(0,V)$ is transparent so braids trivially with all others. Therefore, gauging $\tilde{\cA}$ does not result in a trivial theory but instead lands simply on $U(1)_{1}$, with $(0,V)$ becoming the transparent fermion. By a slight abuse of nomenclature, we will continue to refer to such an algebra as Lagrangian in this context.

In general, a Lagrangian algebra $\cA$ in a fermionic TQFT $\cC$ contains only lines with spin-0 (not 0 or $1/2$) and has dimension (cf.\ \eqref{eq:bosonic_lagr} in the bosonic case)
\begin{equation}\label{eq:fermionic_lagr}
    \dim(\cA) = \cD(\cC) / \sqrt{2}
\end{equation}
Then condensing $\cA$ then gives $SO(n)_1$, where $n/2$ is the central charge of $\cC$, and $\cD(SO(n)_1) = \sqrt{2}$.

This procedure of gauging a fermionic algebra (or, equivalently, gauging a bosonic algebra in the larger product theory) will be essential for the examples studied in the following sections. This is simply because the two-dimensional theory we seek to describe is fermionic, so the local operator spectrum includes spin-1/2 excitations. Since the local operator spectrum in the setup \eqref{eq:3d_setup} is encoded in the Lagrangian algebras $\cA_L$ and $\cA_R$, these algebras will be fermionic and we will need to use the procedure given here to proceed. This is another way to see why the lower slab of $\SO(2N)_1$ must be included in the three-dimensional setup: gauging $\cA_{L,R}$ in $G_k$ does not produce a boundary to the trivial theory as a transparent fermion remains after the gauging.

Everything else about the gauging proceeds as in the bosonic case. Namely, when $\tilde{\cA}$ is gauged any bulk lines differing by fusion with a line in $\tilde{\cA}$ are related, so the lines on a topological boundary where $\tilde{\cA}$ is gauged form multiplets under fusion with $\tilde{\cA}$ called $\tilde{\cA}$-modules. Topological interfaces between topological boundaries where algebras $\tilde{\cA}_L$ and $\tilde{\cA}_R$ are condensed are similarly described by $(\tilde{\cA}_L,\tilde{\cA}_R)$-bimodules.

On the topological boundary of the Chern--Simons theory where $\tilde{\cA}$ is gauged lives the symmetry of a fermionic two-dimensional theory. As discussed above, this must be described by a fermionic symmetry category. In particular, it must admit a $\pi$ line which generates a $\bZ_2$ symmetry. In terms of the bulk lines, $\pi$ is nothing other than the boundary line found by pushing the transparent fermion of the bulk to the boundary.

\subsubsection*{SPT classes}

An important feature of the symmetric interfaces in fermionic theories is their \emph{SPT class}. From the perspective of the two-dimensional CFT, we ultimately seek a boundary condition which preserves a given symmetry $G$. A boundary condition is an interface to the trivial theory. In a fermionic theory, one must also keep track of the action of fermion parity $(-1)^F$. It is important, therefore, to ensure that when constructing boundary conditions we truly build an interface to the trivial theory and not to the Arf SPT. The latter is almost the same as the trivial theory, aside from its non-trivial assignment of $(-1)^F$. It was noted in \cite{Arias-Tamargo:2026urw} that, for $N=2$ Weyl fermions, constructing boundary conditions via folding along duality defects can sometimes give rise to interfaces to Arf, instead of true boundary conditions. All is not lost in this situation, as the Arf SPT itself admits a boundary condition, but it carries an extra unpaired Majorana fermion on the boundary, increasing the Affleck--Ludwig boundary central charge by a factor of $\sqrt{2}$.

Those symmetries for which no unpaired Majorana mode is present on the boundary are said to be in the \emph{vector class}, while those afflicted with the Majorana mode are in the \emph{axial class}. Here we discuss how to determine whether a given interface constructed via the methods introduced in previous sections is in the vector or axial class. This requires some refinements to our story.

First we consider one Weyl fermion, which has a single $U(1)$ symmetry that we ask to be preserved at the interface. After folding along the interface, we recover a Dirac fermion for which there are two $U(1)$-preserving boundary conditions: the vector boundary condition $\psi \to \tilde{\psi}$ and the axial (Andreev) boundary condition $\psi \to \tilde{\psi}^\dag$. The former is free of any unpaired Majorana modes, while the latter is not \cite{BoyleSmith:2019jnh}. We expect, therefore, to find two classes of interfaces in our TQFT construction which differ in how they preserve $(-1)^F$.

Since the single complex fermion carries charge 1 under the $U(1)$ symmetry, the corresponding Chern--Simons theory is simply $U(1)_1$, whose anyons we label $W_0$ and $W_1$. Upon condensing $\cA = 1$, the boundary symmetry consists purely of the lines $1$ and $\pi$ (corresponding to bulk anyons $W_0$ and $W_1$ pushed to the boundary). Neither of these lines can describe fermion parity $(-1)^F$ because they both have trivial full braid with $W_1$, the line to which all fermions are attached. So this setup is incapable of probing $(-1)^F$.

Therefore, we need to replace our TQFT with a bigger one. We can do this by bosonising it (summing over spin structures) which converts $U(1)_1 \to U(1)_4$. Now upon condensing the Lagrangian algebra $\hat{\cA} = (0,1) \oplus (2,V)$ in $U(1)_4 \boxtimes SO(2)_{-1}$, we get a boundary symmetry of $\text{Vec}_{\bZ_4}^{\omega=1}$ (where $\omega \in H^3(B\bZ_4; U(1)) = \bZ_4$ labels the cocycle), whose objects are the bulk Wilson lines $W_{0,1,2,3}$ pushed to the boundary. Of these, $W_2 = \pi$ and $W_1 = (-1)^F$. Indeed, the NS sector fermions are attached to $W_2$, whose full braid with $W_1$ is $e^{2 \pi i \frac{2 \cdot 1}{4}} = -1$. This new setup is, therefore, capable of probing $(-1)^F$.

With the new larger TQFT in place, the three-dimensional setup looks like
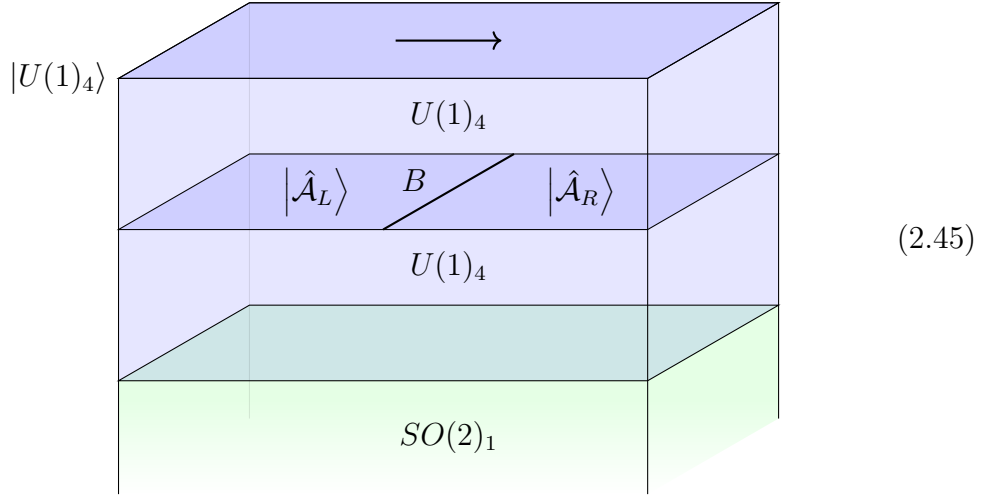
\begin{equation}\label{eq:SPT_determination_special}
\begin{tikzpicture}[z={(30:2cm)},baseline=(current bounding box.center)]
    \newcommand{\x}{7}
    \newcommand{\yB}{2} 
    \newcommand{\yS}{4} 
    \newcommand{\y}{5.5} 

    \shade[top color=green,bottom color=white,middle color=green,opacity=0.1] (0,-\yS,0) -- (0,-\y,0) -- ++(\x,0,0) -- ++(0,0,1) -- (\x,-\yS,1) -- ++(-\x,0,0) -- cycle;
    \fill[blue,opacity=0.1] (0,0,0) -- (0,-\yS,0) -- ++(\x,0,0) -- ++(0,0,1) -- (\x,0,1) -- ++(-\x,0,0) -- cycle;
    \draw[black] (0,0,0) -- ++(\x,0,0) -- ++(0,0,1) -- ++(-\x,0,0) -- cycle;
    \filldraw[black,fill=blue,fill opacity=0.1] (0,0,0) -- ++(\x,0,0) -- ++(0,0,1) -- ++(-\x,0,0) -- cycle;
    \filldraw[black,fill=blue,fill opacity=0.1] (0,-\yB,0) -- ++(\x,0,0) -- ++(0,0,1) -- ++(-\x,0,0) -- cycle;
    \draw[black] (0,-\yS,0) -- ++(\x,0,0) -- ++(0,0,1) -- ++(-\x,0,0) -- cycle;
    \draw[black] (0,0,0) -- ++(0,-\y,0);
    \draw[black] (\x,0,0) -- ++(0,-\y,0);
    \draw[black] (\x,0,1) -- ++(0,-\y,0);
    \draw[black,opacity=0.1] (0,0,1) -- ++(0,-\y,0);
    \draw[thick,black] (\x/2,-\yB,0) -- ++(0,0,1);

    \node at (\x/2,-\yB/2,0.5) {$U(1)_4$};
    \node at (\x/2,-\yB/2-\yS/2,0.5) {$U(1)_4$};
    \node at (\x/2,-\yS/2-\y/2-0.5,0.5) {$SO(2)_1$};
    \node at (\x*0.25,-\yB,0.5) {$\ket{\hat{\cA}_L}$};
    \node at (\x*0.75,-\yB,0.5) {$\ket{\hat{\cA}_R}$};
    \node[above left,xshift=-4pt,yshift=-4pt] at (\x/2,-\yB,0.5) {$B$};

    \draw[->,thick] (\x*0.4,0,0.5) -- (\x*0.6,0,0.5);
    \node[left] at (0,0,0) {$\ket{U(1)_4}$};
\end{tikzpicture}
\end{equation}
Let us interpret the various sections of this picture. The top $U(1)_4$ is the symmetry we want to preserve in the scattering. This might be twisted, for example in the case of the axial boundary condition where $U(1)$ acts as $\psi \to e^{ix} \psi$ on one side of the defect and $\tilde{\psi} \to e^{-ix} \tilde{\psi}$ on the other. The bottom $U(1)_4$, on the other hand, is the symmetry under which the fermions naturally transform. That is, we can understand it as $U(1)_4 \cong \Spin(2)_1$. This is never twisted, and acts either as $e^{ix}$ on both sides of the defect, or as $e^{-ix}$ on both sides, depending on convention.

Suppose, for example, that we want to preserve the axial symmetry, for which the two embeddings $L, R : U(1) \to U(1)$ are
\begin{equation}
    L : e^{ix} \to e^{ix}
    \qquad
    R : e^{ix} \to e^{-ix}
\end{equation}
The two corresponding Lagrangian algebras in $U(1)_4 \boxtimes U(1)_{-4}$ are then
\begin{equation}
    \hat{\cA}_L = (0,0) \oplus (1,1) \oplus (2,2) \oplus (3,3)
    \qc\quad
    \hat{\cA}_R = (0,0) \oplus (1,3) \oplus (2,2) \oplus (3,1)
\end{equation}
This illustrates that although both embeddings give the same Lagrangian algebra in $U(1)_4 \boxtimes SO(2)_{-1}$, they give different Lagrangian algebras in $U(1)_4 \boxtimes U(1)_{-4}$.

Now we claim that the fermion parity operator $(-1)^F$ should be understood as the $W_1$ line of the \emph{bottom} $U(1)_4$.\footnote{We could equally choose $W_3$.} We can write down the possible $(\hat{\cA}_L, \hat{\cA}_R)$-bimodules by inspection, since the set of objects in the bimodule should be closed under fusion with both $\hat{\cA}_L$ and $\hat{\cA}_R$. This yields one possible bimodule
\begin{equation}
\begin{split}
    B &= (0,0) \oplus (1,1) \oplus (2,2) \oplus (3,3) \\
      &\quad\oplus (0,2) \oplus (1,3) \oplus (2,0) \oplus (3,1)
\end{split}
\end{equation}
and then the only other simple bimodule is $B' = B \otimes (1,0)$. From this we see that $B$, regarded as an $\hat{\cA}_L$-module, has internal end algebra
\begin{equation}
    a_L = 0 \oplus 2
\end{equation}
in $\cC_{\cA_L}$ and similarly, as an $\hat{\cA}_R$-module, has internal end
\begin{equation}
    a_R = 0 \oplus 2
\end{equation}
in $\cC_{\cA_R}$. These are simply the sets of lines in $\cC_{\cA_{L,R}}$ that can end on $B$. Note that when we collapse the middle $U(1)_4$ slab to zero, the $2$ line becomes the $\pi$ line. So we could rephrase the above result as $a_L = a_R = 1 \oplus \pi$.

The above result is in stark contrast to what we would have got if we took $\hat{\cA}_R = \hat{\cA}_L$, corresponding to a boundary which preserves the vector symmetry. Then we would have
\begin{equation}
    B = (0,0) \oplus (1,1) \oplus (2,2) \oplus (3,3)
\end{equation}
and thus $a_L = a_R = 1$.

We learn that, in contrast to the direct gauging methods of \cite{Arias-Tamargo:2026urw}, our TQFT setup always produces interfaces $\cT \to \cT$. Equivalently, upon folding along the interface, it always produces genuine boundary conditions of the free fermion theory, and never an interface to the Arf SPT. If we try to apply it to the axial symmetry, for which the naïve symmetry-preserving interface would be an interface $\cT \to \cT \times \Arf$, then the setup automatically produces another factor of $\Arf$ with an unpaired Majorana mode on its boundary, and stacks it with the naïve interface to give an interface $\cT \to \cT$, while increasing its quantum dimension by $\sqrt{2}$. Indeed, as discussed in \cref{sec:2d_descrip}, the dimension of the interface is given by $d_B = \sqrt{\dim(a_L)}$, which is $1$ for the vector case and $\sqrt{2}$ for the axial case. We can also see that the interface is $m$-type simple in the vector case ($\pi$ cannot end on it), while it is $q$-type simple in the axial case ($\pi$ can end on it).

Thus we propose the following diagnostic for the SPT class of the interface which results from our TQFT construction in general. Let $L, R : G_k \to \Spin(2N)_1$ be the two conformal embeddings of the symmetry we wish to preserve.\footnote{Note that we may have to pass to a double cover of $G_k$ to make this embedding exist, which would have the effect of making the $G_k$ Chern--Simons theory bosonic.} Then the branching rules give rise to Lagrangian algebras $\hat{\cA}_{L,R}$ in $G_k \boxtimes \Spin(2N)_{-1}$:
\begin{equation}\label{eq:SPT_determination}
\begin{tikzpicture}[z={(30:2cm)},baseline=(current bounding box.center)]
    \newcommand{\x}{7}
    \newcommand{\yB}{2} 
    \newcommand{\yS}{4} 
    \newcommand{\y}{5.5} 

    \shade[top color=green,bottom color=white,middle color=green,opacity=0.1] (0,-\yS,0) -- (0,-\y,0) -- ++(\x,0,0) -- ++(0,0,1) -- (\x,-\yS,1) -- ++(-\x,0,0) -- cycle;
    \fill[blue,opacity=0.1] (0,0,0) -- (0,-\yS,0) -- ++(\x,0,0) -- ++(0,0,1) -- (\x,0,1) -- ++(-\x,0,0) -- cycle;
    \draw[black] (0,0,0) -- ++(\x,0,0) -- ++(0,0,1) -- ++(-\x,0,0) -- cycle;
    \filldraw[black,fill=blue,fill opacity=0.1] (0,0,0) -- ++(\x,0,0) -- ++(0,0,1) -- ++(-\x,0,0) -- cycle;
    \filldraw[black,fill=blue,fill opacity=0.1] (0,-\yB,0) -- ++(\x,0,0) -- ++(0,0,1) -- ++(-\x,0,0) -- cycle;
    \draw[black] (0,-\yS,0) -- ++(\x,0,0) -- ++(0,0,1) -- ++(-\x,0,0) -- cycle;
    \draw[black] (0,0,0) -- ++(0,-\y,0);
    \draw[black] (\x,0,0) -- ++(0,-\y,0);
    \draw[black] (\x,0,1) -- ++(0,-\y,0);
    \draw[black,opacity=0.1] (0,0,1) -- ++(0,-\y,0);
    \draw[very thick,black] (\x/2,-\yB,0) -- ++(0,0,1);

    \node at (\x/2,-\yB/2,0.5) {$G_k$};
    \node at (\x/2,-\yB/2-\yS/2,0.5) {$\Spin(2N)_1$};
    \node at (\x/2,-\yS/2-\y/2-0.5,0.5) {$\SO(2N)_1$};
    \node at (\x*0.25,-\yB,0.5) {$\ket{\hat{\cA}_L}$};
    \node at (\x*0.75,-\yB,0.5) {$\ket{\hat{\cA}_R}$};
    \node[above left,xshift=-4pt,yshift=-4pt] at (\x/2,-\yB,0.5) {$B$};

    \draw[->,thick] (\x*0.4,0,0.5) -- (\x*0.6,0,0.5);
    \node[left] at (0,0,0) {$\ket{G_k}$};
\end{tikzpicture}
\end{equation}
Then given a simple bimodule $B$, the diagnostic for its SPT class is whether the $V$ line of $\Spin(2N)_1$ can end on $B$ from below. This is equivalent to saying that $\pi$ is in the internal end of $B$ as an $\hat{\cA}_L$- or an $\hat{\cA}_R$-module, and to saying that the corresponding two-dimensional defect is $q$-type.

Coming back to the case of a single Weyl fermion, as commented above, in the axial case we can either have a $q$-type interface $\cT \to \cT$, or an $m$-type interface $\cT \to \cT \times \Arf$. The above construction naturally gave us the first. The second can be found instead by considering the setup
\begin{equation}
\begin{tikzpicture}[z={(30:2cm)},baseline=(current bounding box.center)]
    \newcommand{\x}{7}
    \newcommand{\yB}{2} 
    \newcommand{\yS}{4} 
    \newcommand{\y}{5.5} 
    \newcommand{\rN}{1.4} 

    \shade[top color=green,bottom color=white,middle color=green,opacity=0.1] (0,-\yS,0) -- (0,-\y,0) -- ++(\x,0,0) -- ++(0,0,1) -- (\x,-\yS,1) -- ++(-\x,0,0) -- cycle;
    \fill[blue,opacity=0.1] (0,0,0) -- (0,-\yS,0) -- ++(\x,0,0) -- ++(0,0,1) -- (\x,0,1) -- ++(-\x,0,0) -- cycle;
    \draw[black] (0,-\yS,0) -- ++(\x,0,0) -- ++(0,0,1) -- ++(-\x,0,0) -- cycle;
    \fill[blue!10!white!85!red,opacity=0.85] (\x/2,-\yB,0) arc[start angle=180,end angle=270,radius=\rN] -- (\x,-\yB-\rN,0) -- ++(0,0,1) -- (\x/2+\rN,-\yB-\rN,1) arc[start angle=270,end angle=180,radius=\rN] -- cycle;
    \draw[black] (0,0,0) -- ++(\x,0,0) -- ++(0,0,1) -- ++(-\x,0,0) -- cycle;
    \filldraw[black,fill=blue,fill opacity=0.1] (0,0,0) -- ++(\x,0,0) -- ++(0,0,1) -- ++(-\x,0,0) -- cycle;
    \filldraw[black,fill=blue,fill opacity=0.1] (0,-\yB,0) -- ++(\x,0,0) -- ++(0,0,1) -- ++(-\x,0,0) -- cycle;
    \draw[black] (0,0,0) -- ++(0,-\y,0);
    \draw[black] (\x,0,0) -- ++(0,-\y,0);
    \draw[black] (\x,0,1) -- ++(0,-\y,0);
    \draw[black,opacity=0.1] (0,0,1) -- ++(0,-\y,0);
    \draw[red] (\x/2,-\yB,0) arc[start angle=180,end angle=270,radius=\rN] -- (\x,-\yB-\rN,0) -- ++(0,0,1) -- (\x/2+\rN,-\yB-\rN,1) arc[start angle=270,end angle=180,radius=\rN];
    \draw[very thick,black] (\x/2,-\yB,0) -- ++(0,0,1);

    \node at (\x/2,-\yB/2,0.5) {$G_k$};
    \node at (\x/6,-\yB/2-\yS/2,0.5) {$\Spin(2N)_1$};
    \node at (\x/2,-\yS/2-\y/2-0.5,0.5) {$\SO(2N)_1$};
    \node at (\x*0.25,-\yB,0.5) {$\ket{\hat{\cA}_L}$};
    \node at (\x*0.75,-\yB,0.5) {$\ket{\hat{\cA}_R}$};
    \node[above left,xshift=-4pt,yshift=-4pt] at (\x/2,-\yB,0.5) {$B$};
    \node[red] at (\x*0.75,-\yB-\rN,0.5) {$N_V$};

    \draw[->,thick] (\x*0.4,0,0.5) -- (\x*0.6,0,0.5);
    \node[left] at (0,0,0) {$\ket{G_k}$};
\end{tikzpicture}
\end{equation}
where we insert the condensation defect $N_V$ of $V$. (Recall that $V$ can end on $B$ by virtue of $B$ being in the axial class, so $N_V$ can end on $B$ as well.) Since $N_V$ exchanges $S \leftrightarrow C$, when pushed to the bottom interface it implements the desired fermion parity flip $(-1)^F \leftrightarrow \pi (-1)^F$, which is equivalent to stacking with Arf \cite{Bhardwaj:2024ydc}. Alternatively, when pushed to the top interface, it modifies the embedding $R : G_k \to \Spin(2N)_1$ by composing it with the outer automorphism $S \leftrightarrow C$ of $\Spin(2N)_1$.

We can also see that the effect of $N_V$ is stacking with Arf from the fact $\Spin(2N)_1$ can be regarded as $SO(2N)_1$ with a gauged spin structure, and in this presentation $N_V(\Sigma)$ is $(-1)^{\Arf(\rho_{\text{3d}}|_\Sigma)}$ with $\rho_{\text{3d}}$ the fluctuating spin structure in the bulk. Thus when $N_V$ is pushed to the bottom interface, it stacks the two-dimensional partition function with $(-1)^{\Arf(\rho_{\text{2d}})}$ \cite[Eq.\ (5.3)]{bb}.

\section{Abelian symmetries}\label{sec:abelian}

Here we apply the TQFT methods of the previous section to the case where there are $N$ Weyl fermions and a defect which preserves an abelian $G=U(1)^N$ symmetry. In doing so, we will be adding another entry in the ever-growing catalogue of ways to understand such interfaces, which otherwise includes the explicit construction of corresponding Cardy states \cite{Han_2017,BoyleSmith:2019jnh}, quantum mechanical rotor models \cite{Polchinski:1984uw,Loladze:2024ayk,Loladze:2025jsq,davidandgeorge}, and half-space gauging understood purely in two dimensions \cite{Arias-Tamargo:2026urw,Wei:2026fsn}.

In this case, it is known that scattering a local excitation from the left through the defect results in a twisted operator on the right, where the twist line is invertible and belongs to a finite abelian group symmetry. We will first see that this result can be established immediately from the OPE monodromies discussed in \cref{sec:monodromies}, and then that the same result can be derived from the TQFT approach.

\subsection{Invertible twist lines from OPEs}\label{sec:abelian_OPEs}

The action of the symmetry on the left and right of the defect is specified by a pair of non-degenerate integer-valued charge matrices $Q_L, Q_R$ \cite{BoyleSmith:2019jnh}. Our convention is that each column of $Q_{L,R}$ represents a $U(1)$ symmetry, and the entries of that column are the charges of the fermions under the $U(1)$ symmetry. The local fields are indexed by $i,j=1,\dots,N$ and are denoted $\psi_i$ on the left of the defect and $\tilde{\psi}_i$ on the right. The $U(1)$ symmetries are indexed by $\alpha=1,\dots,N$. Matching the anomaly on the left and right of the defect requires $Q_L^T Q_L = Q_R^T Q_R$. The $U(1)^N$ chiral algebra of this RCFT, and the corresponding Chern--Simons theory, has a $K$-matrix
\begin{equation}
    K = Q_L^T Q_L = Q_R^T Q_R
\end{equation}
$K$ is said to be even if all the elements on the diagonal of $K$ are even, and odd otherwise. There is a freedom to redefine $Q_L \mapsto Q_L M$, $Q_R \mapsto Q_R M$ for any non-degenerate matrix $M$ that preserves the integrality of $Q_{L,R}$, since such a change gives a different parametrisation of the same infinitesimal symmetry, which does not affect the set of symmetry-preserving defects $\cN$. It is useful to introduce the rational orthogonal matrix
\begin{equation}\label{eq:Rdef}
    \cR = Q_R Q_L^{-1}
\end{equation}
which is unaffected by such reparametrisations, so the physically inequivalent $U(1)^N$ symmetries which can be preserved by the defect are labelled by the matrix $\cR \in O(N)_\bQ$.

We consider scattering the local operator $\psi_i$ through the defect from the left. As discussed in \cref{sec:monodromies}, we can probe the twist line attached to $\psi_i$ by considering the OPE with a local operator $\tilde{\psi}_j$ on the right. As the chiral algebra is abelian, the OPE contains only a single primary operator $\cO_{ij}$ with charge $(Q_L)_{i \alpha} + (Q_R)_{j \alpha}$ under $U(1)_\alpha$. The OPE therefore takes the form
\begin{equation}
    \psi_i(z) \tilde{\psi}_j(0) \sim z^{\cR_{ji}} \cO_{ij}(0)
\end{equation}
We see that passing $\tilde{\psi}_j$ through the twist line attached to $\psi_i$, by sending $z\to e^{2\pi i}z$, produces a phase
\begin{equation}\label{eq:abelian_phase}
    \tilde{\psi}_j \to \exp(2\pi i \cR_{ji}) \tilde{\psi}_j
\end{equation}
Hence the twist line corresponds to the element $\cR e_i \in \bR^N / \bZ^N$ of the $U(1)^N$ symmetry of the right fermions $\tilde{\psi}_j$, with $e_i$ the $i$th standard basis vector. Taking the collection of all such twist lines for varying $i = 1, \dots, N$, we find that these lines close into the finite abelian group
\begin{equation}
    \cG_\cR \coloneqq \cR \bZ^N / (\cR \bZ^N \cap \bZ^N) \cong (\cR \bZ^N + \bZ^N) / \bZ^N \subseteq \bR^N / \bZ^N \label{eq:abelianGRdef}
\end{equation}
where the equivalence is by the second isomorphism theorem. This matches recent results of \cite{Wei:2026fsn}, for example, which observed that such a defect can be understood as a duality defect associated with gauging this $\cG_\cR$ symmetry.

\textbf{Example:} We will illustrate many of the results in this section using the 3450 model. The version discussed in \cref{sec:int} has charge matrices $Q_L = \begin{psmallmatrix} 3 & -4 \\ 4 & 3 \end{psmallmatrix}$ and $Q_R = \begin{psmallmatrix} 5 & 0 \\ 0 & 5 \end{psmallmatrix}$. Using a reparametrisation $M$, we can bring this to the minimal form
\begin{equation}\label{eq:3450minimal}
    Q_L = \begin{pmatrix} 2 & -1 \\ 1 & 2 \end{pmatrix}
    \qquad
    Q_R = \begin{pmatrix} 2 & 1 \\ -1 & 2 \end{pmatrix}
    \quad \implies \quad
    K = \begin{pmatrix} 5 & 0 \\ 0 & 5 \end{pmatrix}
\end{equation}
which has lowest possible $\det(K)$. We can compute $\cR = \frac{1}{5} \begin{psmallmatrix} 3 & 4 \\ -4 & 3 \end{psmallmatrix}$. The group $\cG_\cR \cong \bZ_5$ can be taken to be generated by $\cR \begin{psmallmatrix} 1 \\ 2 \end{psmallmatrix} = \frac{1}{5} \begin{psmallmatrix} 1 \\ 2 \end{psmallmatrix}$ mod $\bZ^2$, giving the result \eqref{eq:basictwist} claimed for the basic twist line.

\subsection{TQFT approach}\label{eq:abeliantqftapproach}

We now proceed to the description of the twisted scattering from Chern--Simons theory. The Chern--Simons theory in question is $U(1)^N_K$, which is bosonic or fermionic depending on whether $K$ is even or odd (see \cref{app:MTCs}). However, we can make use of the freedom to reparametrise $Q_{L,R}$ to ensure $K$ is always even. This will be useful not only to reduce the number of cases, but will also be necessary once we reach \cref{sec:abeliansptclass}. From now on we will assume this has been done. The anyons of $U(1)^N_K$ are then labelled by $q \in \bZ^N / K \bZ^N$.

\textbf{Example:} In the 3450 model, the minimal charge matrices \eqref{eq:3450minimal} lead to $K$ odd. But reparametrising the charge matrices by $Q_{L,R} \mapsto Q_{L,R} M$ with $M = \begin{psmallmatrix} 1 & 1 \\ 1 & -1 \end{psmallmatrix}$ yields the new charge matrices
\begin{equation}\label{eq:3450even}
    Q_L = \begin{pmatrix} 1 & 3 \\ 3 & -1 \end{pmatrix}
    \qquad
    Q_R = \begin{pmatrix} 3 & 1 \\ 1 & -3 \end{pmatrix}
    \quad \implies \quad
    K = \begin{pmatrix} 10 & 0 \\ 0 & 10 \end{pmatrix}
\end{equation}
with $K$ even. In general we can always make $K$ even by passing to a basis where only the first diagonal entry of $K$ is odd, then doubling the charge vector associated to the first $U(1)$ symmetry, which corresponds to bosonising $U(1)^N_K$.

\subsubsection*{Topological boundary conditions}

The first step of our programme is to determine the topological boundaries $\ket{\cA_{L,R}}$. More precisely, these are topological interfaces between $U(1)^N_K$ and $\SO(2N)_1$, which are equivalent to topological boundaries of the folded theory
\begin{equation}
    \cC = U(1)^N_K \boxtimes \SO(2N)_{-1}
\end{equation}
The simple lines of this theory can be labelled by pairs $(q, \epsilon)$ with $q \in \bZ^N / K\bZ^N$ and $\epsilon \in \bZ_2$, corresponding to the element $q \boxtimes V^\epsilon$ where $V$ is the transparent fermion of $SO(2N)_{-1}$. The anyons are therefore labelled by the abelian group
\begin{equation}
    G(\cC) = (\bZ^N / K \bZ^N) \times \bZ_2
\end{equation}
with the group multiplication corresponding to the fusion of anyons. Because our theory $\cC$ is an abelian topological order, we will be able to phrase all our results in terms of the group $G(\cC)$, eschewing the more cumbersome categorical formalism.

The local operators of the left Weyl fermions are $\psi^n = \prod_i \psi_i^{n_i}$ for $n \in \bZ^N$, suitably regularised, e.g.\ by point-splitting. Such an operator is attached to the $U(1)^N_K$ line $q = Q_L^T n$, which continues through the interface as $V^{n \cdot \rho}$ with $\rho = (1, \dots, 1)$, since $\psi^n$ has fermion parity $n \cdot \rho$ and all fermionic operators are attached to $V$:
\begin{equation}\label{eq:u1nlocalops}
    \begin{tikzpicture}[z={(20:1.5cm)},baseline=(current bounding box.center)]
    \newcommand{\x}{3.5}
    \newcommand{\y}{3.5}

    \shade[top color=green,bottom color=white,middle color=green,opacity=0.1] (0,-\y/2,0) -- ++(0,-\y/2,0) -- ++(\x,0,0) -- ++(0,0,1) -- ++(0,\y/2,0) -- ++(-\x,0,0) -- cycle;
    \fill[blue,opacity=0.1] (0,0,0) -- ++(0,-\y/2,0) -- ++(\x,0,0) -- ++(0,0,1) -- ++(0,\y/2,0) -- ++(-\x,0,0) -- cycle;
    \draw[black] (0,-\y/2,0) -- ++(\x,0,0) -- ++(0,0,1) -- ++(-\x,0,0) -- cycle;
    \filldraw[black,fill=blue,fill opacity=0.1] (0,0,0) -- ++(\x,0,0) -- ++(0,0,1) -- ++(-\x,0,0) -- cycle;
    \draw[black] (0,0,0) -- ++(0,-\y,0);
    \draw[black] (\x,0,0) -- ++(0,-\y,0);
    \draw[black] (\x,0,1) -- ++(0,-\y,0);
    \draw[black,opacity=0.1] (0,0,1) -- ++(0,-\y,0);

    \begin{scope}[every node/.style={fill=red,shape=circle,inner sep=1pt}]
        \draw[thick,red,decoration = {markings, mark=at position .2 with {\arrowreversed[scale=1.5]{stealth}}, mark=at position .7 with {\arrowreversed[scale=1.5]{stealth}}}, postaction=decorate] (\x*0.4,0,0.5) node{} -- ++(0,-\y/2,0) node{} -- ++(0,-\y/2,0);
    \end{scope}

    \node[left] at (0,-\y*0.75,0) {$\SO(2N)_1$};
    \node[left] at (0,-\y*0.5,0) {$\ket{\cA_L}$};
    \node[left] at (0,-\y*0.25,0) {$U(1)^N_K$};

    \node[red,above,yshift=0.3cm] at (\x*0.4,0,0.5) {$\psi^n$};
    \node[red,right,xshift=0.1cm] at (\x*0.4,-\y*0.25,0.5) {$Q_L^T n$};
    \node[red,right,xshift=0.1cm] at (\x*0.4,-\y*0.75,0.5) {$V^{n \cdot \rho}$};
\end{tikzpicture}
\end{equation}
The interface $\ket{\cA_L}$ is equivalent to a boundary of the folded theory $\cC$. It follows from the above picture that the set of lines which can end on this boundary, and hence make up the Lagrangian algebra $\cA_L$, is
\begin{equation}\label{eq:AL_abelian}
    \cA_L = \{ (Q_L^T n, n \cdot \rho) : n \in \bZ^N / Q_L \bZ^N \} \subseteq G(\cC)
\end{equation}
which in this abelian setting we can view simply as a subgroup of $G(\cC)$. The boundary conditions on the right of the defect are defined similarly with $Q_L$ replaced by $Q_R$. Namely, the algebra of condensed lines on the right of the defect is
\begin{equation}\label{eq:AR_abelian}
    \cA_R = \{ (Q_R^T n, n \cdot \rho) : n \in \bZ^N / Q_R \bZ^N \} \subseteq G(\cC)
\end{equation}
One can indeed check these subgroups have order $|\cA_{L,R}| = |\det(K)|^{1/2} = \cD(\cC)/\sqrt{2}$, as is required for a Lagrangian algebra in a fermionic TQFT $\cC$.

\textbf{Example:} For the 3450 model \eqref{eq:3450even}, the possible anyons are labelled by $G(\cC)=\Z_{10}\times \Z_{10}\times \Z_2$, and the Lagrangian algebras are $\cA_L\cong \cA_R\cong \Z_{10}$.

\subsubsection*{Boundary symmetry categories}

The next step is to determine the symmetry categories $\cS_{L,R}$ of lines supported on the topological boundaries $\ket{\cA_{L,R}}$. As discussed in \cref{sec:scattering_from_bulk}, these are $\cA_L$-modules over $\cC$, which form the category $\cS_L = \cC_{\cA_L}$, and similarly for the right boundary.

To understand what this means in the abelian context, note that an $\cA_L$-module is simply a collection of lines in $G(\cC)$ which is closed under fusion with $\cA_L$. It therefore forms a coset of $\cA_L$ in $G(\cC)$. The set of all such boundary lines is the quotient group
\begin{equation}\label{eq:SLgroupdef}
    \cS_L = G(\cC) / \cA_L
\end{equation}
which fuse according to the same group law. In the categorical language, we would denote this category more explicitly as $\cS_L = \text{Vec}^\omega_{G(\cC) / \cA_L}$, which also depends on a group cocycle $\omega \in H^3(G(\cC) / \cA_L; U(1))$ that we should determine. However, for this section it will suffice to work at the level of objects, for which $\omega$ does not matter.

For $[(q,\epsilon)] \in G(\cC) / \cA_L$, we will denote the corresponding boundary line by $M_{[(q,\epsilon)]}^L$. As a collection of bulk lines in $\cC$, it is given by
\begin{equation}\label{eq:M_induced}
    M_{[(q,\epsilon)]}^L = (q,\epsilon) \otimes \cA_L
\end{equation}
The interpretation is that $M_{[(q,\epsilon)]}^L$ is the bulk line $(q,\epsilon)$ pushed to the boundary, or mathematically, the \emph{induced $\cA_L$-module} from the simple object $(q,\epsilon) \in \cC$.

To compute $\cS_L$ in practice, we note that by fusing with the condensate, we can write any boundary line $M^L_{[(q,1)]} = M^L_{[(q',0)]}$ for some $q'$. Two lines $M^L_{[(q,0)]}$ and $M^L_{[(q',0)]}$ are identified if they differ by an element of the condensate $\cA_L$, which is the case if $q-q'=Q_L^T n$ for some $n\in\bZ^N_\text{even}$ (where $\bZ^N_\text{even}$ is the even sublattice of $\bZ^N$). Therefore,
\begin{equation}
   \cS_L = \{ M^L_{[(q,0)]} : q \in \bZ^N / Q_L^T \bZ^N_\text{even} \}
\end{equation}
and similarly for $\cS_R$ with $Q_L$ replaced by $Q_R$.

\textbf{Example:} For the 3450 model \eqref{eq:3450even}, this gives $\cS_R \cong \bZ^2 / 2 \begin{psmallmatrix} 2 & 1 \\ -1 & 2 \end{psmallmatrix} \bZ^2 \cong \bZ_5 \times \bZ_2^2$. The $\bZ_5$ factor corresponds to the basic twist line \eqref{eq:basictwist}, while the two $\bZ_2$ factors correspond to $(-1)^F$ and $\pi$ (see \cref{sec:SPTs_general}).

\subsubsection*{Symmetric interfaces}

We would now like to understand the interfaces $B$ between the two boundaries $\ket{\cA_{L,R}}$. Ultimately this interface corresponds to the $G$-symmetric defect of the two-dimensional CFT. As discussed in \cref{sec:CS_basics}, since the lines in $\cA_L$ are condensed on one side of the interface and the lines in $\cA_R$ are condensed on the other, the interfaces must admit a left-action of $\cA_L$ and a right-action of $\cA_R$ in a compatible manner. The possible interfaces then form the category $\bimod{\cA_L}{\cC}{\cA_R}$ of $(\cA_L,\cA_R)$-bimodules over $\cC$.

Similarly to the situation for modules, in the abelian context an $(\cA_L, \cA_R)$-bimodule is a collection of lines in $G(\cC)$ closed under both fusion with $\cA_L$ and $\cA_R$. It therefore forms a double coset, and the set of all simple bimodules is
\begin{equation}
    \bimod{\cA_L}{\cC}{\cA_R} = \cA_L \backslash G(\cC) / \cA_R
\end{equation}
in group-theory terms. Of these, there is a unique simplest, or canonical bimodule, which is the double coset containing zero. As a subset of bulk lines of $\cC$, this is
\begin{equation}\label{eq:abelianB}
    B = \cA_L + \cA_R \subseteq G(\cC)
\end{equation}
In this abelian case, all other simple bimodules are found by fusing $B$ with simple lines.

Flitting back to categorical language, note that $B$ occurs in the decomposition of the bimodule induced from the identity line $\1 \in \cC$ into simples. However this induced bimodule is generically not simple itself, but rather a sum of $|\cA_L \cap \cA_R|$ simples.\footnote{An alternative approach is to choose different charge matrices $Q_L,Q_R$ representing the same $\cR$ with trivial overlap, such that the induced bimodules are simple. However, this `minimal' choice of charge matrices may give rise to an odd $K$ matrix, so we would then have to consider this case separately. Here we instead give the computation for generic charge matrices.} To compute the intersection in practice, we use $\cA_L \cap \cA_R = (Q_L^T \bZ^N \cap Q_R^T \bZ^N) / K \bZ^N$.

\textbf{Example:} For the 3450 model, one computes that $|\cA_L \cap \cA_R| = 2$, so the induced bimodule is indeed not simple. The double coset then gives that the set of possible bimodules is $\Z_2\times \Z_2$. Only one of them is the multiplet of lines that contains the identity.

\subsubsection*{The interface as a duality defect}

Our aim is to understand the interfaces between the $\cA_L$ and $\cA_R$ boundaries from a two-dimensional perspective. One way to do this concretely is to realise a given interface as a duality defect, such that the left and right sides are related by the gauging of a particular finite symmetry \cite{Arias-Tamargo:2026urw,Wei:2026fsn}. In the abelian case $G=U(1)^N$, the symmetry to be gauged is a finite abelian group symmetry, as discussed in \cref{sec:abelian_OPEs}. This can be recovered from the TQFT setup via the internal end construction.

As discussed in \cref{sec:2d_descrip}, we begin the internal end construction with a particular $(\cA_L,\cA_R)$-bimodule $B$. Fusion of a boundary line $M^L$ into the interface from the left, or a boundary line $M^R$ from the right, will generically result in a different interface:
\begin{equation}
    \begin{tikzpicture}[scale=0.7,baseline=(current bounding box.center)]
    \fill[gray!15] (-8,-2) rectangle (-2,2);

    \draw[very thick] (-5,-2) -- (-5,2);
    \node[right] at (-5,-1.7) {$B$};
    \node[right] at (-5, 1.7) {$B'$};

    \draw[thick, red] (-7,-2) -- (-5,0);
    \node[above, red] at (-7,-1.8) {$M^{L}$};

    \node[below right] at (-8,2) {$\cS_L$};
    \node[below left] at (-2,2) {$\cS_R$};

    \fill[gray!15] (6,-2) rectangle (0,2);

    \draw[very thick] (3,-2) -- (3,2);
    \node[left] at (3,-1.7) {$B$};
    \node[left] at (3, 1.7) {$B'$};

    \draw[thick, red] (5,-2) -- (3,0);
    \node[above, red] at (5.2,-1.8) {$M^{R}$};

    \node[below right] at (0,2) {$\cS_L$};
    \node[below left] at (6,2) {$\cS_R$};
    \end{tikzpicture}
\end{equation}
The category $\bimod{\cA_L}{\cC}{\cA_R}$ of $(\cA_L,\cA_R)$-bimodules thus admits an action of $\cS_L$ from the left, denoted $B' = M^L \triangleright B$, and a compatible action of $\cS_R$ from the right, denoted $B' = B \triangleleft M^R$, so becomes a $(\cS_L,\cS_R)$-bimodule category. The right internal end algebra $\underline{\End}_{\cS_R}(B)$ of a given bimodule is then formed by boundary lines which can be absorbed by the interface; that is, those $M^R$ for which $B \triangleleft M^R = B$. The left end construction proceeds analogously.

At the level of objects, a bimodule is a double coset $B \in \cA_L \backslash G(\cC) / \cA_R$, and a boundary line is an ordinary coset $M^R \in G(\cC) / \cA_R$. The decomposition of $B$ into ordinary cosets then gives the set of boundary lines $M^R$ which can end on $B$. Hence the right internal end of $B$ is
\begin{equation}\label{eq:abelian_end_general}
    \underline{\End}_{\cS_R}(B) = \frac{\cA_L + \cA_R}{\cA_R} \subseteq G(\cC) / \cA_R
\end{equation}
Since all simple bimodules are found from $B$ by fusion with an invertible line, this result is the same for any simple bimodule.
In terms of concrete data involving the charge matrices, this can be rewritten as the explicit set of lines
\begin{equation}\label{eq:abelian_int_end}
    \underline{\End}_{\cS_R}(B) = \bigoplus_{n \in \cR^{-1} \cG_\cR} M^R_{[(Q_L^T n, n \cdot \rho)]}
\end{equation}
using the first realisation of $\cG_\cR$ in \eqref{eq:abelianGRdef}, which gives $\cR^{-1} \cG_\cR = \bZ^N / (\bZ^N \cap \cR^{-1} \bZ^N)$. This gives an algebra of lines in $\cS_R$ on the right of the defect such that gauging this algebra results in the symmetry $\cS_L$ on the other side of the defect.

The algebra \eqref{eq:abelian_int_end} corresponds to gauging the finite group symmetry $\cG_\cR$ as found already in \cref{sec:abelian_OPEs}. Indeed, we can see that the action of this group is the same as \eqref{eq:abelian_phase} as follows. The local operator $\tilde{\psi}_j$ is attached to the bulk anyon $(Q_R^T e_j, 1) \in \cA_R$, while the generator $\cR e_i \in \cG_\cR$ of the group we have gauged is implemented by the line in \eqref{eq:abelian_int_end} with $n = e_i$. The charge of $\tilde{\psi}_j$ under this line is the full braiding of the corresponding bulk anyons $(Q_R^T e_j, 1)$ and $(Q_L^T e_i, 1)$. The $\SO(2N)_{-1}$ factors play no role since $V$ is transparent, so the full braid comes only from the $U(1)^N_K$ factor,
\begin{equation}
    \exp(2\pi i (Q_L^T e_i)^T K^{-1} (Q_R^T e_j)) = \exp(2\pi i \cR_{ji})
\end{equation}
matching \eqref{eq:abelian_phase}.

\textbf{Example:} In the 3450 model, recall that $\cA_L \cong \bZ_{10}$ and $\cA_R \cong \Z_{10}$, but $\cA_L \cap \cA_R \cong \Z_2$, so the internal end \eqref{eq:abelian_end_general} gives that the symmetry to half-space-gauge is $\Z_5$, as expected. We immediately find the boundary central charge,
\begin{align}
    g = \sqrt{\dim a_R} = \sqrt{5}
\end{align}
where following \cref{sec:2d_descrip} we have identified $a_R = \underline{\End}_{\cS_R}(B)$.

\subsection{SPT class}\label{sec:abeliansptclass}

Let us now turn our attention to the SPT class of the interfaces constructed above preserving $U(1)^N$ symmetries. In \cref{sec:SPTs_general}, a method to determine the class was proposed wherein we expand the SymTFT setup to a larger TQFT as shown in \eqref{eq:SPT_determination}. That is, instead of the Lagrangian algebras in $U(1)^N_K \boxtimes SO(2N)_{-1}$ considered above, we consider Lagrangian algebras in $U(1)^N_K \boxtimes \Spin(2N)_{-1}$. We then ask whether the vector line $V$ of $\Spin(2N)_1$ can end on the bimodule $B$ under consideration. If it can, then the resulting two-dimensional defect is $q$-type and belongs in the axial class.

In \cite{Arias-Tamargo:2026urw}, for $N=2$ complex fermions and a $U(1)^2$-preserving defect, the SPT class of each interface was determined by other methods. Namely, direct orbifold computations were performed keeping track of fermion parity which then indicate whether the assignment of $(-1)^F$ in the Ramond sector differs before and after the orbifold, corresponding to stacking $\Arf$. A generalisation of those methods to $N$ complex fermions is presented in \cref{app:arf_abelian}, the result of which is the following. First we note that there always exist integer vectors $v, w \in \bZ^N$ such that
\begin{equation}\label{eq:useful_result}
    Q_L^T (\rho/2 + v) = Q_R^T (\rho/2 + w)
\end{equation}
Then the SPT class of the interface which preserves a $U(1)^N$ symmetry acting with charges $Q_L$ on the left and $Q_R$ on the right is
\begin{equation}\label{eq:sptclassvw}
    (v + w) \cdot \rho \mod 2
\end{equation}
with 0 indicating the vector class and 1 indicating the axial class.

\textbf{Example:} In the 3450 model we find $Q_L^T \left(\rho/2 + \begin{psmallmatrix} 0 \\ 1 \end{psmallmatrix} \right) = Q_R^T \left(\rho/2 + \begin{psmallmatrix} 1 \\ 0 \end{psmallmatrix} \right)$, thus $\left( \begin{psmallmatrix} 0 \\ 1 \end{psmallmatrix} + \begin{psmallmatrix} 1 \\ 0 \end{psmallmatrix} \right) \cdot \rho = 2$ and the symmetry is in the vector class.

We now demonstrate that this result is recovered by the method introduced in \cref{sec:SPTs_general}. The charge matrices $Q_L, Q_R$ describe group embeddings $U(1)^N \to \Spin(2N)$; this is where we use our assumption that $K$ is even, as foreshadowed in \cref{eq:abeliantqftapproach}. The weight lattices of the $1, V, S, C$ representations of $\Spin(2N)$ are $\bZ^N_\text{even}, \bZ^N_\text{odd}, \bZ^N_\text{even} + \tfrac{\rho}{2}, \bZ^N_\text{odd} + \tfrac{\rho}{2}$ respectively. By applying $Q_L^T$, we obtain the pulled-back charges under $U(1)^N$. This leads to the Lagrangian algebra in $U(1)^N_K \boxtimes \Spin(2N)_{-1}$
\begin{equation}
\begin{split}
    \hat{\cA}_L &= Q_L^T \bZ^N_\text{even} / K\bZ^N \boxtimes 1 \\
    &\oplus Q_L^T \bZ^N_\text{odd} / K\bZ^N \boxtimes V \\
    &\oplus Q_L^T (\bZ^N_\text{even} + \tfrac{\rho}{2}) / K\bZ^N \boxtimes S \\
    &\oplus Q_L^T (\bZ^N_\text{odd} + \tfrac{\rho}{2}) / K\bZ^N \boxtimes C
\end{split}
\end{equation}
The physics of this Lagrangian algebra is that the four rows describe respectively the NS bosons, NS fermions, R bosons, and R fermions, as shown by the following TQFT realisation of the left Weyl fermions:
\begin{equation}
\begin{tikzpicture}[z={(60:0.5cm)},baseline=(current bounding box.center)]
    \newcommand{\xa}{2}
    \newcommand{\xb}{2.5}
    \newcommand{\xc}{4.5}
    \newcommand{\y}{5}

    \shade[right color=green,left color=white,middle color=green,opacity=0.1] (0,0,0) -- ++(0,-\y,0) -- ++(\xa,0,0) -- ++(0,0,1) -- ++(0,\y,0) -- ++(-\xa,0,0) -- cycle;
    \fill[blue,opacity=0.1] (\xa,0,0) -- ++(0,-\y,0) -- ++(\xb+\xc,0,0) -- ++(0,0,1) -- ++(0,\y,0) -- ++(-\xb-\xc,0,0) -- cycle;
    \filldraw[black,fill=blue,fill opacity=0.1] (\xa+\xb+\xc,0,0) -- ++(0,-\y,0) -- ++(0,0,1) -- ++(0,\y,0) -- cycle;
    \filldraw[black,fill=blue,fill opacity=0.1] (\xa+\xb,0,0) -- ++(0,-\y,0) -- ++(0,0,1) -- ++(0,\y,0) -- cycle;
    \draw[black] (\xa,0,0) -- ++(0,-\y,0) -- ++(0,0,1) -- ++(0,\y,0) -- cycle;
    \draw[black] (0,0,0) -- ++(\xa+\xb+\xc,0,0);
    \draw[black] (0,-\y,0) -- ++(\xa+\xb+\xc,0,0);
    \draw[black] (0,0,1) -- ++(\xa+\xb+\xc,0,0);
    \draw[black,opacity=0.1] (0,-\y,1) -- ++(\xa+\xb+\xc,0,0);

    \node[above] at (\xa/2,0,1) {$\SO(2N)_1$};
    \node[below] at (\xa,-\y,0) {$\bra{V}\vphantom{\hat{\cA}}$};
    \node[above] at (\xa+\xb/2,0,1) {$\Spin(2N)_1$};
    \node[below] at (\xa+\xb,-\y,0) {$\langle\hat{\cA}_L|$};
    \node[above] at (\xa+\xb+\xc/2,0,1) {$U(1)^N_K$};

    \begin{scope}[yshift=-0.25cm]
        \foreach \i in {1,2,3,4} {
            \begin{scope}[every node/.style={fill=red,shape=circle,inner sep=1pt}]
                \draw[thick,red] (0,-\i,0.5) -- ++(\xa,0,0) node{} -- ++(\xb,0,0) node{} -- ++(\xc,0,0) node{};
            \end{scope}
        }

        \node[right,xshift=0.2cm] at (\xa+\xb+\xc,-1,0.5) {NS bosons};
        \node[right,xshift=0.2cm] at (\xa+\xb+\xc,-2,0.5) {NS fermions};
        \node[right,xshift=0.2cm] at (\xa+\xb+\xc,-3,0.5) {R bosons};
        \node[right,xshift=0.2cm] at (\xa+\xb+\xc,-4,0.5) {R fermions};

        \node[above] at (\xa/2,-1,0.5) {$1$};
        \node[above] at (\xa/2,-2,0.5) {$V$};
        \node[above] at (\xa/2,-3,0.5) {$1$};
        \node[above] at (\xa/2,-4,0.5) {$V$};

        \node[above] at (\xa+\xb/2,-1,0.5) {$1$};
        \node[above] at (\xa+\xb/2,-2,0.5) {$V$};
        \node[above] at (\xa+\xb/2,-3,0.5) {$S$};
        \node[above] at (\xa+\xb/2,-4,0.5) {$C$};

        \node[above] at (\xa+\xb+\xc/2,-1,0.5) {$Q_L^T \bZ^N_\text{even} / K \bZ^N$};
        \node[above] at (\xa+\xb+\xc/2,-2,0.5) {$Q_L^T \bZ^N_\text{odd} / K \bZ^N$};
        \node[above] at (\xa+\xb+\xc/2,-3,0.5) {$Q_L^T (\bZ^N_\text{even} + \tfrac{\rho}{2}) / K \bZ^N$};
        \node[above] at (\xa+\xb+\xc/2,-4,0.5) {$Q_L^T (\bZ^N_\text{odd} + \tfrac{\rho}{2}) / K \bZ^N$};

        \foreach \i in {3,4} {
            \draw[red] (\xa,-\i,0) node[black,below left,xshift=3pt,yshift=6pt] {$\scriptscriptstyle (-1)^F$} -- ++(0,0,0.5);
        }
    \end{scope}
\end{tikzpicture}
\end{equation}
Here we have arbitrarily chosen $(-1)^F$ to be the $S$ line of $\Spin(2N)_1$ pushed to the boundary; making the opposite choice $C$ would correspond to exchanging the R sector bosons and fermions. Entirely analogous results hold for $\hat{\cA}_R$ with $Q_L$ replaced by $Q_R$.

Firstly, we need to find all the simple $(\hat{\cA}_L, \hat{\cA}_R)$-bimodules. The Lagrangian algebras $\hat{\cA}_{L,R}$ can be viewed as subgroups of the abelian group
\begin{equation}
    G(\widehat{\cC}) = \bZ^N / K \bZ^N \times
    \begin{cases}
        \bZ_4 & N \text{ odd} \\
        \bZ_2 \times \bZ_2 & N \text{ even}
    \end{cases}
\end{equation}
controlling the fusion rules of the folded theory $\widehat{\cC} = U(1)^N_K \boxtimes \Spin(2N)_{-1}$ (see \cref{app:MTCs}). As before, simple bimodules are double cosets, and without loss of generality we may focus on the canonical bimodule containing the zero element,
\begin{equation}
    B = \hat{\cA}_L + \hat{\cA}_R \subseteq G(\widehat{\cC})
\end{equation}
To determine the SPT class of $B$, we need to determine whether this subgroup contains $0 \boxtimes V$. To answer this, first note that
\begin{equation}\label{eq:twolines}
\begin{aligned}
    Q_L^T (v + \tfrac{\rho}{2}) \boxtimes SV^{v \cdot \rho} &\in \hat{\cA}_L \\
    Q_R^T (w + \tfrac{\rho}{2}) \boxtimes SV^{w \cdot \rho} &\in \hat{\cA}_R
\end{aligned}
\end{equation}
From \eqref{eq:useful_result}, taking the difference gives
\begin{equation}
    0 \boxtimes V^{(v + w) \cdot \rho} \in \hat{\cA}_L + \hat{\cA}_R
\end{equation}
Thus we learn that in the axial case, $\hat{\cA}_L + \hat{\cA}_R$ does indeed contain $0 \boxtimes V$. Conversely, in the vector case, the two lines in \eqref{eq:twolines} are the same line, which therefore belongs to $\hat{\cA}_L \cap \hat{\cA}_R$, hence braids trivially with every line in $\hat{\cA}_L + \hat{\cA}_R$. This rules out $0 \boxtimes V$ belonging to $\hat{\cA}_L + \hat{\cA}_R$, because $0 \boxtimes V$ braids non-trivially with \eqref{eq:twolines} since the braiding of $S$ and $V$ gives $-1$. This establishes that
\begin{equation}
    0 \boxtimes V \in B \iff (v+w) \cdot \rho = 1 \mod 2
\end{equation}
which states that the TQFT criterion for the SPT class agrees with the criterion \eqref{eq:sptclassvw}, as claimed.

\section{A non-abelian example}\label{sec:non-abelian}

We now apply our machinery to a family of examples where the preserved symmetry is non-abelian, not extendable to a symmetry of equal rank, but where the theory is still rational with respect to the symmetry. As opposed to the abelian case above, no alternative construction of symmetry preserving boundaries was previously known.

\subsection{The model, and some motivation}\label{sec:mpole}

We consider $(N+1)^2$ right-moving Weyl fermions. As always, our aim is to construct topological defects which preserve some $G_k$ symmetry which acts in a different way on the left and on the right. In particular, we take
\begin{equation}\label{eq:Gk_def}
    G_k = SU(N)_{N+2} \times SU(N+2)_N \times U(1)_{2N(N+1)} \times U(1)_{2(N+1)(N+2)}
\end{equation}
The $(N+1)^2$ independent fermions are assembled into four species which transform in the following representations of $G_k$,
\begin{align}\label{eq:nonabelian_reprs}
\begin{array}{c|cccc}
    &\, \psi_{1}\, &\, \psi_{2}\,  &\, \tilde\psi_1 \, & \tilde\psi_2 \\ \hline
    SU(N)_{N+2} 	& \ydiagram{2} 		& 1  	& \ydiagram{1}   & 1  \\
    SU(N+2)_{N}		& 1	 & \ydiagram{1,1}	& \ydiagram{1} & 1 \\
    U(1)_{2N(N+1)} & 2 & 0 & 1 & -N \\
    U(1)_{2(N+1)(N+2)} 				& 0  	&  2		& 1  & N+2
\end{array}
\end{align}
where $\psi_{1,2}$ live on the left and $\tilde\psi_{1,2}$ on the right of the defect. Explicitly, this matter content corresponds to the following embeddings of $G_k$ into $SO(2(N+1)^2)_1$,
\begin{align}\label{eq: non-abelian L and R}
    L &= (\,\ydiagram{2}\,\,,1 )_{2,0} \,\oplus\,(1, \,\ydiagram{1,1}\, )_{0,2} \nn \\
    R &= (\,\ydiagram{1}\,\,,\,\ydiagram{1}\, )_{1,1} \,\oplus \, (1, 1 )_{-N,N+2}
\end{align}
which are indeed conformal and thus our general story applies. However we are also in the non-equal rank case, with $\text{rank}(G_k) = 2(N+1)$ strictly less than $\text{rank}(SO(2(N+1)^2)_1) = (N+1)^2$, and so the desired interfaces are not accessible through abelian boundary states \cite{BoyleSmith:2019jnh}.

The folded version of this theory appeared previously in \cite{Tong:2021phe}. Indeed, this reference proposed a method of symmetric mass generation in this folded version, which if valid could in principle be used to dynamically generate a $G_k$-preserving conformal boundary following the same logic explored in the abelian case in \cite{Mouland:2025dct,Arias-Tamargo:2026urw}.

Interestingly, this theory arises as the two-dimensional BCFT modelling the low energy scattering of fermions off a 't Hooft line in a particular four-dimensional abelian gauge theory. This theory consists of four left-handed complex Weyl fermions coupled to a $U(1)$ gauge field and transforming as follows under an $H=SU(N)\times SU(N+2)\times U(1)$ global symmetry,
\begin{align}\label{eq: 4d theory general}
\begin{array}{c|cccc}
    &\Psi_{1}&\Psi_{2} & \Psi_3 & \Psi_4 \\ \hline
    U(1)_\text{gauge} 	& 1 		& 1  	& -1   & -1  \\
    SU(N)				& \ydiagram{2}	 & 1	& \overline{\ydiagram{1}} & 1 \\
    SU(N+2) & 1 & \ydiagram{1,1} & \overline{\ydiagram{1}} & 1 \\
    U(1) 				& 1  	&  -1		& 0   & N+1   \\\hdashline
    U(1)_\text{bonus} & 1 & 0 & 0 & -\frac{N(N+1)}{2}
\end{array}
\end{align}
We also give charges under an additional global symmetry $U(1)_\text{bonus}$. The ABJ anomaly of $H\times U(1)_\text{bonus}$ vanishes. Following the general procedure of \cite{vanBeest:2023dbu}, one can check that the problem of scattering low energy fermions off the 't Hooft line of unit magnetic charge reduces to finding a conformal boundary condition for $(N+1)^2$ two-dimensional Dirac fermions that preserves the symmetry \eqref{eq: non-abelian L and R}, where $U(1)_\text{gauge}\times H$ becomes $G_k$ after a suitable change of basis. In particular, all excitations of this two-dimensional effective theory are singlets under the $SU(2)_\text{rot}$ rotational symmetry about the 't Hooft line.

In contrast, the additional symmetry $U(1)_\text{bonus}$ is broken by the 't Hooft line, due to its 't Hooft anomaly in two dimensions, or equivalently its non-vanishing 2-group coefficient with the magnetic 1-form symmetry in four dimensions \cite{vanBeest:2023dbu}. Regular local fermions are scattered into outgoing excitations that generically carry fractional charge under $U(1)_\text{bonus}$.

Finally, note that in the special case of $N=2$ the theory in fact arises as the two-dimensional effective description of scattering off a 't Hooft line in a \textit{different} abelian gauge theory. The theory consists again of four left-handed Weyl fermions coupled to a $U(1)$ gauge field, and carrying charges under global symmetry $H=SU(4) \times U(1)$,
\begin{align}\label{eq: 4d theory lucky}
\begin{array}{c|cccc}
    &\Psi_{1}&\Psi_{2} & \Psi_3 & \Psi_4 \\ \hline
    U(1)_\text{gauge} 	& 3 		& 1  	& -2   & -1  \\
    SU(4) & 1 & \ydiagram{1,1} & \overline{\ydiagram{1}} & 1 \\
    U(1) 				& 0  	& 2 		& -1   & 4  \\ \hdashline
    U(1)_\text{bonus}	& 1  	& 2 		& -1   & -5
\end{array}
\end{align}
Again, all ABJ anomalies for $H\times U(1)_\text{bonus}$ vanish. Once more considering the 't Hooft line of unit charge, one finds once again the theory \eqref{eq: non-abelian L and R} as a low energy effective description of fermion-monopole scattering. The difference here is that fermions in this two-dimensional description sit in non-trivial representations of $SU(2)_\text{rot}$, and indeed it is $SU(2)_\text{rot}\times U(1)_\text{gauge}\times  H$ that is identified with $G_k$. Once again, we have the bonus symmetry $U(1)_\text{bonus}$, charge under which is fractionalised by generic scattering.

\subsection{Twist lines from OPEs}\label{sec:nonabelian_OPEs}

As a first approach to the problem, we would like to determine some properties of the twist lines that get attached to the fermions from the OPEs, as explained in \cref{sec:monodromies} and analogously to what we did for the abelian case in \cref{sec:abelian_OPEs}. For simplicity, we set $N=2$, as it is enough to illustrate the problems that arise in the non-abelian case.

To compute the OPEs, we need two ingredients. The first are the conformal dimensions of each operator, which are straightforwardly determined by the representations of the symmetry group that the fermions live in. One readily checks that
\begin{equation}
    h_{\psi_1}=h_{\psi_2}=h_{\tilde{\psi}_1}=h_{\tilde{\psi}_2}=\frac{1}{2}
\end{equation}
The second are the fusion rules of the relevant representations, which give rise to the fusion channels in the OPE. This is where the difference with the abelian case arises. If we consider scattering $\tilde{\psi}_2$ towards the defect, things are easy as it's only charged under an abelian subgroup of $G_k$. The OPEs then give
\begin{align}
\begin{split}\label{eq:psi2tilde_OPEs}
    \psi_1(z)\tilde{\psi}_2(0) &\sim z^{-1/3}\cO_{(\,\ydiagram{2}\,,1,0,4)}(0) \\
    \psi_2(z)\tilde{\psi}_2(0) &\sim z^{1/3}\cO_{(1\,,\,\ydiagram{1,1}\,,-2,6)}(0)
\end{split}
\end{align}
where we index the operator $\cO$ in the OPE by the representation of $G_k$ it transforms in. The action of the twist line on $\psi_1$ and $\psi_2$ can be probed by studying the monodromy under $z\to e^{2\pi i}z$, giving a phase $\exp(\pm2\pi i/3)$ in either case. This OPE is consistent with a monodromy caused by a $\bZ_3$ line, which can be further checked by computing the OPE involving $(\tilde{\psi}_2)^2$ and $(\tilde{\psi}_2)^3$, for example, although we won't do so here.

On the other hand, if we scatter $\tilde{\psi}_1$, there are several allowed fusion channels, leading to
\begin{align}
\begin{split}\label{eq:psi1tilde_OPEs}
    \psi_1(z)\tilde{\psi}_1(0) &\sim z^{-1/6} \cO_{(\,\ydiagram{1}\,,\,\ydiagram{1}\,,3,1)}(0) + z^{1/3} \cO_{(\,\ydiagram{3}\,,\,\ydiagram{1}\,,3,1)}(0) \\
    \psi_2(z) \tilde{\psi}_1(0)&\sim  z^{-1/3} \cO_{(\,\ydiagram{1}\,,\,\ydiagram{1,1,1}\,,1,3)}(0) + z^{1/6} \cO_{(\,\ydiagram{1}\,,\,\ydiagram{2,1}\,,1,3)}(0)
\end{split}
\end{align}
The immediate consequence of this result is that the twist lines cannot be generators of an invertible symmetry. Studying further properties of the twist lines (such as their fusion or $F$-symbols) from the two-dimensional perspective becomes inaccessible, which is why in the non-abelian case the three-dimensional setup becomes necessary. As anticipated in \cref{sec:monodromies}, the monodromies arise as braiding of anyons.

\subsection{TQFT description}

We now proceed to apply the strategy outlined in \cref{sec:theory}, namely to identify all the bulk anyons and the possible Lagrangian algebras, in order to later compute the bimodules that classify the symmetry-preserving defects. The algorithm to do this is straightforward and can in principle be brute-forced, however in our example of interest there are some simplifications available that we can use to make our lives easier.

Firstly, instead of working directly with the chiral algebra $G_k$ in \eqref{eq:Gk_def}, we start by noting that the $n$-alities of the representations occurring in the embeddings $L, R$ in \eqref{eq: non-abelian L and R} allow for combining $SU(N)_{N+2}\times U(1)_{2N(N+1)}$ into $U(N)$ and $SU(N+2)_N \times U(1)_{2(N+1)(N+2)}$ into $U(N+2)$. To work out their levels, we use the definition \cite{hsin2016levelrank} that $U(N)_{K,K+NK'} = (SU(N)_K \times U(1)_{N(K+NK')}) / \bZ_N$. Therefore, we can replace the symmetry group $G_k$ with the quotient
\begin{equation}\label{eq:Gk_unitary}
    \tilde{G}_k = U(N)_{N+2,2(N+1)} \times U(N+2)_{N,2(N+1)}
\end{equation}
Since both groups are connected, a boundary preserving $\tilde{G}_k$ will automatically preserve $G_k$ as they have the same Lie algebra, and the exponential is surjective into the full group.
As this symmetry group is smaller, it will be simpler to work with. In particular, its Chern--Simons theory will have fewer lines.

In general, the $U(N)_{K,K+NK'}$ Chern--Simons theory is bosonic for $K+K'$ even and fermionic for $K+K'$ odd. Thus both factors in \eqref{eq:Gk_unitary} are fermionic for $N$ even and bosonic for $N$ odd. Since we are particularly interested in the minimal case $N = 2$, we will therefore restrict to $N$ even from now on for simplicity.

Secondly, we invoke level-rank duality \cite{Naculich:1990pa,Nakanishi:1990hj,hsin2016levelrank}
\begin{equation}\label{eq:levelrank1}
    U(N)_{K,K+N} \leftrightarrow \overline{U(K)_{N,N+K}} \times SO(2(KN+1))_1
\end{equation}
which allows us to rewrite the bulk Chern--Simons theory as
\begin{equation}
    \tilde{G}_k = U(N)_{N+2,2(N+1)} \times \overline{U(N)_{N+2,2(N+1)}} \times SO(2(N+1)^2)_1
\end{equation}
This has the advantage that the same TQFT now occurs twice. To summarise, we have the following three-dimensional setup for the Weyl fermions,
\begin{equation}
\begin{tikzpicture}[scale=0.9,z={(30:2cm)},baseline=(current bounding box.center)]
    \newcommand{\x}{13}
    \newcommand{\y}{6}

    \shade[top color=green,bottom color=white,middle color=green,opacity=0.1] (0,-\y/2,0) -- ++(0,-\y/2,0) -- ++(\x,0,0) -- ++(0,0,1) -- ++(0,\y/2,0) -- ++(-\x,0,0) -- cycle;
    \fill[blue,opacity=0.1] (0,0,0) -- ++(0,-\y/2,0) -- ++(\x,0,0) -- ++(0,0,1) -- ++(0,\y/2,0) -- ++(-\x,0,0) -- cycle;
    \draw[black] (0,-\y/2,0) -- ++(\x,0,0) -- ++(0,0,1) -- ++(-\x,0,0) -- cycle;
    \draw[black] (0,0,0) -- ++(\x,0,0) -- ++(0,0,1) -- ++(-\x,0,0) -- cycle;
    \filldraw[black,fill=blue,fill opacity=0.1] (0,0,0) -- ++(\x,0,0) -- ++(0,0,1) -- ++(-\x,0,0) -- cycle;
    \draw[black] (0,0,0) -- ++(0,-\y,0);
    \draw[black] (\x,0,0) -- ++(0,-\y,0);
    \draw[black] (\x,0,1) -- ++(0,-\y,0);
    \draw[black,opacity=0.1] (0,0,1) -- ++(0,-\y,0);
    \draw[thick,black] (\x/2,-\y/2,0) -- ++(0,0,1);

    \node at (\x/2,-\y*0.85,0.5) {$\SO(2(N+1)^2)_1$};
    \node at (\x/2-0.2,-\y*0.25,0.5) {$U(N)_{N+2,2(N+1)} \boxtimes \overline{U(N)_{N+2,2(N+1)}} \boxtimes \SO(2(N+1)^2)_1$};

    \node at (\x*0.25,-\y/2,0.5) {$\ket{\cA_L}$};
    \node at (\x*0.75,-\y/2,0.5) {$\ket{\cA_R}$};
    \node[below] at (\x*0.5,-\y/2,0) {$B$};

    \draw[->,thick] (\x*0.4,0,0.5) -- (\x*0.6,0,0.5);
\end{tikzpicture}
\end{equation}
The theory $U(N)_{N+2,2(N+1)} \boxtimes \overline{U(N)_{N+2,2(N+1)}}$ has central charge zero and contains Lagrangian algebras $\cA_{L,R}$ corresponding to the representations $L, R$ of the fermions. Condensing these algebras yields topological boundaries $\ket{\cA_{L,R}}$, which we can place along the interface between the upper and lower regions to yield the Weyl fermions transforming in the $L$ and $R$ realisations of the symmetry upon compactifying the upper region.

\subsubsection*{The Lagrangian algebras}\label{sec:lagr_algs}

To progress, we need to know the Lagrangian algebras $\cA_{L,R}$. They are found from the branching rules of $\1\oplus V$ in $\SO(2(N+1)^2)_1$ under the embeddings $L,R$. We defer this computation to \cref{app:lagr}. Let us introduce the definition
\begin{equation}
    \cC = U(N)_{N+2,2(N+1)}
\end{equation}
so that our Lagrangian algebras belong to $\cC \boxtimes \overline{\cC}$. Since $\cC$ is fermionic, we recall that in the product theory we must identify the transparent fermion $f$ of both copies of $\cC$.

We find that $\cA_L$ is a product of separate Lagrangian algebras in $\cC$ and $\overline{\cC}$ which turn out to be the same, namely
\begin{equation}\label{eq:AL}
    \cA_L = \cA \boxtimes \overline{\cA}
\end{equation}
for a Lagrangian algebra $\cA$ in $\cC$.

On the other hand, $\cA_R$ is the diagonal (or canonical) Lagrangian algebra of $\cC \boxtimes \overline{\cC}$. For a fermionic TQFT $\cC$, this takes the form
\begin{equation}\label{eq:AR}
    \cA_R = \text{Diag}(\cC) = \bigoplus_{c \in \cC/f} c \boxtimes \overline{c}
\end{equation}
where the sum is over all simples of $\cC$ modulo the transparent fermion line, $f$. The reason for this is to avoid double counting, since $f \boxtimes \overline{f}$ is trivial. Note that fusion with $f$ has no fixed points, since it increases the spin of a line by $\frac{1}{2}$.

\subsubsection*{Example: $N = 2$}

We first give a detailed discussion of the minimal case $N=2$, so $\cC = U(2)_{4,6}$. The objects of $\cC$ are labelled by pairs $(a, b)$, where $0 \leq a \leq 4$ labels an object of $SU(2)_4$, and $0 \leq b \leq 11$ labels an object of $U(1)_{12}$, subject to the constraint $a = b \mod 2$. There are therefore $(5 \times 12) / 2 = 30$ objects. The quantum dimensions of the anyons are
\begin{equation}\label{eq:N=2_dims}
    d_{(a, b)} = \frac{\sin(2 \pi (a + 1) / 12)}{\sin(2 \pi / 12)}
\end{equation}
and their topological spins are
\begin{equation}\label{eq:N=2_spins}
    \theta_{(a, b)} = \exp( 2\pi i \frac{a(a+2) + b^2}{24} )
\end{equation}
The total quantum dimension of $\cC$ is $\cD(\cC) = \sqrt{\sum_a d_a^2} = 6\sqrt{2}$. We give more anyon data of the $SU(2)_k$ and $U(1)_k$ Chern--Simons theories in \cref{app:MTCs}.

The object $f = (4, 6)$ is a transparent fermion line, so $\cC$ is a fermionic TQFT. Fusion with $f$ acts as $(a, b) \otimes f = (4 - a, b + 6)$. Recall from \cref{sec:SPTs_general} that when we form the stacked theory $\cC \boxtimes \overline{\cC}$, the two transparent fermion lines are identified, $(f, \overline{f}) \sim (0, 0)$, so $\cC \boxtimes \overline{\cC}$ has $30^2/2 = 450$ objects.

The structure of the Lagrangian algebras $\cA_L$ and $\cA_R$ in this case are as follows. $\cA_L$ is a product $\cA \boxtimes \overline{\cA}$ where
\begin{equation}\label{eq:A}
    \cA = (0, 0) \oplus (2, 4) \oplus (2, 8) \oplus (4, 0) \in \cC
\end{equation}
is a Lagrangian algebra in $\cC$. Indeed, \eqref{eq:N=2_spins} implies that all lines in $\cA$ have $\theta_{(a,b)}=1$ and \eqref{eq:N=2_dims} gives $\dim(\cA) = 6$, satisfying \eqref{eq:fermionic_lagr}. The unique gaugeable algebra multiplication on the object $\cA$ (up to isomorphism) is
\begin{equation}
    \begin{array}{c|cccc}
          & x & y & z \\ \hline
        x & 2^{1/4} y & -\mathbf{1}+z & x \\
        y & -\mathbf{1}-z & 2^{1/4} x & -y \\
        z & -x & y & \mathbf{1}
    \end{array}
\end{equation}
where $\mathbf{1} = (0,0)$, $x=(2,4)$, $y=(2,8)$, $z=(4,0)$.\footnote{The fact that this algebra is gaugeable is verified in the ancillary file \texttt{category\_examples.nb}.} The product algebra $\cA \boxtimes \overline{\cA}$ is then a sum of 16 distinct simple lines, all of which have multiplicity 1.

$\cA_R$ is the diagonal Lagrangian algebra of $\cC \boxtimes \overline{\cC}$,
\begin{equation}
    \cA_R = \bigoplus_{c \in \cC/f} c \boxtimes \overline{c}
\end{equation}
which is a sum of 15 distinct lines, all of which also have multiplicity 1. The spins of objects in $\cA_R$ are automatically zero, and its quantum dimension is $\dim(\cA_R) = \cD(\cC)^2 / 2 = \cD(\cC \boxtimes \overline{\cC}) / \sqrt{2}$, in accordance with \eqref{eq:fermionic_lagr}. In a natural choice of basis, the algebra multiplication on $\cA_R$ is particularly simple. Namely, all junctions are either $\pm1$ depending solely on the bosonic/fermionic statistics.

Since the gaugeable algebra structure is unique for both $\cA_L$ and $\cA_R$, we will work mainly at the level of the algebra objects in the following.

\subsection{Two-dimensional symmetry categories}\label{sec:SLSRnonabelian}

We now discuss the symmetries of the boundaries $\ket{\cA_L}$ and $\ket{\cA_R}$ in detail. Recall the Lagrangian algebras in $\cC \boxtimes \overline{\cC}$ given in \eqref{eq:AL} and \eqref{eq:AR}. On the $\ket{\cA_{L,R}}$ boundary, the algebra $\cA_{L,R}$ is condensed. As discussed in \cref{sec:CS_basics}, the symmetries of the boundary theory are generated by topological lines described by $\cA_{L,R}$ modules over $\cC\boxtimes\overline{\cC}$. That is, the boundary symmetry categories are
\begin{equation}
    \cS_L = (\cC \boxtimes \overline{\cC})_{\cA_L} \qquad \cS_R = (\cC \boxtimes \overline{\cC})_{\cA_R}
\end{equation}
We start by simplifying the relevant symmetry categories. First of all, on the left of the interface, the categories describing the anyons and the algebra to be gauged are both products, so we have\footnote{Let us briefly clarify our notation. Recall from \cref{sec:CS_basics} that for a Chern--Simons theory $\cC$ we denote its orientation reversal by $\overline{\cC}$, which has the effect of reversing its braiding. Given a Lagrangian algebra $\cA$ in $\cC$, there is a Lagrangian algebra $\overline{\cA}$ in $\overline{\cC}$ which is simply $\cA$ viewed as an algebra in $\overline{\cC}$. For example, one of the conditions for $\overline{\cA}$ to be Lagrangian is that it is braided-commutative, which follows from the braided-commutativity of $\cA$. In \eqref{eq:SL} we have then used that $(\cC\boxtimes\overline{\cC})_{\cA\boxtimes\overline{\cA}} = \cC_\cA \boxtimes \overline{\cC}_{\overline{\cA}}$. The only difference between $\cC_\cA$ and $\overline{\cC}_{\overline{\cA}}$ is that the fusion of the latter has its order reversed. This is because the fusion of $\cA$-modules makes use of the braiding of $\cC$, which is reversed in $\overline{\cC}$. As in \eqref{eq:SL}, we will write $\overline{\cC}_{\overline{\cA}} = \overline{\cC_\cA}$. We note that this is different to the use of the bar on modular tensor categories (e.g.\ $\overline{\cC}$) which signifies a reversal of the braiding. Since the module category $\cC_\cA$ is not braided this should not lead to confusion.}
\begin{equation}\label{eq:SL}
    \cS_L = (\cC \boxtimes \overline{\cC})_{\cA \boxtimes \overline{\cA}} = \cC_\cA \boxtimes \overline{\cC_\cA}
\end{equation}
The $\pi$ line of the fermionic symmetry category $\cC_\cA$ is found by pushing the transparent fermion $f$ of $\cC$ to the boundary (see \cref{sec:SPTs_general}). Moreover, in $\cC_\cA \boxtimes \overline{\cC_\cA}$ we must associate the $\pi$ lines of both factors.

On the right of the defect, instead we have
\begin{equation}\label{eq:SR}
    \cS_R = (\cC \boxtimes \overline{\cC})_{\text{Diag}(\cC)} \cong \cC
\end{equation}
This equivalence corresponds to the fact that unfolding the $\ket{\Diag(\cC)}$ boundary of $\cC \boxtimes \overline{\cC}$ gives the identity interface in $\cC$. Therefore the boundary lines of the folded theory simply correspond to lines in $\cC$ lying along the identity interface after unfolding. This is described in detail in the following subsection and \cref{app:folding_properly}.

\subsubsection*{Example: $N = 2$}

Returning to our minimal example $N=2$, here we will show that the boundary symmetries found when $\cA$ is condensed in $\cC$ in this particular example realise a fermionic extension of the Tambara--Yamagami category $\text{TY}(\bZ_3)$. The symmetry category on the left of the defect in \eqref{eq:SL} is two copies of this.\footnote{The results presented below are computed explicitly in the ancillary file \texttt{category\_examples.nb}.}

Condensing $\cA$ in $\cC=U(2)_{4,6}$ gives boundary symmetries described by the category $\cC_\cA$ of $\cA$ modules, which has 8 simple objects. All of these can be induced from lines in $\cC$. That is, each of the simple boundary lines can be found by pushing a simple bulk line $a\in\cC$ to the boundary, with the induced modules as in \eqref{eq:induced_module}. In particular, taking the bulk lines $(0,2r)$ for $r=0,\dots,5$, as well as $(1,1)$ and $(1,3)$ in $\cC$ gives simple induced modules $\text{Ind}_\cA(a)$. There are other lines in $\cC$ which give non-simple induced modules but in this case they can all be written as sums of the simple ones above, so no new simples are induced from these bulk lines. For example, $\text{Ind}_\cA((2,0)) = \text{Ind}_\cA((0,4))\oplus \text{Ind}_\cA((0,8))$.

Let us denote these simple modules by
\begin{equation}
    M_r = \text{Ind}_\cA( (0,2r) ) \qc X = \text{Ind}_\cA ( (1,1) ) \qc  Y = \text{Ind}_\cA ( (1,3) )
\end{equation}
Their fusion can be computed easily from the fusion in $\cC$ using that
\begin{equation}
    \text{Ind}_\cA(c_i) \otimes_\cA \text{Ind}_\cA(c_j) = \text{Ind}_\cA(c_i\otimes c_j)
\end{equation}
for induced modules. This relation can be shown simply at the level of objects from \eqref{eq:induced_module} using that the relative tensor product satisfies $\cA \otimes_\cA \cA = \cA$. The fusion rules are found to be
\begin{align}
\begin{split}\label{eq:CA_fusion}
    M_r \otimes_\cA M_{r'} &= M_{r+r'\mod 6} \\
    M_r \otimes_\cA X &= \begin{cases} X , & r \text{ even} \\ Y , & r \text{ odd} \end{cases} \\
    M_r \otimes_\cA Y &= \begin{cases} Y , & r \text{ even} \\ X , & r \text{ odd} \end{cases} \\
    X \otimes_\cA X &= Y \otimes_\cA Y = M_1 \oplus M_3 \oplus M_5 \\
    Y \otimes_\cA X &= X \otimes_\cA Y = M_0 \oplus M_2 \oplus M_4
\end{split}
\end{align}
That is, the lines $M_r$ are invertible and generate $\Z_6$, then there are two non-invertible lines $X,Y$ which are exchanged by $M_1$. The quantum dimensions are
\begin{equation}
    d_{M_r} = 1 \qc d_X = d_Y = \sqrt{3}
\end{equation}
As the condensed algebra is fermionic, we expect $\cC_\cA$ to admit the structure of a fermionic symmetry category, and so should admit a $\pi$ line. It can be identified as the image of the transparent fermion,
\begin{equation}
    \pi = \text{Ind}_\cA((4,6)) = M_3
\end{equation}
We then note that the objects can be grouped into $M_0=\1$, $M_2=g$, $M_4=g^2$, $X$ and their images under fusion with $\pi$, $M_1=\pi g^2$, $M_3=\pi$, $M_5=\pi g$, $Y=\pi X$. Furthermore, $g$ generates $\bZ_3$ and interestingly, we find the fusion rules
\begin{equation}\label{eq:X2_fusion}
    X^2 = \pi(1 + g + g^2)
\end{equation}
which is the usual $\bZ_3$ Tambara--Yamagami fusion relation up to the extra $\pi$. Note that there are no $q$-type objects. This is a fermionic extension of $\text{TY}(\bZ_3)$.

\subsection{Canonical choice of simple bimodule}\label{sec:unfolding}

While above we have only constructed $\cC_\cA$, rather than $\cS_L$ or $\cS_R$, we will now see that all the relevant information can be encoded in the $\cA$-modules and $(\cA,\cA)$-bimodules over $\cC$.

Our goal is to construct interfaces $B$ between the two gapped boundaries $\ket{\cA_{L,R}}$,
\begin{equation}\label{eq:original_bimod}
\begin{tikzpicture}[z={(30:2cm)},baseline=(current bounding box.center),scale=0.7]
    \newcommand{\x}{8}
    \newcommand{\y}{6}

    \shade[top color=white,bottom color=blue,middle color=blue,opacity=0.1] (0,0,0) -- ++(0,-\y/2,0) -- ++(\x,0,0) -- ++(0,0,1) -- ++(0,\y/2,0) -- ++(-\x,0,0) -- cycle;
    \filldraw[black,fill=blue,fill opacity=0.1] (0,-\y/2,0) -- ++(\x,0,0) -- ++(0,0,1) -- ++(-\x,0,0) -- cycle;

    \draw[black] (0,0,0) -- ++(0,-\y/2,0);
    \draw[black] (\x,0,0) -- ++(0,-\y/2,0);
    \draw[black] (\x,0,1) -- ++(0,-\y/2,0);
    \draw[black,opacity=0.1] (0,0,1) -- ++(0,-\y/2,0);

    \draw[very thick,black] (\x/2,-\y/2,0) -- ++(0,0,1);

    \node at (\x/2,-\y*0.25,0.5) {$\cC\boxtimes\overline{\cC}$};

    \node[below] at (\x*0.25,-\y/2,0) {$\ket{\cA\boxtimes\overline{\cA}}$};
    \node[below] at (\x*0.75,-\y/2,0) {$\ket{\Diag(\cC)}$};
    \node[below] at (\x*0.5,-\y/2,0) {$B$};
\end{tikzpicture}
\end{equation}
To do this we will first unfold the picture, using the fact that the diagonal topological boundary unfolds to the identity interface of $\cC$,
\begin{equation}\label{eq:unfold_bimod}
    \begin{tikzpicture}[z={(30:2cm)},baseline=(current bounding box.center),scale=0.65]
        \newcommand{\x}{8}
        \newcommand{\y}{5}

        \shade[top color=white,bottom color=white,middle color=blue,opacity=0.2] (0,0,0) -- ++(0,-\y,0) -- ++(\x,0,0) -- ++(0,0,1) -- ++(0,\y,0) -- ++(-\x,0,0) -- cycle;
        \filldraw[black,fill=blue,fill opacity=0.1] (0,-\y/2,0) -- ++(\x/2,0,0) -- ++(0,0,1) -- ++(-\x/2,0,0) -- cycle;

        \draw[black] (0,0,0) -- ++(0,-\y,0);
        \draw[black] (\x,0,0) -- ++(0,-\y,0);
        \draw[black] (\x,0,1) -- ++(0,-\y,0);
        \draw[black,opacity=0.1] (0,0,1) -- ++(0,-\y,0);

        \draw[very thick,black] (\x/2,-\y/2,0) -- ++(0,0,1);

        \node at (\x/2,-\y*0.25,0.5) {$\cC$};
        \node at (\x/2,-\y*0.75,0.5) {$\cC$};

        \node[below] at (\x*0.25,-\y/2,0) {$\ket{\cA}\bra{\cA}$};
        \node[below] at (\x*0.5,-\y/2,0) {$B$};
    \end{tikzpicture}
\end{equation}
Since $\ket{\cA}$ is an interface from $\cC$ to $\SO(N(N+1))_1$, the interface $\ket{\cA}\bra{\cA}$ on the left of $B$ can be split into a pair of interfaces from $\cC$ to $\SO(N(N+1))_1$. These two topological interfaces can then be unfurled, yielding
\begin{equation}\label{eq:unfurl_bimod}
\begin{tikzpicture}[z={(30:2cm)},baseline=(current bounding box.center),scale=0.65]
    \newcommand{\x}{8}
    \newcommand{\y}{6}

    \shade[top color=green,bottom color=white,middle color=green,opacity=0.1] (0,-\y/2,0) -- ++(0,-\y/2,0) -- ++(\x,0,0) -- ++(0,0,1) -- ++(0,\y/2,0) -- ++(-\x,0,0) -- cycle;
    \shade[top color=white,bottom color=blue,middle color=blue,opacity=0.1] (0,0,0) -- ++(0,-\y/2,0) -- ++(\x,0,0) -- ++(0,0,1) -- ++(0,\y/2,0) -- ++(-\x,0,0) -- cycle;
    \draw[black] (0,-\y/2,0) -- ++(\x,0,0) -- ++(0,0,1) -- ++(-\x,0,0) -- cycle;

    \draw[black] (0,0,0) -- ++(0,-\y,0);
    \draw[black] (\x,0,0) -- ++(0,-\y,0);
    \draw[black] (\x,0,1) -- ++(0,-\y,0);
    \draw[black,opacity=0.1] (0,0,1) -- ++(0,-\y,0);

    \draw[very thick,black] (\x/2,-\y/2,0) -- ++(0,0,1);

    \node at (\x/2,-\y*0.25,0.5) {$\cC$};
    \node at (\x/2,-\y*0.85,0.5) {$\SO(N(N+1))_1$};

    \node[below] at (\x*0.25,-\y/2,0) {$\ket{\cA}$};
    \node[below] at (\x*0.75,-\y/2,0) {$\ket{\cA}$};
    \node[below] at (\x*0.5,-\y/2,0) {$B$};
\end{tikzpicture}
\end{equation}
In mathematical terms, we have established an equivalence
\begin{equation}\label{eq:unfolding_equiv}
    \bimod{\cA \boxtimes \overline{\cA}}{(\cC \boxtimes \overline{\cC})}{\text{Diag}(\cC)} \cong \bimod{\cA}{\cC}{\cA}
\end{equation}
We give a mathematical description of this equivalence in \cref{app:folding_properly}. In particular, there is a natural choice of simple object on the right-hand side of \eqref{eq:unfolding_equiv}, namely the identity line of the fusion category $\bimod{\cA}{\cC}{\cA}$. This is often referred to as the regular $(\cA,\cA)$-bimodule. It is simple as $\cA$ is haploid; that is, contains a single copy of the trivial line. It is simply given by the algebra object $\cA$ with left and right $\cA$-action given by the multiplication of $\cA$. By the equivalence \eqref{eq:unfolding_equiv}, this selects a natural choice of simple $(\cA_L,\cA_R)$-bimodule over $\cC\boxtimes\overline{\cC}$.

In view of the above construction, we could alternatively recast our Chern--Simons realisation of the Weyl fermions as follows:
\begin{equation}\label{eq:sausage}
\begin{tikzpicture}[z={(30:2cm)},scale=0.8,baseline=(current bounding box.center)]
    \newcommand{\x}{10}
    \newcommand{\yB}{2} 
    \newcommand{\yS}{4} 
    \newcommand{\y}{5.5} 
    \newcommand{\rN}{0.8} 

    \shade[top color=green,bottom color=white,middle color=green,opacity=0.1] (0,-\yS,0) -- (0,-\y,0) -- ++(\x,0,0) -- ++(0,0,1) -- (\x,-\yS,1) -- ++(-\x,0,0) -- cycle;
    \fill[blue,opacity=0.1] (0,0,0) -- (0,-\yS,0) -- ++(\x,0,0) -- ++(0,0,1) -- (\x,0,1) -- ++(-\x,0,0) -- cycle;

    \draw[black] (0,-\yS,0) -- ++(\x,0,0) -- ++(0,0,1) -- ++(-\x,0,0) -- cycle;

    \draw[black] (0,0,0) -- ++(\x,0,0) -- ++(0,0,1) -- ++(-\x,0,0) -- cycle;
    \filldraw[black,fill=blue,fill opacity=0.1] (0,0,0) -- ++(\x,0,0) -- ++(0,0,1) -- ++(-\x,0,0) -- cycle;

    \draw[black] (0,0,0) -- ++(0,-\y,0);
    \draw[black] (\x,0,0) -- ++(0,-\y,0);
    \draw[black] (\x,0,1) -- ++(0,-\y,0);
    \draw[black,opacity=0.1] (0,0,1) -- ++(0,-\y,0);

    \begin{scope}[draw=black,line join=round,line cap=round]

        \pgfmathsetmacro{\NVsilx}{\x/2-\rN+\rN*cos(60)}
        \pgfmathsetmacro{\NVsily}{-\yB-\rN*sin(60)}

        \coordinate (fronttangent) at (\NVsilx,\NVsily,0);
        \coordinate (reartangent)  at (\NVsilx,\NVsily,1);

        \fill[green!10]
            (0,-\yB-\rN,0)
            -- (\x/2-\rN,-\yB-\rN,0)
            arc[start angle=-90,end angle=-60,radius=\rN]
            -- (reartangent)
            arc[start angle=-60,end angle=90,radius=\rN]
            -- (0,-\yB+\rN,1)
            -- (0,-\yB+\rN,0)
            -- cycle;

        \draw[opacity=0.1]
            (reartangent)
            arc[start angle=-60,end angle=-90,radius=\rN]
            -- (0,-\yB-\rN,1)
            -- (0,-\yB+\rN,1);

        \draw[opacity=0.1]
            (0,-\yB-\rN,0)
            -- (0,-\yB-\rN,1);

        \draw
            (0,-\yB+\rN,0)
            -- (\x/2-\rN,-\yB+\rN,0)
            arc[start angle=90,end angle=-90,radius=\rN]
            -- (0,-\yB-\rN,0)
            -- cycle;

        \draw
            (fronttangent)
            -- (reartangent)
            arc[start angle=-60,end angle=90,radius=\rN]
            -- (0,-\yB+\rN,1)
            -- (0,-\yB+\rN,0);
    \end{scope}

    \node at (\x/2,-0.9*\yB,0.5) {$\ket{\cA}$};
    \node at (\x*0.75-0.2,-1.3*\yB-0.2,0.5) {$U(N)_{N+2,2(N+1)}$};
    \node at (\x/2,-\yS/2-\y/2,0) {$SO(2(N+1)^2)_1$};
    \node at (\x/4,-\yB,0) {$SO(N(N+1))_1$};

    \draw[->,thick] (\x*0.4,0,0.5) -- (\x*0.6,0,0.5);
    \node[left] at (0,0,0) {$\ket{U(N)_{N+2,2(N+1)}}$};
    \draw[->,thick] (\x*0.4,-\yS,0.5) -- (\x*0.6,-\yS,0.5);
    \node[left] at (0,-\yS,0) {$\ket{U(N+2)_{N,2(N+1)}}$};
\end{tikzpicture}
\end{equation}
where on the top of the slab we put the boundary condition realising gapless right-moving modes charged under $U(N)$, and on the interface between the two slabs, we first use the fact such an interface is equivalent to a boundary condition of
\begin{equation}
    \overline{U(N)_{N+2,2(N+1)}} \boxtimes SO(2(N+1)^2)_1 = U(N+2)_{N,2(N+1)}
\end{equation}
then insert the natural boundary condition of this theory realising gapless right-moving modes charged under $U(N+2)$.

\subsection{Algebras to gauge in two dimensions}\label{sec:gauging}

Here we establish that the symmetry categories $\cS_L$, $\cS_R$ written down in \cref{sec:SLSRnonabelian} are related by gauging, and moreover show that the relevant algebras to gauge are those that arise from the simple bimodule chosen in \cref{sec:unfolding}.

Recall from \cref{sec:2d_descrip} that the defect $B$ can be understood from the two-dimensional perspective as a gauging interface. That is, there is a gaugeable algebra $a_L$ in $\cS_L$ such that gauging $a_L$ gives rise to $\cS_R$, and that the interface between the ungauged and gauged theory is $B$. We have selected a natural choice of interface $B$ in the previous section. Using the internal end construction (see \cref{sec:2d_descrip}) we can derive the algebra $a_L$ from it. Similarly, there is an algebra $a_R$ in $\cS_R$ such that gauging yields $\cS_L$.

In \cref{app:deriving_algebras} we explicitly derive the algebra $a_R$ from $B$, and show that it is given by
\begin{equation}
    a_R = \cA \in \cC \cong \cS_R
\end{equation}
Indeed, we note that under gauging $a_R$ the symmetry of the gauged theory is $\bimod{a_R}{(\cS_R)}{a_R}$ \cite{Bhardwaj:2017xup}, and
\begin{equation}
    \bimod{a_R}{(\cS_R)}{a_R} \cong \bimod{\cA}{\cC}{\cA} = \cC_\cA \boxtimes \overline{\cC_\cA} = \cS_L
\end{equation}
with the second equality following from \cite[Corollary 4.1 (ii)]{Davydov:2010}.\footnote{More precisely, the results of \cite{Davydov:2010} apply to non-degenerate $\cC$. We believe the result holds also in the fermionic case once we recall that $\cC_\cA \boxtimes \overline{\cC_\cA}$ has a $\bZ_2$ identification $(\pi,\overline{\pi})\sim(\1,\1)$. Then the dimensions of the left and right-hand sides match: $\cD(\bimod{\cA}{\cC}{\cA})^2=\cD(\cC)^2$ and $\cD(\cC_\cA \boxtimes \overline{\cC_\cA})^2=\frac{1}{2}\cD(\cC_\cA)^4=\frac{1}{2}(\cD(\cC)^2/\dim(\cA))^2 = \cD(\cC)^2$ where the factor of $1/2$ comes from the $\bZ_2$ quotient and we have used \eqref{eq:fermionic_lagr} since $\cA$ is a fermionic Lagrangian algebra.}

On the other hand, the algebra to gauge from the left is
\begin{equation}
    a_L = \text{Diag}(\cC_\cA) \in \cC_\cA \boxtimes \overline{\cC_\cA}
\end{equation}
which, indeed, satisfies
\begin{equation}
    \bimod{a_L}{(\cS_L)}{a_L} = \bimod{\text{Diag}(\cC_\cA)}{(\cC_\cA \boxtimes \overline{\cC_\cA})}{\text{Diag}(\cC_\cA)} \cong \cC \cong \cS_R
\end{equation}
The second equivalence here is established as follows. Firstly, \cite[Theorem 4.14]{mueger2001} implies that $\cC_\cA \boxtimes \overline{\cC_\cA}$ is Morita equivalent to $\cZ(\cC_\cA)$ where $\cZ(\cC_\cA)$ denotes the Drinfel'd centre of $\cC_\cA$. In particular, $\cZ(\cC_\cA)$ can be obtained from $\cC_\cA \boxtimes\overline{\cC_\cA}$ by gauging an algebra $\Diag(\cC_\cA)$. Gauging $\Diag(\cC_\cA)$ gives a dual symmetry $\bimod{\text{Diag}(\cC_\cA)}{(\cC_\cA \boxtimes \overline{\cC_\cA})}{\text{Diag}(\cC_\cA)}$ which is, therefore, equivalent to $\cZ(\cC_\cA)$. We then use that $\cZ(\cC_\cA) \cong \cC$ \cite[Corollary 4.1 (i)]{Davydov:2010}. The physical picture associated with this statement is the following. We start with anyons in $\cC$, which describes a three-dimensional TQFT, and gauge a Lagrangian algebra $\cA$. This defines a two-dimensional boundary, where the symmetry category is $\cC_\cA$. From this boundary, we can reconstruct the three-dimensional TQFT via the Drinfel'd centre $\cZ(\cC_\cA)$. The centre consists of objects in $\cC_\cA$ which can be moved through all other boundary lines in a consistent manner (in particular, they admit a half-braiding) and so can be pulled off the boundary and become bulk lines of a three-dimensional TQFT. Therefore, condensing $\cA$ and then computing the centre simply takes us back to the TQFT $\cC$ that we started with.

In the fermionic case at hand, this logic needs slight modification since the dimensions do not match, $\cD(\cZ(\cC_\cA))^2 = \cD(\cC_\cA)^4 = 2 \cD(\cC)^2$, so the centre $\cZ(\cC_\cA)$ has twice as many objects as required. Instead of $\cZ(\cC_\cA)$, the construction should proceed via the relative centre $\cZ(\cC_\cA,\text{sVec})$ \cite{davydov2011}. This can be seen as follows. As above, the centre consists of those boundary lines which can be consistently pulled into the bulk. Amongst them, the $\pi$ line of the boundary will be pulled into the bulk to yield some anyon. Recall, however, that the $\pi$ line of the boundary can be found by pushing the transparent fermion $f$ of the bulk $\cC$ to the boundary. We should, therefore, require that the $\pi$ line is transparent when pulled into the bulk via the centre construction. We can do this by first forming the centre $\cZ(\cC_\cA)$ and then restricting to the subset of anyons for which $f$ is transparent (that is, those with trivial monodromy with $f$). We can then use \cite[Corollary 4.6]{davydov2011}, giving $\cZ(\cC_\cA,\text{sVec})\cong \cC \boxtimes \text{sVec} \cong \cC$ where the final equivalence follows since we identify the transparent fermions of the two factors of the product (see \cref{sec:SPTs_general}).

\subsubsection*{Example: $N = 2$}

Returning to the non-abelian example with $N=2$, we comment that instead of describing the two-dimensional gauging as algebras within the larger categories $\cS_L\cong \cC_\cA \boxtimes \overline{\cC_\cA}$ and $\cS_R \cong \cC$, we can simplify the answer by passing to smaller subcategories.

On the left of the defect, we have \eqref{eq:SL} with $\cC_\cA$ as described in \cref{sec:SLSRnonabelian}. The algebra to gauge is $a_L = \Diag(\cC_\cA) = (1,1) \oplus (g,g) \oplus (g^2,g^2) \oplus (X,X)$. Note that we have $d_X = \sqrt{3}$ and thus
\begin{align}
\dim a_L = 1 + 1 + 1 +3 = 6
\end{align}
and so, from \cref{sec:2d_descrip}, the resulting half-space gauging defect has Affleck–Ludwig central charge $g = \sqrt{\dim a_L} = \sqrt{6}$. We claim that a fusion subcategory of $\cS_L$ consists of
\begin{equation}
    \{ (g^i, g^j) : 0 \leq i, j < 3\}\cup \{ (X, X) \}
\end{equation}
which moreover has $\text{TY}(\bZ_3 \times \bZ_3)$ fusion rules. Thus within this subcategory, the algebra to gauge $a_L$ is the diagonal of the $\bZ_3 \times \bZ_3$ plus the Tambara--Yamagami line $(X,X)$.\footnote{Taking a bicharacter $\chi((a_1,a_2),(b_1,b_2)) = \exp(2\pi i(a_1 b_1 - a_2 b_2)/3)$ where $a_i,b_i\in\bZ_3$ and Frobenius--Schur indicator equal to $+1$ ensures the algebra $a_L$ is gaugeable, with all junction morphisms equal to 1.}

On the right of the defect, we have \eqref{eq:SR} and the algebra to gauge is $a_R=\cA = (0, 0) \oplus (2, 4) \oplus (2, 8) \oplus (4, 0)$ in $\cC=U(2)_{4,6}$. As a check, note that we indeed have $\dim a_R = 1 + 2 + 2 + 1 = \dim a_L$. We note that the closure of these objects under fusion is the product
\begin{equation}
    \{ 0, 2, 4 \} \boxtimes \{ 0, 4, 8 \} \subset SU(2)_4 \boxtimes U(1)_{12}
\end{equation}
which is equivalent to $\Rep(\bS_3) \boxtimes \text{Vec}_{\bZ_3}$.\footnote{The fact that the subcategory $\{0,2,4\}$ of $\SU(2)_4$ has $\Rep(\bS_3)$ fusion rules is easily established. There are three fusion categories with these fusion rules \cite{etingof2003} and only one of them, $\Rep(\bS_3)$ itself, admits a braiding \cite{ostrik2005}. The subcategory $\{0,2,4\}$ of $\SU(2)_4$ inherits a braiding from $\SU(2)_4$, so must be equivalent to $\Rep(\bS_3)$.} In particular, the $2$ representation of $SU(2)_4$ corresponds to the fundamental 2-dimensional representation of $\bS_3$ and the $4$ representation of $SU(2)_4$ corresponds to the sign representation of $\bS_3$. Thus within this subcategory, the algebra to gauge consists of $(\text{triv}, 0)$, $(\text{sign}, 0)$, and $(\text{fund}, \pm 1)$. Indeed, one can verify that the dual symmetry found by gauging $a_L$ in $\text{TY}(\bZ_3 \times \bZ_3)$ is $\Rep(\bS_3)\boxtimes \text{Vec}_{\bZ_3}$; that is, the latter is the category of $(a_L,a_L)$-bimodules in $\text{TY}(\bZ_3\times\bZ_3)$.

Looking back to \cref{sec:nonabelian_OPEs}, we can see the signatures of these categorical results in the OPEs found simply by using the RCFT data. In particular, consider the OPEs \eqref{eq:psi2tilde_OPEs} where $\tilde{\psi}_2$ is scattered through the defect. From \eqref{eq:nonabelian_reprs}, $\tilde{\psi}_2$ is uncharged under the non-abelian factors of $G_k$ and the OPEs were found to be consistent with the twist line generating a $\bZ_3$ symmetry. Indeed, this is an avatar of the $\bZ_3$ factor in the $\Rep(\bS_3) \times \bZ_3$ symmetry found above which is gauged on the right of the symmetric interface. The OPEs \eqref{eq:psi1tilde_OPEs}, on the other hand, where $\tilde{\psi}_1$ is scattered through the defect, cannot be produced by any invertible twist line. Indeed, since $\tilde{\psi}_1$ is charged under the non-abelian symmetries, the results above imply that it will be sensitive to the non-invertible $\Rep(\bS_3)$ factor found above.

\subsection{The SPT class}

In \cref{sec:SPTs_general} we introduced a general criterion to determine the SPT class of the defect which preserves $G$. It either lies in the vector class, where it does not host any unpaired Majorana modes, or in the axial class, where it does.

We would like to run this analysis for our non-abelian example, for which the relevant Chern--Simons theory was $U(N)_{N+2,2(N+1)} \times U(N+2)_{N,2(N+1)}$. This is a fermionic TQFT, and as such is incapable of probing the SPT class. We will therefore need to replace it with a larger, bosonic TQFT. This bosonisation interferes with the level-rank duality \eqref{eq:levelrank1} that we relied upon to make calculations tractable. We first work out these details using techniques from \cite{Karch:2016sxi}.

\subsubsection*{Level-rank duality}

Recall that for $N$ even the level-rank duality relevant for us takes the form \cite{hsin2016levelrank}
\begin{equation}\label{eq:LR_bosonic}
\resizebox{\textwidth}{!}{$\displaystyle{
    \frac{SU(N)_{N+2} \times U(1)_{2N(N+1)} \times SO(0)_1}{\bZ_N^{[1]}}
    \sim
    \overline{\frac{SU(N+2)_N \times U(1)_{2(N+1)(N+2)} \times SO(0)_1}{\bZ_{N+2}^{[1]}}}
    \times SO(2(N+1)^2)_1
}$}
\end{equation}
which is a duality of fermionic theories. We would like to bosonise this duality. We can do this by gauging \emph{half} of the 1-form symmetry in the quotients of \eqref{eq:LR_bosonic}. Thus, let us rewrite the above duality by performing the gauging in two steps:
\begin{equation}
\resizebox{\textwidth}{!}{$\displaystyle{
    \frac{\frac{SU(N)_{N+2} \times U(1)_{2N(N+1)}}{\bZ_{N/2}^{[1]}} \times SO(0)_1}{\bZ_2^{[1]}}
    \sim
    \frac{\overline{\frac{SU(N+2)_N \times U(1)_{2(N+1)(N+2)}}{\bZ_{(N+2)/2}^{[1]}}} \times SO(0)_1}{\bZ_2^{[1]}}
    \times SO(2(N+1)^2)_1
}$}
\end{equation}
Given a theory $\cT$, the operation $\cT \to \frac{\cT \times SO(0)_1}{\bZ_2^{[1]}}$ describes fermionisation \cite{Bhardwaj:2024ydc}, so the relation above resembles a duality of a pair of fermionised theories. Including the final $SO(2(N+1)^2)$ factor and using the results of \cite[§4.2]{lorentzfrac}, we have
\begin{equation}\label{eq:bosonic_duality}
    \frac{SU(N)_{N+2} \times U(1)_{2N(N+1)}}{\bZ_{N/2}^{[1]}}
    \sim
    \frac{\overline{SU(N+2)_N \times U(1)_{2(N+1)(N+2)}} \times \Spin(2(N+1)^2)_1}{\bZ_{N+2}^{[1]}}
\end{equation}
This is the desired bosonic version of the level-rank duality. Note that both sides have twice as many objects as the original fermionic theory $U(N)_{N+2,2(N+1)}$.

To perform explicit calculations, we will need to know the mapping of lines in this duality. The lines on the left of \eqref{eq:bosonic_duality} are labelled by
\begin{equation}
    (\mu, p) \in \text{Rep}(SU(N)_{N+2}) \times \bZ_{2N(N+1)}
\end{equation}
such that
\begin{equation}
\begin{aligned}
    |\mu| &= p \mod N/2 \\
    (\mu, p)& \sim (a^2 \mu, p + 4(N+1))
\end{aligned}
\end{equation}
where the constraint and identification are due to the gauging of $(\Sym^{N+2}, 2(N+1))^2$ in the denominator. Here $a$ is the order-$N$ operation on $\SU(N)_{N+2}$ Young diagrams of stacking a row of $N+2$ boxes on top and deleting any length-$N$ columns. Similarly, the lines on the right of \eqref{eq:bosonic_duality} are labelled by
\begin{equation}
    (\lambda, q, r) \in \text{Rep}(SU(N+2)_N) \times \bZ_{2(N+1)(N+2)} \times \bZ_4
\end{equation}
such that
\begin{equation}
\begin{aligned}
    |\lambda| &= q + \tfrac{N+2}{2} r \mod N+2 \\
    (\lambda, q, r) &\sim (a \lambda, q + 2(N+1), r + 2)
\end{aligned}
\end{equation}
due to the gauging of $(\Sym^N, 2(N+1), 2)$ in the denominator. Here $a$ denotes order-$(N+2)$ operation on $\SU(N+2)_N$ Young diagrams analogous to the one described above. Note that for $N$ even, $\Spin(2(N+1)^2)_1$ has $\bZ_4$ fusion rules (see \cref{app:MTCs}), so all lines can be written as $S^r$ for some $r \in \bZ_4$, and this is what $r$ above denotes.

We then claim that the map
\begin{equation}
    (\lambda, q, r) \to (a^{\frac{q + \frac{N+2}{2}r - |\lambda|}{N+2} + q} \lambda^T, q + (N + 1)^2 r + 2(N + 1)q)
\label{eq:line map}
\end{equation}
which preserves spins, maps the identity line to itself, and is a bijection, implements the desired mapping of lines in \eqref{eq:bosonic_duality}. For more details, see \cref{app:lagr}.

Note that an equally good guess would be the same formula with $r \to -r$, however this simply reflects the fact $\Spin(n)_1$ enjoys a self-duality that exchanges the $S$ and $C$ lines, implemented by the condensation defect of $V$.

\subsubsection*{Lagrangian algebras}

The $(N+1)^2$ Weyl fermions can be viewed as living on the upper gapless boundary of $\Spin(2(N+1)^2)_1$. Via the conformal embeddings \eqref{eq: non-abelian L and R}, we obtain Lagrangian algebras $|\hat{\cA}_{L,R}\rangle$ in $\SU(N)_{N+2} \times U(1)_{2N(N+1)} \times \SU(N+2)_N \times U(1)_{2(N+1)(N+2)} \times \Spin(2(N+1)^2)_{-1}$. The $n$-alities of the representations work out such that we may quotient the first two factors by $\bZ_{N/2}^{[1]}$, and the last three factors by $\bZ_{N+2}^{[1]}$. Doing so and invoking the level-rank duality \eqref{eq:bosonic_duality} then converts $|\hat{\cA}_{L,R}\rangle$ to Lagrangian algebras in $\widehat{\cC} \boxtimes \overline{\widehat{\cC}}$, where $\hat{\cC}$ is the bosonic theory
\begin{equation}\label{eq:c_hat_def}
    \widehat{\cC} = \frac{SU(N)_{N+2} \times U(1)_{2N(N+1)}}{\bZ_{N/2}^{[1]}}
\end{equation}
In \cref{app:lagr}, we derive the form of these Lagrangian algebras. We find that $\hat{\cA}_R$ is again the diagonal algebra $\hat{\cA}_R = \sum_{c \in \widehat{\cC}} c \boxtimes \overline{c}$, while
\begin{equation}\label{eq:AL_hat_def}
    \hat{\cA}_L = (\cA \boxtimes \overline{\cA}) \otimes \big( \1 \oplus (\ell \boxtimes \overline{\ell}) \oplus (f \boxtimes \overline{f}) \oplus (\ell f \boxtimes \overline{\ell f}) \big)
\end{equation}
with $\ell = (1, \frac{N(N+1)}{2}) \in \widehat{\cC}$, and $f = (\Sym^{N+2}, 2(N+1)) \in \widehat{\cC}$, and $\cA$ as given in \eqref{eq:Adef}. Note that $\cA$ is the same Lagrangian algebra we encountered in the non-$(-1)^F$-refined case \eqref{eq:AL}, but now regarded as a gaugeable but non-Lagrangian algebra in the larger theory $\widehat{\cC}$.

\subsubsection*{Simple bimodule}

We are thus in a situation where the physics of the scattering through $\cN$ is dictated by a choice of simple bimodule $B$ occurring at the interface between the boundaries $|\hat{\cA}_{L,R}\rangle$, as depicted in the following diagram:
\begin{equation}\label{eq:refined_bimod}
\begin{tikzpicture}[z={(30:2cm)},baseline=(current bounding box.center),scale=0.7]
    \newcommand{\x}{8}
    \newcommand{\y}{6}

    \shade[top color=white,bottom color=blue,middle color=blue,opacity=0.1] (0,0,0) -- ++(0,-\y/2,0) -- ++(\x,0,0) -- ++(0,0,1) -- ++(0,\y/2,0) -- ++(-\x,0,0) -- cycle;
    \filldraw[black,fill=blue,fill opacity=0.1] (0,-\y/2,0) -- ++(\x,0,0) -- ++(0,0,1) -- ++(-\x,0,0) -- cycle;

    \draw[black] (0,0,0) -- ++(0,-\y/2,0);
    \draw[black] (\x,0,0) -- ++(0,-\y/2,0);
    \draw[black] (\x,0,1) -- ++(0,-\y/2,0);
    \draw[black,opacity=0.1] (0,0,1) -- ++(0,-\y/2,0);

    \draw[very thick,black] (\x/2,-\y/2,0) -- ++(0,0,1);

    \node at (\x/2,-\y*0.25,0.5) {$\widehat{\cC} \boxtimes \overline{\widehat{\cC}}$};

    \node[below] at (\x*0.25,-\y/2,0) {$|\hat{\cA}_L\rangle$};
    \node[below] at (\x*0.75,-\y/2,0) {$|\hat{\cA}_R\rangle$};
    \node[below] at (\x*0.5,-\y/2,0) {$B$};
\end{tikzpicture}
\end{equation}
We will show that a canonical choice of simple bimodule again exists, and is in the vector class according to the diagnostic of \cref{sec:SPTs_general}.

We will do so by performing the same unfolding-unfurling manoeuvre as in \cref{sec:unfolding}. The form \eqref{eq:AL_hat_def} of $\widehat{\cA}_L$ tells us that the lines $\cA \boxtimes \1$ and $\1\boxtimes\bar{\cA}$ in $\widehat{\cC} \boxtimes \bar{\widehat{\cC}}$ can end on the $\ket{\hat{\cA}_L}$ boundary. Moreover, since $\widehat{\cA}_L$ contains $x \boxtimes \overline{x}$ for $x = 1, \ell, f, \ell f$, these lines can also end there. Since $\widehat{\cA}_R$ is the diagonal algebra, the lines $c\boxtimes\bar{c}$ can end there for any $c\in\widehat{\cC}$. Thus the lines which end on the two boundaries are
\begin{equation}
\begin{tikzpicture}[z={(30:2cm)},baseline=(current bounding box.center),scale=0.7]
    \newcommand{\x}{8}
    \newcommand{\y}{6}

    \shade[top color=white,bottom color=blue,middle color=blue,opacity=0.1] (0,0,0) -- ++(0,-\y/2,0) -- ++(\x,0,0) -- ++(0,0,1) -- ++(0,\y/2,0) -- ++(-\x,0,0) -- cycle;
    \filldraw[black,fill=blue,fill opacity=0.1] (0,-\y/2,0) -- ++(\x,0,0) -- ++(0,0,1) -- ++(-\x,0,0) -- cycle;

    \draw[black] (0,0,0) -- ++(0,-\y/2,0);
    \draw[black] (\x,0,0) -- ++(0,-\y/2,0);
    \draw[black] (\x,0,1) -- ++(0,-\y/2,0);
    \draw[black,opacity=0.1] (0,0,1) -- ++(0,-\y/2,0);

    \draw[very thick,black] (\x/2,-\y/2,0) -- ++(0,0,1);

    \begin{scope}[every node/.style={fill=red,shape=circle,inner sep=1pt}]
        \draw[thick,red,decoration = {markings, mark=at position .5 with {\arrow[scale=1.5]{stealth}}}, postaction=decorate] (\x*0.1,0,0.5) -- ++(0,-\y/2,0) node{};
        \draw[thick,red,decoration = {markings, mark=at position .5 with {\arrow[scale=1.5]{stealth}}}, postaction=decorate] (\x*0.25,0,0.5) -- ++(0,-\y/2,0) node{};
        \draw[thick,red,decoration = {markings, mark=at position .5 with {\arrow[scale=1.5]{stealth}}}, postaction=decorate] (\x*0.4,0,0.5) -- ++(0,-\y/2,0) node{};
        \draw[thick,red,decoration = {markings, mark=at position .5 with {\arrow[scale=1.5]{stealth}}}, postaction=decorate] (\x*0.75,0,0.5) -- ++(0,-\y/2,0) node{};
    \end{scope}

    \node[left, red] at (\x*0.1,0,0.5) {\scriptsize$\cA\boxtimes \1$};
    \node[above, red] at (\x*0.25,0,0.5) {\scriptsize$\1\boxtimes \bar{\cA}$};
    \node[right, red] at (\x*0.4,0,0.5) {\scriptsize$x\boxtimes \bar{x}$};
    \node[right, red] at (\x*0.75,0,0.5) {\scriptsize$c\boxtimes \bar{c}$};

    \node at (\x*0.6,-\y*0.25,0.5) {$\widehat{\cC} \boxtimes \overline{\widehat{\cC}}$};

    \node[below] at (\x*0.25,-\y/2,0) {$|\hat{\cA}_L\rangle$};
    \node[below] at (\x*0.75,-\y/2,0) {$|\hat{\cA}_R\rangle$};
    \node[below] at (\x*0.5,-\y/2,0) {$B$};
\end{tikzpicture}
\end{equation}
We now unfold the diagram along the lower boundary. Since $\hat{\cA}_R$ is the diagonal algebra, it unfolds to the trivial interface in $\widehat{\cC}$. Therefore, we find a setup
\begin{equation}\label{eq:lines_ending}
\begin{tikzpicture}[z={(30:2cm)},baseline=(current bounding box.center),scale=0.65]
    \newcommand{\x}{8}
    \newcommand{\y}{6}

    \shade[top color=white,bottom color=white,middle color=blue,opacity=0.2] (0,0,0) -- ++(0,-\y,0) -- ++(\x,0,0) -- ++(0,0,1) -- ++(0,\y,0) -- ++(-\x,0,0) -- cycle;
    \filldraw[black, fill=blue, fill opacity=0.1] (0,-\y/2,0) -- ++(\x/2,0,0) -- ++(0,0,1) -- ++(-\x/2,0,0) -- cycle;

    \draw[black] (0,0,0) -- ++(0,-\y,0);
    \draw[black] (\x,0,0) -- ++(0,-\y,0);
    \draw[black] (\x,0,1) -- ++(0,-\y,0);
    \draw[black,opacity=0.1] (0,0,1) -- ++(0,-\y,0);

    \begin{scope}[every node/.style={fill=red,shape=circle,inner sep=1pt}]
        \draw[thick,red,decoration = {markings, mark=at position .5 with {\arrow[scale=1.5]{stealth}}}, postaction=decorate] (\x*0.1,0,0.5) -- ++(0,-\y/2,0) node{};
        \draw[thick,red,decoration = {markings, mark=at position .5 with {\arrow[scale=1.5]{stealth}}}, postaction=decorate] (\x*0.25,-\y/2,0.5) node{} -- ++(0,-\y/2,0);
        \draw[thick,red,decoration = {markings, mark=at position .25 with {\arrow[scale=1.5]{stealth}}, mark=at position .75 with {\arrow[scale=1.5]{stealth}}}, postaction=decorate] (\x*0.4,0,0.5) -- ++(0,-\y/2,0) node{} -- ++(0,-\y/2,0);
    \end{scope}

    \filldraw[very thick,black] (\x/2,-\y/2,0) -- ++(0,0,1);

    \node at (\x/2,-\y*0.25,0.5) {$\widehat{\cC}$};
    \node at (\x/2,-\y*0.85,0.5) {$\widehat{\cC}$};

    \node[left, red] at (\x*0.1,0,0.5) {\scriptsize$\cA$};
    \node[left, red] at (\x*0.25,-\y,0.5) {\scriptsize$\cA$};
    \node[right, red] at (\x*0.4,0,0.5) {\scriptsize$x$};
    \node[right, red] at (\x*0.4,-\y,0.5) {\scriptsize$x$};

    \node[below] at (\x*0.2,-\y/2,0) {$\ket{\hat{\cA}_L}$};
    \node[right, below] at (\x*0.5,-\y/2,0.5) {$B$};
\end{tikzpicture}
\end{equation}
Following the set of operations in \cref{sec:unfolding}, we are now instructed to \emph{unfurl} the $\ket{\hat{\cA}_L}$ interface. In order to do so, we must realise it as a factorised interface $\ket{\cA}\bra{\cA}$, where $\ket{\cA}$ denotes an interface where $\cA$ is condensed. This is less obvious here than in the situation in \cref{sec:unfolding} since $\cA$ is not Lagrangian in the larger TQFT $\widehat{\cC}$ and $\hat{\cA}_L$ in \eqref{eq:AL_hat_def} is not simply a product $\cA \boxtimes \bar{\cA}$. The fact that $\ket*{\hat{\cA}_L}$ does take this factorised form stems from the fact that the lines $x=f,\ell$ which pass through the $\ket*{\hat{\cA}_L}$ interface braid trivially with $\cA$. Therefore, when $\cA$ is gauged, these lines survive. Stacking two such interfaces $\ket{\cA}\bra{\cA}$ then gives an interface from $\widehat{\cC}$ to $\widehat{\cC}$ with precisely the properties shown in \eqref{eq:lines_ending}.

In order to unfurl this factorised interface, we must understand the theory which results from gauging $\cA$ in $\widehat{\cC}$. As $\cA$ is not Lagrangian, gauging it gives rise to a smaller TQFT generated by the simple lines $f,\ell$ which survive the gauging. This smaller TQFT is denoted $\cC^\text{loc}_{\cA}$ as the lines $f,\ell$ are said to give local modules. Identifying $\cC^\text{loc}_{\cA}$ is not difficult in this case: it is $\Spin(N(N+1))_1$. This can be evidenced by computing the dimension $\cD(\widehat{\cC}^\text{loc}_\cA) = \frac{\cD(\widehat{\cC})}{\dim(\cA)} = 2$, and the chiral central charge $c(\widehat{\cC}^\text{loc}_\cA) = c(\widehat{\cC}) = \frac{N(N+1)}{2}$. Furthermore, the lines $f,\ell$ generating $\widehat{\cC}^\text{loc}_\cA$ satisfy the relations
\begin{equation}
    f^2 = 1 \qquad \ell^2 = f^{N/2}
\end{equation}
modulo fusion with lines in the condensate $\cA$. This implies they have fusion rules $\bZ_4$ for $N$ = 2 mod 4 and $\bZ_2 \times \bZ_2$ for $N$ = 0 mod 4, which matches the fusion rules of $\Spin(N(N+1))_1$ (see \cref{app:MTCs}). Therefore, when the lines $f, \ell$ of $\widehat{\cC}$ pass through the condensation interface $\ket{\cA}$, they become the lines $V, S$ of $\Spin(N(N+1))_1$.

Finally, then, we unfurl the $\ket{\hat{\cA}_L}=\ket{\cA}\bra{\cA}$ interface to yield 
\begin{equation}\label{eq:refined_unfurl}
\begin{tikzpicture}[z={(30:2cm)},baseline=(current bounding box.center),scale=0.65]
    \newcommand{\x}{8}
    \newcommand{\y}{6}

    \shade[top color=white,bottom color=blue,middle color=blue,opacity=0.1] (0,0,0) -- ++(0,-\y/2,0) -- ++(\x,0,0) -- ++(0,0,1) -- ++(0,\y/2,0) -- ++(-\x,0,0) -- cycle;
    \shade[bottom color=white,top color=green,opacity=0.1] (0,-\y/2,0) -- ++(0,-\y/2,0) -- ++(\x,0,0) -- ++(0,0,1) -- ++(0,\y/2,0) -- ++(-\x,0,0) -- cycle;
    \draw[black] (0,-\y/2,0) -- ++(\x,0,0) -- ++(0,0,1) -- ++(-\x,0,0) -- cycle;

    \draw[black] (0,0,0) -- ++(0,-\y,0);
    \draw[black] (\x,0,0) -- ++(0,-\y,0);
    \draw[black] (\x,0,1) -- ++(0,-\y,0);
    \draw[black,opacity=0.1] (0,0,1) -- ++(0,-\y,0);

    \begin{scope}[every node/.style={fill=red,shape=circle,inner sep=1pt}]
        \draw[thick,red,decoration = {markings, mark=at position .5 with {\arrow[scale=1.5]{stealth}}}, postaction=decorate] (\x*0.15,0,0.5) -- ++(0,-\y/2,0) node{};
        \draw[thick,red,decoration = {markings, mark=at position .5 with {\arrowreversed[scale=1.5]{stealth}}}, postaction=decorate] (\x*0.85,0,0.5) -- ++(0,-\y/2,0) node{};
        \draw[thick,red,decoration = {markings, mark=at position .125 with {\arrow[scale=1.5]{stealth}}, mark=at position .88 with {\arrow[scale=1.5]{stealth}}}, postaction=decorate] (\x*0.25,0,0.5) -- ++(0,-\y/2,0) node{} arc[start angle=-180, end angle=0, radius=0.25*\x] node{} -- ++(0,+\y/2,0);
    \end{scope}

    \draw[very thick,black] (\x/2,-\y/2,0) -- ++(0,0,1);

    \node[left, red] at (\x*0.15,0,0.5) {\scriptsize$\cA$};
    \node[right, red] at (\x*0.85,0,0.5) {\scriptsize$\cA$};
    \node[right, red] at (\x*0.25,0,0.5) {\scriptsize$f,\ell$};
    \node[left, red] at (\x*0.75,0,0.5) {\scriptsize$f,\ell$};
    \node[above, red] at (\x/2,-\y*0.85,0.5) {\scriptsize$V,S$};

    \node at (\x/2,-\y*0.25,0.5) {$\widehat{\cC}$};
    \node at (\x/2,-\y*0.95,0.5) {$\Spin(N(N+1))_1$};

    \node[left] at (0,-\y/2,0) {$\ket{\cA}$};
    \node[right] at (\x,-\y/2,0) {$\ket{\cA}$};
    \node[below] at (\x*0.5,-\y/2,0) {$B$};
\end{tikzpicture}
\end{equation}
This is simply the bosonisation of \eqref{eq:unfurl_bimod}.

Given that both interfaces are now $\ket{\cA}$, there is a canonical choice of bimodule $B$ given by the identity line $\1 \in \widehat{\cC}_\cA$, i.e.\ the regular $\cA$-module. The question of the SPT class of $B$ then becomes the question of whether $f \boxtimes 1$ can end on $B$ in \eqref{eq:refined_bimod}, which is equivalent to whether $f \in \widehat{\cC}$ can end on $\ket{\cA}$ in \eqref{eq:refined_unfurl} from above. That is, $B$ is in the axial class iff $f \in \cA$. Since this is not the case, $B$ is in the vector class, as claimed.

\subsection{Pushing Lagrangian algebras to boundaries}

To finish this section, we make contact with a result of \cite{putrov2024} which says that if $\cA_{L,R}$ are Lagrangian algebras in a generic modular tensor category $\cC$, then pushing $\cA_L$ to the $\ket{\cA_R}$ boundary gives a sum of haploid algebras in $\cC_{\cA_R}$ and vice-versa. Moreover, these algebras give the set of all isomorphism classes of haploid algebras within a single Morita equivalence class; that is, all the algebras $a_L$ in $\cS_L$ which can be gauged to yield $\cS_R$ (see \cref{sec:2d_descrip}). These algebras are also in one-to-one correspondence with the simple bimodules in $\cA_L \otimes \cA_R$. We can therefore glean information about the simple bimodules without directly constructing them. Here we consider this for our $U(2)_{4,6}$ example.

Let us see what we get by pushing $\cA_R$ to the $\ket{\cA_L}$ boundary. We have already determined the structure of $\cC_\cA$ (namely a fermionic extension of $\text{TY}(\bZ_3)$, with fusion rules $X^2 = \pi(1 + g + g^2)$), and from this we can obtain $(\cC \boxtimes \overline{\cC})_{\cA_L} = (\cC_\cA \boxtimes \overline{\cC_\cA})$. Applying the induction functor to $\cA_R$, we find
\begin{align}
\begin{split}\label{eq:AR_decomp}
    \cA_R \to \text{Ind}_{\cA_L}(\cA_R) = 3 & \left[ (1,1) \oplus (g,g) \oplus (g^2,g^2) \oplus (X,X) \right] \\
    \oplus & \left[ (1 + g + g^2, 1 + g + g^2) \oplus 3(X,X) \right]
\end{split}
\end{align}
where we have split the result into 4 haploid algebras. Indeed, \cite[Theorem 4.1]{putrov2024} implies that when pushed to the $\ket{\cA_L}$ boundary, $\cA_R$ should break up into a sum of the internal end algebras corresponding to each object in the module category describing the interface. Let's check this explicitly. As discussed in \cref{sec:gauging}, the interface is described by the module category $\cC_\cA$, whose elements are $\{1,g,g^2,X\}$ and their images under fusion with $\pi$. We now compute the internal end algebra of each element in $\cC_\cA$, finding that they reproduce the four algebras appearing in the decomposition above.

For an object $m \in\cC_\cA$ specifying a particular $\cA_L$-$\cA_R$ interface, its internal end algebra in $\cS_L$ is\footnote{The left action of $S_L=(\cC_\cA\boxtimes\overline{\cC_\cA})/(\pi,\pi)$ on $\cC_\cA$ is $(x \boxtimes \overline{y}) \triangleright m = x \otimes m \otimes \overline{y}$.}
\begin{equation}
    \underline{\End}_{\cS_L}(m) = \bigoplus_{(x,y)\in (\cC_\cA \boxtimes \overline{\cC_\cA})/(\pi,\pi)} \dim \Hom_{\cC_\cA}(x \otimes m , m \otimes y ) (x\boxtimes y)
\end{equation}
Consider first $m=1$, whence $\dim \Hom_{\cC_\cA}(x,y) = \delta_{x,y}$, giving
\begin{equation}
    \underline{\End}_{\cS_L}(1) = (1,1) \oplus (g,g) \oplus (g^2,g^2) \oplus (X,X)
\end{equation}
Using the fusion rules \eqref{eq:CA_fusion} of $\cC_\cA$, it is simple to verify that the same algebra object is found for $m=g$ and $m=g^2$. Finally, for $m=X$ we have $g^i \otimes X \cong X \otimes g^j \cong X$, such that $\dim \Hom_{\cC_\cA}(g^i\otimes X,X\otimes g^j) = 1$ for all $i,j=0,1,2$. Then recall from \eqref{eq:X2_fusion} we have $X^2 = \pi(1\oplus g \oplus g^2)$, thus $\dim \Hom_{\cC_\cA}(X^2,X^2) = 3$, and so
\begin{equation}
    \underline{\End}_{\cS_L}(X) = (1\oplus g\oplus g^2,1\oplus g\oplus g^2) \oplus 3(X,X)
\end{equation}
The internal end construction guarantees that these algebras are gaugeable and that they are all Morita equivalent. Therefore, we see that the decomposition \eqref{eq:AR_decomp} corresponds to the internal end algebras of the objects of $\cC_\cA$, with three copies of the diagonal algebra corresponding to the three invertible elements $1,g,g^2$. Notice that the multiplicity-3 algebra above is the natural algebra $a_L = \text{Diag}(\cC_\cA)$ that we wrote down in \cref{sec:gauging}.

Pushing the lines $\cA_R$ to the $\ket{\cA_L}$ boundary gives the full set of twist lines which can be generated by scattering local operators through a symmetry-preserving defect. However, the results above indicate that for a given choice of defect (i.e.\ a given choice of bimodule $B$) only a subset of these lines will be generated. That is, if the specific defect is found by gauging $\underline{\text{End}}_{\cS_L}(1)$ in half-space, then only the lines in $\underline{\text{End}}_{\cS_L}(1)$ will be generated by the scattering. Using a different internal end algebra to generate a symmetry-preserving defect will generically give rise to a different set of twist lines when local operators are scattered through it. Furthermore, the Affleck--Ludwig central charge of the defects corresponding to gauging different algebras need not agree. In the present example, gauging $\underline{\text{End}}_{\cS_L}(m)$ for $m=1,g,g^2$ gives a defect with $g=\sqrt{6}$ whereas gauging $\underline{\text{End}}_{\cS_L}(X)$ gives a defect with $g=\sqrt{18}$.

We can perform a similar analysis by pushing $\cA_L$ to the $\ket{\cA_R}$ boundary. In that case, we want to decompose $\cA_L\otimes\cA_R$ into haploid algebras in $\cS_R = (\cC\boxtimes\overline{\cC})_{\cA_R} \cong \cC$. This can be done by making use of the folding and unfolding relations discussed in \cref{sec:unfolding} and \cref{app:folding_properly}. Namely, we have
\begin{equation}
    \cA_L \otimes \cA_R = (\cA \boxtimes \overline{\cA})\otimes \Diag(\cC) \cong ( (\cA\otimes\cA)\boxtimes\1) \otimes \Diag(\cC)
\end{equation}
where we have used \eqref{eq:sliding}. This takes the form of an induced $\cA_R$ module, so the equivalence \eqref{eq:unfolding_equiv}, given explicitly in \eqref{eq:unfold_equiv} allows us to relate this to an object in $\cC$. Namely, from \eqref{eq:unfold_equiv}, it is mapped to $\cA\otimes\cA$ in $\cC$. Evaluating this product, it decomposes into four haploid algebras as
\begin{equation}\label{eq:AA_decomp}
\begin{alignedat}{3}
    \cA \otimes \cA = 3 &\big[ &&(0,0) \oplus (2,4) \oplus (2,8) \oplus (4,0) \big] \\
    \oplus &\big[ &&(0,0) \oplus 2(2,0) \oplus (0,4) \oplus (4,0) \\
    &&&\oplus 2(2,4) \oplus (4,4) \oplus (0,8) \oplus 2(2,8) \oplus (4,8) \big]
\end{alignedat}
\end{equation}
The results of \cite{putrov2024} again imply that each haploid algebra in this decomposition should be the internal end algebra for a particular $\cA_L$-$\cA_R$ interface, this time written as an algebra in $\cS_R$. This can be verified explicitly as follows. For an interface $m\in\cC_\cA$, its internal end algebra is
\begin{equation}
    \underline{\End}_{\cS_R}(m) = \bigoplus_{a\in\cC} \dim\Hom_{\cC_\cA}(m\triangleleft a,m) a
\end{equation}
Recall from \cref{sec:SLSRnonabelian} that the simple objects of $\cC_\cA$ can be found by induction from $\cC$. That is, for each $m\in\cC_\cA$ we can choose a simple $c_m \in\cC$ such that $m=\text{Ind}_\cA (c_m) = c_m \otimes \cA$. In particular, a choice of representatives is $c_1=(0,0)$, $c_g=(0,4)$, $c_{g^2}=(0,8)$, and $c_X=(1,1)$. Then by reciprocity \cite{Fuchs:2002cm} we have
\begin{align}
    \Hom_{\cC_\cA}(m\triangleleft a,m) &= \Hom_{\cC_\cA}((c_m \otimes \cA) \triangleleft a, c_m\otimes \cA) \nn\\
    &\cong \Hom_\cC (a \otimes c_m, c_m\otimes \cA) \nn\\
    &\cong \Hom_\cC (a, c_m \otimes \cA \otimes \overline{c_m})
\end{align}
such that
\begin{equation}
    \underline{\End}_{\cS_R}(m) \cong c_m \otimes \cA \otimes \overline{c_m}
\end{equation}
Computing for $m=1$ immediately gives $\underline{\End}_{\cS_R}(1) = \cA$. The same result is found for the other invertible objects $m=g$ and $m=g^2$. For $m=X$ we have $c_X = (1,1)$ and $\overline{c_X} = (1,11)$, using the tensor product of $SU(2)_4$ representations, we find
\begin{equation}
\begin{split}
    \underline{\End}_{\cS_R}(X) &= (0,0) \oplus 2(2,0) \oplus (0,4) \oplus (4,0) \\
    &\qquad \oplus 2(2,4) \oplus (4,4) \oplus (0,8) \oplus 2(2,8) \oplus (4,8)
\end{split}
\end{equation}
Thus, we confirm that the decomposition \eqref{eq:AA_decomp} is made up of the internal end algebras of each simple object in $\cC_\cA$.

\section{Outlook}\label{sec:outlook}

In this work, we have made a significant step in the construction of symmetry preserving boundary conditions for a generic non-anomalous symmetry in two dimensions. The methods we have used can be applied to any case where the boundary preserves a symmetry $G$ whose right- and left-moving sectors both have anomaly $k$, such that $G_k \to SO(2N)_1$ embeds conformally. The non-trivial cases of most interest are where there are two such embeddings so that the $G$ symmetry acts chirally (that is, differently on the left- and right-movers). With sufficient computational effort, from the embeddings one can compute the algebras encoding the relevant boundary conditions of the Chern--Simons theory, the interfaces between them, and the realisation of the interface as a gauging interface.

The most desirable next step is to enhance our methods to tackle the generic case where $G_k$ does not embed conformally. We begin this final section with an outline of such an enhancement and a discussion of the hurdles which remain to be jumped if one is to fully solve the problem.

\subsubsection*{Non-conformally embedded symmetries}

The starting point remains the same as for our discussion. We desire a $G$-symmetric boundary condition of $N$ Dirac fermions, which is equivalent to a $G$-symmetric topological defect of $N$ Weyl fermions after unfolding along the boundary. The preserved symmetry $G$ is necessarily a subgroup of the full invertible symmetry $SO(2N)$ of the fermions, so there are a pair of embeddings $L,R : G_k \to SO(2N)_1$ encoding how $G$ is realised on the left- and right-movers. We now consider the case where these embeddings are \emph{not} conformal. That is, we do not require the central charge of $G_k$ to equal $c(SO(2N)_1)=N$. Equivalently, we do not require the theory to be rational with respect to $G_k$.

A particularly nice example to hang our dreams from is to take $N=15$ and $G=SU(2)\times U(1)^2$ with left-movers $\psi$ transforming in the $\mathbf{1}_{(1,0)} \oplus \mathbf{5}_{(5,0)} \oplus \mathbf{9}_{(9,2)}$ representation and right-movers $\tilde{\psi}$ transforming in the $\mathbf{7}_{(7,2)} \oplus \mathbf{8}_{(8,1)}$ representation, where the subscripts indicate the $U(1)^2$ charges.\footnote{The faithfully acting symmetry is in fact $U(2) \times U(1)$. This example was originally found by and has been emphasised by David Tong since 2017.} The gravitational anomaly clearly vanishes ($c_L=c_R=15$) and one can also check that $G$ has no 't Hooft anomaly. Moreover, the central charge of $G_k$ is less than $N=15$, so the embedding is not conformal.

While the theory is not rational with respect to $G_k$, we can still realise it via the coset construction as in \eqref{eq:coset_CFT}, which we repeat here for convenience:
\begin{equation}\label{eq:repeat_coset}
    \begin{tikzpicture}[z={(30:2cm)},baseline=(current bounding box.center)]
    \newcommand{\x}{4}
    \newcommand{\y}{4}

    \shade[top color=green,bottom color=white,middle color=green,opacity=0.1] (0,-\y/2,0) -- ++(0,-\y/2,0) -- ++(\x,0,0) -- ++(0,0,1) -- ++(0,\y/2,0) -- ++(-\x,0,0) -- cycle;
    \fill[blue,opacity=0.1] (0,0,0) -- ++(0,-\y/2,0) -- ++(\x,0,0) -- ++(0,0,1) -- ++(0,\y/2,0) -- ++(-\x,0,0) -- cycle;
    \draw[black] (0,-\y/2,0) -- ++(\x,0,0) -- ++(0,0,1) -- ++(-\x,0,0) -- cycle;
    \filldraw[black,fill=blue,fill opacity=0.1] (0,0,0) -- ++(\x,0,0) -- ++(0,0,1) -- ++(-\x,0,0) -- cycle;
    \node[left] at (0,0,0) {$\ket{G_k}$};
    \draw[black] (0,0,0) -- ++(0,-\y,0);
    \draw[black] (\x,0,0) -- ++(0,-\y,0);
    \draw[black] (\x,0,1) -- ++(0,-\y,0);
    \draw[black,opacity=0.1] (0,0,1) -- ++(0,-\y,0);

    \node at (\x/2,-\y*0.85,0.5) {$\SO(2N)_1$};
    \node at (\x/2,-\y*0.25,0.5) {$G_k$};

    \node[left] at (0,-\y/2,0) {$\ket{\SO(2N)_1/G_k^{L,R}}$};

    \draw[->,thick] (\x*0.4,0,0.5) -- (\x*0.6,0,0.5);
\end{tikzpicture}
\end{equation}
In particular, there is one such picture for the $L$ embedding and another for the $R$ embedding. The upper boundary hosts gapless modes realising the $G_k$ chiral algebra, while the lower boundary hosts modes realising the coset $\SO(2N)_1/G_k^{L,R}$. As discussed in \cref{sec:CS_basics}, when the embedding is not conformal the coset is not topological, so both boundaries are now gapless. The methods used throughout this work to describe the lower boundary in terms of Lagrangian algebras fail, and it is also not clear how to construct interfaces $B$ between two different such boundaries as in \eqref{eq:3d_setup}.

An alternative approach to construct such interfaces in the generic case proceeds as follows. We begin by constructing an auxiliary non-chiral CFT whose left-moving modes realise the coset $\SO(2N)_1/G_k^L$ and whose right-moving modes realise $\SO(2N)_1/G_k^R$. From the perspective of Chern--Simons theory, each embedding gives rise to an interface between $G_k$ and $SO(2N)_1$ as in \eqref{eq:repeat_coset}, which is equivalent to a boundary of $G_k \boxtimes SO(2N)_{-1}$. Using these boundaries, we assemble the setup
\begin{equation}
\begin{tikzpicture}[z={(30:2cm)},baseline=(current bounding box.center)]
    \newcommand{\x}{6}
    \newcommand{\y}{3}

    \fill[teal,opacity=0.1] (0,0,0) -- ++(0,-\y,0) -- ++(\x,0,0) -- ++(0,0,1) -- ++(0,\y,0) -- ++(-\x,0,0) -- cycle;
    \filldraw[black,fill=teal,fill opacity=0.1] (0,-\y,0) -- ++(\x,0,0) -- ++(0,0,1) -- ++(-\x,0,0) -- cycle;

    \draw[black] (0,0,0) -- ++(0,-\y,0);
    \draw[black] (\x,0,0) -- ++(0,-\y,0);
    \draw[black] (\x,0,1) -- ++(0,-\y,0);
    \draw[black,opacity=0.1] (0,0,1) -- ++(0,-\y,0);

    \filldraw[black,fill=teal,fill opacity=0.1] (0,0,0) -- ++(\x,0,0) -- ++(0,0,1) -- ++(-\x,0,0) -- cycle;

    \node at (\x/2,-\y*0.5,0.5) {$G_k \boxtimes \SO(2N)_{-1}$};
    \node[left] at (0,0,0) {$\ket{\SO(2N)_1/G_k^R}$};
    \node[left] at (0,-\y,0) {$\ket{\SO(2N)_1/G_k^L}$};

    \draw[->,thick] (\x*0.6,0,0.5) -- (\x*0.4,0,0.5);
    \draw[->,thick] (\x*0.4,-\y,0.5) -- (\x*0.6,-\y,0.5);
\end{tikzpicture}
\end{equation}
Since this is a non-chiral ($c = \bar{c}$) fermionic CFT, we expect that it admits conformal boundary conditions. Assuming the existence of such a boundary $B$, the three-dimensional setup becomes
\begin{equation}
\begin{tikzpicture}[z={(30:2cm)},baseline=(current bounding box.center)]
    \newcommand{\x}{7}
    \newcommand{\y}{4}

    \fill[teal,opacity=0.1] (0,0,0) -- ++(0,-\y,0) -- ++(\x,\y/2,0) -- ++(0,0,1) -- ++(-\x,\y/2,0) -- cycle;

    \filldraw[black,fill=teal,fill opacity=0.1] (0,0,0) -- ++(\x,-\y/2,0) -- ++(0,0,1) -- ++(-\x,\y/2,0) -- cycle;

    \filldraw[black,fill=teal,fill opacity=0.1] (0,-\y,0) -- ++(\x,\y/2,0) -- ++(0,0,1) -- ++(-\x,-\y/2,0) -- cycle;

    \draw[very thick] (\x,-\y/2,0) -- ++(0,0,1);

    \draw[black] (0,0,0) -- ++(0,-\y,0);
    \draw[black,opacity=0.1] (0,0,1) -- ++(0,-\y,0);

    \node at (\x/3,-\y*0.5,0) {$G_k \boxtimes \SO(2N)_{-1}$};
    \node[left] at (0,0,0) {$\ket{\SO(2N)_1/G_k^R}$};
    \node[left] at (0,-\y,0) {$\ket{\SO(2N)_1/G_k^L}$};
    \node[below right] at (\x,-\y/2,0.5) {$B$};

    \draw[->,thick] (\x*0.6,-0.3*\y,0.5) -- (\x*0.4,-0.2*\y,0.5);
    \draw[->,thick] (\x*0.4,-0.8*\y,0.5) -- (\x*0.6,-0.7*\y,0.5);
\end{tikzpicture}
\end{equation}
Unfurling this diagram by levering open the wedge, $B$ becomes an interface between the two gapless boundaries $\SO(2N)_1/G_k^{L,R}$. Then unfolding the boundaries of $G_k\boxtimes\SO(2N)_{-1}$ to interfaces between $G_k$ and $\SO(2N)_1$ we have
\begin{equation}
\begin{tikzpicture}[z={(30:2cm)},baseline=(current bounding box.center)]
    \newcommand{\x}{8}
    \newcommand{\y}{5}

    \shade[top color=green,bottom color=white,middle color=green,opacity=0.1] (0,-\y/2,0) -- ++(0,-\y/2,0) -- ++(\x,0,0) -- ++(0,0,1) -- ++(0,\y/2,0) -- ++(-\x,0,0) -- cycle;
    \fill[blue,opacity=0.1] (0,0,0) -- ++(0,-\y/2,0) -- ++(\x,0,0) -- ++(0,0,1) -- ++(0,\y/2,0) -- ++(-\x,0,0) -- cycle;
    \draw[black] (0,-\y/2,0) -- ++(\x,0,0) -- ++(0,0,1) -- ++(-\x,0,0) -- cycle;
    \draw[black] (0,0,0) -- ++(\x,0,0) -- ++(0,0,1) -- ++(-\x,0,0) -- cycle;
    \filldraw[black,fill=blue,fill opacity=0.1] (0,0,0) -- ++(\x,0,0) -- ++(0,0,1) -- ++(-\x,0,0) -- cycle;
    \node[left] at (0,0,0) {$\ket{G_k}$};
    \draw[black] (0,0,0) -- ++(0,-\y,0);
    \draw[black] (\x,0,0) -- ++(0,-\y,0);
    \draw[black] (\x,0,1) -- ++(0,-\y,0);
    \draw[black,opacity=0.1] (0,0,1) -- ++(0,-\y,0);
    \draw[very thick,black] (\x/2,-\y/2,0) -- ++(0,0,1);

    \node at (\x/2,-\y*0.85,0.5) {$\SO(2N)_1$};
    \node at (\x/2,-\y*0.25,0.5) {$G_k$};

    \node[below] at (\x*0.25,-\y/2,0) {$\ket{\SO(2N)_1/G_k^L}$};
    \node[below] at (\x*0.75,-\y/2,0) {$\ket{\SO(2N)_1/G_k^R}$};
    \node[below] at (\x*0.5,-\y/2,0) {$B$};

    \draw[->,thick] (\x*0.4,0,0.5) -- (\x*0.6,0,0.5);
    \draw[->,thick] (\x*0.2,-\y/2,0.5) -- (\x*0.3,-\y/2,0.5);
    \draw[->,thick] (\x*0.7,-\y/2,0.5) -- (\x*0.8,-\y/2,0.5);
\end{tikzpicture}
\end{equation}
where we have placed the standard boundary condition on the upper boundary realising the $G_k$ chiral algebra. This is precisely the setup in \eqref{eq:3d_setup}, with the topological boundaries $\ket{\cA_{L,R}}$ replaced by the gapless coset boundaries. That is, the Chern--Simons realisation of the symmetric boundary condition discussed in this work carries over to the general case, but in the non-conformally embedded case the interface $B$ is no longer described by a bimodule but instead by a conformal boundary condition of a particular heterotic non-chiral CFT.

It remains to understand the scattering process in this generic setup. As always, a generic local operator of the two-dimensional fermion theory is realised in the Chern--Simons construction as a pair of operators on the two gapless boundaries of $G_k$, joined by a bulk anyon. Scattering such an operator through the interface $B$ then proceeds as follows:
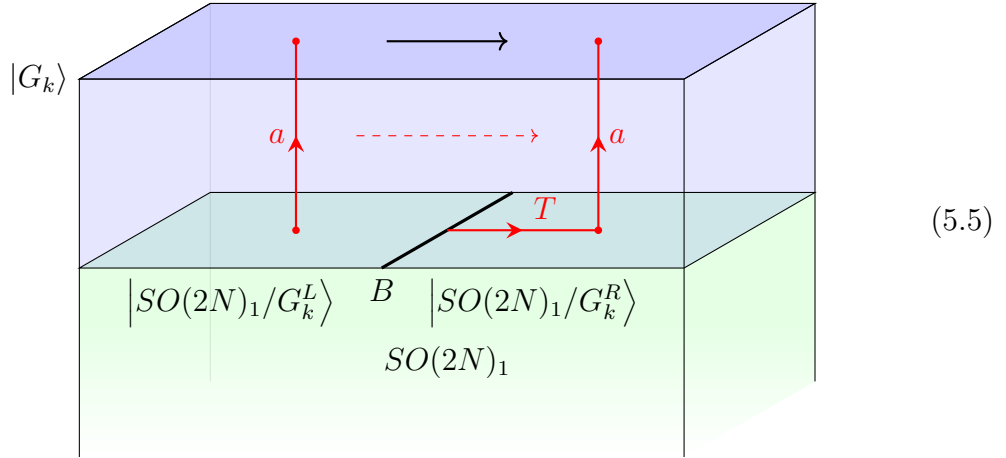
\begin{equation}
\begin{tikzpicture}[z={(30:2cm)},baseline=(current bounding box.center)]
    \newcommand{\x}{8}
    \newcommand{\y}{5}

    \shade[top color=green,bottom color=white,middle color=green,opacity=0.1] (0,-\y/2,0) -- ++(0,-\y/2,0) -- ++(\x,0,0) -- ++(0,0,1) -- ++(0,\y/2,0) -- ++(-\x,0,0) -- cycle;
    \fill[blue,opacity=0.1] (0,0,0) -- ++(0,-\y/2,0) -- ++(\x,0,0) -- ++(0,0,1) -- ++(0,\y/2,0) -- ++(-\x,0,0) -- cycle;
    \draw[black] (0,-\y/2,0) -- ++(\x,0,0) -- ++(0,0,1) -- ++(-\x,0,0) -- cycle;
    \draw[black] (0,0,0) -- ++(\x,0,0) -- ++(0,0,1) -- ++(-\x,0,0) -- cycle;
    \filldraw[black,fill=blue,fill opacity=0.1] (0,0,0) -- ++(\x,0,0) -- ++(0,0,1) -- ++(-\x,0,0) -- cycle;
    \node[left] at (0,0,0) {$\ket{G_k}$};
    \draw[black] (0,0,0) -- ++(0,-\y,0);
    \draw[black] (\x,0,0) -- ++(0,-\y,0);
    \draw[black] (\x,0,1) -- ++(0,-\y,0);
    \draw[black,opacity=0.1] (0,0,1) -- ++(0,-\y,0);
    \draw[very thick,black] (\x/2,-\y/2,0) -- ++(0,0,1);

    \begin{scope}[every node/.style={fill=red,shape=circle,inner sep=1pt}]
        \draw[thick,red,decoration = {markings, mark=at position .5 with {\arrowreversed[scale=1.5]{stealth}}}, postaction=decorate] (\x*0.25,0,0.5) node{} -- ++(0,-\y/2,0) node{};
        \draw[thick,red,decoration = {markings, mark=at position .5 with {\arrowreversed[scale=1.5]{stealth}}}, postaction=decorate] (\x*0.75,0,0.5) node{} -- ++(0,-\y/2,0);
        \draw[thick,red,decoration = {markings, mark=at position .5 with {\arrow[scale=1.5]{stealth}}}, postaction=decorate] (\x*0.5,-\y/2,0.5) -- ++(\x*0.25,0,0) node{};
        \draw[->,dashed,red] (\x*0.35,-\y*0.25,0.5) -- ++(\x*0.3,0,0);
    \end{scope}
    \node[left,red] at (\x*0.25,-\y/4,0.5) {$a$};
    \node[right,red] at (\x*0.75,-\y/4,0.5) {$a$};
    \node[above right, red] at (\x*0.625,-\y/2,0.5) {$T$};

    \node at (\x/2,-\y*0.85,0.5) {$\SO(2N)_1$};

    \node[below] at (\x*0.25,-\y/2,0) {$\ket{\SO(2N)_1/G_k^L}$};
    \node[below] at (\x*0.75,-\y/2,0) {$\ket{\SO(2N)_1/G_k^R}$};
    \node[below] at (\x*0.5,-\y/2,0) {$B$};

    \draw[->,thick] (\x*0.4,0,0.5) -- (\x*0.6,0,0.5);
\end{tikzpicture}
\end{equation}
Namely, the bulk of the anyon simply shifts to the right, as it is topological. The operator on the $SO(2N)_1/G^L_k$ interface, however, moves through the defect $B$ as it is dragged forward. This gives rise to a twist line $T$, a topological line occurring in the fusion $B \otimes \bar{B}$ which trails the outgoing operator. The twist line is confined to the $SO(2N)_1/G^R_k$ interface and cannot venture out into the bulk. This is in contrast with the conformally embedded case, where the lower boundaries are topological and therefore the junction between $T$ and $a$ would be topological also.

A full answer to the scattering problem in general would involve identifying the twist line $T$ for generic embeddings $L,R$, which will depend on the details of the interface $B$. Notably, however, the symmetry properties of the outgoing excitation with respect to $G$ only depend on the operator on the \emph{upper} boundary. This operator is known, and is exactly the operator we sent in. This simple observation, which is true whether the embeddings are conformal or not, ensures that the interface $B$ preserves the $G$ symmetry. The remaining mystery only lies in the operator it is paired with on the lower interface, which lives at the junction between the twist line $T$ and the bulk anyon $a$. It is possible the topological line $T$, assuming it exists, no longer belongs to a \emph{finite} categorical symmetry in this case.

\subsubsection*{'t Hooft lines and the lift to four dimensions}

The non-abelian model we studied in \cref{sec:non-abelian} was motivated, in part, by its utility in modelling the scattering of chiral fermions off certain 't Hooft lines in abelian gauge theories in four-dimensions, following the general strategy set out in \cite{vanBeest:2023dbu,vanBeest:2023mbs}.

The procedure goes as follows. One considers scattering fermions off a Dirac monopole in a tree-level approximation, which amounts to neglecting gauge field fluctuations. To proceed we decompose four-dimensional fermions into angular momentum modes around a given 't Hooft line. All but the lowest modes get bounced back out by a centrifugal barrier. The lowest modes, on the other hand, reduce to a collection of two-dimensional complex Weyl fermions in an effective two-dimensional BCFT. The monopole becomes a conformal boundary, which should preserve any symmetry $G$ with a vanishing 't Hooft anomaly in two dimensions; equivalently, the symmetries which in four dimensions do \emph{not} engage in a 2-group with the magnetic 1-form symmetry. For the minimal charge 't Hooft lines in the theories \eqref{eq: 4d theory general} and \eqref{eq: 4d theory lucky}, we have found precisely these $G$-symmetric conformal boundaries.

In this general procedure, one finally would like to understand how this twisted scattering, encountered only for the lowest angular momentum modes, should be thought about back in four dimensions. The 't Hooft lines tackled in \cite{vanBeest:2023dbu,vanBeest:2023mbs} gave rise to two-dimensional BCFT problems of the equal rank type, which were therefore solved by the emission of twist operators at the end of \emph{invertible} lines. It was found, in every case, that there then exists a codimension-1 topological surface in the four-dimensional theory which corresponds to this twist line upon spherical reduction. One therefore has a purely four-dimensional way to construct the outgoing excitation: it is created by an operator supported not on a 2-sphere like the ingoing excitation, but on a topological 3-ball, with all energy and charge concentrated on its 2-sphere boundary. This topological surface need not be invertible---generically it is not---but it nonetheless reduces to an invertible line in two dimensions.
Furthermore, in this approximation one is in the odd situation that a generic ingoing state, like some localised wavepacket, is scattered into a superposition of a twisted state, along with untwisted states generated by higher incoming angular momentum modes.

Does this four-dimensional formulation of the outgoing excitation in the lowest angular momentum sector generalise to the 't Hooft lines in the theories  \eqref{eq: 4d theory general} and \eqref{eq: 4d theory lucky}? It is far from clear. One would require a codimension-1 surface in four dimensions which reduces to a non-invertible line in two dimensions. As argued in \cite{vanBeest:2023mbs}, such a surface would act non-invertibly on local operators in the four-dimensional theory, which in turn would imply the existence of topological lines in four dimensions; i.e.\ the existence of a 2-form symmetry. No such symmetry of the theories \eqref{eq: 4d theory general} and \eqref{eq: 4d theory lucky} is known.

In lieu of a better understanding of the picture in four dimensions, there is still something robust we can say. If we scatter regular, local fermions off a minimal charge 't Hooft line in one of these theories, the outgoing state will generically carry fractional charge under the symmetry $U(1)_\text{bonus}$. Recall, this is a perfectly good symmetry of the four dimensional theory---its ABJ anomaly vanishes---which is nonetheless broken (and indeed fractionalised) by the 't Hooft line due to its 't Hooft anomaly in two dimensions. In particular, we learn that the outgoing state cannot be built from local fermions of the theory. Exactly how it \emph{can} be built in the four-dimensional theory remains an open question.

\section*{Acknowledgments}

The authors thank George Koundouros and David Tong for providing the inspiration for this paper, for many fruitful discussions, and for communication about upcoming work \cite{davidandgeorge}. PBS thanks Francesco Benini, Antonio Santaniello, and Marco Venuti for discussions leading to much enlightenment, and Kantaro Ohmori and Masataka Watanabe for communications on closely related ideas four years ago. MVCH is grateful to Laurens Lootens for useful discussions. The work of GAT is supported in part by the STFC Consolidated Grant ST/X000575/1, and in part by the R\&D\&I Project ``Uncovering Higher Structures in Supersymmetric and Superconformal Theories'' linked to the  group of Matteo Lotito, funded by CSIC-IFT, PID2024-156043NB-I00 and CEX2025-001574-S, funded by MICIU/AEI/10.13039/501100011033. PBS is supported by the ERC-COG grant NP-QFT No.~864583 and the MUR-FARE2020 grant No.~R20E8NR3HX. RM is supported by UK Engineering and Physical Sciences grant EP/Z000106/1. MVCH was supported by a Research Fellowship from King's College, Cambridge, David Tong's Simons Investigator Award, STFC consolidated grant ST/X000664/1, and UK Research \& Innovation (UKRI) grant UKRI4224.

\appendix
\crefalias{section}{appendix}

\part*{Appendices}
\addcontentsline{toc}{section}{Appendices}
\etocsettocdepth.toc{part}
\etocsetlocaltop.toc{part}
\begingroup
\etocsettocstyle{}{}
\etocignoretoctocdepth
\etocsetnexttocdepth{section}
\localtableofcontents
\endgroup

\section{Chern--Simons anyon data}\label{app:MTCs}

In this appendix we record various formulae for the relevant data of the Chern--Simons theories (i.e.\ (super-)modular tensor categories) used throughout the text.

\subsection{$\SO(2N)_1$}

For all $N$, $\SO(2N)_1$ has two simple lines, $\1$ and $V$, corresponding to the trivial and vector representations. The fusion is invertible
\begin{equation}
    V^2 = \1
\end{equation}
Their spins are
\begin{equation}
    \theta_\1 = 1 \qc \theta_V = -1
\end{equation}
so $V$ is a transparent fermion. The dependence on $N$ is not manifest in the fusion rules and twists alone. $\SO(2N)_1$ is the subcategory of $\Spin(2N)_1$ generated by its $\1$ and $V$ lines (see below).

\subsection{$\Spin(2N)_1$}

For any $N$, $\Spin(2N)_1$ has four simple lines denoted $\1,V,S,C$, corresponding to the trivial, vector, spinor and co-spinor representations. They are all invertible, so all have quantum dimension 1. Their topological spins are
\begin{equation}
    \theta_\1 = 1\qc \theta_V = -1\qc \theta_S = \theta_C = \exp(2\pi i N/8)
\end{equation}

\subsubsection*{$N$ even}

For even $N$, the fusion rules are those of $\bZ_2 \times \bZ_2$,
\begin{equation}
    V^2 = S^2 = C^2 = \1\qc  VS = C\qc VC = S\qc SC = V
\end{equation}
where we suppress the $\otimes$ symbol for brevity. All objects are self dual, with Frobenius--Schur indicators $1,1,(-1)^{N/2},(-1)^{N/2}$.

A consistent choice of $F$-symbols is
\begin{equation}
    F^{SSS}_S = F^{SSC}_C = F^{SCS}_C = F^{SCC}_S = F^{CSS}_C = F^{CSC}_S = F^{CCS}_S = F^{CCC}_C = (-1)^{N/2}
\end{equation}
with all others equal to 1, and a consistent choice of $R$-matrices is
\begin{equation}
    R^{VV}_\1 = R^{VC}_S = R^{SV}_C = -1\qc  R^{SS}_\1 = R^{CC}_\1 = R^{CS}_V = e^{\frac{2 \pi i}{8} N}\qc  R^{SC}_V = -e^{\frac{2 \pi i}{8} N}
\end{equation}
with all others equal to 1.

\subsubsection*{$N$ odd}

For odd $N$, on the other hand, the fusion rules are those of $\bZ_4$,
\begin{equation}
    V^2 = \1\qc VS = C\qc VC = S\qc S^2 = C^2 = V\qc SC = \1
\end{equation}
The $\1$ and $V$ lines are self-dual, both with Frobenius--Schur indicator equal to 1, whereas $S$ and $C$ are dual to each other.

A choice of $F$-symbols is
\begin{equation}
    F^{VSV}_S = F^{VCV}_C = F^{SVC}_V = F^{CVS}_V = F^{SSS}_C = F^{CCC}_S = -1
\end{equation}
with all others equal to 1, and a choice of $R$-matrices is
\begin{gather}
    R^{VV}_\1 = -1\qc R^{VS}_C = R^{SV}_C = R^{VC}_S = R^{CV}_S = -i^{N}\qc R^{SS}_V = R^{CC}_V = e^{\frac{2 \pi i}{8} N}\,,\nn\\
    R^{SC}_\1 = R^{CS}_\1 = e^{-\frac{2 \pi i}{8} N}
\end{gather}

\subsection{$\SU(N)_k$}

The integrable representations of $\SU(N)_k$ can be labelled by the corresponding irreducible representations of the (finite) Lie algebra $\su(N)$, and are those whose standard Young tableau has fewer than or equal to $k$ columns \cite{DiFrancesco:1997nk}. There are $k+N-1 \choose N-1$ such representations. Fusion is then the level-$k$ truncation of the standard $\su(N)$ tensor product.

In terms of Dynkin labels, the integrable highest weights are
\begin{equation}
    P_k^+ = \left\{ \lambda = (\lambda_1,\dots,\lambda_{N-1}) : \lambda_i \in \bZ_{\geq0}, \sum_{i=1}^{N-1}\lambda_i \leq k \right\}
\end{equation}
The dual of the representation $\lambda$ has its Dynkin label reversed. The conformal dimensions are
\begin{equation}
    h_\lambda = \frac{(\lambda,\lambda+2\rho)}{2(k+N)}
\end{equation}
with $\rho=(1,\dots,1)$, which is related to the topological spin via \eqref{eq:theta=exp(h)}. The central charge is
\begin{equation}
    c = \frac{k(N^2-1)}{k+N}
\end{equation}

Further details, including a choice of $F$-symbols and $R$-matrices for $SU(2)_k$ can be found in \cite{Ardonne_2010}.

\subsection{$U(1)^N_K$}

Let us start with the case $N=1$, which is of particular use to us since it is used in the non-abelian example of \cref{sec:non-abelian}. The spin of a primary with charge $q\in\Z$ is
\begin{equation}\label{eq:abelian_conf_dim}
    \theta_q = \exp(\frac{\pi i q^2}{k})
\end{equation}
and the fusion is simply the addition of charges.

There are two cases to distinguish.
\begin{itemize}
    \item \textbf{Even $k$.} In this case, we have only $\abs{k}$ independent anyons $q\in\{0,1,\dots,k-1\}$, with fusion $q \otimes q' = (q+q')\mod k$. This is because with $k$ even \eqref{eq:abelian_conf_dim} is a well-defined quadratic form on $\bZ_k$.

    The $R$-matrices are
    \begin{equation}\label{eq:abelian_R_even}
        R^{q_1 q_2} = \exp(2\pi i\frac{q_1 q_2}{2k})
    \end{equation}
    Since the fusion is abelian we do not include the full set of $R$-matrix labels as in \eqref{eq:R_def} since the internal vertices and outgoing anyon are fully determined by the incoming anyons $q_1$ and $q_2$.

    Shifting $q_2 \to q_2 + k$ changes $R^{q_1 q_2}_{q_3}$ by a sign depending on the parity of $q_1$. The $R$-matrix is not required to be bilinear, but it is required to satisfy the hexagon equations. These involve both the $R$-matrices and $F$-symbols and can be satisfied by choosing the $F$-symbols to be \cite[Example 2.1 (3)]{galindo2025note} (see also \cite[Theorem 8.4.9]{etingof})
    \begin{equation}
        F^{q_1 q_2 q_3} = \begin{cases}
            (-1)^{q_1} & q_2 + q_3 \geq k \\ 1 & \text{else}
        \end{cases}
    \end{equation}
    Again we do not display the other labels on the $F$-symbols as in \eqref{eq:F_def} as they are fixed by the three incoming lines. Moreover, these $F$-symbols satisfy the pentagon equation. This is a bosonic TQFT.

    \item \textbf{Odd $k$.} In this case we see that \eqref{eq:abelian_conf_dim} is not a well-defined quadratic form on $\bZ_k$. Instead, we must now have $2\abs{k}$ independent anyons $q \in \{0,1,\dots,2k-1\}$ with fusion $q \otimes q' = (q+q') \mod 2k$. Then \eqref{eq:abelian_conf_dim} is well-defined on $\bZ_{2k}$.

    The $R$-matrices are again taken as \eqref{eq:abelian_R_even}, which are now well-defined on $\bZ_{2k}$. We can then simply take trivial $F$-symbols
    \begin{equation}
        F^{q_1 q_2 q_3} = 1
    \end{equation}
    which then satisfy both the pentagon and hexagon equations.

    The catch in this case is that the anyon with charge $k$ is now transparent. That is,
    \begin{equation}
        \frac{\theta_k \theta_q}{\theta_{q\otimes k}} = 1
    \end{equation}
    for all $q$. Furthermore, $\theta_k=-1$, so the charge $k$ anyon is a transparent fermion. This is a fermionic TQFT.
\end{itemize}

For $N>1$, the (super-)modular tensor category is now labelled by a symmetric integral non-degenerate matrix $K$ generalising the level $k$ above. Again the primaries are labelled by integer charges $q\in\Z^N$, with topological spin
\begin{equation}\label{eq:abelian_spin_N>1}
    \theta_q = \exp(\pi i q^T K^{-1} q)
\end{equation}
Similarly to the $N=1$ case, we distinguish two cases.
\begin{itemize}
    \item \textbf{Even $K$.} The $K$-matrix is said to be even if all its diagonal entries are even, otherwise it is odd. In this case there are $\abs{\det(K)}$ anyons labelled by $\bZ^N / K\bZ^N$. In this case similar signs can be introduced in the $F$-symbols as above in order to satisfy both the pentagon and hexagon equations. This is a bosonic TQFT.

    \item \textbf{Odd $K$.} In this case \eqref{eq:abelian_spin_N>1} is not well-defined on $\bZ^N/ K\bZ^N$ as
    \begin{equation}
        \theta_{q+Kn} \theta_q^{-1} = \exp(\pi i n^T K n) = \exp(\pi i \sum_i K_{ii} n_i^2)
    \end{equation}
    where we have used that $K$ is symmetric. Hence if any of the diagonal elements of $K$ are odd, then shifting $q\to q+Kn$ can change the spin by a sign.

    We can, however, associate primaries $q \sim q + Kn$ whenever $n^T K n = 0 \mod 2$. Defining $\Lambda = \{ n \in \bZ^N : n^T K n = 0 \mod 2 \}$, the set of distinct anyons is then $\bZ^N / K \Lambda$. There are $2\abs{\det(K)}$ distinct anyons.

    As for $N=1$, there is a transparent fermion in this case which can be represented by the anyon with charge $K n_f$ where $n_f\in\bZ^N$ satisfies $n_f^T K n_f = 1\mod 2$. This is a fermionic TQFT.
\end{itemize}

\section{SPT classes for abelian symmetries}\label{app:arf_abelian}

In this appendix we justify the formula \eqref{eq:sptclassvw} for the SPT class of the abelian boundary states by means of a half-space gauging calculation.

Let $\cR \in O(N)_\bQ$ be a rational orthogonal matrix. As reviewed in the main text, there exist integer vectors $v, w \in \bZ^N$ such that
\begin{equation}\label{eq:appRvw}
    \cR(\rho/2 + v) = \rho/2 + w
\end{equation}
where $\rho = (1, \dots, 1)$.\footnote{We recall the proof here. For any $n \in \bZ^N \cap \cR \bZ^N$, we have mod 2 the equality $n \cdot \rho \equiv n^2 = (\cR^{-1} n)^2 \equiv n \cdot \cR \rho$. Thus $(\cR \rho - \rho) / 2 \in (\bZ^N \cap \cR \bZ^N)^* = \bZ^N + \cR \bZ^N$, hence can be written as $w - \cR v$ for some integer vectors $v, w$. This proves the claim.}

\noindent
\textbf{Claim:} The abelian boundary state labelled by $\cR$ has SPT class
\begin{equation}\label{eq:appvwp}
    (v + w) \cdot \rho \mod 2
\end{equation}
where 0 indicates the vector class and 1 indicates the axial class.

\noindent
\textit{Proof.} We will construct the abelian boundary state, or rather its unfolded version, as the half-space gauging defect of the finite abelian symmetry
\begin{equation}\label{eq:grdef}
    \cG_\cR \coloneqq \cR \bZ^N / (\cR \bZ^N \cap \bZ^N)
\end{equation}
This extends calculations in \cite{Arias-Tamargo:2026urw,Wei:2026fsn}.

\paragraph{Coupling to $U(1)^N$}
Let $\cT$ be the theory of $N$ Weyl fermions. We start by writing down the partition function on the torus with an arbitrary $U(1)^N$ background and spin structure. To do this we introduce the \emph{twisted theta series}
\begin{equation}
    \theta_3(\tau)_a^b \coloneqq \sum_{n \in \bZ^N + a} q^{n^2/2} e^{2 \pi i n \cdot b}
\end{equation}
where $\tau$ is the modular parameter of the torus, $a, b \in \bR^N$, and we define $q = e^{2 \pi i \tau}$. This series has the following periodicity properties in $a$ and $b$ under shifts by an integer vector $n \in \bZ^N$:
\begin{equation}
\begin{aligned}
    \theta_3(\tau)_{a+n}^b &= \theta_3(\tau)_a^b \\
    \theta_3(\tau)_a^{b+n} &= e^{2 \pi i a \cdot n} \theta_3(\tau)_a^b
\end{aligned}
\end{equation}
It also has the following modular properties in the variable $\tau$:
\begin{equation}
\begin{aligned}
    \theta_3(-1/\tau)_a^b &= (-i\tau)^{N/2} e^{2 \pi i a\cdot b} \theta_3(\tau)_b^{-a} \\
    \theta_3(\tau+1)_a^b &= e^{-2\pi i (a^2 + a \cdot \rho) / 2} \theta_{3}(\tau)_a^{b+a+\rho/2}
\end{aligned}
\end{equation}
In terms of the twisted theta series, the partition functions of the Weyl fermions are then, up to a phase convention,
\begin{equation}
\begin{aligned}
    \cZ_\NS^\NS[\tau]_a^b &= \frac{\theta_3(\tau)_a^b}{\eta(\tau)^N}
    &
    \cZ_\NS^\R[\tau]_a^b &= e^{-2\pi i (a^2 + a\cdot \rho)/2} \frac{\theta_3(\tau)_a^{b+\rho/2}}{\eta(\tau)^N}
    \\
    \cZ_\R^\NS[\tau]_a^b &= e^{-2\pi i b^2/2}\frac{\theta_3(\tau)_{a + \rho/2}^b}{\eta(\tau)^N}
    &
    \cZ_\R^\R[\tau]_a^b &= e^{-2\pi i (a^2 + b^2 + a \cdot \rho)/2} \frac{\theta_3(\tau)_{a+\rho/2}^{b+\rho/2}}{\eta(\tau)^N}
\end{aligned}
\end{equation}
Here the lower scripts indicate the backgrounds around the spatial circle, and the upper scripts indicate the backgrounds around the temporal circle.

The above partition functions are sick as a result of the $U(1)^N$ symmetry being anomalous, and this is reflected in two ways. First, the partition functions do not just depend on the equivalence classes of $a, b$ in $\bR^N / \bZ^N = U(1)^N$, but on the choice of representatives $a, b \in \bR^N$ themselves. Specifically, they change in the following way upon changing the choice of representatives:
\begin{itemize}

\item Periodicity under shifting $a$ by $n \in \bZ^N$:
\begin{equation}
\begin{aligned}
    \cZ_\NS^\NS[\tau]_{a+n}^b &= \cZ_\NS^\NS[\tau]_a^b \\
    \cZ_\NS^\R [\tau]_{a+n}^b &= e^{-2 \pi i a\cdot n} \cZ_\NS^\R[\tau]_a^b \\
    \cZ_\R ^\NS[\tau]_{a+n}^b &= \cZ_\R^\NS[\tau]_a^b \\
    \cZ_\R ^\R [\tau]_{a+n}^b &= e^{-2 \pi i a\cdot n} \cZ_\R^\R[\tau]_a^b
\end{aligned}
\end{equation}

\item Periodicity under shifting $b$ by $n \in \bZ^N$:
\begin{equation}
\begin{aligned}
    \cZ_\NS^\NS[\tau]_a^{b+n} &= e^{2 \pi i a \cdot n} \cZ_\NS^\NS[\tau]_a^b \\
    \cZ_\NS^\R [\tau]_a^{b+n} &= e^{2 \pi i a \cdot n} \cZ_\NS^\R[\tau]_a^b \\
    \cZ_\R ^\NS[\tau]_a^{b+n} &= e^{2 \pi i (a-b) \cdot n} \cZ_\R^\NS[\tau]_a^b \\
    \cZ_\R ^\R [\tau]_a^{b+n} &= e^{2 \pi i (a-b) \cdot n} \cZ_\R^\R[\tau]_a^b
\end{aligned}
\end{equation}

\end{itemize}
Second, they are not consistent with fermionic modular invariance:
\begin{itemize}

\item Modular $\cS$-transformations of $\tau$:
\begin{equation}
\begin{aligned}
    \cZ_\NS^\NS[-1/\tau]_a^b &= \textcolor{bad}{e^{2 \pi i a\cdot b}} \cZ_\NS^\NS[\tau]_b^{-a} \\
    \cZ_\NS^\R [-1/\tau]_a^b &= \textcolor{bad}{e^{2 \pi i a\cdot b}} \cZ_\R ^\NS[\tau]_b^{-a} \\
    \cZ_\R ^\NS[-1/\tau]_a^b &= \textcolor{bad}{e^{2 \pi i a\cdot b}} \cZ_\NS^\R [\tau]_b^{-a} \\
    \cZ_\R ^\R [-1/\tau]_a^b &= \textcolor{good}{e^{-2 \pi i N/4}} \textcolor{bad}{e^{2 \pi i a \cdot b}} \cZ_\R^\R[\tau]_b^{-a}
\end{aligned}
\end{equation}

\item Modular $\cT$-transformations of $\tau$:
\begin{equation}
\begin{aligned}
    \cZ_\NS^\NS[\tau+1]_a^b &= \textcolor{good}{e^{-2 \pi i N/24}} \cZ_\NS^\R[\tau]_a^{b+a} \\
    \cZ_\NS^\R [\tau+1]_a^b &= \textcolor{good}{e^{-2 \pi i N/24}} \textcolor{bad}{e^{-2 \pi i a^2}} \cZ_\NS^\NS[\tau]_a^{b+a} \\
    \cZ_\R ^\NS[\tau+1]_a^b &= \textcolor{good}{e^{2 \pi i N/12}} \textcolor{bad}{e^{2 \pi i a \cdot b}} \cZ_\R^\NS[\tau]_a^{b+a} \\
    \cZ_\R ^\R [\tau+1]_a^b &= \textcolor{good}{e^{2 \pi i N/12}} \textcolor{bad}{e^{2 \pi i a \cdot b}} \cZ_\R^\R[\tau]_a^{b+a}
\end{aligned}
\end{equation}

\end{itemize}
The anomalous phases in green are \textcolor{good}{good}, because they are gravitational anomalies and should be present for any theory with central charge $c - \bar{c} = N$. The remaining anomalous phases are \textcolor{bad}{bad}, and reflect the $U(1)^N$ anomaly.

\paragraph{Coupling to $\cG_\cR$}
We now specialise to the case where $a, b$ belong to $\cG_\cR$, namely $\cR \bZ^N / (\cR \bZ^N \cap \bZ^N)$. It can be checked that, on account of the fact $a \cdot b \in \bZ$, this renders the partition functions well-defined (independent of the equivalence class representatives of $a, b$) and modular covariant (all \textcolor{bad}{bad} phases vanish). This shows that $\cG_\cR$ is non-anomalous.\footnote{Working at genus 1 is sufficient for abelian symmetries.} Furthermore, in writing down $\cZ$, we have implicitly made a specific choice of discrete torsion for the coupling to $\cG_\cR$, different choices of which would differ by $H^2(B\cG_\cR; U(1))$.

\paragraph{Gauging $\cG_\cR$}
It remains to establish that with this choice of discrete torsion, $\cT / \cG_\cR \cong \cT \times \Arf^{(v+w) \cdot \rho}$. We do this by computing gauged partition functions.

To warm up, we begin with the NS-NS sector. Including a fugacity $x \in \bR^N$ to keep track of the operators under the duality, the gauged partition function is
\begin{equation}
\begin{aligned}
    (\cZ / \cG_\cR)_\NS^\NS[\tau, x] &= \frac{1}{|\cG_\cR|} \sum_{a, b \in \cG_\cR} \cZ_\NS^\NS[\tau, x]_a^b \\
    &= \frac{1}{|\cG_\cR|} \frac{1}{\eta(\tau)^N} \sum_{a, b \in \cG_\cR} \sum_{n \in \bZ^N + a} q^{n^2/2} e^{2 \pi i n \cdot b + 2 \pi i n \cdot x}
\end{aligned}
\end{equation}
The sum over $b$ acts as a Lagrange multiplier that restricts $n$ to $(\bZ^N \cap \cR \bZ^N) + a$ and inserts a factor of $|\cG_\cR|$. Then $n$ absorbs $a$ and becomes $\cR \bZ^N$-valued. Thus
\begin{equation}
    (\cZ / \cG_\cR)_\NS^\NS[\tau, x]
    = \frac{1}{\eta(\tau)^N} \sum_{n \in \cR \bZ^N} q^{n^2/2} e^{2 \pi i n \cdot x}
    = \cZ_\NS^\NS[\tau, \cR^{-1} x]
\end{equation}
which establishes that $\cT \cong \cT / \cG_\cR$ in the NS-NS sector, with the shift in fugacities $x \to \cR^{-1} x$ indicating a nontrivial remapping of operators.

By modular covariance, the above result also holds in the NS-R and R-NS sectors. It therefore remains to consider the R-R sector. This is where our key sign \eqref{eq:appvwp} finally appears. The gauged partition function is
\begin{equation}
    (\cZ / \cG_\cR)_\R^\R[\tau, x]
    =
    \frac{1}{|\cG_\cR|}
    \sum_{a,b \in \cG_\cR}
    e^{-2 \pi i (a^2 + b^2 + a \cdot \rho) / 2}
    \frac{1}{\eta(\tau)^N}
    \sum_{n \in \bZ^N + a + \rho/2}
    q^{n^2/2} e^{2 \pi i n \cdot (b + \rho/2 + x)}
\end{equation}
Redefining $(a,b,n) \to (\cR a, \cR b, \cR n)$, this becomes
\begin{multline}
    \frac{1}{|\cG_\cR|}
    \sum_{a,b \in \bZ^N / (\bZ^N \cap \cR^{-1} \bZ^N)}
    e^{-2 \pi i (a \cdot \rho + b \cdot \rho + a \cdot \cR^{-1} \rho) / 2} \\
    \frac{1}{\eta(\tau)^N}
    \sum_{n \in \cR^{-1} \bZ^N + a + \cR^{-1} \rho / 2}
    q^{n^2/2} e^{2 \pi i n \cdot (b + \cR^{-1} \rho / 2 + \cR^{-1} x)}
\end{multline}
As before, the sum over $b$ acts as a Lagrange multiplier, adding a factor of $|\cG_\cR|$ and constraining $n$ to lie in $\bZ^N + \rho/2$. Invoking \eqref{eq:appRvw}, we obtain
\begin{equation}
    \sum_{a \in \bZ^N / (\bZ^N \cap \cR^{-1} \bZ^N)}
    \frac{1}{\eta(\tau)^N}
    \sum_{n \in (\bZ^N \cap \cR^{-1} \bZ^N) + a + v + \rho/2}
    q^{n^2/2} e^{2 \pi i n \cdot (\rho/2 + \cR^{-1} x)} e^{2 \pi i (n-a) \cdot (v-\cR^{-1}w)}
\end{equation}
Then we calculate modulo 1
\begin{equation}
\begin{aligned}
    (n - a) \cdot (v - \cR^{-1} w) &= (v + \rho/2) \cdot (v - \cR^{-1} w) \\
    &= (v + \rho/2) \cdot v - (w + \rho/2) \cdot w \\
    &= (v + w) \cdot \rho/2
\end{aligned}
\end{equation}
The variable $a$ can now be absorbed into an enlarged variable $n$ with range $\bZ^N + \rho/2$. This gives the desired answer
\begin{equation}
    (\cZ / \cG_\cR)_\R^\R[\tau, x]
    =
    (-1)^{(v + w) \cdot \rho}
    \cZ_\R^\R[\tau, \cR^{-1} x]
\end{equation}
Just as in the NS-NS sector, we see the same remapping of fugacities $x \to \cR^{-1} x$, but now with an overall sign. This implies the presence of an additional $\Arf^{(v+w)\cdot\rho}$ on the right-hand side of the duality.

\paragraph{Half-space gauging} Via the half-space gauging construction, the duality proved above, which we will denote as
\begin{equation}
    D_\cR : \cT / \cG_\cR \to \cT \times \Arf^{(v+w) \cdot \rho}
\end{equation}
defines a half-space gauging defect $\cN_\cR$ via the setup
\begin{equation}
    \cN_\cR \; = \quad
    \begin{tikzpicture}[scale=0.75,baseline=-1.3cm]
        \fill[blue!5] (0,0) rectangle (4,-3);
        \fill[green!5] (4,0) rectangle (8,-3);
        \fill[blue!5] (8,0) rectangle (12,-3);
        \draw[thick] (4,0) -- (4,-3);
        \draw[thick,dashed] (8,0) -- (8,-3);
        \node at (2,-1.5) {$\cT$};
        \node[above] at (4,0) {$\bra{\text{Dir}}$};
        \node at (6,-1.5) {$\cT / \cG_\cR $};
        \node[above] at (8,0) {$D_\cR$};
        \node at (10,-1.5) {$\cT \times \Arf^{(v+w)\cdot\rho}$};
    \end{tikzpicture}
\end{equation}
where in the middle we have a dynamical $\cG_\cR$ gauge field, capped off by a Dirichlet boundary condition $\bra{\text{Dir}}$ on the left and the duality interface $D_\cR$ on the right.

Because of the mapping of fugacities $x \to \cR^{-1} x$ implemented by $D_\cR$, the defect $\cN_\cR$ preserves a chiral $U(1)^N$ symmetry, and is identified as the unfolded abelian boundary state labelled by $\cR$. We see that this is a vector-class defect $\cT \to \cT$ or an axial-class defect $\cT \to \cT \times \Arf$ depending on the value of \eqref{eq:appvwp}, as claimed. \hfill $\square$

\section{Lagrangian algebras and branching rules}\label{app:lagr}

Here we explicitly compute the branching rules for the family of non-abelian examples, and extract the Lagrangian algebras $\cA_{L,R}$. We restrict to $N$ even.

\subsection{The right embedding}

We start by computing the branching rules of $1 \oplus V$ under the right embedding, which we recall is given by
\begin{alignat}{4}
    R &\quad=\quad \ydiagram{1} &&\quad\otimes\quad \ydiagram{1} &&\quad\otimes\quad 1 &&\quad\otimes\quad 1 \nn \\
      &\quad\oplus\quad 1 &&\quad\otimes\quad 1 &&\quad\otimes\quad -N &&\quad\otimes\quad N+2
\end{alignat}
which gives a group homomorphism of $SU(N) \times SU(N+2) \times U(1)^2$ into $SO(2(N+1)^2)$.

We begin with the representation
\begin{equation}
\begin{array}{c}
    SO(2(N+1)^2)_1
    \\[0.1em] \hline \\[-0.6em]
    1 \oplus V
\end{array}
\end{equation}
First we branch under the direct sum embedding to
\begin{equation}
\begin{array}{ccc}
    SO(2N(N+2))_1 & \times & SO(2)_1
    \\[0.1em] \hline \\[-0.6em]
    (1 \oplus V) & \otimes & (1 \oplus V)
\end{array}
\end{equation}
and then further to
\begin{equation}
\begin{array}{cccccc}
    & SU(N(N+2))_1 & \times & U(1)_{N(N+2)} & \times & U(1)_1
    \\[0.1em] \hhline{~-----} \\[-0.6em]
    \displaystyle{\bigoplus_{\substack{k \in \bZ_{N(N+2)} \\ s \in \bZ_2}}} & \wedge^k & \otimes & k & \otimes & s
\end{array}
\end{equation}
More non-trivially, branching under the tensor product embedding takes us to \cite{altschuler1990branching}
\begin{equation}
\begin{array}{cccccccc}
    & SU(N)_{N+2} & \times & SU(N+2)_N & \times & U(1)_{N(N+2)} & \times & U(1)_1 \\[0.1em] \hhline{~-------}
    \\[-0.6em]
    \displaystyle{\bigoplus_{\substack{\lambda \in \text{Rep}(SU(N)_{N+2}) \\ l \in \bZ_{N+2} \\ s \in \bZ_2}}} & \lambda & \otimes & a^l \lambda^T & \otimes & |\lambda| + Nl & \otimes & s
\end{array}
\end{equation}
In this sum each $\lambda$ is a Young diagram for $SU(N)_{N+2}$, i.e.\ of width $w(\lambda) \leq N+2$ and height $h(\lambda) < N$. We denote by $|\lambda|$ the box count of $\lambda$. By $\lambda^T$, we mean the transpose of $\lambda$ with any length-$(N+2)$ columns deleted, which is a valid Young diagram for $SU(N+2)_N$. Finally, $a$ denotes the operation of stacking a length-$N$ row on top then deleting any length-$(N+2)$ columns. For example, $\lambda = \ydiagram{6,5,1}$ is a Young diagram for $SU(4)_6$, for which $\lambda^T = \ydiagram{2,1,1,1,1}$ and $a\lambda^T = \ydiagram{3,1}$, both Young diagrams for $SU(6)_4$.

We now need to branch under the embedding of $U(1)^2$ symmetries specified by the charge matrix
\begin{equation}
    Q = \begin{pmatrix} 1 & 1 \\ -N & N + 2 \end{pmatrix}
\end{equation}
Via the $K$-matrix computation $Q^T \begin{psmallmatrix} N(N+2) & 0 \\ 0 & 1 \end{psmallmatrix} Q = \begin{psmallmatrix} 2N(N+1) & 0 \\ 0 & 2(N+1)(N+2) \end{psmallmatrix}$, we learn that the new symmetry is $U(1)_{2N(N+1)} \times U(1)_{2(N+1)(N+2)}$. To find the branching rules, we need to solve the equation
\begin{equation}
    Q^{T,-1} \begin{pmatrix} q_1 \\ q_2 \end{pmatrix} = \begin{pmatrix} |\lambda| + Nl \\ s \end{pmatrix} \mod \begin{pmatrix} N(N+2) & 0 \\ 0 & 2 \end{pmatrix} \bZ^2
\end{equation}
for $q_1 \in \bZ_{2N(N+1)}$ and $q_2 \in \bZ_{2(N+1)(N+2)}$, the set of all such pairs $(q_1, q_2)$ then being the branching rules. Doing so, we find the answer
\begin{equation}\label{eq: right 1+V branching}
\begin{array}{cccccccc}
    & SU(N)_{N+2} & \times & SU(N+2)_N & \times & U(1)_{2N(N+1)} & \times & U(1)_{2(N+1)(N+2)} \\[0.1em] \hhline{~-------}
    \\[-0.6em]
    \displaystyle{\bigoplus_{\substack{\lambda \in \text{Rep}(SU(N)_{N+2}) \\ l \in \bZ_{N+2} \\ x \in \bZ_{2(N+1)}}}} & \lambda & \otimes & a^l \lambda^T & \otimes & |\lambda| + Nx & \otimes & |\lambda| - (N+2)x + 2(N+1)l
\end{array}
\end{equation}
Next we group the representations into those of the unitary groups $U(N)_{N+2,2(N+1)}$ and $U(N+2)_{N,2(N+1)}$. This requires verifying two things: i) that each representation $\lambda_1 \otimes \lambda_2 \otimes q_1 \otimes q_2$ obeys $|\lambda_1| = q_1$ mod $N$ and $|\lambda_2| = q_2$ mod $N+2$, which is immediate, and ii) that the set of representations is invariant under $(\lambda_1, q_1) \to (a \lambda_1, q_1 + 2(N+1))$ and $(\lambda_2, q_2) \to (a \lambda_2, q_2 + 2(N+1))$. This action corresponds to fusion with $f_1 = \Sym^{N+2} \otimes 1 \otimes 2(N+1) \otimes 0$ and $f_2 = 1 \otimes \Sym^N \otimes 0 \otimes 2(N+1)$ which are what will eventually become the transparent fermion line. To verify (ii), we note that $l \to l + 1$ effects fusion with $f_2$, while $(\lambda, l, x) \to (a \lambda, l + c(\lambda), x + c(\lambda) + 1)$ effects fusion with $f_1$, where $c(\lambda)$ is the number of length-$(N-1)$ columns in $\lambda$, and we need the identities $|a \lambda| = |\lambda| + N + 2 - Nc(\lambda)$ and $a^{c(\lambda)} (a \lambda)^T = \lambda^T$. The sum can thus be regrouped as
\begin{equation}
\begin{array}{cccc}
    & U(N)_{N+2,2(N+1)} & \times & U(N+2)_{N,2(N+1)} \\[0.1em] \hhline{~---}
    \\[-0.6em]
    \displaystyle{\bigoplus_{\substack{\lambda \in \text{Rep}(SU(N)_{N+2}) \\ x \in \bZ_{2(N+1)} \\ (\lambda, x) \sim (a \lambda, x + c(\lambda) + 1)}}} & (\lambda, |\lambda| + Nx) & \otimes & (\lambda^T, |\lambda| - (N+2)x)
\end{array}
\end{equation}
where each orbit under $f_1, f_2$ has become a single representation of the unitary groups.

Finally we apply level-rank duality to the second factor. This converts each representation of $U(N+2)_{N,2(N+1)}$ to a representation of $U(N)_{N+2,2(N+1)}$ via the replacement $(\lambda_2, q_2) \to (a^{(q_2 - |\lambda_2|)/(N+2)} \lambda_2^T, q_2)$. One can verify that this produces
\begin{equation}
\begin{array}{ccccc}
    & & U(N)_{N+2,2(N+1)} & \times & U(N)_{N+2,2(N+1)} \\[0.1em] \hhline{~~---}
    \\[-0.6em]
    & \displaystyle{\bigoplus_{\substack{\lambda \in \text{Rep}(SU(N)_{N+2}) \\ x \in \bZ_{2(N+1)} \\ (\lambda, x) \sim (a \lambda, x + c(\lambda) + 1)}}} & (\lambda, |\lambda| + Nx) & \otimes & (\lambda, |\lambda| + Nx) \\[4em]
    = &\displaystyle{\bigoplus_{c \in \text{Rep}(U(N)_{N+2,2(N+1)})}} & c & \otimes & c
\end{array}
\end{equation}
where now $c$ runs over all representations of $U(N)_{N+2,2(N+1)}$.

It remains to extract the corresponding Lagrangian algebra $\cA_R$. This is where a slight fermionic subtlety enters. Recall that each $c$ in the branching rules is a representation of $\SU(N)_{N+2} \times U(1)_{2N(N+1)}$ defined up to the equivalence
\begin{equation}
    c \sim c \otimes (\Sym^{N+2}, 2(N+1))
\end{equation}
However, the corresponding $U(N)_{N+2,2(N+1)}$ Chern--Simons theory is defined as
\begin{equation}\label{eq:UN_CS_def}
    U(N)_{N+2,2(N+1)} = \frac{SU(N)_{N+2} \times U(1)_{2N(N+1)} \times SO(0)_1}{\bZ_N^{[1]}\{(\Sym^{N+2}, 2(N+1), V)\}}
\end{equation}
which features an extra factor of $\SO(0)_1$, with $V$ the transparent fermion of $\SO(0)_1$. This means that the lines of the Chern--Simons theory are instead identified up to the coarser equivalence
\begin{equation}\label{eq:Ob(C)}
    c \sim c \otimes (\Sym^{N+2}, 2(N+1))^2
\end{equation}
Thus the branching rules turns into the set of lines
\begin{equation}
    \widehat{\cA}_R = \bigoplus_{c \in \text{Ob}(\cC)/f} (c + fc) \boxtimes \overline{(c + fc)} \; \big/ \; (f \boxtimes \overline{f})
\end{equation}
in $\cC \boxtimes \overline{\cC}$, where $\cC$ denotes the super modular tensor category $U(N)_{N+2,2(N+1)}$, with $f = \Sym^{N+2} \otimes 2(N+1)$ the transparent fermion. Here the notation indicates the quotient of the set of lines in the numerator by a group action given by fusion with the line $f\boxtimes\bar{f}$. To extract the Lagrangian algebra, we simply need to restrict to the lines of spin zero, rather than spin $1/2$. This yields
\begin{equation}
    \cA_R = \bigoplus_{c \in \text{Ob}(\cC)/f} c \boxtimes \overline{c}
\end{equation}
which is the super-diagonal algebra in $\cC \boxtimes \overline{\cC}$, as claimed.

We remark that our logic in this section is in fact entirely backwards: the algebra $\cA_R$ was guaranteed to be the diagonal algebra, and this then fixes the form of level-rank duality map we used to derive it.

\subsection{The left embedding}

Let us now repeat the previous analysis for the left embedding
\begin{alignat}{4}
    L &\quad=\quad \ydiagram{2} &&\quad\otimes\quad 1 &&\quad\otimes\quad 2 &&\quad\otimes\quad 0 \nn \\
      &\quad\oplus\quad 1 &&\quad\otimes\quad \ydiagram{1,1} &&\quad\otimes\quad 0 &&\quad\otimes\quad 2
\end{alignat}
of $SU(N) \times SU(N+2) \times U(1)^2$ into $SO(2(N+1)^2)$.

We begin, as before, with the representation
\begin{equation}
\begin{array}{c}
    SO(2(N+1)^2)_1
    \\[0.1em] \hline \\[-0.6em]
    1 \oplus V
\end{array}
\end{equation}
First we branch under the direct sum embedding to
\begin{equation}
\begin{array}{ccc}
    SO(N(N+1))_1 & \times & SO((N+1)(N+2))_1
    \\[0.1em] \hline \\[-0.6em]
    (1 \oplus V) & \otimes & (1 \oplus V)
\end{array}
\end{equation}
From this it is clear that the branching rules will remain factorised as above. We can therefore treat the two factors entirely separately:
\begin{itemize}

\item Considering the first factor, we further branch to
\begin{equation}
\begin{array}{cccc}
    & SU(N(N+1)/2)_1 & \times & U(1)_{2N(N+1)}
    \\[0.1em] \hhline{~---} \\[-0.6em]
    \displaystyle{\bigoplus_{\substack{k \in \bZ_{N(N+1)}}}} & \wedge^k & \otimes & 2k
\end{array}
\end{equation}
and then under the symmetric embedding to \cite{LevsteinLiberati1995}
\begin{equation}\label{eq:symbranch}
\begin{array}{cccc}
    & SU(N)_{N+2} & \times & U(1)_{2N(N+1)}
    \\[0.1em] \hhline{~---} \\[-0.6em]
    \displaystyle{\bigoplus_{\substack{s \in \{\pm1\}^N \\ r \in \bZ_N}}}
    & a^r \lambda(s) & \otimes & 2c(s) + 2(N+1)r
\end{array}
\end{equation}
To explain the objects $\lambda(s)$ and $c(s)$ used here, for a sign vector $s \in \{\pm1\}^N$, we first let $(a_1, \dots, a_N)$ be the list $(Ns_1, (N-1)s_2, \dots, s_N)$ sorted in descending order. Then
\begin{align}
    \lambda(s) &= [2N + 1 - a_1 + a_N, a_1 - a_2 - 1, \dots, a_{N-1} - a_N - 1] \nn\\
    c(s) &= (N, \dots, 1) \cdot \frac{1+s}{2}
\end{align}
The object $\lambda(s)$ is to be interpreted as a list of Dynkin labels describing an $SU(N)_{N+2}$ representation, while $c(s)$ is an integer.\footnote{To convert a list of Dynkin labels $[\lambda_0, \dots, \lambda_{N-1}]$ to an $SU(N)_M$ Young diagram, one takes $M = \lambda_0 + \dots + \lambda_{N-1}$ and the row lengths to be $(\lambda_1 + \dots + \lambda_{N-1}, \dots, \lambda_{N-2} + \lambda_{N-1}, \lambda_{N-1})$. Also note that at the level of Dynkin labels, the operation $a$ acts as $[\lambda_0, \dots, \lambda_{n-1}] \to [\lambda_{n-1}, \lambda_0, \dots, \lambda_{n-2}]$.} Note that the spins of these representations are $(c(s) + r)/2$.

Next we group these representations into those of $U(N)_{N+2,2(N+1)}$. Each representation $(\lambda, q)$ in the sum obeys $|\lambda| = q$ mod $N$, and the set of all representations is invariant under $(\lambda, q) \to (a \lambda, q + 2(N+1))$, which corresponds to fusion with the transparent fermion $f = (\Sym^{N+2}, 2(N+1))$. Thus the regrouped sum is
\begin{equation}
\begin{array}{cc}
    & U(N)_{N+2,2(N+1)}
    \\[0.1em] \hhline{~-} \\[-0.6em]
    \displaystyle{\bigoplus_{s \in \{\pm1\}^N}}
    & (\lambda(s), 2c(s))
\end{array}
\end{equation}
However, as lines of $U(N)_{N+2,2(N+1)}$, each of these representations $c$ splits into two lines $c$ and $fc$, and we have to take the one of spin 0. This means we should set $r = -c(s)$ instead of $r = 0$ in \eqref{eq:symbranch}. This leads to the Lagrangian algebra
\begin{equation}\label{eq:Adef}
    \cA = \bigoplus_{s \in \{\pm1\}^N} (a^{-c(s)} \lambda(s), -2Nc(s))
\end{equation}
For the reader's convenience, in the ancillary files we provide the Mathematica notebook \texttt{conf-embeddings.nb} to compute $\cA$ for each $N$. For $N = 2$, we find
\begin{equation}
    \cA = (0, 0) \oplus (2, 4) \oplus (2, 8) \oplus (4, 0)
\end{equation}

\item To deal with the second factor, an entirely analogous analysis leads to the branching rules
\begin{equation}
\begin{array}{cc}
    & U(N+2)_{N,2(N+1)}
    \\[0.1em] \hhline{~-} \\[-0.6em]
    \displaystyle{\bigoplus_{s \in \{\pm1\}^N}}
    & (\lambda'(s), 2c(s))
\end{array}
\end{equation}
where we first let $(a_1, \dots, a_{N+1})$ be the list $(Ns_1, (N-1)s_2, \dots, s_N, 0)$ sorted in descending order, then define
\begin{equation}
    \lambda'(s) = [N - a_1, a_1 - a_2 - 1, \dots, a_N - a_{N+1} - 1, N + a_{N+1}]
\end{equation}
One can verify that after applying the level-rank duality map, these branching rules are transformed into the branching rules for the first factor.

More slickly, we could have simply noted that level-rank duality exchanges the $\ydiagram{2}$ and $\ydiagram{1,1}$ embeddings, which implies the result.

\end{itemize}
It follows from these results that the total Lagrangian algebra $\cA_L$ corresponding to the left embedding is
\begin{equation}
    \cA_L = \cA \boxtimes \overline{\cA}
\end{equation}
in $\cC \boxtimes \overline{\cC}$, where as usual, $\cC = U(N)_{N+2,2(N+1)}$.

\subsection{$(-1)^F$ refinement}

In the $(-1)^F$-refined setting of \cref{sec:SPTs_general}, we regard both embeddings as conformal embeddings
\begin{equation}
    SU(N)_{N+2} \times SU(N+2)_N \times U(1)_{2N(N+1)} \times U(1)_{2(N+1)(N+2)} \hookrightarrow \Spin(2(N+1)^2)_1
\end{equation}
and denote the lines of $\Spin(2(N+1)^2)_1$ by $S^r$ with $r = 0,1,2,3 \leftrightarrow 1,S,V,C$. For both embeddings, a line in the branching with $U(1)$ charges $q_1, q_2$ belongs to the branching of $S^{q_1+q_2}$, since every fermion has $q_1 + q_2 = 2$. Pairing each line in the branching of $S^r$ with $\overline{S^r}$, and applying the duality \eqref{eq:bosonic_duality} via \eqref{eq:line map} to the $SU(N+2)_N \times U(1)_{2(N+1)(N+2)} \times \overline{\Spin(2(N+1)^2)_1}$ part, gives Lagrangian algebras $\hat{\cA}_{L,R}$ in $\widehat{\cC} \boxtimes \bar{\widehat{\cC}}$, where now $\widehat{\cC}$ is the bosonic theory
\begin{equation}
    \widehat{\cC} = \frac{SU(N)_{N+2} \times U(1)_{2N(N+1)}}{\bZ_{N/2}^{[1]}}
\end{equation}
We will show that
\begin{equation}\label{eq:refined_algebras}
\begin{aligned}
    \hat{\cA}_R &= \bigoplus_{c \in \widehat{\cC}} c \boxtimes \bar{c} \\
    \hat{\cA}_L &= (\cA \boxtimes \bar{\cA}) \otimes \big( \1 \oplus (\ell \boxtimes \bar{\ell}) \oplus (f \boxtimes \bar{f}) \oplus (\ell f \boxtimes \overline{\ell f}) \big)
\end{aligned}
\end{equation}
where $\cA$ is as in the left embedding above, $f = (\Sym^{N+2}, 2(N+1))$ and $\ell = (1, \frac{N(N+1)}{2})$. As before, the diagonal form of $\widehat{\cA}_R$ then gives a \textit{post-hoc} justification of the level-rank map \eqref{eq:line map}. This follows from the non-refined results with one additional observation.

Let $\cC_0 \subseteq \widehat{\cC}$ consist of the lines braiding trivially with $f$, i.e.\ $(\lambda, q)$ with $|\lambda| = q \mod N$, which form the non-refined theory $U(N)_{N+2,2(N+1)}$. We denote by $\hat{\cA}_{L,R}|_{\cC_0 \boxtimes \bar{\cC}_0}$ the part of $\hat{\cA}_{L,R}$ lying in $\cC_0 \boxtimes \bar{\cC}_0$. This comes from the branching of $1\oplus V$, for which \eqref{eq:line map} reduces to the level-rank map used above. It is therefore given by the non-refined results, except that $f \boxtimes \bar{f}$ is no longer identified with $\1$, so that each line of the non-refined algebras gives rise to two lines differing by fusion with $f \boxtimes \bar{f}$:
\begin{equation}
    \hat{\cA}_R|_{\cC_0 \boxtimes \bar{\cC}_0} = \bigoplus_{c \in \cC_0} c \boxtimes \bar{c}\ , \qquad\qquad
    \hat{\cA}_L|_{\cC_0 \boxtimes \bar{\cC}_0} = (\cA \boxtimes \bar{\cA}) \otimes (\1 \oplus f \boxtimes \bar{f})
\end{equation}
To get the full $\hat{\cA}_{L,R}$ we have to add additional lines coming from the branching of $S\oplus C$. We can do this efficiently by fusion with a carefully chosen line in the branching of $S\oplus C$ that is non-trivial only in the abelian factors. Indeed, for both embeddings such a line is given by
\begin{equation}
    1 \otimes 1 \otimes \tfrac{N(N+1)}{2} \otimes \tfrac{(N+1)(N+2)}{2} \in S\oplus C
\end{equation}
Since $q_1 + q_2 = (N+1)^2 = 1 \mod 4$, it lies in $S$ in both cases. One can then check that \eqref{eq:line map} sends $(1, \frac{(N+1)(N+2)}{2}, 1) \to \ell$, and so $\ell \boxtimes \bar{\ell} \in \hat{\cA}_{L,R}$.

Then, since $f \boxtimes \bar{f} \in \hat{\cA}_{L,R}$, every line of $\hat{\cA}_{L,R}$ braids trivially with $f \boxtimes \bar{f}$, and so lies in either $\cC_0 \boxtimes \bar{\cC}_0$ or $\cC_1 \boxtimes \bar{\cC}_1$, where $\cC_1 = \ell \otimes \cC_0$ consists of the remaining lines of $\cC$. Since $\ell \boxtimes \bar{\ell} \in \hat{\cA}_{L,R}$ is invertible, fusion with it maps $\hat{\cA}_{L,R}$ to itself, exchanging these two parts. Hence
\begin{equation}
    \hat{\cA}_{L,R} = \hat{\cA}_{L,R}|_{\cC_0 \boxtimes \bar{\cC}_0} \otimes (\1 \oplus \ell \boxtimes \bar{\ell})
\end{equation}
which gives \eqref{eq:refined_algebras}.

\section{Folding and unfolding}\label{app:folding_properly}

We now give a more detailed description of the equivalence \eqref{eq:unfolding_equiv}. Firstly, we remark on a subtlety. In \eqref{eq:unfurl_bimod}, $\ket{\cA}$ is not quite a boundary of $\cC$, but rather it is an interface to $\SO(N(N+1))_1$. Therefore, the bimodule $B$ appearing in \eqref{eq:unfurl_bimod} should be understood as an object in $\bimod{\cA}{(\cC\boxtimes\SO(N(N+1))_{-1})}{\cA}$. Working at the level of the categories of lines, however, we are justified in writing \eqref{eq:unfolding_equiv}. This is because both $\cC$ and $\SO(N(N+1))_1$ are fermionic TQFTs so, following the discussion in \cref{sec:SPTs_general}, their transparent fermion lines are identified in the product $\cC \boxtimes \SO(N(N+1))_{-1}$. Since $\SO(N(N+1))_{-1}$ consists only of the trivial line $\1$ and the transparent fermion $V$, this quotient means that the lines of the product are simply those of $\cC$. As a result, the $(\cA,\cA)$-bimodules in $\cC\boxtimes\SO(N(N+1))_{-1}$ are the same as the $(\cA,\cA)$-bimodules in $\cC$. The presence of the $SO(N(N+1))_1$ factor is important to cancel the gravitational anomaly of $\cC$, but it does not affect the spectrum of lines. We will, therefore, ignore it in this appendix and focus on an equality of the categories of lines \eqref{eq:unfolding_equiv}.

We begin with the bimodule $B$ in \eqref{eq:original_bimod}, which is an $(\cA_L,\cA_R)$-bimodule over $\cC\boxtimes\overline{\cC}$. Such a bimodule starts its life as a right $\cA_R$-module over $\cC\boxtimes\overline{\cC}$. We denote its right action
\begin{equation}
    \beta : B \otimes \cA_R \to B
\end{equation}
This is then equipped with a compatible left $\cA_L$ action, which we denote
\begin{equation}\label{eq:AL_action}
    \alpha : \cA_L \otimes B \to B
\end{equation}

Now we unfold $\cC\boxtimes\overline{\cC}$, as illustrated in \eqref{eq:unfold_bimod}. Mathematically, the equivalence of the identity interface in $\cC$ and the $\Diag(\cC)$ boundary of $\cC\boxtimes\overline{\cC}$ is encoded in a tensor equivalence
\begin{equation}\label{eq:unfold_equiv}
    \begin{split}
        \cK : \cC &\xrightarrow{\sim} (\cC\boxtimes\overline{\cC})_{\cA_R} \\
        M &\mapsto \cK(M) \coloneqq (M \boxtimes \1) \otimes \Diag(\cC)
    \end{split}
\end{equation}
The interpretation of this map is as follows. Consider the identity interface in $\cC$. It is transparent, so any line $c\in\cC$ can run through it unchanged. A line of $\cC$ sat on top of this interface is simply a bulk line in $\cC$. When folded, a line $c$ running through the identity interface is folded into a line $c\boxtimes\overline{c} \in\cC\boxtimes\overline{\cC}$. All such lines can end on the boundary, so the boundary is indeed described by the condensation of the canonical algebra $\Diag(\cC)$ in $\cC\boxtimes\overline{\cC}$. Now consider a line $M\in\cC$ lying along the identity interface before folding. After folding it must become a boundary line on the boundary where $\Diag(\cC)$ is condensed. That is, it must become a right $\Diag(\cC)$-module. If we first regard $M$ to be slightly off of the locus of the identity interface then after folding it becomes a line $(M\boxtimes\1) \in\cC\boxtimes\overline{\cC}$. Bringing it to the boundary then gives an induced right $\Diag(\cC)$-module $(M\boxtimes\1) \otimes \Diag(\cC)$. Its right action is simply inherited from the multiplication of $\Diag(\cC)$.

An important property of induced $\Diag(\cC)$-modules that we will use below is
\begin{equation}\label{eq:sliding}
    (\1 \boxtimes \overline{c}) \otimes \Diag(\cC) \cong (c \boxtimes \1) \otimes \Diag(\cC)
\end{equation}
This corresponds to the fact that, before folding, a line $c\in\cC$ can be moved over the identity interface freely.

We wish to demonstrate that the data encoded in $B$, an $(\cA_L ,\cA_R)$-bimodule over $\cC\boxtimes\overline{\cC}$, is precisely the same as that of an $(\cA,\cA)$-bimodule over $\cC$. As above, we can see the bimodule $B$ as a right $\cA_R$ module first. From the equivalence \eqref{eq:unfold_equiv}, we can write it as $B=\cK(M)$ for some $M\in\cC$. The left $\cA_L$ action \eqref{eq:AL_action} can then be written
\begin{equation}\label{eq:AL_action_K}
    \alpha : (\cA \boxtimes \overline{\cA}) \otimes \cK(M) \to \cK(M)
\end{equation}
Now we have
\begin{align}
    (\cA \boxtimes \overline{\cA}) \otimes \cK(M) &= (\cA \boxtimes \overline{\cA}) \otimes (M\boxtimes \1) \otimes \Diag(\cC) \nonumber \\
    &= ( (\cA \otimes M) \boxtimes \overline{\cA}) \otimes \Diag(\cC) \nonumber \\
    &= ( (\cA \otimes M) \boxtimes \1 ) \otimes (\1 \boxtimes \overline{\cA}) \otimes \Diag(\cC) \nonumber \\
    &\cong ( (\cA \otimes M) \boxtimes \1 ) \otimes (\cA \boxtimes \1) \otimes \Diag(\cC) \nonumber \\
    &= ( (\cA\otimes M \otimes \cA) \boxtimes \1) \otimes \Diag(\cC) \nonumber \\
    &= \cK(\cA\otimes M \otimes \cA)
\end{align}
where we have used \eqref{eq:sliding}. Therefore, there is an isomorphism
\begin{equation}
    \chi: (\cA \boxtimes \overline{\cA} ) \otimes \cK(M) \xrightarrow{\sim} \cK(\cA \otimes M \otimes \cA)
\end{equation}
Combining with \eqref{eq:AL_action_K}, the left action $\alpha$ is encoded in a morphism
\begin{equation}\label{eq:alphatilde}
    \tilde{\alpha} = \alpha \circ \chi^{-1} : \cK(\cA\otimes M\otimes \cA) \to \cK(M)
\end{equation}
Since $\cK$ is fully faithful, the left action $\alpha$ is therefore encoded in a morphism\footnote{A functor $F:M\to N$ gives a map $F_{m,m'}:\Hom_M(m,m') \to \Hom_N (F(m),F(m'))$. $F$ is said to be faithful if this map is injective, full if it is surjective, and fully faithful if it is a bijection. Here, since $\cK$ is full it follows from \eqref{eq:alphatilde} that there is a morphism $\mu$ as in \eqref{eq:mu_map} with $\tilde{\alpha}=\cK(\mu)$, and since $\cK$ is faithful it follows that this $\mu$ is unique.}
\begin{equation}\label{eq:mu_map}
    \mu:  \cA \otimes M \otimes \cA \to M
\end{equation}
with $\tilde{\alpha}=\cK(\mu)$. This should be interpreted as saying that the left action of an element $(a,b)$ of $\cA\boxtimes\overline{\cA}$ on $B$, after unfolding, is mapped to the action of $a$ from the left and $b$ from the right on the object $M\in\cC$ which corresponds to $B$ via $B=\cK(M)$.

The conditions on the original left action $\alpha$ descend to conditions on this map $\mu$. Most importantly, we have the associativity condition
\begin{equation}
    \alpha(\cA\boxtimes\bar\cA,\alpha(\cA\boxtimes\bar{\cA}, B)) = \alpha((\cA\boxtimes\bar\cA)\otimes(\cA\boxtimes\bar\cA),B)
\end{equation}
which is represented diagrammatically by
\begin{equation}
    \raisebox{-4em}{
        \begin{tikzpicture}
        \draw [thick] (0,-2) node[below] {\small $B$} to (0,1);
        \draw [thick, red] (-2,-2) node[left] {\small $\cA\boxtimes\bar{\cA}$} to (0,-1);
        \draw [thick, red] (-2,-1) node[left] {\small $\cA\boxtimes\bar{\cA}$} to (0,0);
        \end{tikzpicture}
    }
    \quad=\quad
    \raisebox{-4em}{
        \begin{tikzpicture}
        \draw [thick] (0,-2) node[below] {\small $B$} to (0,1);
        \draw [thick, red] (-2,-2) node[left] {\small $\cA\boxtimes\bar{\cA}$} to (-1,-0.5);
        \draw [thick, red] (-2,-1) node[left] {\small $\cA\boxtimes\bar{\cA}$} to (0,0);
        \end{tikzpicture}
    }
\end{equation}
Once mapped through $\cK$, this gives a condition on $\mu$
\begin{equation}\label{eq:mu_constraint}
    \mu(a_0,\mu(a_1,m,b_1),b_0) = \mu(a_0 a_1, m, b_1 b_0)
\end{equation}
where $(a_0,b_0),(a_1,b_1)$ are elements of $\cA\boxtimes\overline{\cA}$ and $m$ is a simple object in $M$.

We now demonstrate that the morphism $\mu$, satisfying \eqref{eq:mu_constraint}, encodes the structure of an $(\cA,\cA)$-bimodule over $\cC$. That is, from $\mu$ we can find a commuting left and right action of $\cA$ on $\cC$. The left and right actions are given simply by isolating the relevant parts of $\mu$:
\begin{align}
    \lambda_M : \cA \otimes M \to M \qc &\lambda_M = \mu \circ (\id_\cA \otimes \id_M \otimes \eta_\cA ) \\
    \rho_M : M \otimes \cA \to M \qc & \rho_M = \mu \circ (\eta_\cA \otimes \id_M \otimes \id_\cA)
\end{align}
where $\eta_\cA : \1 \to \cA$ is the unit on $\cA$. In components, these are
\begin{equation}
    \lambda_M(a,m) = \mu(a,m,\1_\cA) \qc \rho_M(m,a) = \mu(\1_\cA,m,a)
\end{equation}
where $a\in\cA$ and $m$ is a simple object in $M$.

The constraint \eqref{eq:mu_constraint} gives conditions on $\lambda_M$ and $\rho_M$. Firstly, setting $b_0=b_1=\1_\cA$ in \eqref{eq:mu_constraint}, we find
\begin{equation}
    \lambda_M(a_0,\lambda_M(a_1,m)) = \lambda_M(a_0 a_1,m)
\end{equation}
which is the condition for $\lambda_M$ to be an associative left $\cA$ action. Similarly, setting $a_0=a_1=\1_\cA$ gives
\begin{equation}
    \rho_M(\rho_M(m,b_1),b_0) = \rho_M(m,b_1 b_0)
\end{equation}
so $\rho_M$ is an associative right $\cA$ action. Now setting $(a_0,a_1,b_0,b_1)=(\1_\cA,a,b,\1_\cA)$, we find
\begin{equation}\label{eq:cond1}
    \rho_M(\lambda_M(a,m),b) = \mu(a,m,b)
\end{equation}
Similarly, setting $(a_0,a_1,b_0,b_1)=(a,\1_\cA,\1_\cA,b)$ we find
\begin{equation}\label{eq:cond2}
    \lambda_M(a,\rho_M(m,b)) = \mu(a,m,b)
\end{equation}
Combining \eqref{eq:cond1} and \eqref{eq:cond2} gives
\begin{equation}
    \rho_M(\lambda_M(a,m),b) = \lambda_M(a,\rho_M(m,b))
\end{equation}
such that the left and right actions commute. Therefore, $(M,\lambda_M,\rho_M)$ defined in this manner give the structure of an $(\cA,\cA)$-bimodule over $\cC$.

\section{Algebras to gauge from the bimodule}\label{app:deriving_algebras}

In \cref{sec:unfolding} and \cref{app:folding_properly} we found an equivalence of bimodule categories
\begin{equation}\label{eq:bimod_equivalence}
    F : \bimod{\cA_L}{(\cC\boxtimes\overline{\cC})}{\cA_R} \xrightarrow{\sim} \bimod{\cA}{\cC}{\cA}
\end{equation}
It corresponds to the unfolding procedure in \cref{sec:unfolding}. A natural choice of a simple object in $\bimod{\cA}{\cC}{\cA}$ is the regular $(\cA,\cA)$-bimodule, which has underlying object $\cA$ and a left and right action given by the multiplication of $\cA$. This can be mapped to a bimodule $B \in \bimod{\cA_L}{ (\cC\boxtimes\overline{\cC})}{\cA_R}$ by $F^{-1}$, which is the corresponding bimodule in the folded picture.

In \cref{sec:gauging} we found an algebra $a_L$ in $\cS_L$ which can be gauged to yield $\cS_R$, and a corresponding algebra $a_R$ in $\cS_R$ whose gauging results in $\cS_L$. These algebras were found based purely on properties of the relevant categories. In this appendix we derive these results directly from the regular bimodule, completing the three-dimensional description of the interface as a duality defect.

We begin in the folded picture \eqref{eq:original_bimod}. The category of $(\cA_L,\cA_R)$-bimodules can be acted on from the right by lines in $\cS_R$. That is, topological defects on the right of the interface can be fused into the interface, and generically this will transform one interface into another. Similarly, it admits a left action of $\cS_L$. Therefore, $\bimod{\cA_L}{(\cC\boxtimes\overline{\cC})}{\cA_R}$ admits the structure of an $(\cS_L,\cS_R)$ bimodule category.

For concreteness, we will aim to find the algebra $a_R$ in $\cS_R$. We therefore need to describe the right action of $\cS_R$ on an $(\cA_L,\cA_R)$-bimodule $B$. Recall that a line $X \in \cS_R=(\cC\boxtimes\overline{\cC})_{\cA_R}$ is a right $\cA_R$ module and, as such, comes equipped with a right action $\rho_X:X \otimes \cA_R \to X$. The braiding of $\cC\boxtimes\overline{\cC}$ can be used to construct a left $\cA_R$-action $\lambda_X: \cA_R \otimes X \to X$ from this right action as $\lambda_X = \rho_X \circ c_{\cA_R,X}$.\footnote{There is another way to construct a left action, where one replaces $c_{\cA_R,X}$ by $c_{X,\cA_R}^{-1}$. We will stick to the convention in the text. This is sometimes referred to as alpha-induction, with the resulting bimodules denoted $\alpha^\pm(X)$.} This promotes $X$ to an $(\cA_R,\cA_R)$-bimodule, which we denote $\hat{X}$. This can then be fused with the $(\cA_L,\cA_R)$-bimodule $B$ using the relative tensor product over $\cA_R$. Thus, we define the right action of $X\in \cS_R$ on $B$ by
\begin{equation}
    B \triangleleft X = B \otimes_{\cA_R} \hat{X}
\end{equation}
The result is an $(\cA_L,\cA_R)$-bimodule which generically differs from $B$.

The algebra in $\cS_R$ which must be gauged in order to realise the interface $B$ as a duality defect can be found by the internal end construction. We begin by fixing a particular interface $B$: in our case we will fix $B=F^{-1}(\cA)$, the preimage of the regular bimodule under $F$. We then find the set of lines $X \in \cS_R$ which can be absorbed into $B$. That is, the set of lines for which $B \triangleleft X = B$. This set can be given the structure of a gaugeable algebra called the internal End algebra of $B$, denoted $\underline{\End}_{\cS_R}(B)$. More specifically, it is defined by the property
\begin{equation}\label{eq:internal_end_def}
    \Hom_{\cS_R}(X,\underline{\End}_{\cS_R}(B)) \cong \Hom_{\bimod{\cA_L}{(\cC\boxtimes\overline{\cC})}{\cA_R}} (B \triangleleft X, B)
\end{equation}
Since this refers to the folded picture \eqref{eq:original_bimod}, we write $a_R^{\text{folded}}=\underline{\End}_{\cS_R}(B)$.

We will find that this algebra has a clean presentation in terms of the unfolded quantities resulting from the procedure outlined in \cref{sec:unfolding} and \cref{app:folding_properly}. In particular, recall from \eqref{eq:unfold_equiv} the equivalence $\cK : \cC \xrightarrow{\sim} (\cC\boxtimes \overline{\cC})_{\cA_R} = \cS_R$ which corresponds to unfolding $\cC\boxtimes\overline{\cC}$. Every line $X\in \cS_R$ can be written uniquely as $X=\cK(x)$ for some $x\in\cC$. We also write $a_R^{\text{folded}} = \cK(a_R^{\text{unfolded}})$. In terms of the unfolded quantities, \eqref{eq:internal_end_def} becomes
\begin{equation}\label{eq:unfolded_internal_end}
    \Hom_{\cC} (x,a_R^{\text{unfolded}}) \cong \Hom_{\bimod{\cA}{\cC}{\cA}} ( F(B \triangleleft X), \cA)
\end{equation}
where we have used that unfolding maps $\bimod{\cA_L}{(\cC\boxtimes\overline{\cC})}{\cA_R}$ to $\bimod{\cA}{\cC}{\cA}$ as in \eqref{eq:bimod_equivalence}, and that under this equivalence $B$ is mapped to the regular bimodule $\cA$.

It remains to understand how $B \triangleleft X$ is transformed under the unfolding. As an object in $\cC\boxtimes\overline{\cC}$, we have
\begin{equation}
    B \triangleleft X = B \otimes_{\cA_R} X = \cK(\cA) \otimes_{\cA_R} \cK(x) = \cK(\cA \otimes x)
\end{equation}
where we have used that $\cK$ is a tensor equivalence in the final equality.\footnote{Alternatively, this can be derived directly using the definition \eqref{eq:unfold_equiv}.} Hence, unfolding with $\cK^{-1}$, $B\triangleleft X$ simply becomes $\cA\otimes x$. Its left $\cA$-action is simply given by multiplication with the first $\cA$ factor, while its right action requires first braiding $\cA$ past $x$ and then using the multiplication. Hence, \eqref{eq:unfolded_internal_end} becomes
\begin{equation}\label{eq:almost_there}
    \Hom_\cC (x,a_R^\text{unfolded}) \cong \Hom_{\bimod{\cA}{\cC}{\cA}} (\cA \otimes x, \cA)
\end{equation}
We then use the fact that\footnote{This can be seen by noting that to each bimodule map $\Phi:\cA \otimes x \to \cA$ one can build a map $f_\Phi : x \to \cA$ by $f_\Phi = \Phi \circ (\eta_\cA \otimes \id_x)$. Similarly, from a map $f:x\to\cA$ we build a map $\Phi_f = m \circ (\id_\cA \otimes f) : \cA \otimes x \to \cA$ which can be shown to be left and right $\cA$-linear. The two operations are inverse, i.e.\ $f_{\Phi_f} = f$, establishing the equivalence \eqref{eq:very_almost_there}.

This can be seen as a special case of the results of \cite[Proposition 3.6 \& Lemma 3.13]{Frohlich:2003hm}. Setting $U=x$ and $V=\1$ there, and noting that when $\cA$ is commutative the left centre $C_l(\cA) \cong \cA$, one finds $\Hom_{\bimod{\cA}{\cC}{\cA}}(\alpha^+(x),\alpha^-(\1)) \cong \Hom_\cC (x,\cA)$. The result \eqref{eq:very_almost_there} then follows by noting that $\alpha^-(\1)=\cA$ is the regular bimodule.}
\begin{equation}\label{eq:very_almost_there}
    \Hom_{\bimod{\cA}{\cC}{\cA}} (\cA \otimes x, \cA) \cong \Hom_\cC (x,\cA)
\end{equation}
Combining \eqref{eq:almost_there} and \eqref{eq:very_almost_there}, we then have
\begin{equation}
    a_R^{\text{unfolded}} \cong \cA
\end{equation}
from the Yoneda lemma \cite{etingof}. This establishes that, when unfolded, the algebra to be gauged on the right to produce the defect is $\cA \in \cC$. The reverse argument to find $a_L$ proceeds similarly.

\bibliographystyle{JHEP}
\bibliography{refs}

@article{vanBeest:2023dbu,
    author = "van Beest, Marieke and Boyle Smith, Philip and Delmastro, Diego and Komargodski, Zohar and Tong, David",
    title = "{Monopoles, scattering, and generalized symmetries}",
    eprint = "2306.07318",
    archivePrefix = "arXiv",
    primaryClass = "hep-th",
    doi = "10.1007/JHEP03(2025)014",
    journal = "JHEP",
    volume = "03",
    pages = "014",
    year = "2025"
}

@article{Nakanishi:1990hj,
    author = "Nakanishi, Tomoki and Tsuchiya, Akihiro",
    title = "{Level rank duality of WZW models in conformal field theory}",
    reportNumber = "NU-MATH-002",
    doi = "10.1007/BF02101097",
    journal = "Commun. Math. Phys.",
    volume = "144",
    pages = "351--372",
    year = "1992"
}

@article{Naculich:1990pa,
    author = "Naculich, Stephen G. and Riggs, H. A. and Schnitzer, H. J.",
    title = "{Group Level Duality in {WZW} Models and {Chern-Simons} Theory}",
    reportNumber = "BRX-TH-293",
    doi = "10.1016/0370-2693(90)90623-E",
    journal = "Phys. Lett. B",
    volume = "246",
    pages = "417--422",
    year = "1990"
}

@article{Elitzur:1989nr,
    author = "Elitzur, Shmuel and Moore, Gregory W. and Schwimmer, Adam and Seiberg, Nathan",
    title = "{Remarks on the Canonical Quantization of the Chern-Simons-Witten Theory}",
    reportNumber = "IASSNS-HEP-89/20",
    doi = "10.1016/0550-3213(89)90436-7",
    journal = "Nucl. Phys. B",
    volume = "326",
    pages = "108--134",
    year = "1989"
}

@article{Witten:1988hf,
    author = "Witten, Edward",
    editor = "Mitra, Asoke N.",
    title = "{Quantum Field Theory and the Jones Polynomial}",
    reportNumber = "IASSNS-HEP-88-33",
    doi = "10.1007/BF01217730",
    journal = "Commun. Math. Phys.",
    volume = "121",
    pages = "351--399",
    year = "1989"
}

@article{Apruzzi:2021nmk,
    author = "Apruzzi, Fabio and Bonetti, Federico and Garc{\'\i}a Etxebarria, I{\~n}aki and Hosseini, Saghar S. and Schafer-Nameki, Sakura",
    title = "{Symmetry TFTs from String Theory}",
    eprint = "2112.02092",
    archivePrefix = "arXiv",
    primaryClass = "hep-th",
    doi = "10.1007/s00220-023-04737-2",
    journal = "Commun. Math. Phys.",
    volume = "402",
    number = "1",
    pages = "895--949",
    year = "2023"
}

@article{Freed:2022qnc,
    author = "Freed, Daniel S. and Moore, Gregory W. and Teleman, Constantin",
    title = "{Topological symmetry in quantum field theory}",
    eprint = "2209.07471",
    archivePrefix = "arXiv",
    primaryClass = "hep-th",
    month = "9",
    year = "2022"
}

@article{altschuler1990branching,
    title={The branching rules of conformal embeddings},
    author={Altschuler, Daniel and Bauer, Michel and Itzykson, Claude},
    journal={Communications in mathematical physics},
    volume={132},
    number={2},
    pages={349--364},
    year={1990},
    publisher={Springer}
}

@article{LevsteinLiberati1995,
    author = "Levstein, Fernando and Liberati, José I.",
    title = "{Branching Rules for Conformal Embeddings}",
    doi = "10.1007/BF02100179",
    journal = "Commun. Math. Phys.",
    volume = "173",
    number = "1",
    pages = "1--16",
    year = "1995"
}

@article{Han_2017,
    title={Boundary conformal field theory and symmetry-protected topological phases in $2+1$ dimensions},
    volume={96},
    ISSN={2469-9969},
    url={http://dx.doi.org/10.1103/PhysRevB.96.125105},
    DOI={10.1103/physrevb.96.125105},
    number={12},
    journal={Physical Review B},
    publisher={American Physical Society (APS)},
    author={Han, Bo and Tiwari, Apoorv and Hsieh, Chang-Tse and Ryu, Shinsei},
    year={2017}
}

@article{BoyleSmith:2019jnh,
    author = "Boyle Smith, Philip and Tong, David",
    title = "{Boundary States for Chiral Symmetries in Two Dimensions}",
    eprint = "1912.01602",
    archivePrefix = "arXiv",
    primaryClass = "hep-th",
    doi = "10.1007/JHEP09(2020)018",
    journal = "JHEP",
    volume = "09",
    pages = "018",
    year = "2020"
}

@article{BoyleSmith:2020nuf,
    author = "Boyle Smith, Philip and Tong, David",
    title = "{What Symmetries are Preserved by a Fermion Boundary State?}",
    eprint = "2006.07369",
    archivePrefix = "arXiv",
    primaryClass = "hep-th",
    month = "6",
    year = "2020"
}

@article{Wei:2026fsn,
    author = "Wei, Pengcheng and Zheng, Yunqin",
    title = "{Non-invertible Symmetries in Weyl Fermions, and Applications to Fermion-Boundary Scattering Problem}",
    eprint = "2605.19363",
    archivePrefix = "arXiv",
    primaryClass = "hep-th",
    month = "5",
    year = "2026"
}

@article{Arias-Tamargo:2026urw,
    author = "Arias-Tamargo, Guillermo and Boyle Smith, Philip and Mouland, Rishi and Vel{\'a}squez Cotini Hutt, Maxwell L.",
    title = "{Non-Invertible Symmetries and Boundaries for Two-Dimensional Fermions}",
    eprint = "2605.13952",
    archivePrefix = "arXiv",
    primaryClass = "hep-th",
    month = "5",
    year = "2026"
}

@article{Bhardwaj:2017xup,
    author = "Bhardwaj, Lakshya and Tachikawa, Yuji",
    title = "{On finite symmetries and their gauging in two dimensions}",
    eprint = "1704.02330",
    archivePrefix = "arXiv",
    primaryClass = "hep-th",
    reportNumber = "IPMU-17-0049",
    doi = "10.1007/JHEP03(2018)189",
    journal = "JHEP",
    volume = "03",
    pages = "189",
    year = "2018"
}

@article{Bhardwaj:2024ydc,
    author = "Bhardwaj, Lakshya and Inamura, Kansei and Tiwari, Apoorv",
    title = "{Fermionic non-invertible symmetries in (1+1)d: Gapped and gapless phases, transitions, and symmetry TFTs}",
    eprint = "2405.09754",
    archivePrefix = "arXiv",
    primaryClass = "hep-th",
    doi = "10.21468/SciPostPhys.18.6.194",
    journal = "SciPost Phys.",
    volume = "18",
    number = "6",
    pages = "194",
    year = "2025"
}

@article{Fuchs:2002cm,
    author = "Fuchs, Jurgen and Runkel, Ingo and Schweigert, Christoph",
    title = "{TFT construction of RCFT correlators 1. Partition functions}",
    eprint = "hep-th/0204148",
    archivePrefix = "arXiv",
    reportNumber = "PAR-LPTHE-02-25",
    doi = "10.1016/S0550-3213(02)00744-7",
    journal = "Nucl. Phys. B",
    volume = "646",
    pages = "353--497",
    year = "2002"
}

@article{Kapustin:2010if,
    author = "Kapustin, Anton and Saulina, Natalia",
    editor = "Sati, Hisham and Schreiber, Urs",
    title = "{Surface operators in 3d Topological Field Theory and 2d Rational Conformal Field Theory}",
    eprint = "1012.0911",
    archivePrefix = "arXiv",
    primaryClass = "hep-th",
    reportNumber = "PI-STRINGS-195",
    pages = "175--198",
    month = "12",
    year = "2010"
}

@article{Davydov:2010,
    author = {{Davydov}, Alexei and {Müger}, Michael and {Nikshych}, Dmitri and {Ostrik}, Victor},
    title = "{The Witt group of non-degenerate braided fusion categories}",
    journal = {arXiv e-prints},
    year = 2010,
    month = sep,
    eid = {arXiv:1009.2117},
    pages = {arXiv:1009.2117},
    doi = {10.48550/arXiv.1009.2117},
    archivePrefix = {arXiv},
    eprint = {1009.2117},
    primaryClass = {math.QA},
    adsurl = {https://ui.adsabs.harvard.edu/abs/2010arXiv1009.2117D}
}

@article{Aasen:2017ubm,
    author = "Aasen, David and Lake, Ethan and Walker, Kevin",
    title = "{Fermion condensation and super pivotal categories}",
    eprint = "1709.01941",
    archivePrefix = "arXiv",
    primaryClass = "cond-mat.str-el",
    doi = "10.1063/1.5045669",
    journal = "J. Math. Phys.",
    volume = "60",
    number = "12",
    pages = "121901",
    year = "2019"
}

@article{Komargodski:2020mxz,
    author = "Komargodski, Zohar and Ohmori, Kantaro and Roumpedakis, Konstantinos and Seifnashri, Sahand",
    title = "{Symmetries and strings of adjoint QCD$_2$}",
    eprint = "2008.07567",
    archivePrefix = "arXiv",
    primaryClass = "hep-th",
    reportNumber = "YITP-SB-20-28",
    doi = "10.1007/JHEP03(2021)103",
    journal = "JHEP",
    volume = "03",
    pages = "103",
    year = "2021"
}

@article{hsin2016levelrank,
    title="{Level/rank Duality and Chern--Simons-Matter Theories}",
    author={Po-Shen Hsin and Nathan Seiberg},
    year={2016},
    eprint={1607.07457},
    archivePrefix={arXiv},
    primaryClass={hep-th},
    doi={https://doi.org/10.1007/JHEP09(2016)095},
    url={https://arxiv.org/abs/1607.07457}, 
}

@article{Ardonne_2010,
    author = {{Ardonne}, Eddy and {Slingerland}, Joost},
    title = "{Clebsch--Gordan and 6j-coefficients for rank 2 quantum groups}",
    journal = {Journal of Physics A Mathematical General},
    year = 2010,
    month = oct,
    volume = {43},
    number = {39},
    eid = {395205},
    pages = {395205},
    doi = {10.1088/1751-8113/43/39/395205},
    archivePrefix = {arXiv},
    eprint = {1004.5456},
    primaryClass = {math.QA}
}

@book{etingof,
    author={Etingof, Pavel and Gelaki, Shlomo and Nikshych, Dmitri and Ostrik, Victor},
    title={Tensor Categories},
    year={2015},
    publisher={American Mathematical Society},
    address={Providence},
    isbn={9781470423490; 1470423499},
    doi={10.1090/surv/205}
}

@article{galindo2025note,
    title="{A Note on the Eilenberg--Mac Lane Isomorphism for Quadratic Forms}", 
    author={César Galindo},
    year={2025},
    eprint={2507.15726},
    archivePrefix={arXiv},
    primaryClass={math.CT},
    url={https://arxiv.org/abs/2507.15726}, 
}

@article{putrov2024,
    author = "Putrov, Pavel and Radhakrishnan, Rajath",
    title = "{Non-anomalous non-invertible symmetries in 1+1D from gapped boundaries of SymTFTs}",
    eprint = "2405.04619",
    archivePrefix = "arXiv",
    primaryClass = "hep-th",
    doi = "10.1007/JHEP03(2026)077",
    journal = "JHEP",
    volume = "03",
    pages = "077",
    year = "2026"
}

@article{lorentzfrac,
    author = "Hsin, Po-Shen and Shao, Shu-Heng",
    title = "{Lorentz Symmetry Fractionalization and Dualities in (2+1)d}",
    eprint = "1909.07383",
    archivePrefix = "arXiv",
    primaryClass = "cond-mat.str-el",
    reportNumber = "CALT-TH-2019-035",
    doi = "10.21468/SciPostPhys.8.2.018",
    journal = "SciPost Phys.",
    volume = "8",
    pages = "018",
    year = "2020"
}

@article{Karch:2016sxi,
    author = "Karch, Andreas and Tong, David",
    title = "{Particle-Vortex Duality from 3d Bosonization}",
    eprint = "1606.01893",
    archivePrefix = "arXiv",
    primaryClass = "hep-th",
    doi = "10.1103/PhysRevX.6.031043",
    journal = "Phys. Rev. X",
    volume = "6",
    number = "3",
    pages = "031043",
    year = "2016"
}

@article{Frohlich:2003hm,
    author = "Frohlich, Jurg and Fuchs, Jurgen and Runkel, Ingo and Schweigert, Christoph",
    title = "{Correspondences of ribbon categories}",
    eprint = "math/0309465",
    archivePrefix = "arXiv",
    reportNumber = "HU-EP-03-31",
    doi = "10.1016/j.aim.2005.04.007",
    journal = "Adv. Math.",
    volume = "199",
    pages = "192--329",
    year = "2006"
}

@article{bb,
    author = "Boyle Smith, Philip and Zheng, Yunqin",
    title = "{Backfiring Bosonisation}",
    eprint = "2403.03953",
    archivePrefix = "arXiv",
    primaryClass = "hep-th",
    doi = "10.1007/JHEP03(2026)221",
    journal = "JHEP",
    volume = "03",
    pages = "221",
    year = "2026"
}

@article{Polchinski:1984uw,
    author = "Polchinski, Joseph",
    title = "{Monopole Catalysis: The Fermion Rotor System}",
    reportNumber = "HUTP-84-A005",
    doi = "10.1016/0550-3213(84)90398-5",
    journal = "Nucl. Phys. B",
    volume = "242",
    pages = "345--363",
    year = "1984"
}

@article{Loladze:2024ayk,
    author = "Loladze, Vazha and Okui, Takemichi",
    title = "{Monopole-Fermion Scattering and the Solution to the Semiton{\textendash}Unitarity Puzzle}",
    eprint = "2408.04577",
    archivePrefix = "arXiv",
    primaryClass = "hep-th",
    reportNumber = "KEK-TH-2644",
    doi = "10.1103/PhysRevLett.134.051602",
    journal = "Phys. Rev. Lett.",
    volume = "134",
    number = "5",
    pages = "051602",
    year = "2025"
}

@article{Loladze:2025jsq,
    author = "Loladze, Vazha and Okui, Takemichi and Tong, David",
    title = "{Dynamics of the fermion-rotor system}",
    eprint = "2508.21059",
    archivePrefix = "arXiv",
    primaryClass = "hep-th",
    doi = "10.1007/JHEP01(2026)052",
    journal = "JHEP",
    volume = "01",
    pages = "052",
    year = "2026"
}

@article{Tong:2021phe,
    author = "Tong, David",
    title = "{Comments on symmetric mass generation in 2d and 4d}",
    eprint = "2104.03997",
    archivePrefix = "arXiv",
    primaryClass = "hep-th",
    doi = "10.1007/JHEP07(2022)001",
    journal = "JHEP",
    volume = "07",
    pages = "001",
    year = "2022"
}

@book{DiFrancesco:1997nk,
    author = "Di Francesco, P. and Mathieu, P. and Senechal, D.",
    title = "{Conformal Field Theory}",
    doi = "10.1007/978-1-4612-2256-9",
    isbn = "978-0-387-94785-3, 978-1-4612-7475-9",
    publisher = "Springer-Verlag",
    address = "New York",
    series = "Graduate Texts in Contemporary Physics",
    year = "1997"
}

@article{jensen,
	archiveprefix = {arXiv},
	author = {Jensen, Kristan and Shaverin, Evgeny and Yarom, Amos},
	doi = {10.1007/JHEP01(2018)085},
	eprint = {1710.07299},
	journal = {JHEP},
	pages = {085},
	primaryclass = {hep-th},
	title = {{{\textquoteright}t Hooft anomalies and boundaries}},
	volume = {01},
	year = {2018}
}

@article{hellerman,
    author = "Hellerman, Simeon and Orlando, Domenico and Watanabe, Masataka",
    title = "{Quantum Information Theory of the Gravitational Anomaly}",
    eprint = "2101.03320",
    archivePrefix = "arXiv",
    primaryClass = "hep-th",
    month = "1",
    year = "2021"
}

@article{Thorngren:2020yht,
	archiveprefix = {arXiv},
	author = {Thorngren, Ryan and Wang, Yifan},
	doi = {10.1007/JHEP09(2021)017},
	eprint = {2012.15861},
	journal = {JHEP},
	pages = {017},
	primaryclass = {hep-th},
	title = {{Anomalous symmetries end at the boundary}},
	volume = {09},
	year = {2021}
}

@article{Wei:2025zyd,
    author = "Wei, Pengcheng and Zheng, Yunqin",
    title = "{On the Absence of Symmetric Simple Conformal Boundary Conditions}",
    eprint = "2512.23976",
    archivePrefix = "arXiv",
    primaryClass = "hep-th",
    month = "12",
    year = "2025"
}

@article{Choi_2023,
   title={Remarks on boundaries, anomalies, and noninvertible symmetries},
   volume={108},
   ISSN={2470-0029},
   url={http://dx.doi.org/10.1103/PhysRevD.108.125005},
   DOI={10.1103/physrevd.108.125005},
   number={12},
   journal={Physical Review D},
   publisher={American Physical Society (APS)},
   author={Choi, Yichul and Rayhaun, Brandon C. and Sanghavi, Yaman and Shao, Shu-Heng},
   year={2023},
   month=Dec
}

@inproceedings{tongscgp,
    author = {Tong, David},
    title = "{What I Don't Understand About Monopole-Fermion Scattering}",
    publisher = {Paths to Quantum Field Theory, SCGP, June 2026, \url{https://scgp.stonybrook.edu/video/video.php?id=7747}}
}

@article{vanBeest:2023mbs,
    author = "van Beest, Marieke and Boyle Smith, Philip and Delmastro, Diego and Mouland, Rishi and Tong, David",
    title = "{Fermion-monopole scattering in the Standard Model}",
    eprint = "2312.17746",
    archivePrefix = "arXiv",
    primaryClass = "hep-th",
    doi = "10.1007/JHEP08(2024)004",
    journal = "JHEP",
    volume = "08",
    pages = "004",
    year = "2024"
}

@article{Callan:1982ac,
    author = "Callan, Jr., Curtis G.",
    title = "{Monopole Catalysis of Baryon Decay}",
    reportNumber = "LPTENS-82-20",
    doi = "10.1016/0550-3213(83)90677-6",
    journal = "Nucl. Phys. B",
    volume = "212",
    pages = "391--400",
    year = "1983"
}

@article{Callan:1982au,
    author = "Callan, Jr., Curtis G.",
    title = "{Dyon-Fermion Dynamics}",
    reportNumber = "Print-82-0476 (PRINCETON)",
    doi = "10.1103/PhysRevD.26.2058",
    journal = "Phys. Rev. D",
    volume = "26",
    pages = "2058--2068",
    year = "1982"
}

@article{Callan:1982ah,
    author = "Callan, Jr., Curtis G.",
    title = "{Disappearing Dyons}",
    reportNumber = "Print-82-0006 (PRINCETON)",
    doi = "10.1103/PhysRevD.25.2141",
    journal = "Phys. Rev. D",
    volume = "25",
    pages = "2141",
    year = "1982"
}

@article{Rubakov:1982fp,
    author = "Rubakov, V. A.",
    title = "{Adler--Bell--Jackiw Anomaly and Fermion Number Breaking in the Presence of a Magnetic Monopole}",
    doi = "10.1016/0550-3213(82)90034-7",
    journal = "Nucl. Phys. B",
    volume = "203",
    pages = "311--348",
    year = "1982"
}

@article{affleckludwig,
    title={Universal noninteger ``ground-state degeneracy'' in critical quantum systems},
    volume={67},
    ISSN={0031-9007},
    url={http://dx.doi.org/10.1103/PhysRevLett.67.161},
    DOI={10.1103/physrevlett.67.161},
    number={2},
    journal={Physical Review Letters},
    publisher={American Physical Society (APS)},
    author={Affleck, Ian and Ludwig, Andreas W. W.},
    year={1991},
    month={Jul},
    pages={161-164}
}

@article{mueger2001,
    author = {{Müger}, Michael},
    title = "{From Subfactors to Categories and Topology II. The quantum double of tensor categories and subfactors}",
    journal = {arXiv Mathematics e-prints},
    year = 2001,
    month = nov,
    eid = {math/0111205},
    pages = {math/0111205},
    doi = {10.48550/arXiv.math/0111205},
    archivePrefix = {arXiv},
    eprint = {math/0111205},
    primaryClass = {math.CT}
}

@article{davydov2011,
    author = {{Davydov}, Alexei and {Nikshych}, Dmitri and {Ostrik}, Victor},
    title = "{On the structure of the Witt group of braided fusion categories}",
    journal = {arXiv e-prints},
    year = 2011,
    month = sep,
    eid = {arXiv:1109.5558},
    pages = {arXiv:1109.5558},
    doi = {10.48550/arXiv.1109.5558},
    archivePrefix = {arXiv},
    eprint = {1109.5558},
    primaryClass = {math.QA}
}

@article{etingof2003,
    author = {{Etingof}, Pavel and {Gelaki}, Shlomo and {Ostrik}, Viktor},
    title = "{Classification of fusion categories of dimension $pq$}",
    journal = {arXiv Mathematics e-prints},
    year = 2003,
    month = apr,
    eid = {math/0304194},
    pages = {math/0304194},
    doi = {10.48550/arXiv.math/0304194},
    archivePrefix = {arXiv},
    eprint = {math/0304194},
    primaryClass = {math.QA}
}

@article{Mouland:2025dct,
    author = "Mouland, Rishi and Tong, David and Zan, Bernardo",
    title = "{Phases of 2d gauge theories and symmetric mass generation}",
    eprint = "2509.12305",
    archivePrefix = "arXiv",
    primaryClass = "hep-th",
    doi = "10.1007/JHEP04(2026)154",
    journal = "JHEP",
    volume = "04",
    pages = "154",
    year = "2026"
}

@article{ostrik2005,
    author = {{Ostrik}, Victor},
    title = "{Pre-modular categories of rank 3}",
    journal = {arXiv Mathematics e-prints},
     year = 2005,
    month = mar,
      eid = {math/0503564},
    pages = {math/0503564},
      doi = {10.48550/arXiv.math/0503564},
    archivePrefix = {arXiv},
    eprint = {math/0503564},
    primaryClass = {math.QA}
}

@article{Antinucci:2026uuh,
    author = "Antinucci, Andrea and Copetti, Christian and Galati, Giovanni and Rizi, Giovanni",
    title = "{A Twist on Scattering from Defect Anomalies}",
    eprint = "2605.13961",
    archivePrefix = "arXiv",
    primaryClass = "hep-th",
    month = "5",
    year = "2026"
}

@article{Tachikawa:2026cxd,
    author = "Tachikawa, Yuji and Tsuji, Keita and Watanabe, Masataka",
    title = "{What happens to wavepackets of fermions when scattered by the Maldacena--Ludwig wall?}",
    eprint = "2603.25508",
    archivePrefix = "arXiv",
    primaryClass = "hep-th",
    month = "3",
    year = "2026"
}

@article{fs,
      title="{Straightening Out the Frobenius--Schur Indicator}", 
      author={Steven H. Simon and Joost K. Slingerland},
      year={2022},
      eprint={2208.14500},
      archivePrefix={arXiv},
      primaryClass={hep-th},
      url={https://arxiv.org/abs/2208.14500}, 
}

@inproceedings{davidandgeorge,
    author = {Koundouros, George and Tong, David},
    title = "{Fermions Scattering Off Non-Abelian Rotors}",
    publisher = {to appear}
}

@inproceedings{yunqinetal,
    author = {Ohmori, Kantaro and Watanabe, Masataka and Wei, Pengcheng and Zheng, Yunqin},
    title = "{Intertwining Defects and Fermion-Boundary Scattering from TQFT}",
    publisher = {to appear}
}

@article{Freed:2012bs,
    author = "Freed, Daniel S. and Teleman, Constantin",
    title = "{Relative quantum field theory}",
    eprint = "1212.1692",
    archivePrefix = "arXiv",
    primaryClass = "hep-th",
    doi = "10.1007/s00220-013-1880-1",
    journal = "Commun. Math. Phys.",
    volume = "326",
    pages = "459--476",
    year = "2014"
}

@article{ostrik2001,
      title="{Module categories, weak Hopf algebras and modular invariants}",
      author={Viktor Ostrik},
      year={2001},
      eprint={math/0111139},
      archivePrefix={arXiv},
      primaryClass={math.QA},
      url={https://arxiv.org/abs/math/0111139}, 
}

@article{Moore:1989yh,
    author = "Moore, Gregory W. and Seiberg, Nathan",
    title = "{Taming the Conformal Zoo}",
    reportNumber = "IASSNS-HEP-89/6",
    doi = "10.1016/0370-2693(89)90897-6",
    journal = "Phys. Lett. B",
    volume = "220",
    pages = "422--430",
    year = "1989"
}

@article{Kitaev:2005hzj,
    author = "Kitaev, Alexei",
    title = "{Anyons in an exactly solved model and beyond}",
    eprint = "cond-mat/0506438",
    archivePrefix = "arXiv",
    doi = "10.1016/j.aop.2005.10.005",
    journal = "Annals Phys.",
    volume = "321",
    number = "1",
    pages = "2--111",
    year = "2006"
}

@article{Gu:2010na,
    author = "Gu, Zheng-Cheng and Wang, Zhenghan and Wen, Xiao-Gang",
    title = "{Classification of two-dimensional fermionic and bosonic topological orders}",
    eprint = "1010.1517",
    archivePrefix = "arXiv",
    primaryClass = "cond-mat.str-el",
    doi = "10.1103/PhysRevB.91.125149",
    journal = "Phys. Rev. B",
    volume = "91",
    number = "12",
    pages = "125149",
    year = "2015"
}

@article{Eliens:2013epa,
    author = {Eli{\"e}ns, I. S. and Romers, J. C. and Bais, F. A.},
    title = "{Diagrammatics for Bose condensation in anyon theories}",
    eprint = "1310.6001",
    archivePrefix = "arXiv",
    primaryClass = "cond-mat.str-el",
    doi = "10.1103/PhysRevB.90.195130",
    journal = "Phys. Rev. B",
    volume = "90",
    number = "19",
    pages = "195130",
    year = "2014"
}

@article{Karch:2019lnn,
    author = "Karch, Andreas and Tong, David and Turner, Carl",
    title = "{A Web of 2d Dualities: ${\bf Z}_2$ Gauge Fields and Arf Invariants}",
    eprint = "1902.05550",
    archivePrefix = "arXiv",
    primaryClass = "hep-th",
    doi = "10.21468/SciPostPhys.7.1.007",
    journal = "SciPost Phys.",
    volume = "7",
    pages = "007",
    year = "2019"
}

@ARTICLE{Johnson2021,
    author = {{Johnson-Freyd}, Theo and {Reutter}, David},
    title = "{Minimal nondegenerate extensions}",
    journal = {arXiv e-prints},
    year = 2021,
    month = may,
    eid = {arXiv:2105.15167},
    pages = {arXiv:2105.15167},
    doi = {10.48550/arXiv.2105.15167},
    archivePrefix = {arXiv},
    eprint = {2105.15167},
    primaryClass = {math.QA}
}

@article{Moore:1988qv,
    author = "Moore, Gregory W. and Seiberg, Nathan",
    title = "{Classical and Quantum Conformal Field Theory}",
    reportNumber = "IASSNS-HEP-88-39",
    doi = "10.1007/BF01238857",
    journal = "Commun. Math. Phys.",
    volume = "123",
    pages = "177",
    year = "1989"
}

@article{Gannon:2026ttf,
    author = "Gannon, Terry and Rayhaun, Brandon C.",
    title = "{Hypergroup Symmetry in Relative Quantum Field Theories and Chiral Algebras}",
    eprint = "2606.05279",
    archivePrefix = "arXiv",
    primaryClass = "hep-th",
    month = "6",
    year = "2026"
}

\end{document}